\documentclass[a4paper,12pt]{article}
\title{Black Hole / String Transition and Rolling D-brane}
\author{Yu Nakayama}
\usepackage{amsmath}
\usepackage{amssymb}
\usepackage{graphicx}
\usepackage{hyperref}
\def\drawbox#1#2{\hrule height#2pt
        \hbox{\vrule width#2pt height#1pt \kern#1pt
              \vrule width#2pt}
              \hrule height#2pt}

\def\Fund#1#2{\vcenter{\vbox{\drawbox{#1}{#2}}}}
\def\Asym#1#2{\vcenter{\vbox{\drawbox{#1}{#2}
              \kern-#2pt       
              \drawbox{#1}{#2}}}}

\def\funda{\Fund{6.5}{0.4}}

\def\symm{\funda\kern-0.4pt\funda}

\newcommand{\om}{\omega}
\newcommand{\al}{\alpha}
\newcommand{\ep}{\epsilon}

\newcommand{\lb}{\lbrack}
\newcommand{\rb}{\rbrack}

\newcommand{\msc}[1]{\mbox{\scriptsize #1}}
\newcommand{\dsp}{\displaystyle}

\newcommand{\bc}{\mathbb{C}}
\newcommand{\br}{\mathbb{R}}
\newcommand{\bz}{\mathbb{Z}}

\newcommand{\bsz}{\mathbb{Z}}

\newcommand{\cA}{{\cal A}}

\newcommand{\cO}{{\cal O}}
\newcommand{\cN}{{\cal N}}
\newcommand{\cM}{{\cal M}}

\newcommand{\cR}{{\cal R}}
\newcommand{\cC}{{\cal C}}
\newcommand{\cQ}{{\cal Q}}

\newcommand{\cH}{{\cal H}}

\newcommand{\ket}[1]{{|#1\rangle}}
\newcommand{\bra}[1]{{\langle#1|}}
\newcommand{\dket}[1]{{\left.\left|#1\right\rangle\right\rangle}}
\newcommand{\dbra}[1]{{\left\langle\left\langle#1\right|\right.}}

\renewcommand{\th}{{\theta}}

\renewcommand{\Im}{\mbox{Im}}

\renewcommand{\Re}{\mbox{Re}}

\newcommand{\nn}{\nonumber\\}

\newcommand{\be}{\begin{align}}
\newcommand{\ee}{\end{align}}
\newcommand {\rmd} {{\rm d}}
\newcommand{\bea}{\begin{align}}
\newcommand{\eea}{\end{align}}
\newcommand{\f}{\frac}
\newcommand{\dd}{\mbox{d}}

\allowdisplaybreaks[0]

\newcommand{\sectiono}[1]{\section{#1}\setcounter{equation}{0}}

\begin{document}

\begin{titlepage}
\thispagestyle{empty}
\begin{flushright}
UT-07-09\\
hep-th/0702221\\
\end{flushright}

\vskip 1.5 cm
\vspace*{2.5cm}
\begin{center}
\noindent{\textbf{\LARGE{ Black Hole - String Transition  \\\vspace{0.5cm}
 and Rolling D-brane
\vspace{0.5cm}\\
}}} 
\vskip 1.5cm
\noindent{\large{Yu Nakayama}\footnote{E-mail: nakayama@hep-th.phys.s.u-tokyo.ac.jp}}\\ 
\vspace{1cm}
\noindent{\small{\textit{Department of Physics, Faculty of Science, University of 
Tokyo}} \\ \vspace{2mm}
\small{\textit{Hongo 7-3-1, Bunkyo-ku, Tokyo 113-0033, Japan}}}
\end{center}
\vspace{1cm}
\newpage

\vspace*{4cm}
\begin{abstract}
We investigate the black hole - string transition in the two-dimensional Lorentzian black hole system from the exact boundary states that describe the rolling D-brane falling down into the two-dimensional black hole. The black hole - string phase transition is one of the fundamental properties of the non-supersymmetric black holes in  string theory, and we will reveal the nature of the phase transition from the exactly solvable world-sheet conformal field theory viewpoint. Since the two-dimensional Lorentzian black hole system ($SL(2;\br)_k/U(1)$ coset model at level $k$) typically appears as near-horizon geometries of various singularities such as NS5-branes in string theory, our results can be regarded as the probe of such singularities from the non-supersymmetric probe rolling D-brane.
The exact construction of boundary states for the rolling D0-brane falling down into the two-dimensional D-brane enables us to probe the phase transition at $k=1$ directly in the physical amplitudes. During the study, we uncover three fundamental questions in string theory as a consistent theory of quantum gravity: small charge limit v.s. large charge limit of non-supersymmetric quantum black holes, analyticity v.s. non-analyticity in physical amplitudes and physical observables, and unitarity v.s. open closed duality in time-dependent string backgrounds. This work is based on the PhD thesis submitted to Department of Physics, Faculty of Science, University of Tokyo, which was defended on January 2007. 

\end{abstract}

\end{titlepage}

\tableofcontents
\newpage

\sectiono{Introduction}\label{sec:1}
{\bf From the Heaven}

{\it A luminous star, of the same density as the Earth, and whose diameter should be two hundred and fifty times larger than that of the Sun, would not, in consequence of its attraction, allow any of its rays to arrive at us; it is therefore possible that the largest luminous bodies in the universe may, through this cause, be invisible} (Laplace: 1798). It was Laplace who first predicted the existence of the black hole from the Newtonian mechanics. More than a hundred years later, in 1915 when he was serving in Russia for World War I, Schwarzshild discovered the exact static black hole solution in Einstein's general relativity. Ever since, the black hole has continued to attract  our broad attention in theoretical physics.

Black holes are fascinating and indeed mysterious. It is remarkable that some properties of the black hole are quite reminiscent of those of the thermodynamics: it has a definite temperature, energy and entropy, and moreover it satisfies the thermodynamical laws. To understand this coincidence, it had been long suggested that the quantum gravity would explain the microscopic statistical origins of the thermodynamic properties of the black hole.
Furthermore black holes challenge the validity of the quantum mechanics. The Hawking radiation, predicted from the quantum mechanics, leads to evaporation of the black hole, which ironically results in the failure of the unitary evolution of the quantum system. These mysterious natures of the black holes have continued to enchant generations of theoretical physicists.

Over this past two decades, theoretical physicists have gained more and more confidence in string theory as a candidate for the final  theory of everything. The theory of everything, from its tacit implication, should include a consistent theory of quantum gravity with sufficient predictive power. The best arena to test the quantum gravity is quantum black hole systems, where the semiclassical analysis leads to the puzzling issues raised above. Whether the string theory resolves these issues or not is a big challenge to string theorists.

One of the greatest achievements of the string theory so far is to yield a microscopic explanation of the entropy for (near) BPS black holes with large charges. The string theory, along with various dualities, has enabled us to ``count" microscopic states forming such black holes. The counting successfully agrees with the classical Bekenstein Hawking entropy formula of the corresponding macroscopic black hole. 

The situations, however, still remain unclear when one studies non-BPS black holes with small charge. The large quantum corrections, both in string coupling constant and large curvature effects, prevent us from the quantitative enumeration of quantum states corresponding to the black hole. Qualitatively, it has been suggested that the so-called black hole - string phase transition occurs when we consider such a small charge black hole. One of the motivation of this thesis is to understand the black hole - string phase transition in exactly solvable string theory backgrounds.

In this thesis, we study the exact dynamics of rolling D-brane in the two-dimensional black hole system. The two-dimensional black hole system not only gives a toy model for the exactly solvable black hole systems in string theory, but also it can be embedded in the full superstring theory as a solution corresponding to black NS5-branes. Although our model is rather specific, we will uncover many important and universal features of quantum gravity such as the black hole - string transition. In particular We would like ask three fundamental questions about the nature  of the quantum gravity, or string theory as a candidate for the theory of everything.

The first problem we would like to ask in this thesis is the small charge limit of the non-supersymmetric black hole and its relation to the black hole - string transition.
By studying the black hole - string transition in the two-dimensional black hole, we would like to explicitly show the phase transition between the large charge limit and the small charge limit of the non-BPS black hole systems. The origin of the phase transition is the existence of two characteristic temperatures in the string theory: the one is the Hawking temperature associated with the Hawking radiation from the black hole, and the other is the Hagedorn temperature of the underlying string theory. The relation between the two temperatures is of utmost importance in understanding the black hole - string phase transition, and we will show that the phase transition occurs exactly when these two temperatures coincide in the two-dimensional black hole system by examining the properties of the exact probe rolling D-brane boundary states. 

A related issue is whether the genuine two-dimensional non-critical string theory (i.e. the target space is two-dimensional) admits a black hole solution. The question has remained long unanswered. Actually, the two-dimensional black hole in the two-dimensional non-critical string theory is located well below the black hole - string phase transition point, suggesting the difficulty of physical interpretations as a black hole. Our study will also support this argument in a negative way.

The second problem we would like to investigate in this thesis is the relation between the analyticity and non-analyticity in amplitudes and physical quantities. It is well-known that in the supersymmetric situations, holomorphy (analyticity) plays a crucial role in determining exact BPS properties of the theory. On the other hand, to discuss phase transitions such as the black hole - string transition, the non-analyticity of the physical quantities is essential. Throughout this thesis, the interplay between the analyticity and non-analyticity appears intermittently. Especially, we highlight the universality of the decaying D-brane and the subtleties associated with Wick rotation in curved spaces in this context.
 
The third problem we would like to study is the consistency between the unitarity and the open-closed duality. The unitarity is one of the crucial ingredients of the quantum theory. In the first quantized string theory, however, the unitarity in time-dependent background is not always manifest, especially in the Euclidean world-sheet formulation.
The simplest consequence of the unitarity is the optical theorem. In the time-dependent physics associated with the D-brane decay, however, it is not apparently obvious whether the analytic continuation involved is consistent with the requirement from the unitarity. Indeed, the abuse of the careless Wick rotation between the Lorentzian world-sheet theory and the Euclidean world-sheet theory, would result in inconsistent results, violating the optical theorem, which will be only fixed after the careful studies of the neglected pole contributions that appear through the process of Wick rotation. The rolling D-brane in the two-dimensional black hole system is an excellent arena to check the validities of proposed prescription for the Wick rotations given in the literatures.

{\bf Down to Earth}

So far, we have stated the celestial motivations of the thesis. What about the terrestrial motivations? In other words, which practical physics can we learn from the study of the rolling D-brane in two-dimensional black holes?

The dynamics of the rolling D-brane in the two-dimensional black holes closely resembles that of the rolling tachyon associated with the D-brane decay in flat space. Indeed, our study suggests that this tachyon - radion correspondence shows rather universal features of closed string radiation rate from the decaying D-brane. The string (particle) production from the time-dependent system such as the dynamical D-brane system itself is an interesting arena of theoretical physics, but it also has potential applications to the quantum cosmology based on the superstring theory.

In the recent observational cosmology, the existence of the inflational epoch of our universe has been confirmed with increasingly great accuracy. It is, therefore, a great challenge for the string theory to provide a natural setup for the inflation. One viable scenario for the string inflation is the so-called brane inflation, where the potential between the D-brane and anti D-brane provides the inflaton field. Recent studies show that the brane inflation could be embedded in the flux compactification of the type II string theory with all moduli fixed.

The end-point of the brane inflation is the pair annihilation between the D-brane and the anti D-brane. This is the point where the effective field theory approximation for the brane inflation breaks down and the stringy effects dominate. The reheating of the universe associated with the inflation decay is astonishingly different in the brane inflation scenario from the conventional field theory scenario. To understand the reheating process with the open string tachyon condensation, the universality of the radiation rate of the D-brane decay we will discuss in this thesis is crucial.

We will also see that large closed string loops will form during the D-brane decay and they will dominate the radiated energy once the fundamental string charge is induced. The subsequent evolution of such macroscopic strings will be of great importance to understand and estimate the relic cosmic strings in our universe, which might be observed in near future by experiment, directly proving the string theory.

In this way, the study of the D-brane decay has potential applications to quantum cosmology. We believe that our results, especially the universal properties of the decaying D-branes will become fundamental backgrounds for the realistic brane inflation models with successful reheating.

{\bf Organization of the Thesis}

The organization of this paper is as follows. In section \ref{sec:2}, we review the two-dimensional black hole from the space-time viewpoint. In section \ref{sec:3}, we review the two-dimensional black hole from the conformal field theory viewpoint. In section \ref{sec:4}, we introduce the concept of the black hole - string transition. In section \ref{sec:5}, we study the rolling tachyon dynamics and introduce the tachyon - radion correspondence conjecture. In section \ref{sec:6}, we study the D-branes in two-dimensional black hole system in the Euclidean signature. In section \ref{sec:7}, we construct the exact boundary states for the rolling D-brane in two-dimensional black hole in the Lorentzian signature. In section \ref{sec:8}, we study the closed string radiation rate from the rolling D-brane and probe the black hole - string transition. In section \ref{sec:9}, we present some discussions and the conclusion of the paper.

In appendices we collect useful facts used in the main part of the thesis. In appendix \ref{sec:A}, we fix our conventions and collect useful formulae. In appendix \ref{sec:B}, we present miscellaneous topics, whose detailed discussions are omitted in the main stream of the thesis.

A part of the thesis is based on the published papers. In particular, a large portion of the discussions in section \ref{sec:7} and \ref{sec:8} is based on \cite{Nakayama:2005pk,Nakayama:2006qm}.

\newpage
\sectiono{Two-dimensional Black Hole: Space-Time Viewpoint}\label{sec:2}
In this section, we review the two-dimensional black hole from the space-time viewpoint. We will see that the string theory is replete with exactly solvable solutions containing the two-dimensional black hole systems. By studying such backgrounds, we can understand the $\alpha'$ exact physics of the string theory near singularities.

The organization of this section is as follows. In section \ref{sec:2-1}, we introduce the black NS5-brane background as a most typical string solution based on the two-dimensional black hole system. In section \ref{sec:2-2}, we generalize the construction to study string theory near various singularities. In section \ref{sec:2-3}, we review the basic aspect of the classical two-dimensional black hole system. In particular we focus on the thermodynamic properties in section \ref{sec:2-4}.
\subsection{(Black) NS5-brane background}\label{sec:2-1}
As is often said, the string theory is not a theory of strings only. It turns out to contain other higher dimensional nonperturbative objects such as D-branes and NS-branes. Stable D-branes are charged under the Ramond-Ramond fields, and defined as objects on which perturbative strings can end. On the other hand, NS-branes are charged under the Kalb-Ramond $B_{\mu\nu}$ field, and do not possess a perturbative definition. They can be constructed as solitonic solutions of the equation of motions of the effective supergravity in ten-dimension.

Historically, all these important ingredients of the string theory are discovered as exact (BPS) solitonic solutions of the effective supergravity in ten-dimension. The tension of the D-branes is proportional to $1/g_s$ while the tension of the NS-branes is proportional to $1/g_s^2$, where $g_s$ denotes the string coupling constant. Hence, in the perturbative limit (i.e. $g_s \to 0$), all these objects are infinitely massive compared with the perturbative string spectrum and could  be neglected as excitations. Rather we regard the existence of such solitonic objects as super-selection sectors of the perturbative string theory.

The moduli spaces of the string theory is connected by various dualities. In particular, one of the most important recent achievements is the advent of the gauge - gravity correspondence. Before this new development, it had been believed that the local quantum field theory cannot realize the gravitational theory (Weinberg-Witten theorem \cite{Weinberg:1980kq}). However, the holographic realization of the gauge theory avoid this no-go theorem in a remarkable manner, and it has enabled us to study the strongly coupled gauge theory from the weakly coupled gravity. Explicit realization in the string theory involves the low-energy decoupling limit (Maldacena limit \cite{Maldacena:1997re}) of the localized excitations: the most famous example is the low-energy field theory limit of open-string theory living on the D3-brane in flat ten-dimensional space, which yields the duality between type IIB string theory on $AdS_5 \times \mathbb{S}^5$ and the $\mathcal{N}=4$ supersymmetric Yang-Mills theory on $\br^{1,3}$ (or $\br^1 \times \mathbb{S}^3$) \cite{Maldacena:1997re,Aharony:1999ti}.

The decoupling limit of the localized degrees of freedom and the gauge - gravity correspondence are not only important for the understanding of the strongly coupled gauge theories, but also essential to understand the quantum gravitational nature of the string theory. What is the microscopic origin of the black hole entropy? What is the fundamental degrees of freedom for the quantum gravity? How does (or does not) string theory solve the information paradox? These questions have been partially answered from the gauge - gravity correspondence of D-branes.
The decoupling limit is essentially the near horizon limit of the corresponding supergravity background, and the properties of black hole can be understood through the gauge - gravity correspondence in this way.

For NS5-brane, situations are more involved. Compared with D-branes, the NS5-brane is more geometrical in its origin. Indeed, as we will see in section \ref{sec:2-2}, it is T-dual to the singular geometry, and it appears not obvious what is the localized degrees of freedom in the decoupling limit. On the other hand, the closed string background for the near horizon limit of the NS5-brane is exactly quantized, so we are able to understand the gauge - gravity correspondence beyond the supergravity approximation.

Our starting point is the supergravity solution for the extremal NS5-brane: the solution contains nontrivial dilaton and the metric\footnote{Throughout this thesis, we use the string frame for supergravity solutions.}
\begin{align}
 \dd s^2   \equiv G_{\mu\nu} \dd x^\mu \dd x^\nu =- \dd t^2
+ \left(1+\frac{k\al'}{r^2}\right) \left(\dd
r^2+r^2 \dd \Omega_3^2\right)+ \dd {\bf
y}^2_{\mathbb{R}^5}~,  ~~~
e^{2\Phi(r)} = g_s^2 \left(1+\frac{k\al'}{r^2}\right) \ ,
\label{exNS5}
\end{align}
along with $k$-units of NS-NS $H_3$-flux penetrating through
$\mathbb{S}^3$:
\begin{align}
H_{mnp} = -\epsilon_{mnp}^{\ \ \ \ q} \partial_q \Phi(r) \ ,
\end{align}
where $x^m$ ($m=6,\cdots,9$) are transverse to the 5-brane.
Thus, $k$ refers to the number of NS5-branes at
$r=0$, ${\bf y}$ are
the spatial coordinates of the planar NS5-brane worldvolume, and
$g_s$ is the string coupling constant at infinity. The background preserves 16 supercharges of the type II (A or B) supergravity.

Following the argument of decoupling limit given above, we take the near horizon limit of the geometry \eqref{exNS5} by zooming in the $r^2 \ll \alpha'$ region. Neglecting the constant term (i.e. $1$) in the harmonic function $\left(1+\frac{k\al'}{r^2}\right)$, we obtain the near horizon limit of the extremal NS5-brane background \cite{Rey:1989xi,Rey:1989xj,Rey:1991uu,Callan:1991dj,Callan:1991at}

\begin{align}
 \dd s^2 = - \dd t^2 + k\al' \dd \rho^2 + k\al' \dd
\Omega_3^2 + \dd {\bf y}_{\mathbb{R}^5}^2 ~, \qquad {\Phi} = -\rho +
\text{constant}\ , \label{NH ext NS5}
\end{align}
where $r  = \sqrt{k\alpha'} \exp \rho$. This near horizon background remarkably admits an exact conformal field theory description involving a linear
dilaton theory and $SU(2)_k$ super Wess-Zumino-Novikov-Witten (WZNW)
model:\footnote{Here, $k$ is the level of total current of super SU(2) WZNW models and $\sqrt{\frac{2}{k}}$ is the amount of background charge for the linear dilaton system.}
\begin{align} \Big[ \mathbb{R}_t \times \mathbb{R}_{\rho , \sqrt{2\over k}}
\times SU(2)_k \Big]_\perp \times \Big[\mathbb{R}^5 \Big]_{||}\ .
\end{align}
The first part describes the five-dimensional curved space-time
transverse to the NS5-brane while the second part describes the
flat spatial directions parallel to the NS5-brane. The criticality
condition for superstring theories is satisfied for any $k$ because
\begin{align}
 \left( 1 + \frac{6}{k} + \frac{1}{2}\right) + 3 \times \left(
\frac{k-2}{k} + \frac{1}{2}\right)+6\times \left(1 + \frac{1}{2}
\right) =15 \label{crit} \ .
\end{align}

Although the background is exactly solvable, the string background is 
singular due to the existence of the linear dilaton direction $\rho$.
In the large negative $\rho$, the string coupling constant effectively diverges 
and the string perturbation theory is ill-defined. Physically, there exists a core of NS5-branes at $r=0$, and the dynamical degrees of freedom on the NS5-brane cannot be neglected.

There are several ways to regularize this linear dilaton singularity so that the string world-sheet perturbation theory makes sense with sufficient predictive power. 
One way to do this is to introduce the non-extremality to the geometry. Let us consider the non-extremal or black NS5-brane solution in the ten-dimensional type II supergravity:
\begin{align}
 \dd s^2 = -\left(1-\frac{r_0^2}{r^2}\right) \dd t^2
+ \left(1+\frac{k\al'}{r^2}\right) \left(\frac{\dd
r^2}{1-\frac{r_0^2}{r^2}} +r^2 \dd \Omega_3^2\right)+ \dd {\bf
y}^2_{\mathbb{R}^5}~,  ~~~
e^{2\Phi(r)} = g_s^2 \left(1+\frac{k\al'}{r^2}\right) \ 
\label{blackNS5}
\end{align}
along with $k$-units of NS-NS $H_3$-flux penetrating through
$\mathbb{S}^3$ again. Here $r=r_0$ is the location of the event horizon of the black NS5-brane.

One type of near-horizon limit
is $r_0 \rightarrow 0$ and $g_s \rightarrow 0$ independently,
leading to the `throat geometry' of extremal NS5-branes that reduce to
\eqref{NH ext NS5}.
Another type of near-horizon limit is $r_0 \rightarrow 0$ and $g_s
\rightarrow 0$ while keeping the energy density above the extremal
configuration $\mu \equiv {r_0^2}/{g_s^2 \al'}$ fixed. It yields
`throat geometry' of the near-extremal NS5-branes \eqref{blackNS5}
\cite{Maldacena:1997cg,Kutasov:2000jp}:
\begin{align}
\hspace{-5mm} \dd s^2 = - \tanh^2\rho \, \dd t^2 + k\al' \dd
\rho^2 + k\al' \dd \Omega_3^2 + \dd {\bf y}_{\mathbb{R}^5}^2 ~,
\qquad e^{2\Phi} = \frac{k}{\mu \cosh^2 \rho} ~, \label{NH black
NS5}
\end{align}
where $r= r_0\cosh \rho$. For $(t,\rho)$-subspace, the metric and
the dilaton coincide with those of the two-dimensional black hole with a Lorentzian signature.
This Lorentzian black hole can be described by Kazama-Suzuki
supercoset conformal field theory $SL(2; \br)_k / U(1)$ (where
$U(1)$ subgroup is chosen to be the non-compact component
(i.e. space-like direction)) of central charge $c=3(1+2/k)$. Likewise,
taking account of the NS-NS $H_3$-flux penetrating through
$\mathbb{S}^3$ which is omitted in (\ref{NH black NS5}), the angular
part $\mathbb{S}^3$ can be described by the (super) $SU(2)$-WZNW model as we have seen in the extremal case.
In this way, the string background of the nonextremal NS5-brane is
reduced to a solvable superconformal field theory system:\footnote
  {Here again, $k$ denotes the level of total current of super WZNW models.
   Namely, $k+2$, $k-2$ are the levels of bosonic
    $SL(2)$ and $SU(2)$ currents.}
\begin{align}
\Big[ {SL(2;\br)_{k} \over U(1)} \times SU(2)_{k} \Big]_\perp \times
\Big[\, \mathbb{R}^5 \, \Big]_{||}~. \label{SCFT black NS5}
\end{align}
Here, the first part describes the five-dimensional curved space-time
(including the time direction) transverse to the NS5-brane, while
the second part describes the flat spatial directions parallel to
the NS5-brane. The criticality condition is satisfied for any $k$ as
in \eqref{crit}.

As we will review in the next section, the classical geometry of the two-dimensional black hole itself is not singularity free. This is because although in the Schwarzshild-like coordinate used in \eqref{NH black NS5} there is no singularity at all, the event horizon exists at $\rho=0$, and we can extend the coordinate inside the horizon. In the maximally extended geometry, we observe a curvature and dilaton singularity as is the case with the usual Schwarzshild black hole. It is interesting, however, despite the appearance of the singularity, the exact SCFT formulation \eqref{SCFT black NS5} appears perfectly well-defined, at least formally.

Another way to regularize the linear dilaton singularity, while keeping the space-time supersymmetry in contrast with the above non-extremal resolution, is to separate the position of the NS5-branes in a ring-like manner and study the smeared solution \cite{Sfetsos:1998xd} (see also \cite{Israel:2004ir,Itzhaki:2005zr}). The background is described by the coset model 
\begin{align}
\frac{\Big[ {SL(2;\br)_{k} \over U(1)} \times \frac{SU(2)_{k}}{U(1)} \Big]_\perp}{\bz_k} \times
\Big[\, \mathbb{R}^{1,5} \, \Big]_{||}~. \label{SCFT sNS5}
\end{align}
Here $\bz_k$ orbifold serves as a GSO projection\footnote{With the abuse of convention, the Gliozzi-Scherk-Olive (GSO) projection has a two-fold meaning in this thesis (and in many literatures). The one is the summation over the spin structure \cite{Gliozzi:1976qd}, and the other is the restriction to the integral $U(1)_R$ charge sector for the internal SCFT. Both are imperative to preserve the target-space supersymmetry.} that restricts the spectrum to the sector with integral $U(1)_R$ charge so that the space-time supercharge is well-defined. Intuitively, we have extracted a particular $U(1)$ direction from the $SU(2)$ WZNW model and combined it with the linear dilaton direction to construct  the Euclidean ${SL(2;\br)_{k} \over U(1)}$ coset model by a marginal deformation.\footnote{$U(1)$ subgroup here is chosen to be the compact direction.} The linear dilaton direction together with the $U(1)$ direction is deformed to the ${SL(2;\br)_{k} \over U(1)}$ coset model that does not possess a dilaton singularity.

To see the geometrical meaning of this deformation, we write the coset part $\Big[ {SL(2;\br)_{k} \over U(1)} \times \frac{SU(2)_{k}}{U(1)} \Big]_\perp $ as
\begin{align}
\dd s^2 = \alpha'k(\dd \theta^2 + \tan^2\theta \dd\tilde{\phi}^2_2 + \dd\rho^2 + \tanh^2\rho \dd\tilde{\phi}_1^2) \ ,  \ \ e^{2\Phi} = \frac{1}{\cos^2\theta\cosh^2\rho} \ . \label{startm}
\end{align}
It is interesting to note that this geometry does {\it not} admit any Killing spinor needed for an apparent supersymmetry: the supersymmetry will be recovered after taking the $\bz_k$ orbifold \cite{Israel:2004ir} (see also \cite{Bakas:1994ba,Bergshoeff:1994cb,Bakas:1995hc} for earlier discussions). The $\bz_k$ orbifold is defined as $(\tilde{\phi}_1,\tilde{\phi}_2) \sim (\tilde{\phi}_1+2\pi/k,\tilde{\phi}_2+2\pi/k)$. We define new coordinates 
\begin{align}
\tilde{\phi}_1 = \phi_1+\phi_2/k \ , \ \ \tilde{\phi}_2 = \phi_2/k
\end{align}
so that the $\bz_k$ orbifold simply acts as $(\phi_1,\phi_2) \sim (\phi_1, \phi_2 + 2\pi)$. In the new coordinates, the metric reads
\begin{align}
\dd s^2 = \alpha'k(\dd\theta^2 +\dd\rho^2 \tanh^2\rho \dd\phi_1^2) + 2\alpha'\tanh^2\rho \dd\phi_1\dd\phi_2 + \frac{\alpha'}{k}(\tan^2\theta + \tanh^2\rho) \dd\phi_2^2 \ .
\end{align}
Since $\phi_2$ direction has a usual $2\pi$ periodicity, one can perform the T-duality along the $\phi_2$ direction. Applying Buscher's rule (see appendix \ref{busc}), we obtain
\begin{align}
\dd s^2 &= \alpha'k\left(\dd\theta^2 + \dd\rho^2 + \frac{\tan^2\theta\tanh^2\rho}{\tan^2\theta+\tanh^2\rho} \dd\phi_1^2 + \frac{1}{\tan^2\theta + \tanh^2\rho}{\dd\hat{\phi}_2} \right)  \cr
B&= \frac{\alpha'k\tanh^2\rho}{\tan^2\theta+ \tanh^2\rho}\dd\phi_1\wedge \dd\hat{\phi}_2 \ , \ \ e^{2\Phi} = \frac{1}{\cos^2\theta\cosh^2\rho(\tan^2\theta+\tanh^2\rho)} \ . \label{rmet}
\end{align}

\begin{figure}[htbp]
   \begin{center}
    \includegraphics[width=0.5\linewidth,keepaspectratio,clip]{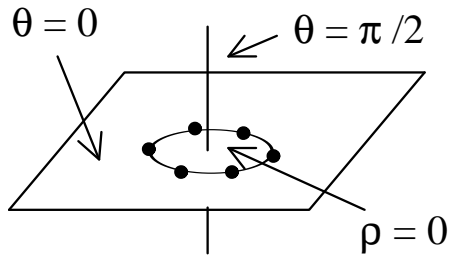}
    \end{center}
    \caption{NS5-branes are localized along the ring in the $(x_8,x_9)$ plane with $\theta = \rho = 0$.}
    \label{fig:ring}
\end{figure}

In the asymptotic region $\rho \to \infty$, we recover the NS5-brane solution \eqref{NH ext NS5}, and we can also see that the NS5-branes are now localized along the ring $\theta = \rho=0$, where the dilaton diverges (see figure \ref{fig:ring} for a description of our coordinate system). In other words, the NS5-branes are located along the ring in the $(x^8,x^9)$ plane.\footnote{Our parametrization is $x^6 = \rho_0 \sinh\rho \sin \theta \cos \phi_1$, $x^7 = \rho_0 \sinh\rho \sin \theta \sin\phi_1$, $x^8 = \rho_0 \cosh\rho\cos\theta \cos \hat{\phi}_2$, $x^9 = \rho_0\cosh\rho \cos \theta \sin\hat{\phi}_2$. } In this sense, the geometry still appears singular, but as we will discuss later, this is just an artefact of loose applications of T-duality: the trumpet singularity in \eqref{rmet} will be resolved by the ``winding tachyon condensation". Another quick way to see the absence of singularity is to revisit our starting point \eqref{startm}: it does not possess any dilaton singularity. It is also clear that the coset \eqref{SCFT sNS5} is manifestly singularity free as an SCFT up to a harmless orbifold structure.

Although we will not explicitly do it here, we can begin with the appropriate (smeared) harmonic function ansatz for the ring-likely distributed NS5-branes and reproduce the metric \eqref{rmet} purely from the supergravity solution by taking a suitable near horizon limit \cite{Sfetsos:1998xd}. In this approach, the space-time supersymmetry is manifest.

\subsection{Noncritical superstring and LST}\label{sec:2-2}
In section \ref{sec:2-1}, we discussed the relation between the two-dimensional black hole systems and the near horizon NS5-brane configurations. It is possible to generalize this construction to describe the singular limit of the geometry from exactly solvable conformal field theories. The construction is similar to the Gepner construction  for compact Calabi-Yau spaces \cite{Gepner:1987vz,Gepner:1987qi}, and it can be named ``non-compact Gepner construction" \cite{Ghoshal:1995wm,Ooguri:1995wj,Giveon:1999zm,Lerche:2000uy,Hori:2002cd,Eguchi:2004ik}. In this subsection, we would like to review this construction. Thanks to this generalized ``non-compact Gepner construction", most of the results we will present in later sections can be applied to various singular Calabi-Yau spaces.

\subsubsection{noncompact Calabi-Yau and wrapped NS5-branes}\label{sec:2-2-1}
Our starting points to construct exactly solvable world-sheet conformal field theories for singular Calabi-Yau spaces from two-dimensional black hole and minimal models are the following two claims.

{\bf Calabi-Yau / Landau-Ginzburg correspondence} \cite{Vafa:1988uu,Lerche:1989uy,Martinec:1988zu,Witten:1993yc}

Let us consider the algebraic varieties defined by
\begin{align}
\sum_{i=1}^{n+2} x_i^{r_i} = 0 \label{nocv}
\end{align}
in the weighted projective space $\mathbb{WCP}_{n+1}\left(\frac{1}{r_1},\cdots
,\frac{1}{r_{n+2}}\right)$. The Calabi-Yau condition reads $\sum_{i=1}^{n+2}\frac{1}{r_i} = 1$. The Calabi-Yau / Landau-Ginzburg correspondence says that the (quantum) sigma model defined on the $n$-dimensional algebraic varieties \eqref{nocv} is (weakly) equivalent\footnote{The precise meaning of the weak equivalence can be found e.g. in \cite{Hori:2000kt}.} to the $\mathcal{N} = 2$ supersymmetric Landau-Ginzburg orbifold theory with the superpotential
\begin{align}
W(X_i) = \sum_{i=1}^{n+2} X_i^{r_i}  \ . \label{lgo}
\end{align}
The orbifold projection serves as a GSO projection demanding the integrality of the $U(1)_R$-charge of the total model. The Calabi-Yau condition can be understood as the criticality condition for the SCFT with the central charge $\hat{c} = c/3 = n$.

When all $r_i$ are positive, the resulting model is nothing but the Gepner construction for compact Calabi-Yau spaces (see also \cite{Greene:1988ut,Witten:1993yc}). When some of $r_i$ are negative, the Calabi-Yau manifold is non-compact and the definition of the Landau-Ginzburg orbifold needs extra care as we will do it momentarily.

{\bf Landau-Ginzburg / minimal model correspondence} \cite{Lerche:1989uy}

The $\mathcal{N}=2$ supersymmetric Landau-Ginzburg model with the superpotential $W(X) = X^{k}$ is equivalent to the $\mathcal{N}=2$ $(k-2)$-th minimal model with the central charge $\hat{c} = c/3 = 1 - \frac{2}{k}$. The minimal model has an algebraic formulation, but an alternative construction is based on the Kazama-Suzuki coset $SU(2)_k/U(1)$ associated with the level $k$ $SU(2)$ current algebra.\footnote{We always stick to the convention where $k$ denotes the {\it total} level of the current algebra: the bosonic $SU(2)$ current algebra has the level $\kappa = k-2$ and the bosonic $SL(2;\br)$ current algebra has the level $\kappa = k+2$.} Kazama-Suzuki construction guarantees that the coset CFT associated with the $\mathcal{N}=1$ current algebra actually possesses $\mathcal{N}=2$ superconformal symmetry when the target space is a special Kahler manifold (the hermitian symmetric manifold) \cite{Kazama:1988uz,Kazama:1988qp}. In our simplest case, it is indeed the case and the theory is equivalent to the $\mathcal{N}=2$ minimal model.\footnote{Actually, if the denominator $H$ in the coset $G/H$ is a Cartan subgroup of $G$, the coset admits the $\mathcal{N}=2$ superconformal symmetry even if it is a non-symmetric space \cite{Kazama:1988va}.}

We can formally generalize the above discussion to define the $\mathcal{N}=2$ supersymmetric Landau-Ginzburg model with the negative power superpotential $W(X) = X^{-k}$. The analytic continuation of the central charge for the positive power superpotential yields $\hat{c} = c/3 = 1 + \frac{2}{k}$. The Kazama-Suzuki coset construction has a natural generalization in this case as well. Instead of considering $SU(2)_k/U(1)$ supercoset model, we consider $SL(2;\br)_k/U(1)$ supercoset model whose central charge is also given by $\hat{c} = c/3 = 1 + \frac{2}{k}$. This CFT will be reviewed in section \ref{sec:3}. Since the Lagrangian formulation based on the Landau-Ginzburg model with the negative power superpotential does not seem to be well-defined while the $SL(2;\br)_k/U(1)$ coset does, the precise claim of the non-compact Gepner construction is that the Landau-Ginzburg orbifold appearing in \eqref{lgo} should be understood as the $SL(2;\br)_k/U(1)$ coset model.

At this point, it would be interesting to mention that the formal Landau-Ginzburg description suggests a duality between the $\mathcal{N}=2$ Liouville theory and the $SL(2;\br)_k/U(1)$ coset model. We begin with the topological path integral for the partition function on the sphere: 
\begin{align}
Z &= \int \dd X \dd \bar{X} \frac{1}{g_s^2} e^{-W(X)-\bar{W}(\bar{X})}  \cr
&=\int \dd X \dd \bar{X} \frac{1}{g_s^2}e^{-X^{-k}-\bar{X}^{-k}} \ . \label{pi}
\end{align}
Introducing the $\mathcal{N}=2$ Liouville coordinate $X^{-k} = e^{\frac{1}{\mathcal{Q}}\Phi}$ with $\mathcal{Q}^2 = \frac{2}{k}$, we can rewrite the path integral \eqref{pi} as
\begin{align}
Z = \int \dd\Phi \dd\bar{\Phi} \frac{1}{g_s^2} \exp\left({-\mathcal{Q}\mathrm{Re}\Phi -e^{\frac{1}{\mathcal{Q}}\Phi}- e^{\frac{1}{\mathcal{Q}}\bar{\Phi}}}\right) \ .
\end{align}
The important step is to regard the measure factor $\exp\left({-\mathcal{Q}\mathrm{Re}\Phi}\right)$ as a space-dependent coupling constant, namely, linear dilaton background with the slope $\mathcal{Q}$. The remaining action shows the structure of the $\mathcal{N}=2$ Liouville superpotential $W(\Phi) = e^{\frac{1}{\mathcal{Q}}\Phi}$. This heuristic equivalence between the $SL(2;\br)/U(1)$ coset model and the $\mathcal{N}=2$ Liouville theory at the topological level will be made more precise in later section \ref{sec:3-4}.

Now combining these two facts, we can construct the equivalent description for  (non-compact) Calabi-Yau spaces by considering tensor products of $SL(2;\br)/U(1)$ coset models ($\mathcal{N}=2$ Liouville theory) and $SU(2)/U(1)$ coset models ($\mathcal{N}=2$ minimal models) with appropriate GSO projections. We call such a construction a generalized (non-compact) Gepner construction.

Let us discuss some simple examples.

1) $A_{k-1}$ type ALE spaces

We take $n=2$, and set $-r_1 = r_2 = k$, $r_3=r_4=2$. From the projective invariance, we can set $x_1= - \mu$ without loss of generality.\footnote{Note that we are considering a noncompact space, so the domain of the projective coordinate $x_1$ is in $\mathbb{C}^* \equiv \mathbb{C} - \{0\}$. } The resulting algebraic variety is given by
\begin{align}
x_2^k + x_3^2 + x_4^2 = \mu^k \label{deak}
\end{align}
in $\mathbb{C}^3$, which is nothing but the $A_{k-1}$ type ALE space with a deformation parametrized by $\mu$. On the other hand, the noncompact Gepner construction yields

\begin{align}
\frac{\Big[ {SL(2;\br)_{k} \over U(1)} \times \frac{SU(2)_{k}}{U(1)} \Big]}{\bz_k}
\end{align}
because the massive theory with the quadratic superpotential $W(X) = X^2$  will decouple under the renormalization group flow. We now recognize that the resulting theory is same as the near horizon limit of the $k$ NS5-brane solutions discussed in section \ref{sec:2-1}. This shows an equivalence between the near horizon limit of $k$ NS5-brane solutions and the $A_k$ type ALE spaces. They are indeed related with each other via the T-duality. The deformation parameter $\mu$ in the ALE space corresponds to the separation of NS5-branes. We can easily generalize the construction for other ALE spaces with ADE singularities.

2) Generalized conifolds

We next consider the case of Calabi-Yau three-fold $(n=3)$. We take $r_2,r_3,r_4,r_5> 0$ and set $r_1 = 1-\frac{1}{\sum_{i=2}^5 r_i^{-1}} < 0$. After fixing the projective invariance by eliminating $x_1$, the resultant Calabi-Yau space is the so-called generalized (deformed) conifold  
\begin{align}
x_2^{r_2} + x_3^{r_3} + x_4^{r_4} + x_5^{r_5} = \mu \ \label{dgc}
\end{align}
in $\bc^4$. Mathematically, we can regard it as a complex structure deformation of the Brieskorn-Pham type singularity (see section \ref{sec:2-2-3}). The Gepner construction leads to the orbifolded tensor products of $\mathcal{N}=2$ minimal models with one $SL(2;\br)_{-r_1}/U(1)$ coset model.
 The simplest example is the case with $r_1=-1$ and $r_2 = r_3= r_4=r_5=2$. The geometry is the deformed conifold:
\begin{align}
x_2^2 + x_3^2 + x_4^2 + x_5^2 = \mu \ .
\end{align}
The noncompact Gepner construction is given by $SL(2;\br)_1/U(1)$ coset model with the level 1 parent current algebra. This is the famous Ghoshal-Vafa duality \cite{Ghoshal:1995wm}.

3) ALE($A_{k-1})$ fibration over weighted projective spaces

We finally consider the model with two negative charges: $n=3$, $r_1=-k(1+\frac{k_1}{k_2})$, $r_2 = -k(1+\frac{k_2}{k_1})$, $r_3 = k$, and $r_4=r_5=2$. The Landau-Ginzburg superpotential is given by
\begin{align}
W(X_i) = X_1^{-k(1+\frac{k_1}{k_2})} + X_2^{-k(1+\frac{k_2}{k_1})} + X_3^{k} \ .
\end{align}
By introducing new variables: $Z= \log X_1 + \log X_2$, $Y=kk_1\log X_1 - kk_2 \log X_2$ and $X^k = e^{kZ} X_3^k $, we can rewrite the Landau-Ginzburg superpotential as
\begin{align}
W = e^{-kZ}(X^k + e^{Y/k_1} + e^{-Y/k_2}) \ ,
\end{align}
with the linear dilaton $\Phi = -\mathrm{Re} Z$. After integrating over $Z$, the topological path integral is localized along the locus\footnote{We have added the superpotential term $W_1^2 + W_2^2$ by hand to match the dimensionality.} 
\begin{align}
e^{y/k_1} + e^{-y/k_2} + x^k + w_1^2 + w_2^2 = 0 \ ,
\end{align}
which shows a structure of ALE($A_{k-1}$) fibration over 
 $\mathbb{WCP}^1(k_1,k_2)$. The simplest example with $k_1=k_2$, we obtain the ALE($A_{k-1}$) fibration over $\mathbb{CP}_1$. The geometry of the two $SL(2;\br)/U(1)$ coset and one $SU(2)/U(1)$ coset can be analysed in a similar way as we did in section \ref{sec:2-1}, and the result is given by the wrapped NS5-brane solution around $\mathbb{CP}_1$, where we have chosen $k_1=k_2=1$ for simplicity (see \cite{Hori:2002cd} for details). This is expected from the fact that the $A_{k-1}$ singularity is T-dual to flat $k$ NS5-branes and we could perform the fiber-wise T-duality for the ALE($A_{k-1}$) fibration over $\mathbb{CP}_1$.

The partition functions and elliptic genera of these noncompact Gepner models have been studied in \cite{Eguchi:2004yi,Eguchi:2004ik,Eguchi:2006tu}. In this section we restricted ourselves to the Landau-Ginzburg construction where the theory is defined as (an orbifold of) the direct product of the Landau-Ginzburg models with monomial superpotentials. Geometrically, they corresponded to the (deformations of) the Brieskorn-Pham type singularities. It is possible to construct more general Landau-Ginzburg orbifolds with generic polynomial superpotentials. The generalized models have a potential applications to the singular locus of the $\mathcal{N}=2$ supersymmetric Yang-Mills theories (Argyres-Douglas point) and their deformations. The exact quantization of the world-sheet theory beyond the topological subsector, however, is a difficult task and we would not pursue these generalizations any further in this thesis.

\subsubsection{singular limit and LST}\label{sec:2-2-2}
In section 2.1, we have discussed that the coinciding $k$ NS5-branes superstring solution corresponds to the linear dilaton background while the supersymmetric deformation (separation of NS5-branes in a ring-like manner) corresponds to the $SL(2;\br)/U(1)$ coset background (i.e. two-dimensional Euclidean black hole). Here we would like to take the similar singular limit in more general noncritical superstring backgrounds discussed in section \ref{sec:2-2-1}.\footnote{What we mean by ``noncritical" here is that the SCFTs involved does not necessarily possess an apparent 10-dimensional background as is the case with the Gepner construction for compact Calabi-Yau spaces. In a more specific narrower sense, we sometimes call a theory ``noncritical" when it possesses a Liouville direction.}

It is particularly easy to see the singular limit if we start with the $\mathcal{N}=2$ Liouville description: it has a superpotential
\begin{align}
W(\Phi) = \mu e^{\frac{1}{\mathcal{Q}}\Phi} \ ,
\end{align}
and the parameter $\mu$ directly corresponds to the deformation parameter appearing e.g. in \eqref{deak},\eqref{dgc}. Thus the singular limit $\mu \to 0$ is equivalent to switching off the Liouville potential so that we are left with the linear dilaton theory. The duality between the $\mathcal{N}=2$ Liouville theory and $SL(2;\br)/U(1)$ coset theory then confirms the statement that the singular limits of the non-compact Gepner models correspond to replacing $SL(2;\br)/U(1)$ coset part by the $\mathcal{N}=2$ supersymmetric linear dilaton theory with the same central charge and the same asymptotic dilaton gradient.

Let us formulate the proposal discussed above in a more precise way \cite{Giveon:1999zm}. We begin with the type II string theory on a singular Calabi-Yau varieties $X^{2n}$ with  the complex dimension $n$ defined as the vicinity of a hypersurface singularity 
\begin{align}
F(z_1,\cdots, z_{n+1}) = 0 \label{defeq}
\end{align}
in $\mathbb{C}^{n+1}$, where $F$ is a quasi-homogeneous polynomial on $\mathbb{C}^{n+1}$. This means that $F$ has degree $1$ under the $\mathbb{C}^{*}$ action:
\begin{align}
z_i \to \lambda^{r_i} z_i \ . \label{Cac}
\end{align}

Now we can define a locally holomorphic $n$-form $\Omega$ as
\begin{align}
\Omega = \frac{\dd z_1\wedge \dots \wedge \dd z_n}{\partial F/\partial z_{n+1}} \ 
\end{align}
on the patch $\partial F/\partial z_{n+1}\neq 0$. It can be extended to other patches where $\partial F\partial z_i \neq 0$ with the similar expressions and glued together to form a globally well-defined holomorphic $n$-form with the charge $r_{\Omega} = \sum_i r_i -1$ under the $\mathbb{C}^{*}$ action \eqref{Cac}. Such constructed varieties $X^{2n}$ are Gorenstein\footnote{Gorenstein means that  $X^{2n}-\{0\}$ admit a nowhere vanishing holomorphic $n$-form.} equipped with a natural $\mathbb{C}^{*}$ action \eqref{Cac} by construction.

We consider the type II string theory on the singular Calabi-Yau varieties $\br^{d-1,1}\times X^{2n}$ in the vicinity of the isolated singular point $y_0$ in the decoupling limit $g_s \to 0$. The proposed dual theory is the type II string theory on $\br^{d-1,1} \times \br_{\phi} \times \mathcal{N}$. Here $\mathcal{N}$ is the infrared limit of the sigma model on the manifold $\mathcal{N} = X^{2n}/\br_+$, where the division by $\br_+$ is an action on $z_i$ as \eqref{Cac} with $\lambda \in \br_+$, The infrared limit\footnote{We assume $r_{\Omega}>0$ so that $\mathcal{N}$ is Fano meaning that the curvature of the Einstein metric on it is positive.} of the sigma model on $\mathcal{N}$ is given by a Landau-Ginzburg model with superpotential $F(Z_i)$ and an additional $\mathbb{S}^1$ circle.\footnote{As a CFT, they are not a simple direct product but an orbifold. We need to impose the GSO projection to preserve the target-space supersymmetry.} Here $\mathbb{S}^1$ direction corresponds to the $U(1)_R$ symmetry associated with the residual $U(1)$ action \eqref{Cac} with $|\lambda| =1$. The quotient space $\mathcal{N}/U(1)$ is equivalent to the Landau-Ginzburg model with superpotential $W= F(Z_i)$ from the standard Landau-Ginzburg / non-linear sigma model correspondence. The linear dilaton slope is determined by the total criticality condition of the string theory.

This construction is equivalent to the one discussed above by turning off the $\mathcal{N}=2$ Liouville superpotential or $SL(2;\br)/U(1)$ deformation. For later purposes, it is worthwhile studying the normalizability of such deformations.
 Consider the variation of the complex structure of $X^{2n}$:
\begin{align}
F(z_i) + \sum_a t_a A_a(z_i)  = 0 \ .
\end{align}
Here $t_a$ are complex deformation parameters, and $A_a(z_i)$ are complex structure deformations of the defining equation \eqref{defeq}. The Kahler potential of the Weil-Petersson metric for such complex structure deformations is given by the formula \cite{Candelas:1990pi}
\begin{align}
K = - \log \int_{X^{2n}} \Omega \wedge \bar{\Omega} \ .
\end{align}
To discuss the normalizability of the perturbation associated with $A_a$, we have to evaluate
\begin{align}
\frac{\partial^2}{\partial t_a \partial \bar{t}_a} \Omega \wedge \bar{\Omega} \ . \label{metd}
\end{align}
This can be done by the simple scaling argument \cite{Gukov:1999ya}: if $A_a$ scales under \eqref{Cac} as $\lambda^{r_a}$, $t_a$ should scale as $\lambda^{1-r_a}$, so \eqref{metd} scales with a weight 
\begin{align}
\omega_a = 2\left(\sum_{i}r_i + r_a - 2\right) \ .
\end{align}
The modes satisfying $r_a > 1-r_{\Omega}$ are non-normalizable deformations as $|z_i| \to \infty$ while the modes satisfying $r_a < 1-r_{\Omega}$ are normalizable deformations. 

The non-normalizable deformations should be regarded as boundary conditions we have to impose at infinity to define a theory. Different boundary conditions would give rise to different theories. On the other hand, the normalizable deformations should be regarded as fluctuating fields after the quantization. Their values can be varied within a given theory. As we will discuss later in section \ref{sec:3-1}, the normalizability of the deformations from the space-time viewpoint presented here is deeply connected with the normalizability of the corresponding operators in the world-sheet $\mathcal{N}=2$ linear dilaton theory. Indeed, one can regard this agreement as a nontrivial support for the duality proposed in \cite{Giveon:1999zm} and reviewed here.

As an example, let us consider a class of generalized conifolds defined by the hypersurface
\begin{align}
F(z_i) = H(z_1,z_2) + z_3^2 + z_4^2 \ 
\end{align}
in $\bc^4$. It can be regarded as an NS5-brane wrapped around the Riemann surface $H(z_1,z_2)= 0$ along the line of arguments reviewed at the end of section \ref{sec:2-2-1}. We begin with the $A_{n-1}$ type Brieskorn-Pham singularity with $H(z_1,z_2) = z_1^n + z_2^2$, and consider the perturbations of the form $z_1^a$ $(a=0,1,\cdots,n-2)$. $U(1)_R$-charges are given by $r_{\Omega} = \frac{1}{n}+\frac{1}{2}$ and $r_a = \frac{a}{n}$. From the condition $r_a > 1-r_{\Omega}$, we conclude that the deformations with
\begin{align}
a > \frac{n}{2}-1 \label{normc}
\end{align}
are non-normalizable \cite{Gukov:1999ya,Giveon:1999zm}. We can also understand the normalizability of these deformations from the dual $\mathcal{N}=2$ supersymmetric four-dimensional field theory viewpoints by studying the Seiberg-Witten theory near the Argyres-Douglas points \cite{Argyres:1995jj,Argyres:1995xn,Eguchi:1996vu}. 

To conclude this section, we would like to revisit the question: what is the decoupling limit of the theory defined on the singularities \eqref{defeq}? We have reviewed the proposed {\it dual} string theory defined as (deformations of) the $\mathcal{N}=2$ linear dilaton theory coupled with the Landau-Ginzburg model. We here summarize the low-energy decoupled physics from the original NS5-brane construction in flat ten-dimensional Minkowski space. The decoupled theory has a conventional name ``little string theory (LST)"  \cite{Losev:1997hx}.\footnote{See \cite{Aharony:1999ks} for an earlier review.}

\begin{itemize}
	\item As a decoupled six-dimensional theory, it has $\mathcal{N}=(2,0)$  (type IIA) or $\mathcal{N}=(1,1)$ (type IIB) supersymmetry. The theory is non-local.
	\item They are classified by the ADE classification.
	\item IR limit is the six-dimensional super Yang-Mills theory in type IIB and the six-dimensional interacting (2,0) SCFT in type IIA \cite{Seiberg:1997ax}.
	\item BPS excitation includes a string with tension $l_s$ (little string) and the theory shows a Hagedorn-like thermodynamics with the Hagedorn temperature $\beta_{\mathrm Hg} \sim 2\pi\sqrt{2k}$ (see section \ref{sec:2-4}). Most of these high-energy states are nonperturbative in nature.
\end{itemize}

After compactification (by wrapping NS5-branes on projective spaces for instance), we have four- (or two-) dimensional effective theory, which includes Seiberg-Witten theory near the Argyres-Douglas singularities, corresponding to the Calabi-Yau 3-fold singularities.

The properties of these theories, known as the LST, are less known than the field theory living on D-branes. However, the closed string dual theory is  exactly quantizeable in many cases unlike the R-R background in the near horizon limit of the D-branes. The study of $\alpha'$ exact information is an interesting subject of its own, besides the application to the dual theories, and we will pursue this direction in the following sections.

\subsubsection{obstruction for conical metrics}\label{sec:2-2-3}
So far, we have assumed the existence of the Calabi-Yau varieties defined on the hypersurface singularity \eqref{defeq}. In the compact Calabi-Yau case such as the curve defined in \eqref{nocv}, Calabi-Yau theorem guarantees the existence of the unique Ricci-flat Kahler metric once the Calabi-Yau condition is satisfied. The existence of the Calabi-Yau metric on the hypersurface singularities \eqref{defeq}, however, is an open problem (see \cite{SE} for a review).

From the $\mathbb{C}^{*}$ action \eqref{Cac} on the complex variables $z_i$, it is natural to assume the conical metric
\begin{align}
\dd s^2_{X^{2n}} = \dd r^2  + r^2 \dd s^2_L \ ,\label{metc}
\end{align}
where $r\le 0$ denotes the radial direction $\br_+$ and $\dd s^2_L$ is the Sasaki-Einstein metric of the link $L$ associated with the non-compact Calabi-Yau variety $X^{2n}/\{0\} = \br_+ \times L $. The Sasaki condition is equivalent to the statement that the total metric is Kahler, and the Einstein condition is equivalent to the statement that the total metric is Ricci flat.

It turns out to be extremely difficult to give necessary and sufficient conditions for the existence of such conical metric (or alternatively existence of the Sasaki-Einstein metric on the link $L$) while infinitely many examples of explicit metrics have been constructed quite recently \cite{Gauntlett:2004yd,Cvetic:2005ft,Cvetic:2005vk}.

For definiteness we restrict ourselves to the Brieskorn-Pham type singularities defined by the particular polynomial
\begin{align}
F(z_i) = \sum_{i=1}^{n+1} z_i^{a_i} \ . \label{bpsin}
\end{align}
The corresponding hypersurface singularities $X^{2n}$ are always isolated and Gorenstein. However from the following physical reasoning, we believe that not every singularity possesses a conical metric.

Assuming the existence of such a conical metric, we can compute the volume of such a hypothetical Sasaki-Einstein link $L$ by the formula \cite{Bergman:2001qi}
\begin{align}
\mathrm{Vol}(L) = \frac{r_{\Omega}^n}{n^n\prod_{i=1}^{n+1}r_i} \mathrm{Vol}(\mathbb{S}^{2n-1}) \ ,
\end{align}
where $\mathrm{Vol}(\mathbb{S}^{2n-2}) = \frac{2\pi^n}{(n-1)!}$.\footnote{We have assumed that the Reeb-vector (conformal $U(1)_R$-symmetry) is given by the natural $\mathbb{C}^{*}$ action \eqref{Cac}. If this is not the case, we have to determine the ``correct" Reeb-vector by using the $Z$-minimization \cite{Martelli:2005tp,Martelli:2006yb} ($a$-maximization \cite{Intriligator:2003jj}) principle. }
Via the AdS-CFT correspondence, the central charge $a$ of the dual SCFT living on D3-branes placed at the tip of the cone is related to the volume  of the Sasaki-Einstein link $L$ \cite{Gubser:1998vd} as 
\begin{align}
a \propto \frac{1}{\mathrm{Vol}(L)} \ .
\end{align}
On the other hand, the conjectured $a$-theorem states that $a$ is a decreasing function during a  renormalization group flow from UV to IR. Geometrically speaking, the relevant deformations to \eqref{bpsin} should increase the volume.\footnote{From the pure gravity viewpoint, this is a consequence of the weaker energy condition \cite{Freedman:1999gp}.} However, this statement is clearly violated in some explicit examples such as the series $F = z_1^k + z_2^2 + z_3^2 + z_4^2$ with $k>4$, where $\mathrm{Vol}(k+1) > \mathrm{Vol}(k)$ contradicting with the $a$-theorem.

Recently, \cite{Gauntlett:2006vf} has given two mathematical obstructions for the existence of conical Calabi-Yau metric for such varieties.

{\bf The Bishop obstruction}

For the existence of the conical Calabi-Yau metric \eqref{metc}, $\mathrm{Vol}(L) < \mathrm{Vol}(\mathbb{S}^{2n-1})$ is necessary. From the dual gauge theory viewpoint,  the condition corresponds to the fact that by appropriate Higgsing that decreases $a$, we can reach $\mathcal{N}=4$ SYM theory.

{\bf The Lichnerowicz obstruction}

When $X^{2n}$ admits a holomorphic function with $U(1)_R$-charge $\lambda <1$, the conical Calabi-Yau metric does not exist. For the Brieskorn-Pham type singularities, it is equivalent to the statement $r_{\Omega} \le n r_a$ for any deformation. From the dual gauge theory viewpoint, the condition corresponds to the unitarity bound of dual operators for the deformations. It can be shown that the Lichnerowicz obstruction also eliminate a possible violation of $a$-theorem for the Brieskorn-Pham type singularities (see appendix \ref{Lic}).

As an example let us consider the Calabi-Yau four-fold defined by 
\begin{align}
F = z_1^k + z_2^2 + z_3^2 + z_4^2 + z_5^2 = 0 \ . \label{fourc}
\end{align}
The conical Calabi-Yau metric only exists for $k=2$, and other varieties are obstructed from the Lichnerowicz bound. Thus the $AdS_3\times L_7$ compactification of M-theory is only possible for $k=2$. This should be so because otherwise the $a$-theorem would be violated or the weaker energy-condition would be spoiled. 

On the other hand, one can consider the two-dimensional string compactification on such a hypothetical conical Calabi-Yau manifold \eqref{fourc} and add $N_f$ fundamental strings on the noncompact $\br^{1,1}$ directions at the tip of the cone. Due to the gravitational backreaction, we can see the near horizon geometry would be $AdS_3 \times \mathcal{N}$ with the constant string coupling $g_s^2 \sim \frac{1}{N_f}$, where $\mathcal{N}$ has been introduced in section \ref{sec:2-2-2} denoting the infrared limit of the sigma model on the hypothetical Sasaki-Einstein link $L$ associated with \eqref{fourc}. As discussed in this section, the Sasaki-Einstein link $L$ is obstructed, but the string theory on $AdS_3  \times \mathbb{S}^1 \times (LG(F)\sim \mathcal{N}/U(1))$ has a well-defined perturbative description based on the $SL(2;\br)$ current algebra with level $k$ times Landau-Ginzburg orbifold defined by $F(Z_i)$ (or $\mathcal{N}=2 $ minimal model) \cite{Aharony:1999ti}.\footnote{This does not mean the existence of such Sasaki-Einstein metric because we are sitting at the Gepner-point of the sigma model, where the geometrical description is questionable due to large $\alpha'$ corrections.} Interestingly, unlike the $a$-theorem associated with the M-theory compactification on $AdS_3 \times \mathcal{N}$, the $c$-theorem for the dual CFT of $AdS_3 \times \mathbb{S}^1 \times (LG(F)\sim \mathcal{N}/U(1))$ is always satisfied because the dual CFT central charge, which is determined from the curvature of the $AdS_3 \sim SL(2;\br)$, is given by $c=6kQ_1$.\footnote{For instance, $k=\frac{n}{n+1}$ for $A_{n-1}$ type Calabi-Yau four-folds.}

 In a similar fashion, the near horizon geometry of  every Brieskorn-Pham singularities admit the noncritical string construction based on the non-compact Gepner models as we have presented in this section irrespective of the above-mentioned obstructions. It would be interesting to understand the obstructions of the existence of conical metrics for such singularities from the noncritical string theory viewpoint. For instance, we can translate the Lichnerowicz obstruction as the claim that every relevant deformations up to $z_i^{a_i-3}$ should be normalizable.

\subsection{Classical two-dimensional black hole}\label{sec:2-3}
In section \ref{sec:2-1}, we have introduced the two-dimensional black hole geometry as a near horizon limit of the black NS5-brane solutions in the type II superstring theory:
\begin{align}
\dd s^2 = k\al'(- \tanh^2\rho \, \dd t^2 +  \dd 
\rho^2) \ , \label{tbhm}
\end{align}
with nontrivial dilaton gradient $e^{2\Phi} = \frac{k}{\mu \cosh^2 \rho}$.\footnote{We have rescaled the normalization of $t$ for simplicity of notation. We will sometimes do this in the following without notice, for it would be convenient to stick to $2\pi$ periodicity in the Euclidean time direction after the Wick rotation.}
It has been claimed in the literature that the background is $\alpha'$ exact perturbatively in the type II superstring theory while the bosonic two-dimensional black hole might receive perturbative world-sheet $\alpha'\sim \frac{1}{k}$ corrections \cite{Tseytlin:1992ri}. We will discuss physical importance of the nonperturbative corrections later in section \ref{sec:3-4}.

\begin{figure}[htbp]
   \begin{center}
    \includegraphics[width=0.5\linewidth,keepaspectratio,clip]{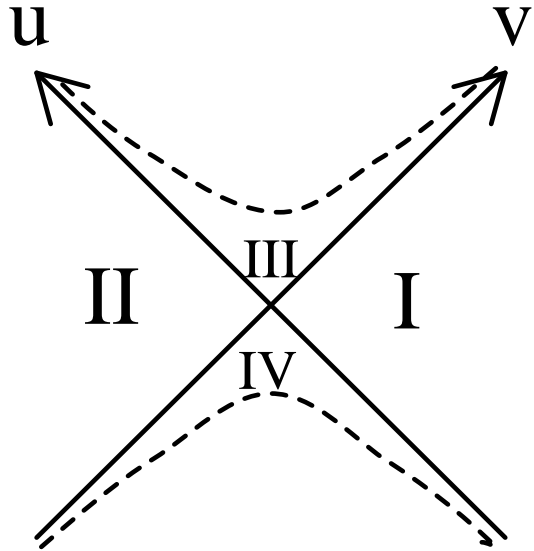}
    \end{center}
    \caption{Kruscal diagram for the two-dimensional black hole system.}
    \label{fig:kruscal}
\end{figure}

In this subsection, we review the classical geometry of the two-dimensional black hole. First of all, the metric \eqref{tbhm} has an event horizon at $\rho = 0$, but the $(t,\rho)$ coordinate does not cover the whole causal region of the black hole. One can maximally extend the geometry \eqref{tbhm} by introducing the Kruscal coordinate
\begin{align}
u = \sinh\rho e^{t} \ , \ \ v= -\sinh\rho e^{-t} \ , \ \ \dd s^2 = -2k\frac{\dd u\dd v}{1-uv} \ , \ \  e^{2\Phi} = \frac{k}{\mu(1-uv)} \ .
\end{align}
Note that in two-dimension, it is always possible to introduce the conformal coordinate $(u,v)$ locally with the conformally flat metric $\dd s^2 = f(u,v)\dd u\dd v$.
In this coordinate, the event horizon is located at $uv = 0$, and inside the horizon, we encounter singularity at $uv=1$, where the curvature and the dilaton diverges. The Kruscal diagram can be found in figure \ref{fig:kruscal}. Causal region of the
Lorentzian black hole background has four boundaries: past and
future horizons ${\cal H}^\pm$, and past and future asymptotic
infinities ${\cal I}^\pm$.

\begin{figure}[htbp]
   \begin{center}
    \includegraphics[width=0.5\linewidth,keepaspectratio,clip]{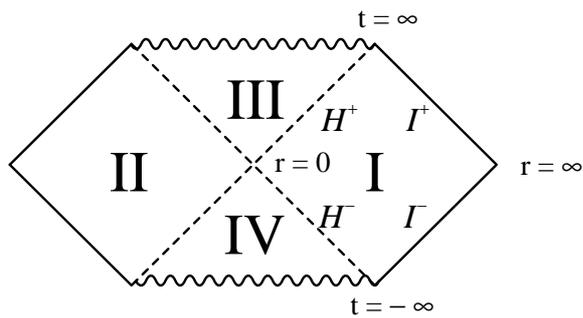}
    \end{center}
    \caption{Penrose diagram for the two-dimensional black hole.}
    \label{fig:bh}
\end{figure}

We can also study the global structure of the metric by using the Penrose coordinates, and one can write down the Penrose diagram (see figure \ref{fig:bh}) of the two-dimensional black hole system, which looks exactly same as that for the four-dimensional Schwarzshild black hole system (upon neglecting $\mathbb{S}^2$ angular directions). This is one of the motivations to study the two-dimensional black hole system as an exactly solvable toy model for four-dimensional Schwarzshild black hole.

The geodesic motion of a particle with the minimal coupling interaction to the geometry
\begin{align}
S = \int \dd s \  = \int \dd\tau \sqrt{\frac{\dot{u}\dot{v}}{1-uv}} \ ,
\end{align}
where dot denotes the derivative with respect to $\tau$, is given  by
\begin{align}
\begin{cases} \ddot{u}(1-uv) = -v\dot{u}^2 \cr
\ddot{v}(1-uv) = -u \dot{v}^2 \ .
\end{cases}
\end{align}
Later, we will compare this with the string motion and D-particle motion, both of which show quite distinct properties.

Finally, we would like to study the ``mass" of the two-dimensional black hole.
 For this purposes, it is convenient to use the Schwarzshild(-like) coordinate \eqref{blackNS5}:
\begin{align}
\dd s^2 = -\left(1-\frac{2M}{r}\right)\dd t^2 +  \frac{k}{1-\frac{2M}{r}}\frac{\dd r^2}{r^2} \ ,  \ \ e^{2\phi} = r \ ,
\end{align}
where $ r_0 = 2M$ is the location of the horizon. From the expression, it is clear that we can easily shift the value of $M$ multiplicatively $M\to aM$ by the scaling of $r$ as $r \to r/a$. Therefore the physical meaning of the ``mass" of the two-dimensional black hole is solely determined by the value of the string coupling constant (dilaton) at the horizon $r=2M$. It corresponds to the fact that the mass parameter $M$ is related to the world-sheet $\mathcal{N}=2$ Liouville cosmological constant $\mu$ in the dual $\mathcal{N}=2$ Liouville theory, where $\mu$ can be shifted by the shift of the Liouville coordinate (see section \ref{sec:3-4} for more about the duality).  

Because of this property, the Hawking temperature of the two-dimensional black hole is independent of $M$ unlike the case with higher-dimensional black-holes. Similarly many features of the string theory in the two-dimensional black hole background such as scattering amplitudes also show rather trivial dependence on $M$.\footnote{It is customary to set $M=1$ as we will do in most part of the thesis.} It is related to the Knizhnik-Polyakov-Zamolodchikov (KPZ) scaling law of the dual Liouville theory \cite{Knizhnik:1988ak}.
 
\subsection{Wick rotation: thermodynamic properties}\label{sec:2-4}

In this section, we would like to study thermodynamic properties of the two-dimensional black hole (and hence LST on the black NS5-branes). It is a well-known but profound fact that the black hole system shows a thermodynamic properties \cite{Bekenstein:1973ur,Bardeen:1973gs}:
\begin{itemize}
	\item Surface gravity $\kappa$ (temperature $T$) is constant over horizon of stationary black hole (the zeroth law: $T=\frac{\kappa}{2\pi}$)
	\item $\dd M = \frac{1}{8\pi} \kappa \dd A + \Omega \dd Q$ (the first law: $S= \frac{A}{4}$)	
	\item $\delta A \ge 0$ in any physical process (the second law)
	\item It is impossible to achieve $\kappa = 0$ by any physical process (the third law)
\end{itemize}
Here $M$ is the mass of the black hole, $A = 4S$ is the area of the event horizon ($=$ entropy), $Q$ is the charge, and $\Omega$ is its chemical potential.
One of the biggest motivations to study the quantum gravity such as the string theory is to understand the black hole thermodynamics from the microscopic viewpoint.

Let us begin with the temperature of the two-dimensional black hole. One convenient way to compute the Hawking temperature of black hole systems \cite{Hawking:1974sw}
 is to use the Euclidean path integral formalism \cite{Gibbons:1976ue}. In our case, we can study the Wick rotation (i.e. $ t\to i\tau_E$) of the Lorentzian two-dimensional black hole \label{tbhm}:
\begin{align}
\dd s^2_E =  \tanh^2\rho \, \dd \tau_E^2 + k\al' \dd 
\rho^2 \ . \label{tbhmE}
\end{align}
To avoid a conical singularity at the origin $\rho=0$, we have to set the periodicity of the Euclidean time direction by $\beta_{\rm Hw} = {2\pi}{\sqrt{k\alpha'}}$: $\tau_E \sim \tau_E + \beta_{\rm Hw}$. In the Euclidean path integral formulation, we regard this periodicity as the inverse of the Hawking temperature of the black hole: $T_{\rm Hw} = \frac{1}{\beta_{\rm Hw}} = \frac{1}{2\pi\sqrt{k\alpha'}}$ because in the Matsubara formalism, the periodicity of the Euclidean time corresponds to the inverse temperature. Note that in the large $k$ semi-classical limit, we have vanishing Hawking temperature so that the back-reaction associated with the Hawking radiation is negligible and the black hole geometry is infinitely long-lived (i.e. eternal). As we mentioned in section \ref{sec:2-3}, it is also interesting to note that the Hawking temperature does not depend on the mass $m$ of the two-dimensional black hole. In the context of the dual LST defined in section \ref{sec:2-3}, we will regard this temperature as the (non-perturbative) Hagedorn temperature of the LST.

There are several derivations of the Hawking temperature other than the Euclidean path integral method. Recently \cite{Robinson:2005pd,Iso:2006wa} proposed a new derivation based on the gravitational anomaly in the vicinity of the event horizon.\footnote{See also \cite{Christensen:1977jc} for a derivation based on the trace anomaly.} We briefly review their derivation focusing on the two-dimensional case (see \cite{Iso:2006ut,Murata:2006pt,Vagenas:2006qb,Setare:2006hq} for related works). 

Let us take the very near-horizon limit (Rindler-limit)\footnote{Although there is nothing wrong with taking the Rindler-limit in the general relativity, the limit is a little bit subtle in the string theory because the string theory introduces a ``minimal size" (string scale) to the geometry. In our example, the central charge of the original two-dimensional black-hole and its very near horizon limit is different, so we need an extra compensation of the central charge in order to preserve the criticality condition. Since we are only interested in the classical thermodynamics, we will neglect this subtlety for a time being. See also the discussion of stretched horizon of the two-dimensional black holes in section \ref{sec:4}.} of the two-dimensional black hole:
\begin{align}
\dd s^2 = -\frac{2(r-r_0)}{\sqrt{k\alpha'}} \dd t^2 + \frac{\sqrt{k\alpha'}}{2(r-r_0)} \dd r^2 \ , \ \ \Phi = \mathrm{const} \ .
\end{align}
Now let us suppose we neglect the classically irrelevant in-falling modes of any scalar field propagating in the vicinity of the horizon $r=r_0$. The massless scalar fields with this boundary condition are effectively chiral, so it will show a gravitational anomaly:
\begin{align}
\nabla_\mu T^{\mu}_{\ \nu} = \frac{1}{\sqrt{-g}}\partial_\mu N^\mu_{\ \nu} 
\end{align}
with the explicit component expression
\begin{align}
N_{t}^t = N_{r}^r = 0 \ , \ \ N_t^r = \frac{1}{192\pi} \frac{4}{k\alpha'} \ , \ \ N_{r}^t = -\frac{1}{192\pi (r-r_0)^2}  \ . 
\end{align}
Especially, it shows a pure flux contribution
\begin{align}
\Phi = N_t^r|_{r=r_0} = \frac{1}{192\pi} \frac{4}{k\alpha'}  \ . \label{fl}
\end{align}

To cancel the gravitational anomaly, we need a quantum contribution that can be attributed to the Hawking radiation from the black hole. The black body radiation with the temperature $T_{\rm Hw}$ gives rise to the flux
\begin{align}
\Phi = \frac{\pi}{12}T^2_{\rm Hw} \ , \label{flt}
\end{align}
and the comparison between \eqref{fl} and \eqref{flt} establishes the Hawking temperature $T_{\rm Hw} = \frac{1}{2\pi\sqrt{k\alpha'}}$. We will later see similar effects related with the choice of boundary conditions of wavefunction at the horizon when we study D-brane motions in the two-dimensional black hole geometries.

Let us move on to the other thermodynamic quantities. Since the temperature does not depend on the mass of the two-dimensional black hole, we see that the black hole entropy is given by
\begin{align}
S(m) =  \beta_{\rm Hw} m = 2\pi\sqrt{\alpha' k}  m \ . \label{entr}
\end{align}
In higher dimensions, we could identify the entropy of the black hole as the area of the event horizon (i.e. the Bekenstein formula $S=\frac{A}{4\pi}$), but in the two-dimensional space-time, the event horizon is just a point and we cannot apply the Bekenstein formula. Instead, we have defined the entropy through the thermodynamic relation $\beta = \frac{\partial S}{\partial m} $. This formula predicts the high energy density of states in the LST. Assuming the microscopic explanation of the black hole entropy \eqref{entr} from the LST, the density of states of LST should be given by
\begin{align}
\rho(E) \sim e^{2\pi\sqrt{\alpha' k} E } \ ,
\end{align}
in the high energy limit $E\to \infty$.
\newpage
\sectiono{Two-dimensional Black Hole: CFT Viewpoint}\label{sec:3}
In this section, we review the two-dimensional black hole from the exactly solvable CFT viewpoint. We begin with the Euclidean version of the two-dimensional black hole  and then we move on to the Lorentzian two-dimensional black hole from an appropriate Wick rotation. This is because the Euclidean version is much better understood than the Lorentzian counterpart.

The organization of this section is as follows. In section \ref{sec:3-1}, We begin with the classical geometries for the $SL(2;\br)/U(1)$ coset model that yields an exact CFT model for the two-dimensional black hole system. In section \ref{sec:3-2}, we review the Euclidean spectrum of the $SL(2;\br)/U(1)$ coset model. In section \ref{sec:3-3}, we deal with the Lorentzian case in detail. In section \ref{sec:3-4}, we comment the duality between $SL(2;\br)/U(1)$ coset model and the $\mathcal{N}=2$ Liouville theory and discuss implications of the associated winding tachyon condensation.

\subsection{Classical geometries for $SL(2;\br)/U(1)$ coset}\label{sec:3-1}
From the world-sheet viewpoint, the reason why we are interested in the two-dimensional black hole system is that we can quantize the string theory on it by using the $SL(2;\br)/U(1)$ coset CFT. In this subsection, we would like to overview the correspondence between the $SL(2;\br)/U(1)$ coset model and the two-dimensional black hole system from the gauged Wess-Zumino-Novikov-Witten (WZNW) model construction \cite{Witten:1991yr,Elitzur:1991cb,Mandal:1991tz}.

It is possible to define the coset CFT such as $SL(2;\br)/U(1)$ model purely from the algebraic viewpoint (at least at the level of the left-right chiral $SL(2;\br)/U(1)$ representations: the most difficult point is to construct the modular invariant combinations), but we would like to begin with the Lagrangian construction based on \cite{Witten:1991yr}. This is because the construction directly gives the geometric interpretation of the model as the non-linear sigma model on the two-dimensional black hole in the semi-classical limit $(k \to \infty)$. The path integral formulation based on the Lagrangian  can also be used to derive the (formally) modular invariant partition function of the Euclidean two-dimensional black hole \cite{Hanany:2002ev} (see appendix \ref{part}).

The ungauged WZNW model for a general Lie group $G$ has the following action:
\begin{align}
S_{WZNW}(g) = \frac{\kappa}{8\pi} \int_{\Sigma} \dd^2 x \sqrt{|\gamma|}\gamma^{ij} \mathrm{tr} (g^{-1}\partial_i g g^{-1} \partial_j g) + i\kappa\Gamma(g) \ . \label{WZWa} 
\end{align}
The Wess-Zumino term $\Gamma(g)$ is given by 
\begin{align}
\Gamma(g) = \frac{1}{12\pi} \int_B \mathrm{tr} (g^{-1}\dd g\wedge g^{-1}\dd g\wedge g^{-1}\dd g ) \ ,
\end{align}
where $B$ is a three-dimensional manifold whose boundary is $\Sigma$. $\kappa$ denotes the level of the current algebra realized by the WZNW model.
 When the Lie group $G$ is compact, the level $\kappa$ should be quantized so that the Wess-Zumino term contributes to the path-integral uniquely with an arbitrary choice of $B$. In our case, however, since the Lie group is non-compact and $H^3(SL(2;\br),\br) = 0$, the quantization condition of the level $\kappa$ is not necessary.

Let $G$ be $SL(2;\br)$ for our discussion in the following. 
The action \eqref{WZWa} possesses a global $SL(2;\br)\times SL(2;\br)$ symmetry $g \to a gb^{-1}$, with $a,b \in SL(2;\br)$. Quantum mechanically, it will be elevated to a current algebra with the level $\kappa$: the chiral current
\begin{align}
j^A(z) = \kappa \mathrm{Tr}(T^A \partial g g^{-1}) \ ,
\end{align}
with $T^3= \frac{1}{2}\sigma_2$, $T^{\pm} = \pm \frac{1}{2}(\sigma_3 \pm i\sigma_1)$, satisfies the OPE of the affine $\widehat{SL(2;\br)_{\kappa}}$  current algebra
\begin{align}
\begin{cases} j^3(z)j^3(0) \sim -\frac{\kappa}{2z^2} \cr
j^3(z) j^{\pm}(0) \sim \pm \frac{1}{z}j^{\pm}(0) \cr
j^+(z)j^{-}(0) \sim \frac{\kappa}{z^2} - \frac{2}{z}j^{3}(0) 
\end{cases} \ .
\end{align}
The  bosonic $SL(2;\br)$ WZNW model has the central charge $c = \frac{3\kappa}{\kappa-2}$. 
We will gauge the anomaly free subgroup of the global symmetry of the $SL(2;\br)$ WZNW model to obtain the Lagrangian formulation for the coset CFT.

Let us first begin with the Euclidean coset. The $SL(2;\br)$ WZNW model has a negative-signature direction in $J^3 \sim i\sigma_2$ and the target space is a Lorentzian manifold. We gauge the (compact) $U(1)$ subgroup generated by
\begin{align}
\delta g = \epsilon (i\sigma_2 \cdot g + g \cdot (i\sigma_2)) \ , 
\end{align}
or by setting
\begin{align}
a = b^{-1} = h= \begin{pmatrix} \cos\epsilon & \sin\epsilon \\ -\sin\epsilon  & \cos\epsilon \end{pmatrix}  \ . \label{gauget}
\end{align}

To promote the global axial symmetry $g \to hgh $ to the gauge symmetry, we have to introduce the gauge connection $A_i$ that transforms as $\delta A_i = -\partial_i \epsilon$ under the gauge transformation \eqref{gauget} with a space varying gauge parameter $\epsilon(x_i)$. The covariantized action reads
\begin{align}
S_{\mathrm{gauged}} &= S_{\mathrm{WZNW}}(g) + \cr &+ \frac{\kappa}{2\pi} \int \dd^2z \bar{A} \mathrm{Tr}\left(i\sigma_2 g^{-1}\partial g\right) + A\mathrm{Tr}\left(i\sigma_2 \bar{\partial} g ^{-1}\right) + A\bar{A}\left(-2 + \mathrm{Tr}\left(i\sigma_2 g i\sigma_2 g^{-1}\right) \right) \ .  \label{gWZW}
\end{align}
We call this gauged WZNW model as $SL(2;\br)^{(A)}/U(1)$ axial coset.
To obtain the classical geometry of the $SL(2;\br)^{(A)}/U(1)$ coset CFT, we will integrate out the gauge field by fixing the gauge \footnote{In terms of the Euler angle parametrization (see appendix \ref{a-1-2}), $g = e^{i\sigma_2 \frac{t-\phi}{2}} e^{r\sigma_1}e^{i\sigma_2\frac{t+\phi}{2}}$ and we set $t=0$, where $\phi = \theta - \frac{\pi}{2}$. It is clear that this gauge fixing is always possible and unique.}
\begin{align}
g = \cosh r + \sinh r \begin{pmatrix} \cos\theta & \sin\theta \\ \sin\theta & -\cos\theta \end{pmatrix} \ .
\end{align}
The resulting sigma model for the gauge fixed coordinate $(r,\theta)$ is given by the action
\begin{align}
S = \frac{\kappa}{2\pi} \int \dd^2z \left(\partial r \bar{\partial} r + \tanh^2 r \partial \theta \bar{\partial} \theta \right) \ , \label{axcea}
\end{align}
with the dilaton gradient $e^{2\Phi} = \frac{k}{\mu \cosh^2 r}$, which originates from the one-loop determinant factor for the gauge field $A_i$. The geometry one can read from the sigma model action is the Euclidean two-dimensional black hole we have introduced in section \ref{sec:2}.

In the bosonic coset model, we expect a perturbative (and nonperturbative) $\alpha'$ corrections for this gauge fixing procedure and the corrected sigma model was proposed in \cite{Dijkgraaf:1992ba}. In the supersymmetric Kazama-Suzuki coset, it is believed that there is no perturbative $\alpha'$ corrections to the metric. Nonperturbative corrections which will be reviewed in section \ref{sec:3-4}, however, are present and they are one of the key elements to  understand the ``black hole - string transition".

The Lorentzian coset is obtained by gauging the non-compact subgroup
\begin{align}
\delta g = \epsilon\left(\sigma_3 g + g \sigma_3\right) \ .
\end{align}
We fix the gauge by setting
\begin{align}
g = \begin{pmatrix} a & u \\ -v & a \end{pmatrix}
\end{align}
with the determinant constraint $uv = 1- a^2$. The gauge fixing condition is valid for $1-uv>0$ and we can see that $u$ and $v$ are gauge invariant coordinates. With this gauge fixing condition,\footnote{Strictly speaking, the coset is a double cover of the $(u,v)$ plane, where the two-sheets are distinguished by the signature of $a$. We will neglect this small subtlety throughout the thesis.} the target space is spanned by the $(u,v)$ plane. After integrating out the gauge field $A_i$, the resulting sigma model is given by
\begin{align}
S = -\frac{\kappa}{4\pi} \int \dd^2x \sqrt{|\gamma|}\frac{\gamma^{ij}\partial_i u\partial_j v}{1-uv} \ ,
\end{align}
which reproduces the classical two-dimensional black hole system discussed in section \ref{sec:2}. Here we have used the Lorentzian signature world-sheet so that the sigma model with a Lorentzian  signature target space is well-defined.

In this thesis, we mainly focus on the supersymmetric generalization of the two-dimensional black hole based on the supersymmetric $SL(2;\br)_k/U(1)$ coset model. The starting point is the bosonic $SL(2;\br)_\kappa /U(1)$ coset mode with the level $\kappa = k+2$ bosonic current algebra. In addition to the bosonic action \eqref{gWZW}, we introduce the fermionic part:
\begin{align}
S_f = \frac{1}{2\pi} \int \dd^2z \left(\psi^+(\bar{\partial}-\bar{A})\psi^- + \psi^- (\bar{\partial}+\bar{A}) \psi^{+} + \tilde{\psi}^+ (\partial -A) \tilde{\psi}^-  + \tilde{\psi}^-(\partial +A)\tilde{\psi}^+ \right) \ , 
\end{align}
with the OPE $\psi^+(z)\psi^-(0) \sim 1/z$, $\psi^{\pm}(z)\psi^{\pm}(0) \sim 0$.
Let us concentrate on the Euclidean case for definiteness. From the Kazama-Suzuki construction, we can realize the $\mathcal{N}=2$ superconformal symmetry on the supersymmetric $SL(2;\br)/U(1)$ coset model. The explicit realization is given by
\begin{align}
T(z) & = \frac{1}{k}(j^{1}j^{1} +j^2j^2) - \frac{1}{2}(\psi^+\partial \psi^- - \partial \psi^+ \psi^-) \cr
G^{\pm}(z) &= \frac{1}{\sqrt{k}} \psi^{\pm} j^{\mp} \cr
J(z) & = \psi^+\psi^- + \frac{2}{k}(j^{3} + \psi^+\psi^-) \ ,
\end{align}
whose central charge is given by $c = 3(1+\frac{2}{k}) $.
The gauging current is defined by $J^3 = j^3 + \psi^+ \psi^-$, which commutes with all the elements of the $\mathcal{N}=2$ superalgebra. The fermionic part of the Lorentzian case is obtained from the analytic continuation by formally replacing $\psi^+ = \frac{1}{\sqrt{2}}(\psi^1 + i \psi^2) \to \frac{1}{\sqrt{2}}(\psi^1 + \psi^3)$ and $\psi^- = \frac{1}{\sqrt{2}}(\psi^1 - i \psi^2) \to \frac{1}{\sqrt{2}}(\psi^1 - \psi^3)$ .

From the path integral viewpoint, instead of treating the (Euclidean) $SL(2;\br)^{(A)}/U(1)$ coset, it is more convenient to study the equivalent description based on the $\mathbb{H}_3^+/\br$ coset model. The $\mathbb{H}_3^+ = SL(2;\mathbb{C})/SU(2)$ model is defined by the sigma model on the upper sheet
\begin{align}
g =\begin{pmatrix} a & u \\ \bar{u}  & b \end{pmatrix} =  \begin{pmatrix} e^{\phi} & e^{\phi}\bar{\gamma} \\ e^{\phi}\gamma  & e^{\phi}\gamma\bar{\gamma} + e^{-\phi} \end{pmatrix} \ ,
\end{align}
where we have introduced a real field $\phi$ and a complex field $\gamma$ with its complex conjugation $\bar{\gamma}$. The sigma model has the action
\begin{align}
S = -\frac{\kappa}{2\pi} \int \dd^2z \left(\partial\phi \bar{\partial}\phi + e^{2\phi}\partial\gamma\bar{\partial}\gamma \right)  \ .
\end{align}
The model has a positive definite action and the path integral is well-defined (unlike the ungauged $SL(2;\br)$ WZNW model).\footnote{However, the model has an imaginary $H_3$ flux, so the physical interpretation is unclear.} The two-dimensional Euclidean black hole is obtained by axially gauging the noncompact $U(1)$ direction $\sigma_2$ (e.g. one can choose a gauge $a = b$ and obtain the metric $ds^2 = \kappa\frac{dud\bar{u}}{1+|u|^2}$). This construction has the advantage that the parent sigma model has a definite Euclidean path integral while the $SL(2;\br)/U(1)$ coset does not because the parent WZNW model has a Lorentzian signature. 

So far, we have studied the axial coset of the $SL(2;\br)$ WZNW model, but it is possible to gauge the vector symmetry 
\begin{align}
\delta g = \epsilon (i\sigma_2 \cdot g - g \cdot (i\sigma_2)) \ .
\end{align}
This symmetry has a fixed point, and the corresponding effective action
\begin{align}
S = \frac{\kappa}{2\pi} \int \dd^2z \left(\partial \rho \bar{\partial} \rho + \frac{1}{\tanh^2 \rho} \partial \tilde{\theta} \bar{\partial} \tilde{\theta} \right) \ , \label{efftru}
\end{align}
which is also known as the trumpet model, has a singularity at $\rho=0$. However, from the algebraic coset viewpoint, there is no singularity at all. We have just replaced right moving $\bar{J}_3$ current of the $SL(2;\br)$ WZNW model with $-\bar{J}_3$.

In order to specify the model, we have to determine the periodicity of the variable $\tilde{\theta}$ in the effective action \eqref{efftru}. The angular variable $\theta$ in the cigar geometry has a natural periodicity $2\pi$ coming from the $SL(2;\br)$ WZNW model. In the trumpet case, there is no apriori natural periodicity of $\tilde{\theta}$ because it is non-contractible loop in $SL(2;\br)$ and any periodicity is allowed if we study the universal cover of the $SL(2;\br)$. In terms of the Euler angle parametrization (see appendix \ref{a-1-2}), $g = e^{i\sigma_2 \frac{t-\phi}{2}} e^{r\sigma_1}e^{i\sigma_2\frac{t+\phi}{2}}$ and we set $\phi=0$. In this sense, the natural periodicity for $t = \tilde{\theta}$ is $2\pi$. We {\it define} that the $SL(2;\br)^{(V)}/U(1)$ vector coset model has $2\pi$ periodicity in $\tilde{\theta}$.

From the algebraic construction, the vector coset merely changes the sign convention of the right-moving current $\bar{J}_3$, which reminds us of the T-duality or mirror symmetry. Along this line of reasoning, an alternatively good definition of the vector coset is to take $ 2\pi/k$ periodicity in $\tilde{\theta}$.\footnote{This convention is the one given in \cite{Dijkgraaf:1992ba}.} This is indeed motivated by Bucsher's T-duality rule: if we perform the T-duality to our original cigar model \eqref{axcea} with $2\pi$ periodicity in $\theta$, we obtain the trumpet model with $ 2\pi/k$ periodicity. We call this model as ``$\bz_k$ orbifold of the vector coset $SL(2;\br)^{(V)}/U(1)$". The $\bz_k$ orbifold of the $SL(2;\br)^{(V)}/U(1)$ or $\bz_k$ orbifold of the trumpet model is same as the cigar model as a CFT (up to a GSO projection in the supersymmetric case).\footnote{Let us mention the similar structure in the $SU(2)$ case. It is known that the axial coset $SU(2)^{(A)}/U(1)$ is the $\bz_k$ orbifold of the vector coset $SU(2)^{(V)}/U(1)$. However, in this case, $\bz_k$ orbifold of the vector coset $SU(2)^{(V)}/U(1)$ is the same model as the original $SU(2)^{(V)}/U(1)$. Therefore, we can also say that the T-dual of the $SU(2)^{(A)}/U(1)$ coset is the $SU(2)^{(V)}/U(1)$. In the $SL(2;\br)$ case, although the asymptotic spectrum of the $\bz_k$ orbifold of the $SL(2;\br)^{(V)}/U(1)$ coincides with that of the $SL(2;\br)^{(V)}/U(1)$, the (un-regularized) partition function is different from each other. Here we implicitly assumed that $k$ is an integer, but the situation is more involved when $k$ is not an integer because the meaning of the $\bz_k$ orbifold is obscure. Note that we have defined the $\bz_k$ orbifold of the $SL(2;\br)^{(V)}/U(1)$ as the T-dual of the $SL(2;\br)^{(A)}/U(1)$, which perfectly makes sense even for irrational $k$.}

Since the vector coset model with the trumpet geometry is related to the T-duality of the cigar geometry, there should be no singularity at all in the vector coset model as a CFT. What happens to the apparent singularity of the classical geometry? We will come back to this problem in section \ref{sec:3-4}.

The equivalence between the axial coset and the ($\bz_k$ orbifold of) vector coset leads to a remarkable observation made in \cite{Dijkgraaf:1992ba} --- the duality between the singularity and the horizon. If we gauge the vector symmetry for the Lorentzian coset, we end up with the same Lorentzian two-dimensional black hole.\footnote{This is up to global duplications. For instance, if one considers an axial coset of the universal cover of the $SL(2;\br)/U(1)$ coset, we have an infinite copies of the Lorentzian two-dimensional black holes.} However, the analytic continuation of the trumpet geometry (vector coset) has a natural interpretation as the region inside the singularity:
\begin{align}
\dd s^2 = \alpha' k\left(- \frac{1}{\tanh^2\rho} \dd t^2 + \dd\rho^2\right) \ .
\end{align}

In this way, the axial-vector duality suggests the duality of the region parametrized $uv$ and $1-uv$ while keeping $t$. In particular, it exchanges the region outside the horizon and the one inside the singularity. This duality would shed a new light on the physics of the black hole and especially, relation between the T-duality and ``winding tachyon" in the Lorentzian two-dimensional black hole. It is, however, fair to say that the precise physical meaning of the duality is far from being well-understood in the Lorentzian signature.

To close this section, we discuss the pure two-dimensional background by setting $k=1/2$ or $\kappa = 9/4$. Then the model is a critical string theory by itself and one can regard it as a pure two-dimensional background \cite{Witten:1991yr}. In the Euclidean signature, there is a proposed dual matrix model \cite{Kazakov:2000pm} and the theory is supposedly exactly solvable. In the context of more general dilaton gravity in the two-dimensional space-time, the classical solutions are investigated in \cite{Grumiller:2005sq}.

\subsection{Euclidean spectrum}\label{sec:3-2}
Let us study the spectrum of the Euclidean $SL(2;\br)/U(1)$ coset model. 

\subsubsection{algebraic coset}\label{sec:3-2-1}
We begin with the algebraic structure of the coset model from the noncompact para-fermion construction. 

Let $\Phi_{jm}(z)$ be holomorphic part of the primary fields of $SL(2;\br)$ WZNW model with bosonic level $\kappa$. It has a (bosonic) left-moving $j^3_0$ eigenvalue $m$:
\begin{align}
j^3(z) \Phi_{jm}(0) = m \frac{\Phi_{jm}}{z} \ .
\end{align}
 It is also a holomorphic part of the primary for the supersymmetric $SL(2;\br)$ WZNW model with the same eigenvalue for $J^3_0$. Here, the supersymmetric current $J^3$ is defined by
\begin{align}
J^3 = j^3 -\psi^-\psi^+ \ .
\end{align}
For later convenience, we introduce the bosonized current\footnote{Note that the gauging current must be time-like in the Euclidean coset.}
\begin{align}
\partial H &= i\psi^-\psi^+ \cr
J^3 &= -\sqrt{\frac{k}{2}}\partial X_3  \cr
j^3 &= J^3 + i\partial H = -\sqrt{\frac{\kappa}{2}}\partial x_3 \ .
\end{align}
Using $X_3$, or $x_3$, we can decompose $\Phi_{jm}$ as
\begin{align}
\Phi_{jm} = U_{jm} e^{m\sqrt{\frac{2}{k}}X_3} = V_{jm} e^{m\sqrt{\frac{2}{\kappa}}x_3} \ . \label{decb}
\end{align}
The para-fermion fields $U_{jm}$ and $V_{jm}$ have the conformal dimension
\begin{align}
\Delta(U_{jm}) &= \frac{-j(j+1)+m^2}{k} \cr
\Delta(V_{jm}) &= -\frac{j(j+1)}{\kappa-2} + \frac{m^2}{\kappa} \ , 
\end{align}
and they are (holomorphic) primaries of the supersymmetric Euclidean coset $SL(2;\br)/U(1)$ and the bosonic coset respectively. 

For the supersymmetric case, we should also decompose the $U(1)$ fermion current as 
\begin{align}
e^{inH} = e^{n\sqrt{\frac{2}{k}}X_3} e^{i\sqrt{\frac{c}{3}} X_R} \ , \label{decf}
\end{align}
where the bosonized $U(1)_R$-current (of the $\mathcal{N}=2$ SUSY algebra: see appendix \ref{SCA2}) is defined by
\begin{align}
J = i\sqrt{\frac{c}{3}}\partial X_R \ ,
\end{align}
where
\begin{align}
iH = \sqrt{\frac{2}{k}}X_3 + i\sqrt{1+\frac{2}{k}}X_R \ .
\end{align}

Under the decomposition \eqref{decb} and \eqref{decf}, (holomorphic) primary fields of the supersymmetric $SL(2;\br)/U(1)$ coset takes the form
\begin{align}
V^{n}_{jm} = V_{jm} e^{i(\frac{2m}{k+2} +n)\sqrt{\frac{c}{3}}X_R} \ ,
\end{align}
which has the conformal dimension
\begin{align}
\Delta(V_{jm}^n) = -\frac{j(j+1) + (m+n)^2}{k} + \frac{n^2}{2} \ ,
\end{align}
and the $U(1)_R$ charge
\begin{align}
R(V_{jm}^n) = \frac{2m}{k} + \frac{nc}{3} \ .
\end{align}
The (half) integer $n$ is the amount of the spectral flow of the $\mathcal{N} = 2 $ superconformal algebra. The structure of the descendants, depending on quantum numbers $(j,m,n)$, are completely fixed from that of the $SL(2;\br)$ (see appendix \ref{a-1}), or alternatively from the representation of the $\mathcal{N}=2 $ superconformal algebra.

\subsubsection{spectrum from partition function}\label{sec:3-2-2}

To obtain the full spectrum of the CFT from the holomorphic data discussed in section \ref{sec:3-2-1}, we need to combine left-moving parts and right-moving parts in a consistent way. For example, in the compact $SU(2)$ WZNW model, the complete classification of the modular invariant partition function (and hence the spectrum) is given by the so-called ADE classification. In the non-compact case, we have not yet achieved such a systematic classification. Physically, however, we are primarily interested in the two-dimensional black hole interpretation of the $SL(2;\br)/U(1)$ coset model, so we will only consider the simplest realization from the gauged WZNW model as we have focused in section \ref{sec:3-1}.

We begin with the partition function for the bosonic Euclidean two-dimensional black hole \cite{Hanany:2002ev}
\begin{align}
Z_{\mathbb{H}^{3(A)}_{+}/\br} = \int_{\Sigma} \frac{\dd u^2}{\tau_2} \frac{e^{\frac{u_2^2}{\tau_2}}}{\sqrt{\tau_2}|\theta_1(\tau,u)|^2 } \sqrt{\tau_2}|\eta(\tau)|^2 \sum_{m,\omega\in \bz} e^{-\frac{\pi \kappa}{\tau_2}|\omega\tau - m + u|^2} \ . \label{ppo}
\end{align}
See appendix \ref{part} for a summary of various partition functions. Unfortunately, the partition function is divergent, and the leading divergence comes from the integration near $u_1 = u_2 = 0$. The diverging factor could be attributed to the volume divergence in the radial $\rho$ direction of the cigar model. Thus, at the leading order, we have
\begin{align}
Z_{\mathbb{H}^{3(A)}_{+/\br}} \sim \frac{1}{2\pi} (\log\epsilon) Z_{\mathrm{free}}(\tau) Z_{\sqrt{\kappa}}(\tau) + \text{finite part} \ , \label{lpf}
\end{align}
which gives the asymptotic degrees of freedom realized by a free non-compact boson (with the linear dilaton):
\begin{align}
Z_{\mathrm {free}}(\tau) = \frac{1}{\sqrt{\tau_2}|\eta(\tau)|^2} \ ,
\end{align}
 and a compact boson with radius $R^2 = \kappa$. 
\begin{align}
Z_{\sqrt{\kappa}} = \frac{\sqrt{\kappa}}{\sqrt{\tau_2}|\eta(\tau)|^2} \sum_{m,\omega\in \bz} e^{-\frac{\pi \kappa}{\tau_2}|\omega\tau - m|^2} \ .
\end{align}
The appearance of the compact boson is due to the summation over the lattice $(n,\omega)$, which arises from the zeros of $\theta_1(\tau,u)$ in \eqref{ppo}.

A more precise (but formal) manipulation \cite{Hanany:2002ev} leads to the following decomposition of the partition function:
\begin{align}
Z_{\mathbb{H}^{3(A)}_{+/\br}} = \int^{-1/2}_{-(\kappa-1)/2} \dd j \mathrm{Tr}_{\hat{D}^+_j\otimes\hat{D}^+_j} q^{L_0}q^{\bar{L}_0} + \sum_{\omega,n}\int_0^\infty \dd p 2\rho(p) \mathrm{Tr}_{\hat{C}_{-\frac{1}{2}+ip}\otimes\hat{C}_{-\frac{1}{2}+ip}} q^{L_0}q^{\bar{L}_0} + \dots \ , \label{decompp}
\end{align}
where the Hilbert space $\hat{D}^+_j\otimes\hat{D}^+_j$ is the discrete representations of the $SL(2;\br)$ with the constraints $J_0^3 -\bar{J}_0^3 = n$, $J_0^3 + \bar{J}_0^3 = \kappa \omega $ and no contribution from the $J^3_{n<0}$ oscillators.
The same restriction is imposed on the continuous representations $\hat{C}_{-\frac{1}{2}+ip}\otimes\hat{C}_{-\frac{1}{2}+ip}$. The density of states for the continuous representations is given by
\begin{align}
\rho(p) = \frac{1}{2\pi} 2 \log \epsilon + \frac{1}{2\pi i} \frac{\partial}{2\partial p} \log \frac{\Gamma(-ip+\frac{1}{2}-m)\Gamma(-ip+\frac{1}{2}+\bar{m})}{\Gamma(+ip+\frac{1}{2}+\bar{m})\Gamma(+ip+\frac{1}{2}-m)} \ , \label{dnsi}
\end{align}
where $m = \frac{1}{2}(n+\kappa \omega)$, $\bar{m} = -\frac{1}{2}(n-\kappa\omega)$ are the eigenvalues of $J_0^3$ and $\bar{J}_0^3$. The density of states appearing here is consistent with the reflection amplitude (or sphere two-point function) of the two-dimensional black hole as we will review in section \ref{sec:3-2-3}. The leading diverging part proportional to $\log \epsilon$ agrees with \eqref{lpf}. 

There are, however, several subtleties associated with the decomposition \eqref{decompp}. First of all, the expression \eqref{decompp} is {\it not} modular invariant although our starting point \eqref{ppo} is formally invariant. The failure is due to the nontrivial $p$ dependent density of states \eqref{dnsi} and the contribution from the discrete series. In other words, the regularization rule for the character decomposition \eqref{decompp} does not preserve the modular invariance. The only one can say is that the leading part, (or the partition function per unit volume as $\epsilon \to 0$) is modular invariant \eqref{lpf}. Another subtlety is related to the omitted terms in  \eqref{decompp}. The regularization procedure proposed in \cite{Hanany:2002ev} (see also \cite{Maldacena:2000hw} for related models) actually leaves us with finite terms that could not be written as the character appearing in \eqref{decompp} \cite{Israel:2004ir}. Again this depends on the regularization scheme and one natural (but not unique) solution is to omit this part as we will implicitly assume in the following.

Despite all these subtleties, the decomposition \eqref{decompp} seems to capture important physics of the two-dimensional black hole. In particular, it predicts the existence of the discrete spectrum localized near the tip of the cigar. Indeed the range of the discrete representations $\frac{-\kappa+1}{2}<j<-\frac{1}{2}$ will be independently checked by the Cardy analysis of the boundary states for $SL(2;\br)/U(1)$ coset model as we will see in section \ref{sec:6-2-3}. Also, the minisuperspace analysis for the two-dimensional black hole reproduces the zero-slope limit of the results given here including the density of states \eqref{dnsi}. We will review the mini-superspace analysis in section \ref{sec:3-2-3}.

Before moving on to the mini-superspace analysis, we will briefly present a generalization to the supersymmetric $SL(2;\br)/U(1)$ coset model. The partition function is given by
\begin{align}
Z^{(NS)}(\tau) = \int_{\Sigma} \frac{\dd u^2}{\tau_2} \frac{|\theta_3(\tau,u)|^2}{\sqrt{\tau_2}|\theta_1(\tau,u)|^2 } \sqrt{\tau_2}|\eta(\tau)|^2 \sum_{m,\omega\in \bz} e^{-\frac{\pi k}{\tau_2}|\omega\tau - m + u|^2} \ 
\end{align}
and the decomposition to the character is obtained as 
\begin{align}
\int^{-1/2}_{-(k+1)/2} \dd j \mathrm{Tr}_{\hat{D}^+_j\otimes\hat{D}^+_j} q^{L_0}q^{\bar{L}_0} + \sum_{\omega,n \in \bz}\int_0^\infty \dd p 2 \rho(p) \mathrm{Tr}_{\hat{C}_{-\frac{1}{2}+ip}\otimes\hat{C}_{-\frac{1}{2}+ip}} q^{L_0}q^{\bar{L}_0} + \dots \ , \label{decomps} \ .
\end{align}
In this case, the trace should be taken over the (NS-NS) Hilbert space of the supersymmetric coset instead of the bosonic one. Explicitly
\begin{align}
\mathrm{Tr}_{\hat{C}_{-\frac{1}{2}+ip}\otimes\hat{C}_{-\frac{1}{2}+ip}} q^{L_0}q^{\bar{L}_0} = q^{\frac{p^2+m^2}{k}}\bar{q}^{\frac{p^2+\bar{m}^2}{k}} \frac{|\theta_3(\tau)|^2}{|\eta(\tau)|^6} \ , 
\end{align}
and 
\begin{align}
 \int^{-1/2}_{-(k+1)/2} \dd j \mathrm{Tr}_{\hat{D}^+_j\otimes\hat{D}^+_j} q^{L_0}q^{\bar{L}_0} &=  \sum_{\omega,n \in {\mathbf{Z}}} \sum_{j \in \mathcal{J}_{\omega,n}} \chi_{\mathrm{dis}, 1+j+\frac{k}{2}, m +\frac{k}{2}}(\tau) \chi_{\mathrm{dis}, 1+j+\frac{k}{2}, \bar{m} + \frac{k}{2}}(\bar{\tau}) \cr
\mathcal{J}_{\omega,n} &= \left.\left[-\frac{k+1}{2},-\frac{1}{2}\right.\right) \cap \left(\frac{k\omega-n}{2} + \mathbb{Z}\right) \cr
\chi_{\mathrm{dis}, j,j+n}({\tau}) &= \frac{q^{\frac{(j+n)^2}{k}-\frac{1}{4k}}}{1+q^{n+1/2}}\frac{\theta_3(\tau,0)}{\eta(\tau)^3} \ .
\end{align}
The partition functions of the other sectors are readily obtained by performing the spectral flow symmetry of the $\mathcal{N}=2$ SCA. We note that in order to obtain a superstring compactification with other sectors (such as $\mathcal{N}=2$ minimal models), we have to project down to the sectors with integral $U(1)_R$ -charge so that the space-time supersymmetry is well-defined (GSO projection).

We can read some important physics from the spectrum of the Euclidean two-dimensional black hole:
\begin{itemize}
	\item The continuous representations have a mass gap, which is consistent with the (asymptotic) linear dilaton background. Due to the mass gap, would-be graviton is massive, which is again consistent with the statement that the LST is non-gravitational theory.
	\item The discrete representations correspond to local dynamical degrees of freedom that has a winding quantum number being localized near the tip of the cigar. From the space-time point of view, they are normalizable deformation of the background localized in the vicinity of the singularity. The improved unitarity bound perfectly agrees with the geometrical normalizability condition discussed in section \ref{sec:2-2-2}.
\end{itemize}

The improved unitarity bound has an important application to obtain the normalizable deformations of the LST.
As promised, we will derive the geometrical bound \eqref{normc} from the improved unitarity bound of the $SL(2;\br)/U(1)$ coset model. The dual string theory for the generalized conifold
\begin{align}
z_1^{n} + z_2^2 + z_3^2 + z_4^2 = 0 \ 
\end{align}
is given by the $(n-2)$-th $\mathcal{N}=2$ minimal model coupled with $SL(2;R)/U(1)$ coset with the level $k = \frac{2n}{n+2}$. 

The vertex operators corresponding to massless deformations of the geometry can be obtained by combining (anti-)chiral primary operators of the $\mathcal{N}=2$ minimal model and the $SL(2;\br)/U(1)$ coset model restricted to $h=\bar{h}=\frac{1}{2}$. Labeling the chiral primaries of the minimal model by $l$ ($0 \le l \le n-2)$ with the $U(1)_R$ charge $Q_R = \frac{l}{n}$, we obtain the conformal condition:
\begin{align}
\frac{l}{n} + \frac{2m}{k} = 1 \ .
\end{align}

On the other hand, the improved unitarity constraint is 
\begin{align}
1 \le 2m \le 1+k \ ,
\end{align}
which gives the constraint
\begin{align}
l = 0 , 1 , \cdots , \left[\frac{n-2}{2}\right] \ .
\end{align}
The bound is in perfect agreement with \eqref{normc}.

\subsubsection{minisuperspace analysis}\label{sec:3-2-3}
For a complementary method to read the spectrum of the sigma model is to use the point particle approximation known as the mini-superspace approximation.

Let us consider the Euclidean two-dimensional black hole background, known
as `cigar geometry':
\begin{equation}
\dd s^2 \equiv G_{ij} \dd x^i \dd x^j = 2k (\dd \rho^2 + \tanh^2\rho
\dd \theta^2) \qquad \mbox{and} \qquad e^{\Phi} =
\frac{e^{\Phi_0}}{\cosh\rho} ~. \label{Euclidean cigar}
\end{equation}
Recall that $k$ sets characteristic curvature radius in unit of the
string scale and hence string world-sheet effects, while $e^{\Phi_0}$
sets the maximum value of the string coupling at the tip $\rho=0$ of
the cigar geometry. We shall assume the limit $k \gg 1$ and
$e^{-\Phi_0} \gg 1$: this limit suppresses both string world-sheet
and space-time quantum effects and facilitates to truncate closed
string spectrum to zero-modes, viz. to mini-superspace approximation.

In the mini-superspace approach, difference between bosonic strings
(with no world-sheet supersymmetry) and fermionic strings (with
${\cal N}=2$ world-sheet supersymmetry) becomes unimportant. The
closed string Hamiltonian $L_0 + \overline{L}_0$ is reduced in the
mini-superspace approximation to the target space Laplacian
$\Delta_0$, where:
\begin{align}
\Delta_0 &= \frac{1}{e^{-2\Phi} \sqrt{G}} \partial_i
\left(e^{-2\Phi} \sqrt{G} G^{ij} \partial_j\right) \equiv
-\frac{1}{2k}[\partial_\rho^2 +2\coth2\rho\partial_\rho +
\coth^2\rho\partial_\theta^2] ~. \label{Laplacian}
\end{align}
The Hamiltonian is defined with respect to the volume element:
\begin{align}
\dd \mbox{Vol} = e^{-2\Phi}\sqrt{G} \dd \rho \dd \theta := {2k}
\sinh \rho \cosh \rho \dd \rho \, \dd \theta \equiv k \sinh 2\rho
\dd \rho \, \dd \theta~, \label{vol cigar}
\end{align}
inherited from the Haar measure on the $SL(2;\br)$ group manifold.
In the volume element, the dilaton factor $e^{-2\Phi}$ is taken into
account, as the inner product for closed string states is defined by
the world-sheet two-point correlators on the sphere. The normalized
eigenfunctions are obtained straightforwardly \cite{Dijkgraaf:1992ba,Ribault:2003ss}. They
are:
\begin{align}
 \phi_n^j(\rho,\theta) & =  -\frac{\Gamma^2(-j+\frac{|n|}{2})}{\Gamma(|n|+1)\Gamma(-2j-1)}
e^{in\theta} \times\cr
& \times \left[ \sinh^{|n|}\rho \cdot
F\left(j+1+\frac{|n|}{2},-j+\frac{|n|}{2};|n|+1;-\sinh^2\rho
\right)\right] ~ , \label{ef}
\end{align}
where $F(\alpha, \beta; \gamma; z)$ is the Gaussian
hypergeometric function. These eigenfunctions correspond to the
primary state vertex operators of conformal weights
\begin{align}
&& h= \bar{h}= -\frac{j(j+1)}{k-2} + \frac{n^2}{4k} \qquad
\mbox{or} \qquad h= \bar{h}= -\frac{j(j+1)}{k} + \frac{n^2}{4k}
\end{align}
for bosonic\footnote{The eigenvalue is actually proportional to
$-\frac{j(j+1)}{k}+ \frac{n^2}{4k}$. We will return to this small mismatch at the end of this subsection.} and fermionic strings,
respectively. We shall focus on the continuous series, parametrise
the radial quantum number $j$ as $j= -\frac{1}{2}+ i\frac{p}{2}$
$(p\in \br)$, and label the eigenfunctions as
$\phi^p_n(\rho,\theta)$ instead of $\phi^j_n(\rho,\theta) $. We adopt
the convention that, in the asymptotic region $\rho \sim \infty$,
the vertex operators with $p>0$ corresponds to the incoming waves
and those with $p<0$ corresponds to the outgoing waves. The
eigenfunctions \eqref{ef} are then normalized as
\begin{align}
\Big(\phi^p_n, \phi^{p'}_{n'} \Big) = \delta_{n,n'} \Big\lb 2 \pi
\delta(p-p')+ \cR_0(p',n) \, 2 \pi \delta(p+p') \Big\rb
~,\label{inner product}
\end{align}
where the inner product is defined with respect to the volume
element \eqref{vol cigar}. Here, $ \cR_0(p,n)$ refers to the
reflection amplitude of the mini-superspace analysis:
\begin{align}
\cR_0(p,n) =
\frac{\Gamma(+ip)\Gamma^2(\frac{1}{2}-\frac{ip}{2}+\frac{n}{2})}
{\Gamma(-ip)\Gamma^2(\frac{1}{2}+\frac{ip}{2}+\frac{n}{2})} \ .
\label{cref amp}
\end{align}
That is, from the definition \eqref{ef}, the reflection amplitude is
seen to obey the mini-superspace reflection relation:
\begin{align}
\phi^{-p}_n(\rho,\theta) = \cR_0(-p,|n|) \,
\phi^{+p}_n(\rho,\theta)~. \label{cref rel}
\end{align}
We shall refer $\cR_0(p,n)$ as `mini-superspace' reflection
amplitude, valid strictly within mini-superspace approximation at $k
\rightarrow \infty$, and anticipate string world-sheet effects at
finite $k$. Notice that no winding states wrapping around
$\theta$-direction are present since by definition the
mini-superspace approximation retains states with zero winding only.

Utilizing the analytic continuation formula of the hypergeometric
functions:
\begin{align}
F(\alpha,\beta;\gamma;z) &=
\frac{\Gamma(\gamma)\Gamma(\beta-\alpha)}
{\Gamma(\beta)\Gamma(\gamma-\alpha)}
(-z)^{-\alpha}F(\alpha,\alpha+1-\gamma;
\alpha+1-\beta;1/z) \cr
&+ \frac{\Gamma(\gamma)\Gamma(\alpha-\beta)}
{\Gamma(\alpha)\Gamma(\gamma-\beta)}(-z)^{-\beta}
F(\beta,\beta+1-\gamma;\beta+1-\alpha;1/z)
\label{eq:inv} \ ,
\end{align}
the eigenfunction \eqref{ef} can be decomposed into
\begin{align}
\phi^p_n(\rho,\theta) = \phi^p_{L,n}(\rho,\theta) + \cR_0(p,|n|)
\phi^p_{R,n}(\rho,\theta) ~, \label{decomp ef} \end{align}
where
\begin{align}
\phi^p_{L,n}(\rho,\theta) &\equiv e^{in\theta} (\sinh
\rho)^{-1-ip}\,
F\Big(\frac{1}{2}+\frac{ip+n}{2},\frac{1}{2}+\frac{ip-n}{2}; 1+ip;
-\frac{1}{\sinh^2\rho} \Big) ~,\nn & \sim 
e^{-\rho}e^{-ip\rho+in\theta} \qquad \mbox{at} \qquad \rho \,
\rightarrow\, +\infty~ \label{phiL} \end{align}
and
\begin{align}
\phi^p_{R,n}(\rho,\theta) &\equiv e^{in\theta} (\sinh
\rho)^{-1+ip}\,
F\Big(\frac{1}{2}-\frac{ip+n}{2},\frac{1}{2}-\frac{ip-n}{2}; 1-ip;
-\frac{1}{\sinh^2\rho} \Big) \nn & \sim 
e^{-\rho}e^{ip\rho+in\theta} \qquad \mbox{at} \qquad
   \rho \, \rightarrow\, +\infty
\label{phiR}
\end{align}
refer to the left- and the right-movers, respectively, at $\rho
\rightarrow +\infty$, and $\cR_0(p, |n|)$ is defined in \eqref{cref
amp}. Obviously, they are related to each other under the reflection of
radial momentum: $\phi^{+p}_{R, n} = \phi^{-p}_{L, n}$, which is
also evident from \eqref{decomp ef} and \eqref{cref amp}. These
mini-superspace wave functions \eqref{decomp ef} constitute the
starting point of constructing boundary states of D-brane in the
Euclidean two-dimensional black hole background.

We close the mini-superspace analysis with remarks concerning Wick
rotation of the results to the Lorentzian background and string
world-sheet effects present at finite $k$.
\begin{enumerate}
 \item The decomposition of $\phi^p_n$ into $\phi^p_{L,n}$ and $\phi^p_{R,n}$ cannot
 globally defined over the entire cigar geometry. They
 are ill-defined around the tip $\rho =0$, and the reflection relation
\eqref{cref rel} implies that $\phi^{-p}_n$ is not independent of
$\phi^{+p}_n$. Therefore, of the continuous series, only the
eigenfunctions $\phi^p_n$ with $p>0,~ n\in \bz$ span the physical
Hilbert space of the closed strings on the Euclidean two-dimensional
black hole. On the other hand, the situation will become further
complicated once Wick rotated to the Lorentzian two-dimensional
black hole.
 \item Notice that $\phi^p_n$ is not analytic
with respect to the angular quantum number $n$ as it depends on its
absolute value, $|n|$. This leads to the ambiguity for Wick rotation
from Euclidean to Lorentzian background, under which roughly
speaking $i n$ is replaced by energy $\omega$. As for the
mini-superspace reflection amplitude $\cR_0(p, n)$, since
$\cR_0(p,-n)= \cR_0(p,n)$ holds for all $n \in \bz$, it is
unnecessary to take absolute value $|n|$ in \eqref{cref rel},
\eqref{decomp ef}. When taking Wick rotation, we will start from the
expression $\cR_0(p,|n|)$. In other words, we analytically continue
$\cR_0(p,n)$ if $n > 0$ and $\cR_0(p,-n)$ if $n<0$.
 \item It is evident that $|\cR_0(p,n)|=1$, viz, the mini-superspace
 reflection
 amplitude is purely a phase shift in the Euclidean black hole
 background. It is of
 utmost importance that, in the Lorentzian black hole background,
 $n$ is analytically continued to pure imaginary value,
 and the modulus of the reflection amplitude becomes less than unity.
 \item For the fermionic Euclidean $SL(2; \br)/U(1)$ conformal field
 theory, exact result for the reflection amplitude (i.e. taking account
 of all string world-sheet effects) is known
 \cite{Teschner:1997ft,Teschner:1999ug,Giveon:1999px,Giveon:1999tq}. In our notations, it is
\begin{align}
\cR(j,m,\bar{m}) = \nu(k)^{-2j-1}\,
\frac{\Gamma(1+\frac{2j+1}{k})}{\Gamma(1-\frac{2j+1}{k})}
\frac{\Gamma(2j+1)\Gamma(-j+m)\Gamma(-j-\bar{m})}{\Gamma(-2j-1)
\Gamma(j+1+m)\Gamma(j+1-\bar{m})}, \label{qref amp}\end{align}
where
\begin{align} \nu(k)\equiv \frac{1}{\pi}\frac{\Gamma(1-\frac{1}{k})}
{\Gamma(1+\frac{1}{k})}~, \qquad m=\frac{kw+n}{2}~, \qquad \bar{m} =
\frac{kw-n}{2}~.
 \nn
\end{align}
Denoting by $\Phi_{j;m,\bar{m}}$ the vertex operator with conformal weights
$h= \frac{m^2-j(j+1)}{k}$, $\bar{h}= \frac{\bar{m}^2-j(j+1)}{k}$,
the exact reflection relation reads
\begin{align}
\Phi_{-(j+1);m,\bar{m}} = \cR(-(j+1),m,\bar{m}) \Phi_{j;m,\bar{m}}~, \label{qref rel}
\end{align}
The mini-superspace reflection amplitude $\cR_0(p,n)$ is
then related to the exact one $\cR(j,m,\bar{m})$
by taking the $k\,\rightarrow\,\infty$ limit as mentioned above
(up to overall constant):
\begin{align}
\cR_0(p,n) = \lim_{k\rightarrow + \infty}\,
\cR(j=-\frac{1}{2}+\frac{ip}{2}, m=\frac{n}{2}, \bar{m}=-\frac{n}{2})~.
\end{align}
\end{enumerate}

Although the mini-superspace approximation can only describe the momentum mode of the full spectrum, it is possible to study the winding mode by using the T-duality even if we restrict ourselves to the mini-superspace approximation. In the remaining part of this subsection, we will study the mini-superspace spectrum for the T-dualized trumpet geometry (i.e. $\bz_k$ orbifold of the vector coset $SL(2;\br)^{(V)}/U(1)$).
T-dualized classical geometry is given by the trumpet geometry
\begin{align}
\dd s^2 = 2\left(k \dd\rho^2 + \frac{\dd\tilde{\theta}^2}{k\tanh^2\rho}\right) \ , \ \ e^{\Phi} = \frac{\mu}{\sinh\rho} \ . 
\end{align}
Note that we have a curvature and dilaton singularity at $\rho = 0$.
For later purposes, let us discuss the minisuperspace analysis for the bulk spectrum. The minisuperspace spectrum is determined by the eigenfunctions of the string Laplacian
\begin{align}
\Delta &= - \frac{1}{e^{-2\Phi}\sqrt{\det G}} \partial_i e^{-2\Phi} \sqrt{\det{G}} G^{ij} \partial_j \cr
 &= -\frac{2}{k}\left[\partial_\rho^2 + (\coth\rho+ \tanh\rho) \partial_\rho + k^2 \tanh^2\rho\partial_{\tilde{\theta}}^2 \right]
\end{align}
The (delta-function normalizable) eigenfunctions are given by
\begin{align}
\phi_{p,w}(\rho,\tilde{\theta}) &= C_1 e^{iw\theta}(\cosh\rho)^{-1-ip} F\left(\frac{1}{2}-\frac{kw}{2}+\frac{ip}{2},\frac{1}{2}+\frac{kw}{2}+ \frac{ip}{2};1+ip;\frac{1}{\cosh^2\rho}\right) \cr
 &+ C_2 e^{iw\tilde{\theta}}(\cosh\rho)^{-1+ip} F\left(\frac{1}{2}-\frac{kw}{2}-\frac{ip}{2},\frac{1}{2}+\frac{kw}{2}- \frac{ip}{2};1-ip;\frac{1}{\cosh^2\rho}\right) \ .
\end{align}
It is not apriori clear which boundary condition one should impose because the trumpet geometry has a singularity at $\rho = 0$. Our natural guess would be to impose $\phi_{p,w}(\rho = 0,\tilde{\theta}) \equiv 0$. With our convenient normalization $C_1= 1$, this boundary condition amounts to 
\begin{align}
C_2 = \mathcal{R}_0(p,\omega) = \frac{\Gamma(ip)\Gamma(\frac{1}{2}-\frac{ip}{2}+\frac{kw}{2})\Gamma(\frac{1}{2}-\frac{ip}{2}-\frac{kw}{2})}{\Gamma(-ip)\Gamma(\frac{1}{2}+\frac{ip}{2}+\frac{kw}{2})\Gamma(\frac{1}{2}+\frac{ip}{2}-\frac{kw}{2})} \ , \label{ref dual}
\end{align}
which is consistent with the semiclassical limit of the exact reflection amplitude that is descended from the $SL(2;\br)$  WZNW model (or $\mathbb{H}_3^+$ model). Here we have used the formulae in the appendix A to evaluate the behavior of the hypergeometric function near the singularity.

Before we close our study of the mini-superspace Euclidean two-dimensional black hole system, we introduce the so-called ``exact string background" for the bosonic two-dimensional black hole. As we have seen, for the bosonic string case, the spectrum shows $1/k$ corrections as
\begin{align}
h = \bar{h} = -\frac{j(j+1)}{k-2} + \frac{n^2}{4k} \ , \label{extst}
\end{align}
compared with the mini-superspace results
\begin{align}
h_0 = \bar{h}_0 =  -\frac{j(j+1)}{k} + \frac{n^2}{4k} \ .
\end{align}
To cure this small mismatch, \cite{Dijkgraaf:1992ba} introduced the following improved Laplacian
\begin{align}
\Delta'_0 = -\frac{1}{k-2}\left(\frac{\partial^2}{4\partial \rho^2} + \coth 2\rho \frac{\partial}{2\partial \rho} + (\coth^2\rho-\frac{2}{k})\frac{\partial^2}{\partial \theta^2}\right) \ , \label{imlap}
\end{align}
to reproduce the exact $1/k$ corrected spectrum \eqref{extst}. The corresponding metric is
\begin{align}
ds^2 = 2(k-2)\left(\dd\rho^2 + \frac{\dd\theta^2}{(\coth^2\rho -\frac{k}{2})} \right)
\end{align}
with the dilaton
\begin{align}
e^{2\Phi} = \mu \sinh2\rho\sqrt{\coth^2\rho-\frac{2}{k}} \ .
\end{align}

In the literature, it has been shown that this background is a solution of the bosonic string equation of motion in a particular renormalization scheme \cite{Tseytlin:1992ri}. However, from the modern viewpoint, the reflection amplitude obtained from \eqref{imlap} is the same as the one from \eqref{Laplacian}, and does not capture the nonperturbative $1/k$ corrections that appear in the exact result \eqref{qref amp}. We will study the origin of the non-perturbative corrections to the mini-superspace reflection amplitude in section \ref{sec:3-4}.

\subsection{Lorentzian spectrum}\label{sec:3-3}
\subsubsection{classical string in two-dimensional black hole}\label{sec:3-3-1}
We would like to study the classical string solution in the two-dimensional black hole geometry. For this purposes, we can directly solve the string equation of motion (and Virasoro constraint) on the classical background, or we can study the gauged WZNW model before integrating out the gauge constraint \cite{Bars:1994sv}.

If one takes the axial gauge $A=0$ in the classical gauged WZNW action \eqref{gWZW}, the solution can be constructed as follows.\footnote{Since we are studying the Lorentzian target space, we use the Lorentzian signature world-sheet.} The classical solution of the parent WZNW model is written as 
\begin{align}
g(\sigma_+,\sigma_-) =g_L(\sigma_+)g_R(\sigma_-)^{-1} \ , 
\end{align}
where we parametrize $g_L(\sigma_+)$, $g_R(\sigma_-)^{-1}$ $\in SL(2;\br)$ as
\begin{align}
g_L = \begin{pmatrix} a_L & u_L \\ -v_L  & b_L \end{pmatrix} \ ,  \ \ g_R^{-1} = \begin{pmatrix} b_R & -u_R \\ v_R  & a_R \end{pmatrix} \ , \\
g =\begin{pmatrix} u_Lv_R+a_Lb_R & -a_Lu_R+u_La_R \\ b_Lv_R-v_Lb_R  & v_Lu_R+b_La_R \end{pmatrix}  \ ,
\end{align}
with the determinant constraint $u_Lv_L+a_Lb_L = u_Rv_R+a_Rb_R = 1$.

Now the current constraint $J_2 = \bar{J}_2=0$ reduces to
\begin{align}
\mathrm{Tr}(\sigma_3\partial_+ g_Lg_L^{-1})= v_L \partial_+ u_L + b_L \partial_+ a_L = -u_L\partial_+ v_L - a_L\partial_+ b_L = 0 \ , \label{curcon}
\end{align}
and the Virasoro constraint becomes\footnote{It is interesting to study general solutions of the coset CFT without imposing the Virasoro constraint. We will later come back to this point.}
\begin{align}
(a_L\partial_+ u_L - u_L \partial a_L)(b_L\partial_+v_L - v_L \partial_+ b_L) = 0 \ . \label{vircon}
\end{align}
The right moving part satisfies the similar equations.

Due to the determinant constraint, two equations in \eqref{curcon} are not independent, so we expect an appearance of one arbitrary function in the full solutions. Indeed, \eqref{curcon} and \eqref{vircon} suggests that either $\partial_+ u_L = \partial_+ a_L =0$, or $\partial_+v_L = \partial_+ b_L =0$ should be satisfied. Then the general solutions can be expressed as 
\begin{align}
g_L = \begin{pmatrix} a_L & u_L \\ -v_L(\sigma^+) &\frac{1}{a_L} - \frac{u_L}{a_L} v_L(\sigma^+)    \end{pmatrix} \ \text{or} \ \ \begin{pmatrix}  \frac{1}{b_L} - \frac{v_L}{b_L} u_L(\sigma^+) &u_L(\sigma^+) \\ -v_L &b_L   \end{pmatrix}  \cr
g_R^{-1} =  \begin{pmatrix} \frac{1}{a_R} -\frac{u_R}{a_R} v_R(\sigma^-)  & -u_R \\ v_R(\sigma^-) & a_R\end{pmatrix} \ \text{or} \ \ \begin{pmatrix} b_R & -u_L(\sigma^-) \\ v_R  &\frac{1}{b_R} - \frac{v_R}{b_R} u_R(\sigma^-)  \end{pmatrix} \ .
\end{align}
Combining the left moving part and the right moving part, we totally obtain four possible solutions:
\begin{align}
g_A(\sigma_+,\sigma_-) &= \begin{pmatrix} \frac{1}{b}(1-{u}(\sigma^+){v}(\sigma^-)) & {u}(\sigma^+) \\ -{v}(\sigma^-) & b \end{pmatrix}  \cr
g_B(\sigma_+,\sigma_-) &= \begin{pmatrix} a & \bar{u}(\sigma^-) \\ -\bar{v}(\sigma^+) & \frac{1}{a}(1-\bar{u}(\sigma^-)\bar{v}(\sigma^+)) \end{pmatrix} \cr
g_C(\sigma_+,\sigma_-) &= \begin{pmatrix} {a}(\sigma^-) & c_1\\ \frac{1}{c_1}(-1+{a}(\sigma^{-}){b}(\sigma^+)) &{b}(\sigma^+) \end{pmatrix} \cr
g_D(\sigma_+,\sigma_-) &= \begin{pmatrix} \bar{a}(\sigma^+) & \frac{1}{c_2}(1-\bar{a}(\sigma^+)\bar{b}(\sigma^-)) \\ -c_2 & \bar{b}(\sigma^-) \end{pmatrix} \ ,
\end{align}
where $u(\sigma^+), v(\sigma^-), \bar{u}(\sigma^{-}), \bar{v}(\sigma^{+}), a(\sigma^{-}), b(\sigma^+) , \bar{a}(\sigma^{+}), \bar{b}(\sigma^{-})$ are arbitrary real functions, and $(b,a,c_1,c_2)$ are real integration constants. We can read off the gauge invariant motion of strings from $u$ and $v$ components: 
\begin{align}
&A: &  u&=u(\sigma^+) \ , & v&= v(\sigma^-) \cr
&B: &  u&=\bar{u}(\sigma^-) \ ,&  v&= \bar{v}(\sigma^+) \cr
&C: &  u&=c_1 \ , & v&= \frac{1}{c_1}(1-a(\sigma^-)b(\sigma^+)) \cr
&D: &  u&= \frac{1}{c_2}(1-\bar{a}(\sigma^+)\bar{b}(\sigma^-)) \ , & v&= c_2 \  . \label{sols}
\end{align}
It is interesting to note that the solution $A$ and $B$ are actually solutions of the string equation of motion  of any two-dimensional target space metric written in the conformal (Kruscal) coordinate: $\dd s^2 = f(u,v) \dd u\dd v$.

We still have a gauge degree of freedom associated with the conformal transformation: $\sigma^+ \to f(\sigma^+)$ and $\sigma^- \to \bar{f}(\sigma^-)$. By using this gauge symmetry, we can {\it locally} gauge away the arbitrary functions in  \eqref{sols} to make them reduce to the point particle (collapsed) string solution. For example, in the solution $A$ (similarly for $B$), we can expand
\begin{align}
u = u_0 + p_u(\tau + \sigma) + \sum_{n\neq 0} \alpha_n e^{-in(\tau + \sigma)} \ .
\end{align}
As is the case with the flat space, we can gauge away the oscillatory part, and due to the periodicity of $\sigma$ for closed strings, we have to set $p_u = 0$ unless the target space has a periodic directions. In the solution $C$ (similarly for $D$), we can locally set $ v = v(\tau)$ independent of $\sigma$ by using the conformal transformation. The resultant string motion is nothing but the geodesics for the massless point particle in the two-dimensional black hole. 

Thus the classical solution of the string equation collapses to a massless point particle in the two-dimensional black hole background, and it seems to be consistent with the vertex operator analysis, where the only tachyon fields are dynamical classically. However, if one allows a folded string solution, we can construct more general string solutions as we will review in the following section \ref{sec:3-3-2}. This can be done by patching different solutions of \eqref{sols} within the same world-sheet.

\subsubsection{folded strings}\label{sec:3-3-2}
For an illustrative purpose, let us begin with the folded string solution  \cite{Bardeen:1975gx,Maldacena:2005hi} in two-dimensional flat space $(T,X)$. We can fix the world-sheet conformal invariance by choosing the gauge $\tau = T$.\footnote{This static gauge choice is not always possible especially in the curved space-time background.} The classical equation of motion is given by $\partial_+\partial_- X = 0$ with the Virasoro constraint\footnote{If one embeds the two-dimensional Minkowski space in higher dimensional space-time, the Virasoro constraint can be relaxed. For instance, if one introduces a contribution from the other zero modes than $(T,X)$ the resultant two-dimensional motion becomes massive rather than massless.}
\begin{align}
-2 + (\partial_+ X)^2 =  0  \ ,  \ \ -2 + (\partial_- X)^2 = 0 \ ,
\end{align}
which amounts to $\partial_+ X = \pm \sqrt{2}$. If we restrict ourselves to the solutions with continuous first derivative, they reduce to massless particles. To obtain the folded string solution, we can set $X= X_+(\sigma^+) + X_-(\sigma^-)$, where $X_+$ and $X_-$ are periodic functions with the same period and with derivatives $\pm \sqrt{2}$. In other words, we partition the world-sheet and assign different solutions that satisfy $ \partial_+ X = \pm \sqrt{2}$ and patch-work together so that the full solution is periodic in $\sigma$ direction. A simple solution with the static center of motion is given by 
\begin{align}
X = \sqrt{2}(|\sigma^+|_{\mathrm{per}} + |\sigma^-|_{\mathrm{per}}) \ , \label{minfold}
\end{align}
where the periodic absolute value function $|\sigma^+|_{\mathrm{per}}$ is given by $|\sigma^+|_{\mathrm{per}} = |\sigma^+|$ for $-\pi<\sigma^+ \le \pi$ and periodically extended outside this interval. More complicated solutions are possible, and one simple way to obtain them is to perform the target space Lorentz boost to the solution \eqref{minfold}. The resultant string solutions describes the folded string with the motion of the center of mass. Note that the Lorentz boost breaks our original gauge choice. 

Next. we will consider the linear dilaton background $\Phi = QX$. The equation of motion does not change, but the Virasoro constraint is modified as
\begin{align}
- 2 + (\partial_+ X)^2 - Q\partial_+^2 X = 0 \ ,  \ \ - 2 + (\partial_- X)^2 - Q\partial_-^2 X = 0 \ . 
\end{align}
The point is that due to the existence of the second derivatives in the Virasoro constraint, the solutions such as \eqref{minfold} are no more allowed. The most general solutions are given by
\begin{align}
X = X_0 -Q \log\left(\cosh\frac{\sqrt{2}\tau}{Q} + \cosh\frac{\sqrt{2}\sigma}{Q}\right) \ .
\end{align}
Except for the special limit where the string collapse to a point particle, the solution is not periodic in $\sigma$ direction. Thus only the long string stretched to the infinity is the allowed folded string solution. This agrees with the fact that there are no closed string states in the linear dilaton theory other than the massless tachyon. The inclusion of the Liouville potential does not alter the qualitative feature of the classical string solutions.

\begin{figure}[htbp]
   \begin{center}
    \includegraphics[width=0.5\linewidth,keepaspectratio,clip]{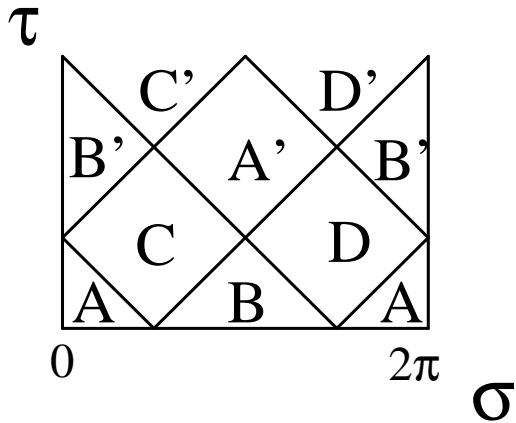}
    \end{center}
    \caption{In order to obtain folded string solution, we partition the world-sheet and assign different solutions on each patch.}
    \label{fig:partition}
\end{figure}

Let us now move on to the folded string solutions in the two-dimensional black hole. First, we partition the string world-sheet as in figure \ref{fig:partition}. We know that the solution should be locally given by \eqref{sols}. We fix the world-sheet conformal invariance by giving the boundary condition:
\begin{align}
&A: &  u&=u_0 +p^+\sigma^+_{\mathrm{per}}\ , & v&=v_0+p^-\sigma^-_{\mathrm{per}} \ ,\cr
&B: &  u&=u_0 +p^+\sigma^-_{\mathrm{per}}\ , & v&=v_0+p^-\sigma^+_{\mathrm{per}}  \ , \label{first}
\end{align}
where $\sigma_{\mathrm{per}}^+ = \sigma^+ - m\pi$ for $\frac{\pi}{2}\le \sigma^+ <\frac{\pi}{2}$ and periodically identified outside this range.
If we had considered a flat Minkowski space $\dd s^2 = \dd u\dd v$, the solution in the region $C$ and  $D$ would be independent of $\sigma$:
\begin{align}
&C_{\mathrm{flat}}: &  u&=u_0 + \frac{\pi p^{+}}{2}\ , & v&=v_0 + \frac{\pi p^-}{2}+ p^-(\sigma_{\mathrm{per}}^++\sigma_{\mathrm{per}}^-) \ \cr
&D_{\mathrm{flat}}: &  u&=u_0 +\frac{\pi p^+}{2} +p^+(\sigma^-_{\mathrm{per}}+\sigma^-_{\mathrm{per}})\ , & v&=v_0+\frac{\pi p^-}{2}  \ .
\end{align}
 so that we have a fold as in \eqref{minfold}. In the two-dimensional black-hole case, the solution in $C$ and $D$ region is given by 
 \begin{align}
 C_{\mathrm{2DBH}}:  u&=u_0 + \frac{\pi p^{+}}{2}\ , \cr  v
 &=\frac{1}{u_0+\frac{\pi p^+}{2}} \left(1-\frac{[1-(u_0+\frac{\pi p^+}{2})(v_0+p^-\sigma_{\mathrm{per}}^+)]\times{[1-(u_0+\frac{\pi p^+}{2})(v_0+p^-\sigma_{\mathrm{per}}^-)]}}{[1-(u_0+\frac{\pi p^+}{2})(v_0-\frac{\pi p^-}{2})]} \right) \ , \cr  
D_{\mathrm{2DBH}}: u& =\frac{1}{v_0+\frac{\pi p^-}{2}} \left(1-\frac{[1-(v_0+\frac{\pi p^-}{2})(u_0+p^+\sigma_{\mathrm{per}}^+)]\times{[1-(v_0+\frac{\pi p^-}{2}(u_0+p^+\sigma_{\mathrm{per}}^-)]}}{[1-(v_0+\frac{\pi p^-}{2})(u_0-\frac{\pi p^+}{2})]} \right) \cr v&=v_0+\frac{\pi p^-}{2}  \ .
\end{align}
We could continue this analysis, but important point is that in the region $A'$ and $B'$ the solution turns out to be linear in $\sigma^+$ and $\sigma^-$ again just as in \eqref{first}. Thus the structure repeats itself and the simple recursion formula to derive the full solution exists (see \cite{Bars:1994sv}). Physically, the pulsating string falls into the black-hole as a massive particle. The folds move with the speed of light because $u$ or $v$ is constant.

The folded string solution in the two-dimensional Liouville background was identified as the {\it open string} attached to the FZZT brane \cite{Fateev:2000ik,Teschner:2000md} in a certain limit \cite{Maldacena:2005hi}. The scattering amplitude computed from the CFT analysis completely matches with the matrix model computation in the non-singlet sector \cite{Fidkowski:2005ck,Kostov:2006dy}. In this sense, it makes sense to regard the non-singlet sector (or winding sector) in the matrix model corresponds to the folded string solution in the Lorentzian target space theory in the Liouville background.

Therefore, one might expect that the folded string solution in the two-dimensional black hole background should play an important role in constructing the dual matrix model for the two-dimensional black hole with the Lorentzian target space signature. At present, we do not have a conclusive argument for or against this direction, but we present some remarks here:
\begin{itemize}
	\item It is important to note that the folded string solution in the Liouville background is in the open string sector and not in the closed string sector. This should be contrasted with the winding sector in the Euclidean two-dimensional black hole (and its hypothetical analytic continuation to the Lorentzian signature black hole).
	\item To obtain the long stretched string solution, the existence of the linear dilaton in the Virasoro constraint has been crucial. In our semiclassical analysis for the two-dimensional black hole, the existence of the nontrivial dilaton was neglected. This is the reason why we obtained a folded string solution that is asymptotically identical to the flat space solution \eqref{minfold}. Since in the linear dilaton case, such a solution was excluded, we might expect that only the long string solution would survive in the full quantization.\footnote{Just adding the dilaton term in the classical energy-momentum tensor is not consistent in the classical treatment because the one-loop correction to the equation of motion together with the classical contribution from the dilaton guarantee the one-loop holomorphic structure of the energy-momentum tensor.}
	\item In order to have a support for such long string solutions, we need a D-brane localized at asymptotic infinity. However, the two-dimensional Lorentzian  black hole does {\it not} admit such a D-brane solution from the classical DBI action analysis (see section \ref{sec:7-1}).
\end{itemize}

We leave the role of the folded strings in the full quantization of the Lorentzian two-dimensional black hole system for future studies. In most of the following sections, we will concentrate on the mini-superspace approximation (point-particle approximation).

Before closing our discussion on the classical string solutions in the two-dimensional black hole, we would like to present the most general solutions of the  classical sigma model without imposing the Virasoro constraint for completeness \cite{deVega:1993pm}. First we introduce arbitrary three-component vectors lying on the hyperboloids
\begin{align}
\vec{A}\cdot\vec{A} &= - (A_0)^2 + (A_1)^2 + (A_2)^2 = 1 \cr
\vec{B}\cdot\vec{B} &= - (B_0)^2 + (B_1)^2 + (B_2)^2 = 1 \ .
\end{align}
Then the most general solutions of the sigma model (in the Schwarzshild-like coordinate) are given by
\begin{align}
\cosh^2 \rho(\sigma,\tau) &= \frac{1}{2}\left(1+\vec{A}(\sigma^+)\cdot \vec{B}(\sigma^-)\right) \cr
t(\sigma,\tau) &= \frac{1}{2}\int^{\sigma^+}_a\left(\frac{\epsilon_{abc}A'_aA_bB_c}{\vec{A}\cdot\vec{B}-1}\right)(x_+,b)dx_+ \cr
&-\frac{1}{2}\int^{\sigma^-}_b\left(\frac{\epsilon_{abc}B'_aA_bB_c}{\vec{A}\cdot\vec{B}-1}\right)(\sigma_+,x_-)dx_- + c
\end{align}
The energy momentum tensor takes the form
\begin{align}
T_{++} = -\frac{A'(\sigma^+)^2}{8} \ ,  \ \ T_{--} = -\frac{B'(\sigma^-)^2}{8} \ .
\end{align}
If one imposes the Virasoro constraint, the solutions reduce to \eqref{sols}.

\subsubsection{spectrum from partition function?}\label{sec:3-3-3}
The partition function for the Lorentzian two-dimensional black hole should be able to determine its spectrum in principle. However, in practice the subtleties associated with the  non-compactness of the target space and the Lorentzian signature make the partition function ill-defined and prevent us from reading the spectrum. 

At the classical level, the Lorentzian two-dimensional black hole is obtained by the Wick rotation $\theta \to i t$ in the cigar geometry. The vertex operator (corresponding to the massive character) should be Wick rotated as well
\begin{align}
\Delta(j,m=-\bar{m}=\frac{n}{2}) &= -\frac{j(j+1)}{k-2} + \frac{n^2}{4k} \cr
\to \Delta(j,\omega) &= -\frac{j(j+1)}{k-2} - \frac{\omega^2}{4k}
\end{align}
based on the naive Wick rotation $n \to i \omega$. Since the time-direction $t$ is non-compact, $\omega$ should be continuous unlike the Euclidean two-dimensional black hole case, where $n$ is quantized. For the same reason, the Lorentzian black hole does not have a winding states along the $t$ direction.\footnote{Since the discrete states that descend from the $SL(2;\br)$ primaries have a winding quantum number in the Euclidean black hole, the corresponding states do not seem to exist in the Lorentzian black hole, but this is a controversial issue.} From the mini-superspace analysis we will review in \ref{sec:3-3-4}, the classical spectrum is obtained from this naive analytic continuation (with some care about the boundary conditions). 

We would like to make an attempt to read the spectrum from the Lorentzian partition function with some hindsight from the mini-superspace analysis.
The proposed partition function (with a suitable analytic continuation) for the Lorentzian $SL(2;\br)/U(1)$ coset is 
\begin{align}
Z_{SL(2;\br)/U(1)} =  \int_{\br^2} \frac{\dd v^2}{\tau_2}\frac{e^{\frac{v_1^2}{\tau_2}-\frac{\pi k}{\tau_2}|v|^2}  }{\sqrt{\tau_2}|\theta_1(\tau,iv)|^2 } \sqrt{\tau_2}|\eta(\tau)|^2  \ . \label{ppl}
\end{align}
Here we study the bosonic case first for simplicity. Since we are integrating over the whole complex plane spanned by $v$, this partition function is formally same as that for the Euclidean axial coset \eqref{ppo}. The conclusion that the spectrum of the Euclidean coset and the Lorentzian coset, nevertheless, has the same spectrum seems too quick given the fact that the mini-superspace approximation gives a totally different answer. 

In the Euclidean case, the divergence near $iv = u = n+ \omega \tau$ gives a bulk contribution of the noncompact boson (with a linear dilaton) coupled to a compact boson with radius $R^2 = k$, where the summation over $n$ and $\omega$ shows the existence of the compact direction. In the Lorentzian case, we propose that the the torus modulus $\tau$ should be Lorentzian, namely $\tau$ and $\bar{\tau}$ and real and mutually independent.

On the Lorentzian torus, the divergence of the partition function \eqref{ppl} only appears at $v=0$, leading to the contribution of the noncompact boson (with a linear dilaton) coupled to a non-compact boson:\footnote{On the Lorentzian torus, the Euclidean coset partition function still has an infinitely many origin of divergence at $ u = n+\omega \tau$, which gives a compact boson.}
\begin{align}
Z \sim \log \epsilon Z_{free}^2 + \text{finite terms}\ .
\end{align}
The leading order partition function seems to agree with the mini-superspace analysis.\footnote{One should note that the partition function on the Lorentzian torus needs a careful $i\epsilon$ prescription. This is an interesting but subtle subject, and we will not delve into the details here.}

More precise discussions are needed to determine the finite part of the partition function. The situation is more involved than in the Euclidean case, and so far no conclusive agreements are available.
One interesting related question is how the target-space supersymmetry is broken in the (world-sheet supersymmetric) Lorentzian two-dimensional black hole background. If we restrict ourselves to the leading order partition function, we can certainly define a GSO projection, and the partition function (at the leading order) vanishes because asymptotically the theory is essentially free and the spectrum coincides with the supersymmetric linear dilaton theory, and the target-space supersymmetry operator can be constructed. Therefore, the target-space supersymmetry should be broken near the horizon, where the curvature effects will be present.

To study this problem, let us consider the type II GSO projected partition function for the two-dimensional Lorentzian black hole (coupled to free transverse sectors that is represented by free CFTs for simplicity). Our original partition function for the Lorentzian $SL(2;\br)/U(1)$ Kazama-Suzuki coset is the diagonal modular invariant one:
\begin{align}
Z^{(NS)}(\tau) = \int_{\br^2} \frac{\dd v^2}{\tau_2} \frac{|\theta_3(\tau,iv)|^2}{\sqrt{\tau_2}|\theta_1(\tau,iv)|^2 } \sqrt{\tau_2}|\eta(\tau)|^2 e^{-\frac{\pi k}{\tau_2}|v|^2} \ 
\end{align}
for NS-NS sectors (other sectors can be obtain by replacing $\theta_3(\tau,u)$ with $\theta_a(\tau,u)$ $(a=0,1,2)$ from the spectral flow). A natural candidate for the type II partition function is 
\begin{align}
Z(\tau) 
= \int_{\br^2} \frac{\dd v^2}{\tau_2} \frac{|\theta_3(\tau,iv)\theta_3^3 - \theta_2(\tau,iv)\theta_2^3 \pm \theta_1(\tau,iv)\theta_1^3 - \theta_0(\tau,iv)\theta_0^3)|^2}{\sqrt{\tau_2}|\theta_1(\tau,iv)|^2 } \sqrt{\tau_2}|\eta(\tau)|^{-14} e^{-\frac{\pi k}{\tau_2}|v|^2} \ , \label{typii}
\end{align}
where $\theta_a^3 = \theta_a(\tau,0)^3$ have been introduced from the free CFT contributions. The expression is almost supersymmetric: for example, if we first take the leading diverging part at $v=0$, then the fermionic oscillator part gives zero after summing over the spin structure due to the abstruse identity of Jacobi.

A remaining uncancelled part, which describes the breaking of the target-space supersymmetry, is of particular interest.
By using the Riemann quartic identity \eqref{rqi}, we can rewrite \eqref{typii} as
\begin{align}
Z(\tau) = \int_{\br^2} \frac{\dd v^2}{\tau_2} \frac{|\theta_1(\tau,i\frac{v}{2})|^8}{\sqrt{\tau_2}|\theta_1(\tau,iv)|^2 } \sqrt{\tau_2}|\eta(\tau)|^{-14} e^{-\frac{\pi k}{\tau_2}|v|^2} \ . \label{typeiit}
\end{align}
Now one can see that the leading order divergence near the origin $(v=0)$ is indeed removed, which suggests that the bulk part of the spectrum is supersymmetric.\footnote{Unfortunately, for complex $\tau$, the partition function still shows a volume divergence at $iv = n + \omega \tau$ with a pair of {\it odd} integers $(n,\omega)$, where the GSO projection acts {\it oppositely}. We do not fully understand the origin of this failure of bulk cancellation. Since these divergences seem unphysical in the Lorentzian partition function if we stick to the Lorentzian torus, we do not see their physical relevance.}

It is still difficult to evaluate \eqref{typeiit} to uncover the non-supersymmetric spectrum of the two-dimensional Lorentzian black hole partially because the formal $q$ expansion of \eqref{typeiit} gives a divergent series. Let us suppose that the major part of the $v$ integral comes near the origin $v=0$ since in the large $k$ limit, the Gaussian factor would provide a strong convergence factor for the integral (with no good justification because of the oscillatory nature of the $\theta$ functions on the Lorentzian torus). The subsequent integration over $v$ would lead to
\begin{align}
Z(\tau) \sim |\eta(\tau)|^4 \ .
\end{align}
The partition function looks like a free 0-dimensional bosonic string (probably localized near the horizon). The support of the (non-supersymmetric) bosonic degrees of freedom should be localized because we do not have a diverging volume factor.

It seems likely that the supersymmetry is broken only locally in the vicinity of the horizon. One naively guess that this should correspond to the Lorentzian version of the winding condensation near the tip of the cigar in the Euclidean two-dimensional black hole as we will see in section \ref{sec:3-4}. A further study of this subject is of great interest and the more precise definition of the (almost) supersymmetric partition function \label{typeiit} and its evaluation is highly desirable.

\subsubsection{minisuperspace approximation}\label{sec:3-3-4}
Since it is difficult to read the spectrum of the Lorentzian two-dimensional black hole directly from the partition function, it is very important to study the classical spectrum based on the mini-superspace approximation.
The Wick rotation of the mini-superspace eigenfunctions in the
Euclidean cigar geometry \eqref{ef} is not so trivial. Fortuitously,
the Lorentzian eigenfunctions are already classified thoroughly in
\cite{Dijkgraaf:1992ba}. The complete basis for waves outside the black hole
horizon are spanned by the following four types of
eigenfunctions\footnote{Here we adopt slightly different
normalization  from \cite{Dijkgraaf:1992ba}.} of the Lorentzian
Klein-Gordon operator. For those with the eigenvalue
$\frac{p^2}{4k}-\frac{\om^2}{4k}+ \frac{1}{4k}$ of the Klein-Gordon
operator, the four eigenfunctions are
\begin{align}
U^p_{\om}(\rho,t) &= -
\frac{\Gamma^2(\nu_+)}{\Gamma(1-i\om)\Gamma(-ip)} e^{-i\om t} (\sinh
\rho)^{-i\om} F(\nu_+,\nu^*_-;1-i\om;-\sinh^2\rho) \nn &\sim
e^{-i\om t - i\om \ln \rho} \qquad \mbox{as} \qquad
\rho\,\rightarrow\,0~,
\label{U} \\
V^p_{\om}(\rho,t) &=  -
\frac{\Gamma^2(\nu^*_+)}{\Gamma(1+i\om)\Gamma(ip)} e^{-i\om t}
(\sinh \rho)^{i\om} F(\nu^*_+,\nu_-;1+i\om;-\sinh^2\rho) \nn &\sim
e^{-i\om t + i\om \ln \rho} \qquad \mbox{as} \qquad
\rho\,\rightarrow\,0~,
\label{V} \\
L^p_{\om} (\rho,t ) &= e^{-i\om t} (\sinh \rho)^{-1-ip} F(\nu^*_+,
\nu^*_-;1+ip; -\frac{1}{\sinh^2 \rho}) \nn &\sim e^{-\rho}
e^{-ip\rho -i\om t} \qquad \mbox{as} \qquad
\rho\,\rightarrow\,\infty~,
\label{L} \\
R^p_{\om} (\rho,t )
&= e^{-i\om t} (\sinh \rho)^{-1+ip} F(\nu_+, \nu_-;1-ip;
-\frac{1}{\sinh^2 \rho}) \nn &\sim e^{-\rho} e^{+ip\rho -i\om t}
\qquad \mbox{as} \qquad \rho\,\rightarrow\,\infty \label{R}
\end{align}
with the notations
\begin{align}
\nu_{\pm} = \frac{1}{2} - i\left(\frac{p}{2}\pm
\frac{\om}{2}\right)~. \nonumber
\end{align}
These eigenfunctions are defined by the following analytic
continuations of the mini-superspace Euclidean eigenfunctions:
\begin{align}
 U^p_{\om}(\rho,t)& =
\left\{
\begin{array}{ll}
\phi^p_{n=+i\om}(\rho,\theta = +it)~  ~~ (\om>0,~n<0)  \\
\phi^p_{n=-i\om}(\rho,\theta = -it)~  ~~ (\om<0,~n>0)
\end{array}
\right. \nn
 V^p_{\om}(\rho,t&)=
\left\{
\begin{array}{ll}
\phi^{-p}_{n=-i\om}(\rho,\theta = - it)~  ~~ (\om>0,~n<0)  \\
\phi^{-p}_{n=+i\om}(\rho,\theta = +it)~  ~~ (\om<0,~n>0)
\end{array}
\right. \nn  L^p_{\om}(\rho,t)&= \phi^p_{L,n=i\om} (\rho,
\theta=+it) \nn  R^p_{\om}(\rho,t)&= \phi^p_{R,n=i\om} (\rho,
\theta=+it) ~, \label{ac UVLR}
\end{align}
where the $n<0$ and $n>0$ ranges are mapped to $\om>0$ and $\om<0$,
respectively.

\begin{figure}[htbp]
    \begin{center}
   \includegraphics[width=13cm,height=10cm]
      {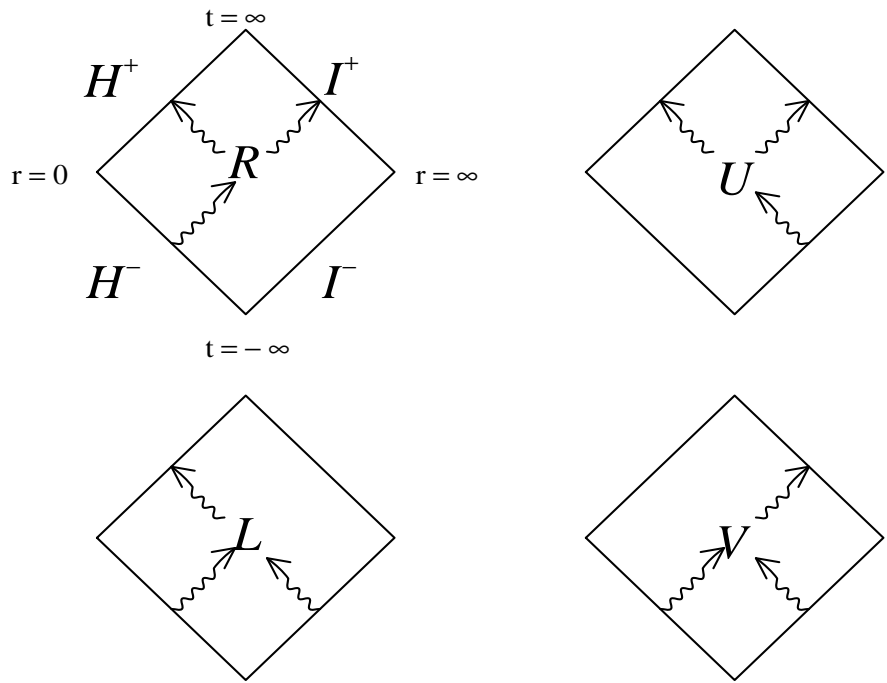}
    \end{center}
    \caption{The boundary conditions of the Lorentzian
       eigenfunctions ($\om>0$ sector). For $\om<0$, the
        figures for $L$ and $R$ should be interchanged.}
    \label{wave}
\end{figure}

As discussed in \cite{Dijkgraaf:1992ba}, only two out of the four eigenfunctions
are linearly independent. In particular,
\begin{align}
V_\omega^p(\rho, t) = U^{p*}_\omega(\rho, -t) \qquad \mbox{and}
\qquad R_\omega^p(\rho, t) = L^{p*}_\omega (\rho, -t). \nonumber
\end{align}
The reason why we introduce the above four eigenfunctions is because
they encode four possible boundary conditions (We here assume $p>0$)
in the Lorentzian black hole background. Recall that, for the region
outside the horizon of the eternal black hole, the boundaries
consist of four segments: `future (past) horizon' $t=+\infty, \,
\rho=0$ ($t=-\infty, \, \rho=0$) by ${\cal H}^+$ (${\cal H}^-$), and
the `future (past) infinity' $t=+\infty, \, \rho=+\infty$
($t=-\infty, \, \rho=+\infty$) by ${\cal I}^+$ (${\cal I}^-$). The
four eigenfunctions $U, V, L, R$ are the ones obeying boundary
conditions:
\begin{align}
 U^p_{\om} &= 0 ~~ \mbox{at}~ {\cal H}^-~,~~~ V^p_{\om} = 0 ~~
\mbox{at}~~ {\cal H}^+~, \cr ~~~ L^p_{\om} &= 0~ (R^p_{\om} = 0)~~ 
\mbox{at}~~ {\cal I}^+~, ~~~ R^p_{\om} = 0 ~(L^p_{\om} = 0) ~~
\mbox{at}~~ {\cal I}^- 
\end{align}
for $\om>0$ ($\om <0$). See Figure \ref{wave}.

%


By Wick rotating the mini-superspace reflection relations \eqref{cref
rel}, we obtain linear relations among the Lorentzian
eigenfunctions:
\begin{align}
 U^p_{\om} = L^{p}_{\om} + \cR_0(p,\om) R^p_{\om} \qquad
\mbox{and} \qquad V^p_{\om} = R^{p}_{\om} + \cR^*_0(p,\om)
L^p_{\om}~. \label{decomp ef 2}
\end{align}
Equivalently,
\begin{align}
 L^p_{\om} &= \frac{1}{1-|\cR_0(p,\om)|^2} \left\{ U^p_{\om} -
\cR_0(p,\om) V^p_{\om} \right\} \cr \quad \mbox{and} \quad R^p_{\om} &=
\frac{1}{1-|\cR_0(p,\om)|^2} \left\{ V^p_{\om} - \cR^*_0(p,\om)
U^p_{\om} \right\}. \label{decomp ef 2-2}
\end{align}
Here, the mini-superspace reflection amplitude $\cR_0(p,\om)$ in
Lorentzian theory is given by
\begin{align}
\cR_0(p,\om) = \frac{\Gamma(+ip)\Gamma^2(\nu_+)}
{\Gamma(-ip)\Gamma^2(\nu^*_-)}
 \equiv - \frac{B(\nu_+,\nu_-)}{B(\nu_+^*,\nu_-^*)} \cdot
\frac{\cosh \pi \left(\frac{p-\om}{2}\right)}
{\cosh \pi \left(\frac{p+\om}{2}\right)}~.
\label{cref amp 2}
\end{align}
Notice that, in sharp contrast to the Euclidean black hole, the
reflection amplitude is less than unity due to the second factor:
\begin{align} |\cR_0(p,\om)|^2 = {\cosh^2 \pi \left({p-\om \over 2}\right)
\over \cosh^2 \pi \left({p+\om \over 2} \right)} \leq 1.
\label{inequality} \end{align}
The inequality is saturated at $p=\omega = 0$. The inequality
\eqref{inequality} shall play a prominent role for understanding string
dynamics in the Lorentzian black hole background. The
mini-superspace reflection relations for $U^p_{\om}$, $V^p_{\om}$
are also expressible in a form similar to the Euclidean ones.
Recalling that $\cR_0(-p, \om) \cR_0(+p, \om) = 1$,
\begin{align}
U^{-p}_{\om}(\rho,t)= \cR_0(-p,\om) U^{p}_{\om}(\rho,t) \qquad
\mbox{and} \qquad V^{-p}_{\om}(\rho,t)= \cR^*_0(-p,\om)
V^{p}_{\om}(\rho,t)~, \label{cref rel UV}
\end{align}
while $L^p_{\om}$ and $R^p_{\om}$ are simply related by reflection:
\begin{align}
 L^{-p}_{\om} (\rho,t) = R^{+p}_{\om} (\rho,t)~. \label{cref rel
LR}
\end{align}
Moreover, $U^p_{\om}$ and $V^p_{\om}$ are linearly independent
except for the special kinematic regime, $\om=0$. Notice also, in
the relation \eqref{inequality}, the reflection amplitude involves the
mini-superspace contribution only, not the full-fledged stringy one.

Before proceeding further, we shall here collect explicitly
relations among inner products of Lorentzian primary fields, where
the inner product is defined with respect to the Lorentzian measure
$\dd v_L = k\sinh 2\rho \dd \rho \dd t$. Taking quantum
numbers $p$, $\om$ fixed and dropping off delta function factors
$2\pi\delta(p-p')$, $2\pi\delta(\om-\om')$ for notational
simplicity, we have
\begin{align}
 & (U^p_{\om}, U^p_{\om}) = (V^p_{\om}, V^p_{\om})
   = N_0(p,\om)~, \qquad N_0(p,\om)
\equiv \frac{1+|\cR_0(p,\om)|^2}{2}
\nn
 & (U^p_{\om},V^p_{\om}) = \cR_0^*(p,\om)~,
\nn
 & (L_{\om}^p,L_{\om}^p)= (R^p_{\om}, R^p_{\om})= \frac{1}{2}~,
 \qquad \quad (L^p_{\om}, R^p_{\om})= 0 ~, \nn
 & (U^p_{\om},L^p_{\om}) = (V^p_{\om}, R^p_{\om}) =\frac{1}{2}~,
 \qquad \quad (R^p_{\om},U^p_{\om}) = (V^p_{\om}, L^p_{\om})
   =\frac{\cR_0(p,\om)}{2}~.
\label{inner product UVLR}
\end{align}
The inner products involving $L^p_\om$ and $R^p_\om$ are readily
evaluated since dominant contributions are supported in the
asymptotic region $\rho \gg 0$, yielding the volume factor
$2\pi\delta(0)$. The remaining inner products ca be extracted
from the linear relations \eqref{decomp ef 2}, \eqref{decomp ef 2-2}.\footnote
{We checked these inner products numerically using
MATHEMATICA.} We also fixed the overall normalization factors from
consistency with the Euclidean inner product \eqref{inner product}
under the $\om\,\rightarrow\, 0$ limit. Notice also that
\begin{align}
 N_0(-p,\om) = \left|\cR_0(-p,\om)\right|^2 \, N_0(+p,\om)~,
\end{align}
as is consistent with the mini-superspace reflection relation
\eqref{cref rel UV}.

It is easy to construct the exact string vertex operators or primary
states corresponding to the mini-superspace eigenfunctions $U$, $V$,
$L$, $R$.
To be specific, we shall consider primarily the fermionic $SL_k (2,
\br)/U(1)$ supercoset conformal field theory.\footnote
    {For the bosonic $SL(2;\br)_{\kappa}/U(1)$ coset conformal field theory,
  we instead have
$h= \bar{h}= \frac{p^2}{4(\kappa-2)}-\frac{\om^2}{4\kappa}
 +\frac{1}{4(\kappa-2)}$, and
$\cR(p,\om)\equiv
\cR_0(p,\om) \frac{\Gamma\left(1+\frac{ip}{\kappa-2}\right)}
{\Gamma\left(1-\frac{ip}{\kappa-2}\right)}$.
}
The primary states $\ket{U^p_{\om}}$, $\ket{V^p_{\om}}$ are the ones
of conformal weights $h= \bar{h}= \frac{p^2}{4k}-\frac{\om^2}{4k}
+\frac{1}{4k}$ and obey the exact reflection relations
\begin{align}
 \ket{U^{-p}_{\om}}= \cR(-p,\om) \ket{U^{p}_{\om}}~, ~~~
\ket{V^{-p}_{\om}}= \cR^*(-p,\om) \ket{V^{p}_{\om}}~, \label{qref rel
UV}
\end{align}
and the exact reflection amplitude is given by
\begin{align} \cR(p,\om)\equiv \cR_0(p,\om)
\frac{\Gamma\Big(1+\frac{ip}{k}\Big)}
{\Gamma\Big(1-\frac{ip}{k}\Big)}. \label{exactra} \end{align}
Notice that the string world-sheet effect entering through the
$1/k$-correction is a pure phase. Thus, the exact reflection
probability $\vert {\cal R} (p, \om) \vert^2$ remains unmodified
from the mini-superspace approximation result $\vert {\cal R}_0(p,
\om) \vert^2$ given in \eqref{inequality}. We shall normalize the
primary states $\ket{U^p_{\om}}$, $\ket{V^p_{\om}}$ ($p>0$) as
\begin{align}
& \bra{U^p_{\om}} U^{p'}_{\om'}\rangle = \bra{V^p_{\om}}
V^{p'}_{\om'}\rangle = N(p,\om) \, 2\pi\delta(p-p')
2\pi\delta(\om-\om') ~, \nonumber \\
& \bra{V^p_{\om}} U^{p'}_{\om'} \rangle = \cR^* (p,\om) \,
2\pi\delta(p-p') 2\pi\delta(\om-\om') ~, \label{norm UV}
\end{align}
where the new normalization factor $N(p,\om)$ is simply defined by
replacing $\cR_0$ with $\cR$ in $N_0(p,\om)$. The primary states
$\ket{L^p_{\om}}$, $\ket{R^p_{\om}}$ are also definable by using the
linear relations \eqref{decomp ef 2} or \eqref{decomp ef 2-2} but now
with $\cR_0$ replaced by $\cR$. Notice that $\ket{U^p_{\om}}$,
$\ket{V^p_{\om}}$ are the ones analytically continuable to the
Euclidean primary states $\ket{\phi^{\pm p}_n}$, so often referred
as the `Hartle-Hawking vacua'. On the other hand, the states
$\ket{L^p_{\om}}$, $\ket{R^p_{\om}}$ does not have Euclidean
counterparts. Recall that, over the Euclidean black hole background,
$\phi^p_{L,n}$, $\phi^p_{R,n}$ behave badly in the vicinity of $\rho
= 0$ and hence ill-defined.


We also find it useful to introduce the dual basis
$\widehat{\bra{U^p_{\om}}}$, $\widehat{\bra{V^p_{\om}}}$ ($p,p'>0$)
with inner products
\begin{align}
 & \widehat{\bra{U^p_{\om}}} U^{p'}_{\om'} \rangle
   = \widehat{\bra{V^p_{\om}}} V^{p'}_{\om'} \rangle
   = 2\pi\delta(p-p')2\pi\delta(\om-\om')~, \qquad
 \widehat{\bra{U^p_{\om}}} V^{p'}_{\om'} \rangle
   = \widehat{\bra{V^p_{\om}}} U^{p'}_{\om'} \rangle
   = 0~.
\label{hat U V}
\end{align}
Explicitly, they are given by
\begin{align}
 \widehat{\bra{U^p_{\om}}} &= \frac{2}{1-\left|\cR(p, \om)
\right|^2} \left\{ \bra{L^p_{\om}} - \cR^* (p, \om) \bra{R^p_{\om}}
\right\}~, \cr \qquad \widehat{\bra{V^p_{\om}}} &=
\frac{2}{1-\left|\cR(p, \om) \right|^2} \left\{ \bra{R^p_{\om}} -
\cR(p, \om) \bra{L^p_{\om}} \right\}~. \label{def hat U V}
\end{align}
As such, these dual basis obey the following exact reflection
relations:
\begin{align}
 \widehat{\bra{U^{-p}_{\om}}} = \cR(p,\om)
 \widehat{\bra{U^{p}_{\om}}} \qquad \mbox{and} \qquad
\widehat{\bra{V^{-p}_{\om}}} = \cR(p,\om)^*
 \widehat{\bra{V^{p}_{\om}}}~.
\label{ref hat U V}
\end{align}

A remark is in order. The dual basis $\widehat{\bra{U^{p}_{\om}}}$,
$\widehat{\bra{V^{p}_{\om}}}$ are {\em not\/} Wick rotatable to the
Euclidean dual basis $\bra{\phi^{+p}_n}$, $\bra{\phi^{-p}_n}$, since
$|\cR(p,\om)|=1$ for $\om \in i\br$. The correct procedure would be
that we first define Wick rotations for the `ket' states, and then
define their dual states within the Lorentzian Hilbert space.
Nevertheless, one-point correlators in the Lorentzian theory, from
which a set of physical observables can be computed, ought to be
always analytically continuable to the one-point correlators in the
Euclidean theory. Roughly speaking, ambiguities inherent to the Wick
rotation of dual states drop out upon taking inner product.
%
\subsection{Duality and winding tachyon condensation}\label{sec:3-4}
One of the salient features of the (Euclidean) $SL(2;\br)/U(1)$ coset model is the so-called Fateev-Zamolodchikov-Zamolodchikov (FZZ) duality \cite{FZZ}. Mathematically speaking, this duality has enabled us to compute exact two-, and three-point functions of the $SL(2;\br)/U(1)$ coset model and revealed their pole structures. Physically speaking, on the other hand, it has established a duality between the winding tachyon condensation and the singularity resolution of the geometry, and uncovered, from an exact CFT perspective, the importance of the winding tachyon condensation near the classical singularities.

Let us formulate the FZZ duality in the $\mathcal{N}=2$ supersymmetric case. The FZZ duality states:

{\bf FZZ duality} ($\mathcal{N}=2$) supersymmetric $SL(2;\br)/U(1)$ coset model with level $k$ is equivalent (up to chirality) to the $\mathcal{N}=2$ Liouville field theory (see e.g. \cite{Nakayama:2004vk} for reference): 
\begin{align}
L = \int \dd^4\theta \Phi^\dagger \Phi + \int \dd^2\theta W(\Phi) + h.c. \cr
W(\Phi) = \mu e^{\frac{1}{Q}\Phi}  \ ,
\end{align}
where $Q^2 = \frac{2}{k}$.

The appearance of the chirality flip suggests the T-dual nature of the duality. Indeed, the FZZ duality can be proved in a more general context of the mirror symmetry. Physically, the appearance of the $\mathcal{N}=2$ Liouville potential in the T-dualized set up can be anticipated as follows. As we studied in section \ref{sec:3-1}, the T-dual of the $SL(2;\br)^{(A)}/U(1)$ axial coset model, whose classical geometry is the cigar, is classically described by the trumpet geometry. However, the trumpet geometry has a singularity coming from the fixed point of the (T-dualized) $U(1)$ angular direction. To avoid the existence of a naked singularity of the space-time, the (T-dualized) winding tachyon will condensate. From the world-sheet viewpoint, the (winding) tachyon condensation is nothing but the $\mathcal{N}=2$ Liouville superpotential.\footnote{To avoid a possible confusion, we note that the original winding tachyon condensation becomes non-winding tachyon with $\tilde{\theta}$ momentum after taking the T-duality.}

The operator correspondence of the FZZ duality is almost clear. In the asymptotic region, one can write the vertex operators of primary states in the $SL(2;\br)^{(A)}/U(1)$ coset model by using those of the linear dilaton times $U(1)$ angular direction. We then perform the T-duality, to write them down as asymptotic vertex operators of primary states in the $\mathcal{N}=2$ Liouville theory. The descendant structure is completely fixed by the $\mathcal{N}=2$ superconformal algebra.

There are several ``proofs" of the FZZ duality available in the literature. In the original work, FZZ has given a direct computation of the two- and three-point functions of the both models (including winding violating correlation functions) and has shown the equivalence between the two models when the computation based on the screening operator is available. In \cite{Hori:2001ax}, the duality has been established rigorously at the level of the topological field theory from the viewpoint of the mirror symmetry (T-duality of the linear sigma model that flows to $SL(2;\br)/U(1)$ coset in the infrared). As is the case with the usual mirror symmetry, it is natural to expect that the full conformal field theory is dual with each other, and indeed there is much supporting evidence for that. In another interesting derivation of the FZZ duality \cite{Tong:2003ik}, the domain wall dynamics of  a certain $2+1$ dimensional gauge theory has been studied, resulting in two complementary descriptions --- $SL(2;\br)/U(1)$ coset model on one hand and $\mathcal{N}=2$ Liouville theory on the other hand.

We will not review the derivation of the FZZ duality (see any of the references above, or consult \cite{Nakayama:2004vk} for a brief summary of the related discussions). Instead, we will see some physical consequences of the duality in the remaining part of this section. Let us begin with the comparison between the classical two-point function of the $SL(2;\br)/U(1)$ coset CFT from the minisuperspace approximation and the exact one. The mini-superspace result (see \eqref{cref amp} and \eqref{ref dual}) is 
\begin{align}
\cR_0(j,m,\bar{m}) =
\frac{\Gamma(2j+1)\Gamma(-j+m)\Gamma(-j-\bar{m})}{\Gamma(-2j-1)
\Gamma(j+1+m)\Gamma(j+1-\bar{m})}, \end{align}
where
\begin{align}  \qquad m=\frac{kw+n}{2}~, \qquad \bar{m} =
\frac{kw-n}{2}~ ,
 \nn
\end{align}
while the exact result is 
\begin{align}
\cR(j,m,\bar{m}) &= \nu(k)^{-2j-1}\,
\frac{\Gamma(1+\frac{2j+1}{k})}{\Gamma(1-\frac{2j+1}{k})}
\frac{\Gamma(2j+1)\Gamma(-j+m)\Gamma(-j-\bar{m})}{\Gamma(-2j-1)
\Gamma(j+1+m)\Gamma(j+1-\bar{m})}, \cr
\nu(k) &\equiv \frac{1}{\pi}\frac{\Gamma(1-\frac{1}{k})}
{\Gamma(1+\frac{1}{k})}~, \label{strt}
\end{align}
The effects of the winding tachyon condensation can be seen in the $1/k$ suppressed factor in the exact formula as $\frac{\Gamma(1+\frac{2j+1}{k})}{\Gamma(1-\frac{2j+1}{k})}$. As is well-known in Liouville field theory, the poles in the correlation function appear when the screening interaction coming from the $\mathcal{N}=2$ superpotential $W= \mu e^{\frac{1}{Q}\Phi}$ satisfies the screening condition for the Liouville momenta $\phi$. Indeed, the perturbative Liouville insertion predicts poles in the two-point functions exactly as indicated by the factor $\frac{\Gamma(1+\frac{2j+1}{k})}{\Gamma(1-\frac{2j+1}{k})}$.

Another important aspect of the FZZ duality is that it has provided a perspective on the winding number non-conservation process. In the $SL(2;\br)^{(A)}/U(1)$ axial coset model, one can define an asymptotic winding quantum number by $\omega$. However, since the cigar geometry has a trivial fundamental group, the winding number is not a conserved quantity. In the free field construction of the $SL(2;\br)^{(A)}/U(1)$ coset model (such as the one based on the Wakimoto construction of the $SL(2;\br)$ current algebra), it is difficult to compute the winding number violating correlation functions. Indeed this was the first motivation of FZZ to propose the dual description.\footnote{At the same time, FZZ has also given an ingenious way to compute the winding violating correlation function within the $SL(2;\br)/U(1)$ coset model by introducing the dual operators.} 

Situations are worse in the naive T-dualized trumpet geometry. In the trumpet metric, it appears that we have a $U(1)$ isometry along $\tilde{\theta}$ that is the dual coordinate for $\theta$, suggesting that the momentum quantum number as well as the winding quantum number are well-defined conserved quantities. The breaking of the winding number (or momentum mode in the T-dual picture) is quite obscure: the origin of the winding non-conservation, i.e. the fixed point of the $U(1)$ action, has now become the singularity of the target space.\footnote{It is well-known that when we gauge the axial symmetry, the vector current has an anomaly and vice versa, and this is indeed the origin of this apparent paradox. In the same token, the $U(1)$ isometry of the vector coset is broken down to $\bz_k$. We should be, therefore, careful when we talk about the ``T-duality" of the trumpet.}

The resolution of this puzzle is given by the FZZ duality. In the T-dualized picture, the singularity is removed by the tachyon condensation, or $\mathcal{N}=2$ Liouville superpotential. At the same time, the $\mathcal{N}=2$ superpotential explicitly breaks the translation invariance along the $\theta$ direction, which gives the origin of the momentum non-conservation in the T-dual picture. Actually, the explicit breaking of the momentum conservation is quite useful to compute the winding number violating process in $SL(2;\br)/U(1)$ coset model: by a direct insertion of the $\mathcal{N}=2$ Liouville superpotential, the winding number violating process can be computed perturbatively.

We end this section with three remarks
\begin{itemize}
	\item Supersymmetric $SL(2;\br)/U(1)$ coset model has a conserved $U(1)_R$ current. By taking quotient of the theory with this $U(1)_R$  current, we obtain the duality between the bosonic $SL(2;\br)/U(1)$ coset model and the sine-Liouville theory \cite{Karczmarek:2004bw}. The sine-Liouville theory has the potential
\begin{align}
V(\phi,Y) &= \mu (S^+ + S^-) \cr
S^{\pm} &=  e^{-\frac{1}{\cQ}(\phi \pm \sqrt{1+\cQ^2}iY)}
\equiv e^{-\sqrt{\frac{\kappa-2}{2}} \phi \mp
\sqrt{\frac{\kappa}{2}} iY}~,
~~~ (\cQ=\sqrt{2/(\kappa-2)})~.
\end{align}
We note that the potential preserves the $W_{\infty}$ symmetry as a side remark, which makes the model integrable \cite{Baseilhac:1998eq,Lukyanov:2003nj}. In their original work (FZZ), they proposed the duality between the bosonic $SL(2;\br)/U(1)$ coset model and the bosonic sine-Liouville theory.

	\item There is a small controversial issue in the interpretation of the FZZ duality. Our standpoint has been that the dual description of the cigar geometry is given by the $\mathcal{N}=2 $ Liouville theory, and the $\mathcal{N}=2$ Liouville superpotential does not appear in the original cigar geometry explicitly (otherwise the source of the winding number non-conservation is two-fold). The other common interpretation of the FZZ duality is that the winding tachyon condensation ($\mathcal{N}=2$ Liouville superpotential written in the dual coordinate) also appears in the original cigar geometry. This interpretation is natural in the sense that it gives a natural explanation about the coexistence of the poles coming from the geometry part and the Liouville insertion part. Whichever interpretation one may take, we believe that what we call the supersymmetric $SL(2;\br)/U(1)$ theory and the $\mathcal{N}=2$ Liouville theory is identical, and the structure constant, e.g. the two-point function is uniquely given by formulae like \eqref{strt}.

	\item So far, we have focused on the Euclidean $SL(2;\br)/U(1)$ coset model. However, things are unclear in the Lorentzian $SL(2;\br)/U(1)$ coset model, where the dual $\mathcal{N}=2$ Liouville theory is unavailable. A naive analytic continuation of the $\mathcal{N}=2$ superpotential gives a wrong Liouville wall, which is localized near the weakly coupled region \cite{Hikida:2004mp}. Furthermore, the Lorentzian coset does not have a winding mode, so the interpretation of the winding tachyon condensation is not evident. Nevertheless, we believe that the exact structure constants are given by the analytic continuation of the exact results for the Euclidean coset since the analytic continuation correctly reproduces the mini-superspace part. The clear explanation of the origin of the extra poles in the Lorentzian coset is still an open question.\footnote{The origin might be given by the degrees of freedom near the horizon on which we mentioned in section \ref{sec:3-2-3}. A related interpretation based on the idea of stretched horizon has been given in \cite{Kutasov:2005rr}.}
\end{itemize}

\newpage
\sectiono{Black Hole - String Transition}\label{sec:4}
In this section, we review ``black hole - string transition". The transition is believed to be a fundamental property of the quantum black hole in the non-BPS regime. We will also see that the transition is related to the thermal winding tachyon condensation.

The organization of the section is as follows. In section \ref{sec:4-1}, we formulate the ``black hole - string transition" in general dimensions. In section \ref{sec:4-2}, we specialize in the two-dimensional case, where $\alpha'$ exact treatment is possible. In section \ref{sec:4-3}, we briefly summarize the current status of the black hole - string transition in other solvable backgrounds.

\subsection{In general dimensions}\label{sec:4-1}

One of the most profound results in (semi-)classical gravity is the thermodynamics of the black hole. Thus one of the most significant benchmarks of any theory of quantum gravity is to provide a satisfactory understanding of the thermodynamics of the black hole. Especially, understanding of the black hole entropy from the microscopic viewpoint has been one of the greatest achievements of the string theory as a quantum theory of gravity \cite{Strominger:1996sh}. 

Let us consider the Schwarzshild black hole in the string theory as a simplest example of the non-extremal black hole system.\footnote{The exact quantization of the string in the Schwarzshild black hole is not known. However, since one can make the curvature of the Schwarzshild black hole arbitrarily small outside the horizon, it is natural to assume the existence of string solutions asymptotically given by the Schwarzshild black hole. The existence of the $SL(2;\br)/U(1)$ two-dimensional black hole strongly supports this assumption.} 
The Schwarzshild black hole in any dimension is completely determined by the parameter $r_h$ that determines the horizon size. When $r_h \gg l_s $, the classical supergravity description is good (at least outside of the horizon), and we can trust the effective supergravity action to discuss the properties of the black hole. 

If one gradually decreases the horizon size $r_h$, the effects of higher derivative corrections coming from the underlying quantum gravity will become important. Within the superstring theory, some higher derivative corrections are known, and it has been shown that these corrections will beautifully explain the apparent mismatch between the macroscopic derivation of the small charge BPS black hole entropy and the microscopic derivation from the string theory (see e.g. \cite{Dabholkar:2004yr}). In the non-extremal cases we are discussing now, we have not yet completely grasped the structure of the higher derivative corrections and the quantitative match of the black hole entropy, but the guiding principle is summarized by the so-called ``black hole - string transition" or ``black hole - string crossover" introduced in \cite{'tHooft:1987tz,Holzhey:1991bx,Susskind:1993ws,Horowitz:1996nw,Sen:1995in}.

When $r_h \le l_s$, the geometrical description of the black hole breaks down and it should be replaced with the microscopic description based on the quantum strings. This is natural because the string theory has a natural cutoff  given by the string length $l_s$ as a length scale, and the objects smaller than $l_s$ do not possess an ordinary geometrical meaning. The principle of the ``black hole - string transition" is that the black hole can be understood either as the higher excitation of the strings or as the classical solution of the (higher derivative) gravities. Especially, the crossover is parametrically smooth as a function of the coupling constant $g_s$ and $l_s$. 

In the Schwarzshild black hole example, we can roughly estimate the transition point and the ``black hole - string crossover" as follows. Let us assume the four dimensional Schwarzshild black hole for definiteness. The four-dimensional Newton constant $G$ is given by $G \sim g_s^2 l_s^2$, so the Schwarzshild radius of the string is estimated as $r_{0} = m_{\rm str} G \sim m_{\rm str} g_s^2 l_s^2$ with the mass of excited string $m^2_{\rm str} \sim \frac{n}{l_s^2}$, where $n$ denotes the oscillator level. At the black hole - string transition point $r_0 \sim l_s$, we have
\begin{align}
\frac{l_s^2}{G} \sim \frac{1}{g_s^2} \sim \sqrt{n} \ .
\end{align}
Thus, the classical Bekenstein entropy is given by $S_{Bek} \sim \frac{r_0^2}{G} \sim \frac{l_s^2}{G} \sim \sqrt{n}$, which indeed agrees with the entropy of the perturbative string expected from the Cardy formula up to a numerical factor. Alternatively speaking, one can say that the requirement of the smooth overlap of the entropy demands that $r_0 \sim l_s$ should be the ``black hole - string transition" point.

Another important concept associated with the $\alpha'$ corrections to the geometry is the stretched horizon \cite{Susskind:1993aa,Susskind:1993ws,Kutasov:2005rr}. We can formulate the stretched horizon based on the local temperature of the geometry. As is the case with the two-dimensional black hole, any neutral black hole has an intrinsic temperature determined by the periodicity of the Euclidean time (Hawking temperature). From the Lorentzian viewpoint, the temperature is defined by the observer at spacial infinity. From an observer at a fixed proper distance $R$ from the horizon, the Hawking radiation is observed with much higher temperature
\begin{align}
T_{u}(R) = \frac{T_{\rm Hw}}{\sqrt{g_{00}(R)}} \ , \label{uhw}
\end{align}
due to the gravitational red-shift. On the other hand, the string theory has the ``highest temperature" determined by the Hagedorn temperature. Since the number of perturbative string states grows exponentially as a function of energy (mass): \begin{align}
Z(\beta) = \mathrm{Tr} e^{-\beta E} \sim \int \dd M \rho(M) e^{-\beta M} \ . \label{hgr}
\end{align}
with the density of states given by $\rho(M) \sim e^{\beta_{\rm Hg} M} $, the partition function of the perturbative string theory is ill-defined beyond the Hagedorn temperature $\beta < \beta_{\rm Hg}$. There we expect that the string interactions are much more important and the strings will disentangle. 

Now let us return to the Hawking radiation. From \eqref{uhw}, one can see that the (red-shifted) temperature becomes infinite at the classical horizon. Actually, before reaching the event horizon  we will encounter the radius when the local temperature exceeds the Hagedorn temperature. The local Hagedorn transition blurs the local geometry near the black hole horizon. This is what we call the stretched horizon. Note that we can make the curvature at the horizon arbitrarily small, and in this regime, the size of the stretched horizon is of order one in the string unit.

It is interesting to consider some extreme limits of the above discussions. The first example is the large $T_{\rm Hw}$ limit: what happens if the Hawking temperature in the asymptotic infinity is larger than $T_{\rm Hg}$? We expect that the stretched horizon completely blur the black hole geometry. Indeed in the leading order estimation of the four-dimensional Schwarzshild black hole, we have $\beta_{\rm Hw} = r_0$ and  $\beta_{\rm Hg} = \text{const}$, and such scenario occurs when $r_0 \sim l_s$. It is interesting to note that the condition roughly coincides with that for the ``black hole - string transition". In section \ref{sec:4-2}, we will see this coincidence is exact (after taking $\alpha'$ corrections into account) in the two-dimensional black hole that is an exactly solvable string background. 

Another limit is the (extremal) charged black hole solution. In the charged black hole examples, the above discussion based on the Hagedorn temperature and the Hawking temperature should be generalized. This is because, as pointed out in \cite{Horowitz:1996nw},  we can arbitrarily lower the Hawking temperature while keeping possible $\alpha'$ corrections large. In other worlds, one can make the transition temperature arbitrarily lower than the Hagedorn temperature. The most extreme case is the (BPS) extremal black hole, where the Hawking temperature is zero. The generalization proposed in \cite{Giveon:2005jv} states that the ``black hole - string transition" occurs when the Hawking temperature coincides with the temperature of the free-string with the same mass {\it and} charge. We will briefly review their discussions later in section \ref{sec:4-3}.

It is also instructive to recapitulate the problem from the Euclidean approach. In the flat Minkowski space, the Hagedorn divergence of the partition function can be attributed to the thermal winding tachyon condensation \cite{Polchinski:1985zf,Sathiapalan:1986db,Kogan:1987jd,O'Brien:1987pn,Atick:1988si}. 
We begin with the more precise version of \eqref{hgr}.
\begin{align}
\beta F &= \mathrm{Tr}_{\mathrm{phys}} \log (1- e^{-\beta E}) \cr
    &= \int_{-\infty}^{\infty}\dd\tau_2 \int_{-1/2}^{1/2} \dd\tau_1 \frac{1}{\tau_2} \mathrm{Tr}_{\mathrm{CFT}} q^{L_0}q^{\bar{L}_0} \ . \label{hgg}
\end{align}
In the second line, we have introduced the Schwinger parameter $\tau_2$ and the level matching condition by 
\begin{align}
\int_{-1/2}^{1/2} \dd\tau_1 e^{2\pi i \tau_1 (L_0-\bar{L}_0)} \ .
\end{align}
The trace is taken over the original space-like CFT with an additional free $\mathbb{S}_1$ CFT whose radius is $\beta$ {\it restricted to the momentum mode}. The Hagedorn divergence appears in the ultraviolet region $\tau_2 \to 0$. Now let us use the Polchinski's trick \cite{Polchinski:1985zf} to rewrite the thermal partition function \eqref{hgg} as the string 1-loop partition function
\begin{align}
\beta F &= \int_{\mathcal{F}}\frac{\dd\tau^2}{\tau_2} \mathrm{Tr}_{\mathrm{CFT}\times \mathbb{S}^1} q^{L_0}q^{\bar{L_0}} \ ,
\end{align}
where $\mathcal{F}$ is the fundamental domain of the torus, and the trace is taken over the original CFT with the free $\mathbb{S}^1$ CFT {\it including winding modes}. The Hagedorn divergence is now translated to the IR instability $\tau_2 \to \infty$. Apart from the ground state tachyon that should be GSO-projected out in the supersymmetric theory, a possible instability comes from the thermal winding tachyon whose mass is given by
\begin{align}
m(\beta)^2 = -1 + \beta^2 \ .
\end{align}
When $m(\beta)^2 < 0$, the Hagedorn instability occurs. In this way, we can understand the Hagedorn divergence as the appearance of the winding tachyon in the Euclideanized thermal string theory.

The argument above suggests that when the thermal direction shrinks enough to admit ``winding tachyon" in the Euclidean spectrum, the Hagedorn phase transition occurs. Assuming a semiclassical quantization of string in the Schwarzshild black hole, a similar situation occurs in the thermal string theory in the Euclidean Schwarzshild black hole background. The thermal winding tachyon has an effective mass
\begin{align}
m^2(r) = -1 + r_0^2\left(1-\frac{r_0}{r}\right) \ .
\end{align}
At the point where $m^2(r)$ becomes negative, the black hole develops a stretched horizon, and when $m^2(\infty) < 0$, we expect the ``black hole - string" phase transition. We will see later that the winding tachyon is crucial in the two-dimensional black hole and its exact ``black hole - string phase transition".

Recently Horowitz \cite{Horowitz:2005vp} studied the real-time winding tachyon condensation in the black hole system. If one considers a compactified black string solution, the extra dimension can show a winding tachyon condensation as the direction shrinks toward the black hole singularity. After the winding tachyon condensation, the black hole evaporates as a bubble of nothing. This process is proposed to be a new interesting end point of the Hawking black hole evaporation (see \cite{Ross:2005ms,Bergman:2005qf,Horowitz:2006mr,Dine:2006we} for related studies). The winding tachyon condensation could also give a solution of the cosmological singularity problems as studied in \cite{McGreevy:2005ci,Nakayama:2006gt}.

\subsection{Two-dimensional black hole case}\label{sec:4-2}
To discuss the ``black hole - string transition" introduced in section \ref{sec:4-1} in a more quantitative manner, it is imperative to study the exact string background rather than the approximate Schwarzshild black hole solution. Especially, the arguments related to the (thermal) winding tachyon condensation is rather speculative, and a demonstration based on the exactly solvable string background would be highly desirable. As we have seen in section \ref{sec:2}, the simplest exactly solvable (non-BPS) black hole background is the two-dimensional black hole. In this subsection, we specialize in the ``black hole - string transition" in the two-dimensional black hole.\footnote{Of course, what we mean by the ``two-dimensional black hole" includes the embedding into the superstring theory such as the black NS5-brane background, so our results have a direct application to the ten-dimensional critical string theories.}

Let us consider the supersymmetric $SL(2;\br)/U(1)$ coset model. As we discussed in section \ref{sec:2-4}, the two-dimensional black hole has the Hawking temperature
\begin{align}
T_{\rm Hw} = \frac{1}{\beta_{\rm Hw}} = \frac{1}{2\pi\sqrt{\alpha'k}} \ .
\end{align}
Since the two-dimensional black hole is asymptotically a linear dilaton theory,  the Hagedorn temperature shows a $1/k$ corrected shift compared with the flat Minkowski theory\footnote{When we mention the Hagedorn temperature of the two-dimensional black hole, we always assume that the criticality condition of the string theory is satisfied by adding {\it non-dilatonic} CFTs. The NS5-brane background is a typical example.}:
\begin{align}
T_{\rm Hg} = \frac{1}{\beta_{\rm Hg}} = \frac{1}{4\pi\sqrt{1-\frac{1}{2k}}} \ . \label{Hg}
\end{align}
To derive this formula, one should first note that the $SL(2;\br)/U(1)$ coset model has a mismatch between the genuine central charge $c^{SL(2;\br)/U(1)} =  3 + \frac{6}{k}$ and the effective central charge $c_{\mathrm{eff}}^{SL(2;\br)/U(1)} = 3$ due to the asymptotic linear dilaton.\footnote{We can also see this directly from the one-loop partition function and the spectrum. See section \ref{sec:3} and appendix A.} Therefore, the total theory has a deficit effective central charge $c_{\mathrm{eff}}^{\mathrm total} = 12 - \frac{6}{k}$ after the subtraction of the ghost contribution. Now we recall the Cardy formula:
\begin{align}
\rho(M) \sim \exp\left(2\pi\sqrt{\frac{c_{\mathrm{eff}}}{12}}M + 2\pi \sqrt{\frac{\bar{c}_{\mathrm {eff}}}{12}}M \right) \ ,
\end{align}
which immediately gives the Hagedorn temperature \eqref{Hg}. 

For later references, we present here a similar formula for the bosonic two-dimensional black hole. The Hawking temperature and the Hagedorn temperature is given by
\begin{align}
T_{\rm Hw} = \frac{1}{\beta_{\rm Hw}} = \frac{1}{2\pi\sqrt{\alpha'\kappa}} \ . \label{hwb}
\end{align}
and
\begin{align}
T_{\rm Hg} =\frac{1}{\beta_{\rm Hg}} = \frac{1}{4\pi\sqrt{2-\frac{1}{2(\kappa-2)}}} \ . \label{Hgb}
\end{align}
There is no apparent reason to exclude possible $\alpha'$ corrections to the Hawking temperature in the bosonic string theory, but the exact string quantization reveals that the formula \eqref{hwb} is the correct one.\footnote{In general, the Hawking temperature is classically determined solely from the information near the event horizon (the Rindler limit), where the curvature and the $\alpha'$ corrections could become large. The effects of such $\alpha'$ corrections and possible renormalization of the Hawking temperature are interesting subjects to study.} We will return to this problem when we discuss the exact boundary states of the probe rolling D-brane in section \ref{sec:8}. On the other hand, the Hagedorn temperature here is obtained from the exact string quantization and trustful.

From the general discussions in section \ref{sec:4-1}, we expect ``black hole -  string" at $k=1$ (or $\kappa=3$ for the bosonic case) when the Hawking temperature and the Hagedorn temperature coincide. At this point, the stretched horizon becomes so large that it will swallow the complete space-time. This ``black hole - string transition" in the two dimension black hole is induced by the strong $\alpha'$ corrections: when $k$ is large (recall $1/k$ correction corresponds to $\alpha'$ correction) the Hawking temperature is much larger than the Hagedorn temperature, and the geometry is not disturbed by the back-reaction of the Hawking radiation. When $k$ becomes smaller, $1/k$ corrections will become more and more important, and at the phase transition point, i.e. at $k=1$, physics changes drastically. One of the main focus of this thesis is to study this transition from the rolling D-brane probe.

At this point, we would like to point out that the ``black hole - string" phase transition of the two-dimensional black hole does not involve the string coupling $g_s$ in the discussion. This is one of the features of the two-dimensional black hole that we can clearly separate the (typically more difficult) problem of the genus expansion from the more tractable $\alpha'$ corrections in order to understand the ``black hole - string transition".

What is the origin of the strong $1/k$ correction? As we have mentioned earlier in section \ref{sec:3-1}, the metric for the supersymmetric two-dimensional black hole does not receive perturbative $1/k$ corrections. The origin of the (nonperturbative) $1/k$ corrections that trigger the ``black hole - string transition" is most clearly seen in the Wick rotated Euclidean two-dimensional black hole for which the dual description is available. 

In the dual description, the two-dimensional black hole is described by the $\mathcal{N}=2$ Liouville theory. The $\mathcal{N}=2$ superpotential
\begin{align}
W(\Phi) = \mu \int d^2\theta e^{\frac{1}{Q}\Phi} \label{Liousup}
\end{align}
can be seen as the localized (winding) tachyon condensation.\footnote{The duality between the $SL(2;\br)^{(A)}/U(1)$ coset and the $\mathcal{N}=2$ Liouville theory is a kind of T-duality as we discussed in section \ref{sec:3-4}. Thus the condensation of the momentum mode in $\mathcal{N}=2$ Liouville theory can be regarded as the condensation of the winding mode in the original $SL(2;\br)^{(A)}/U(1)$ coset model.} The condensation is a localized mode because the Liouville momentum $j$ corresponding to the superpotential \eqref{Liousup} is not given by the continuous series $j=-\frac{1}{2}+ip$, but lies in the discrete series.

The crucial observation has been already made early in \cite{Kutasov:1990ua} in the context of the noncritical superstring theory. The spirit is close to the discussions given in section \ref{sec:2-3} and \ref{sec:3-4}. The superpotential \label{Liousup} is a normalizable perturbation if $\frac{1}{Q}>\frac{Q}{2}$ (i.e. $k >1$) and it is a non-normalizable deformation otherwise. In the language of the noncritical string theory, the $\mathcal{N}=2$ super Liouville potential satisfies the Seiberg bound \cite{Seiberg:1990eb} only when  $\frac{1}{Q}<\frac{Q}{2}$  holds. This directly means that the $\mathcal{N}=2$ Liouville description is good for $k<1$ and the two-dimensional black hole description is good for $k>1$. The transition point is exactly at $k=1$.\footnote
   {Another interesting observation related
    to the $k=1$ transition is the following. If we consider
        a two-dimensional $U(1)$ gauge theory in the ultraviolet that
        flows to $SL(2;\br)/U(1)$ coset theory in the infrared (as was
        introduced in \cite{Hori:2001ax} to prove the mirror duality to the
        $\cN=2$ Liouville theory),
  the central charge of the $U(1)$ gauge theory
  is given by $9$. Since the IR $SL(2;\br)/U(1)$ coset theory
  has a central charge $c=3(1+\frac{2}{k})$, there is an
  apparent contradiction to Zamolodchikov's $c$-theorem
  if the level $k<1$ is considered. However, we should note that
  $SL(2;\br)/U(1)$ coset theory is dilatonic so that the effective
  central charge is always given by $3$.}

We can repeat the same analysis for the bosonic $SL(2;\br)/U(1)$ coset. The duality between the bosonic $SL(2;\br)/U(1)$ and the sine-Liouville theory, together with the Seiberg bound, leads to the conclusion that $\kappa=3$ is the phase transition point. The potential is given by
\begin{align}
V = \mu (S^+ + S^-) \  , \ \ \ S^{\pm} =  e^{-\frac{1}{\cQ}(\phi \pm \sqrt{1+\cQ^2}iY)}
\equiv e^{-\sqrt{\frac{\kappa-2}{2}} \phi \mp
\sqrt{\frac{\kappa}{2}} iY}~,
~~~ (\cQ=\sqrt{2/(\kappa-2)})~,
\end{align}
and the normalizability changes precisely at $\kappa = 3$.
Assuming that this occurs when the Hawking temperature and the Hagedorn temperature coincides, we have verified that the Hawking temperature of the bosonic two-dimensional black hole is not renormalized. We will see another support from the probe rolling D-brane later in section \ref{sec:8}.

We could argue this transition without using the dual Liouville picture \cite{Karczmarek:2004bw}. The black hole perturbation descends from the $SL(2;\br)$ states
\begin{align}
J_{-1}^+ \bar{J}_{-1}^+|j=-1; m = \bar{m} = -1 \rangle \ .
\end{align}
The normalizability of such states (see section \ref{sec:3-2-2}) demand
\begin{align}
-\frac{1+k}{2} < j  < -\frac{1}{2} \ 
\end{align}
with $j=-1$, which suggests the same phase transition point $k=1$ (or $\kappa=3$).

The situation in the Lorentzian two-dimensional black hole is less clear. We cannot perform the Wick rotation to the winding tachyon potential \eqref{Liousup} naively because the time is continuous and there is no apparent winding mode in the Lorentzian two-dimensional black hole. The same thing can be said in the Hagedorn instability of the free string theory in the flat Minkowski space: the existence of the thermal winding tachyon in the Wick rotated theory does not mean the tachyonic instability in the real time physics. Rather it should be understood as the phase transition associated with the thermal dissolution of strings. At the temperature beyond the Hagedorn point, there would be no distinction between the gas of strings and the black hole.

\subsection{Other solvable backgrounds}\label{sec:4-3}
There are many other exactly solvable string theory backgrounds that exhibit the ``string black hole transition". Most of the examples are more or less related to the $SL(2;\br)$ WZNW model. In this subsection, we will briefly review the transition in such backgrounds.

The black hole - string transition across $k=1$ also has a natural
interpretation in terms of the holographic principle, as recently
discussed in \cite{Giveon:2005mi}. Adding $Q_1$ fundamental strings to $k$
NS5-branes (more generally Calabi-Yau singularities) as we reviewed in section 2.4, one obtains the familiar bulk geometry of the
$AdS_3/CFT_2$-duality. In this context, the density of states of the
dual conformal field theory is given by the naive Cardy formula
$S=2\pi\sqrt{\frac{cL_0}{6}}+2\pi\sqrt{\frac{\bar{c}\bar{L}_0}{6}}$
with $c = 6 k Q_1$ for $k>1$, but not for $k<1$. Rather, the central
charge that should be used in the Cardy formula is replaced by an
effective one $c_{\rm eff}= 6Q_1(2-\frac{1}{k})$ \cite{Kutasov:1990ua}.

The origin of the difference between the $c_{\mathrm{eff}}$ and $c$ is again the normalizability of a certain operator. The $SL(2;\mathbb{C})$ vacuum of the dual CFT corresponds to the states
\begin{align}
J_{-1}^+ \bar{J}_{-1}^+|j=-1; m = \bar{m} = -1 \rangle \ 
\end{align}
in the world-sheet $SL(2;\br)$ WZNW model, and as we have seen several times, for $k>1$, the operator is normalizable, and $c_{\mathrm{eff}} = c$. On the other hand, for $k<1$, the operator is non-normalizable, and we expect $c_{\mathrm{eff}} <c$. A short computation based on the string description gives $c_{\rm eff}= 6Q_1(2-\frac{1}{k})$.

We note that for $k>1$, the BTZ black hole excitation is normalizable and the partition function and the entropy is dominated by the Bekenstein-Hawking entropy of the BTZ black hole while for $k<1$, the BTZ black hole excitation is non-normalizable and the entropy is solely explained by the string excitations. This argument is completely in agreement with the ``black hole - string transition" picture at $k=1$.

Another interesting generalization is the two-dimensional charged black hole. We consider the asymmetric coset
\begin{align}
\frac{SL(2;\br)_k\times U(1)_L}{U(1)} \ , \label{chtbk}
\end{align}
where the $U(1)$ gauging acts on one of the (space-like) left-moving current in $SL(2;\br)$ and a linear combination of the right-moving current of $SL(2;\br)$ and $U(1)_L$. After the Kaluza-Klein reduction, the geometry of \eqref{chtbk} is described by the metric ($Q^2 = \frac{2}{k}$)
\begin{align}
\dd s^2 = \dd\phi^2 - \left(\frac{\tanh\frac{Q}{2}\phi}{1-a^2\tanh^2\frac{Q}{2}\phi}\right)^2 \dd\theta^2 \ ,
\end{align}
the dilaton
\begin{align}
\Phi = \Phi_0 - \frac{1}{2}\log\left(1+(1-a^2)\sinh^2\frac{Q}{2}\phi\right) \ ,
\end{align}
and the gauge field
\begin{align}
A = \frac{a\tanh^2\frac{Q}{2}\phi}{1-a^2\tanh^2\frac{Q}{2}\phi} \dd\theta \ .
\end{align}
Here $a^2$ is related to the mass $m$ and the charge $q$ of the black hole as
\begin{align}
a^2 = \frac{m-\sqrt{m^2-q^2}}{m+\sqrt{m^2-q^2}} \ .
\end{align}
At $a = 0$, the model reduces to the undeformed $SL(2;\br)/U(1)$ black hole (and a compact boson).

The Hawking temperature of the black hole (e.g. from the Euclidean geometry) is given by
\begin{align}
\beta_{\rm Hw} = \frac{4\pi}{Q} \frac{1}{1-a^2} \ .
\end{align}
On the other hand, the Hagedorn temperature is given by 
\begin{align}
T_{\rm Hg} =\frac{1}{\beta_{\rm Hg}} = \frac{1}{4\pi\sqrt{1-\frac{1}{2k}}} \ , 
\end{align}
irrespective of the deformation parameter $a$.

From the world-sheet perspective, the ``black hole - string transition" of the charged two-dimensional black hole inherits from the $SL(2;\br)$ WZNW model and the transition point should be $k=1$. This is different from the naive guess based on the relation $\beta_{\rm Hw} = \beta_{\rm Hg}$. A resolution proposed in \cite{Giveon:2005jv} is that the more precise definition of the transition temperature is when the Hawking temperature coincides with the temperature of the string that has the same mass and charge of the black hole. 

In this example, the entropy of the string with charge $q$ is given by\footnote{The shift is due to $q$-amount of right-moving $U(1)$ charge: we are summing over the string states with fixed $U(1)$ charge $q$ instead of summing over all states.}
\begin{align}
S = 2\pi \sqrt{1-\frac{Q^2}{4}} \left(m+\sqrt{m^2-q^2}\right)
\end{align}
resulting in the corresponding string temperature
\begin{align}
\beta_{\rm str} = \left.\frac{\partial S}{\partial m} \right|_{q} = \sqrt{1-\frac{Q^2}{4}}\frac{4\pi}{1-a^2} \ .
\end{align}
It is easy to see that the condition $\beta_{\rm str} = \beta_{\rm Hw}$ exactly reproduces the CFT computation, i.e. $k=1$.
\newpage
\sectiono{Tachyon Radion Correspondence}\label{sec:5}
In this section, we review the tachyon radion correspondence, which is one of the greatest motivations to study the rolling D-brane in the two-dimensional black hole system. The correspondence says that the dynamics of the open-string tachyon condensation may be geometrically realized by the rolling D-brane system. 

The organization of the section is as follows. In section \ref{sec:5-1}, we overview the rolling tachyon problem. In section \ref{sec:5-2}, we study the closed string emission rate from the rolling tachyon boundary states and their variations.\footnote{This part of the thesis is based on \cite{Nakayama:2006qm}.} In section \ref{sec:5-3}, we study the correspondence at the classical level. In section \ref{sec:5-4}, we summarize our results on the quantum correspondence. In section \ref{sec:5-5} some cosmological implications are studied.

\subsection{Rolling tachyon}\label{sec:5-1}
\subsubsection{overview}\label{sec:5-1-1}
In the days of early developments of string theory, tachyon used to be thought of as a nuisance in constructing realistic models for particle physics of our world. In recent years, open-string tachyons have obtained civil rights and have played more and more important roles in acquiring our knowledge on the nonperturbative D-brane physics with (spontaneously) broken SUSY. In addition, they have been even providing phenomenological applications such as brane inflation. More recently, the closed string tachyons (especially localized winding tachyon) have attracted much attention in relation to the topological change \cite{Adams:2001sv,Adams:2005rb} and the resolution of singularities \cite{McGreevy:2005ci}.

One of the important steps in understanding the physics of unstable D-branes is Sen's conjecture with subsequent advancement (see \cite{Sen:2004nf} for a review), which states that the decaying process of the unstable D-branes can be regarded as the open string tachyon condensation. In particular, the energy difference between the false (perturbative) tachyonic vacuum and the true vacuum of the open string tachyon potential should explain the tension of the decaying D-brane exactly. Furthermore, the cohomology of the open string theory at the true vacuum must vanish. In the context of the open string field theory, these conjectures have been analytically proved in \cite{Schnabl:2005gv,Ellwood:2006ba}. 

More interesting aspects of the tachyon dynamics is to study its time evolution \cite{Sen:2002nu,Sen:2002in,Sen:2002an}. Based on the effective field theory analysis (which has been confirmed by the exact boundary states analysis later), it was found that the late time evolution of the open-string tachyon gives rise to the so-called ``tachyon matter", which is a pressureless fluid. Such a ``rolling tachyon" evolution has provided us with novel understanding of the tachyon condensation and time-dependent physics in string theory. The feasibility to construct the exact boundary states enables us to study the highly non-supersymmetric time evolution in a quantitative way.

Let us begin with the effective DBI type action for the rolling tachyon
\begin{align}
S = - \int \dd t V(T)\sqrt{1-\dot{T}^2} \ . \label{tDBI}
\end{align}
Since we will focus on the homogeneous decay, we have assumed the D0-brane action without loss of generality. The effective potential $V(T)$ takes the form
\begin{align}
V(T) = M_0 \frac{1}{\cosh\frac{T}{2x}} \ ,
\end{align}
where the D0-brane tension $M_0 \propto \frac{1}{g_s} $. For the non-BPS D-branes in supersymmetric theory, $x=1$, and for the unstable D-branes in bosonic string theory, $x=1/2$.
The solution of the equation of motion is given by
\begin{align}
\sinh\frac{T}{2x} = a\cosh\frac{t}{2x} \ , \label{seom}
\end{align}
leading to the classical energy momentum tensor:
\begin{align}
T_{\mu\nu} = \frac{V(T)\partial_\mu T\partial_\nu T}{\sqrt{1+\eta^{\mu\nu}\partial_\mu T\partial_\nu T}} - V(T) \eta_{\mu\nu} \sqrt{1+\eta^{\mu\nu}\partial_\mu T\partial_\nu T} \ .
\end{align}
We are interested in the late time behavior of the energy momentum tensor, which is explicitly given by 
\begin{align}
T_{00} & \sim E \cr
T_{ij} & \sim -E \exp(-t/x) \delta_{ij} \ ,
\end{align}
where $E = M_0/\sqrt{1+a^2}$. As we mentioned before, we have obtained the pressureless dust as a final product of the D0-brane decay.

The energy momentum tensor yields the coupling of the rolling D-brane to the gravity. In order to study the coupling to higher string modes, we need the exact boundary state that describes the rolling D-brane. In the boundary conformal field theory approach, we introduce the boundary interaction\footnote{We focus on the bosonic case for simplicity. The generalization to the non-BPS D-branes in superstring theory is straightforward.}
\begin{align}
\delta S_{\mathrm{full}} = \tilde{\lambda} \int \dd s \cosh X^0(s) \ , \label{fulli}
\end{align} 
for the ``full S-brane" model, and
\begin{align}
\delta S_{\mathrm{half}} = \lambda \int \dd s e^{X^0(s)} \ , \label{halfi}
\end{align}
for the ``half S-brane" model. Here $X^0$ denotes the target-space time coordinate and the integration is taken over the boundary of the world-sheet parametrized by $s$.

There are several different ways to obtain the boundary states. Originally Sen \cite{Sen:2002nu} proposed to obtain the boundary states for \eqref{fulli} by starting with the (compactified) space-like model (boundary sine-Gordon model) and performing the Wick rotation. In the ``half S-brane model", Gutperle and Strominger \cite{Gutperle:2003xf} proposed to use the Wick rotation of the Liouville theory in the zero linear dilaton limit (time-like Liouville theory). 

\begin{figure}[htbp]
   \begin{center}
    \includegraphics[width=0.5\linewidth,keepaspectratio,clip]{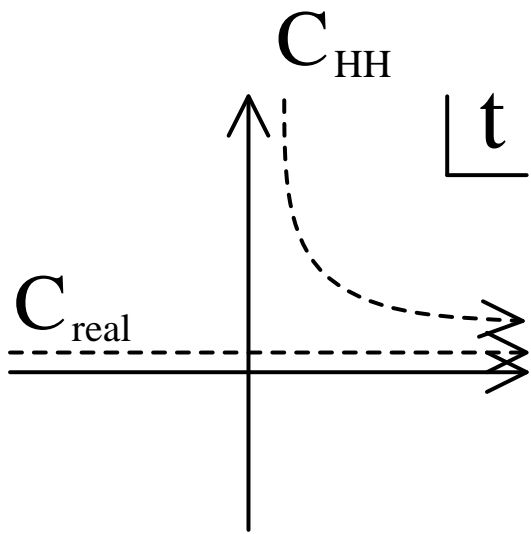}
    \end{center}
    \caption{Different contour integration gives different boundary states.}
    \label{fig:realHH}
\end{figure}

In the coordinate space, the behavior of the boundary states from different prescription shows a different behavior (mainly in the region $ X^0 < 0$), but in the momentum (energy) space, they are related with each other in the zero mode sector. To see this, let us expand the rolling tachyon boundary states as
\begin{align}
|B\rangle &= i\int_C \dd t  \rho (t) |0\rangle  + \sigma (t) \alpha^0_{-1}\bar{\alpha}^0_{-1}|0\rangle + \cdots \cr
 &= i\int \dd\omega \tilde{\rho} (\omega) |\omega \rangle  + \tilde{\sigma} (\omega) \alpha^0_{-1}\bar{\alpha}^0_{-1}|\omega\rangle + \cdots \ .
\end{align}
Since we are dealing with the time-dependent theory based on the analytic continuation, the contour choice will affect the physics. The zero mode density $\rho(\omega)$ has been computed as 
\begin{align}
i \int_{C_{real}}\dd t \rho_{full} (t) e^{i\omega t} &= \left( e^{-i\omega \log\hat{\lambda}} - e^{i\omega \log \hat{\lambda}}\right) \frac{\pi}{\sinh\pi \omega} \cr
i \int_{C_{real}} \dd t \rho_{half} (t) e^{i\omega t} & =  e^{-i\omega \log\hat{\lambda}} \frac{\pi}{\sinh\pi \omega} \cr
i \int_{C_{HH}} \dd t \rho_{full} (t) e^{i\omega t} & =  e^{-i\omega \log\hat{\lambda}} \frac{\pi}{\sinh\pi \omega} \ .
\end{align}
Note that the Hartle-Hawking contour $C_{HH}$ integral of the full S-brane solution coincides with the real contour $C_{real}$ integral of the half S-brane solution (see figure \ref{fig:realHH}). This is intuitively expected because the half S-brane solution describes the later half dynamics of the rolling D-brane (decaying brane) and the Hartle-Hawking contour effectively sets the initial condition at $t=0$ to give a decaying D-brane. We also note that the boundary time-like Liouville field approach directly gives the same zero mode boundary wavefunction for the half S-brane solution. Thus we can conclude that various approaches yield essentially the identical results for the zero mode boundary wavefunction (i.e. coupling to the scalar tachyon mode).

Nevertheless, there are differences in the nonzero mode sectors between the boundary sine-Gordon approach and the boundary time-like Liouville approach. The origin of the difference is that the descendants for the boundary sine-Gordon model is based on the $SU(2)$ current algebra (at the self dual radius) and those for the boundary Liouville theory is based on the Virasoro algebra. For on-shell amplitudes (and energy-momentum tensor) we can gauge away these differences as we will do in section \ref{sec:5-2} to compute the closed string emission rate. However, at least in the two-dimensional noncritical string example, it has been stressed in \cite{Sen:2004yv} that such off-shell boundary states will be important to generate infinitely many conserved charges in addition to the energy momentum tensor. We will revisit the problem later in the discussion of the rolling D-brane, so we will not delve into the details any further at this point and concentrate on the physics associated with the zero mode.

\subsection{Radiation from rolling tachyon boundary states}\label{sec:5-2}
In this subsection, we would like to study the closed string emission rate from the rolling tachyon by using the exact boundary states. We will present rather a technical aspects of the computation for two reasons. One is that we are going to compare the results of the rolling tachyon and rolling D-brane in later sections in detail. The other is to understand the nontrivial relation between the unitarity (optical theorem) and the open - closed duality, which we revisit in the more nontrivial rolling D-brane case in section \ref{sec:8-2}.

Before entering into the computation, we summarize the main physics involved.
\begin{itemize}
	\item At the one-loop level computation, all the energy of the D0-brane is converted into closed string radiation: the radiation rate shows a power-like divergence.
	\item Most of the energy is converted into highly massive strings whose mass is effectively cut off by $M\sim 1/g_s$.
	\item The emitted strings are highly non-relativistic.
	\item If one considers the D$p$-brane as we increase $p$, the divergence becomes milder, but the spectral density is still power-like and the higher moment diverges.
	\item The inclusion of the {\it space-like} linear dilaton makes the divergence disappear due to the exponential suppression for the growth of density of states.
	\item On the contrary, the {\it time-like} linear dilaton (along the rolling tachyon direction) does not affect the divergence. This suggests a first hint of the universality of the decay of unstable D-branes.
\end{itemize}

Now we will begin our study on the closed string emission rate from the unstable D-brane. For a slight generalization of section \ref{sec:5-1}, we consider the unstable D-brane in the linear dilaton background. For the boundary states, we will use the one obtained from the time-like Liouville theory because with a time-like linear dilaton, the corresponding boundary states from the boundary sine-Gordon theory is unavailable. For zero time-like linear dilaton limit, however, the two computation agrees as expected.

 The dilaton gradient is set
by:
\begin{align} \Phi = {1 \over \sqrt{\alpha'}} (Q\, X^0 + {\bf V} \cdot {\bf
X}), \qquad \mbox{where} \qquad Q \equiv \beta - {1 \over \beta} ~
\qquad(\beta \ge 1)~. \label{dilaton}\end{align}
This puts the critical dimension $D$ for the bosonic string theory
to be
\begin{align} 26 = D - 6 Q^2 + 6 {\bf V}^2, \qquad \mbox{so} \qquad c_{\rm
eff} = 6 Q_\beta^2 - 6 {\bf V}^2~, \end{align}
where $Q_\beta \equiv (\beta + 1/\beta)$. The effective central
charge $c_{\rm eff}$ sets the growth of density of closed string
states \cite{Kutasov:1990ua}:
\begin{align}
\rho^{(c)}(M) \sim e^{4\pi \sqrt{\frac{c_{\msc{eff}}}{24}
\alpha' M^2}}
\label{dos}
\end{align}
up to subleading pre-exponential factor of $M$. It grows slower than
the density of states for flat space-time (obtainable by setting
$Q={\bf V} = 0$).

\subsubsection{closed string emission}\label{sec:5-2-1}
Let us consider the decay of an unstable D-brane in the linear dilaton
background.
The radiative transition of a D$p$-brane to a single closed string
state of mass $M$ (set by the integer-valued oscillator level
$N=\widetilde{N}$), whose on-shell energy-momentum $(\omega, {\bf
k})$ is given by
\begin{align} \Big(\omega_E - {i Q \over \sqrt{\alpha'}}\Big)^2 - \Big({\bf
k}_E + {i {\bf V} \over \sqrt{\alpha'}} \Big)^2 =(\omega^2 -{\bf
k}^2) =  M^2 \quad \mbox{where} \quad \frac{1}{4}\, \alpha' M^2 = N
- {c_{\rm eff} \over 24}, \end{align}
where $(\omega_E, {\bf k}_{E})$ and $(\omega, {\bf k})$ are
energy-momenta in the Einstein and the string frame, respectively.
In string loop perturbation theory, the transition amplitude is
computed by the disk one-point function $\langle \exp ((-i \omega +
\frac{Q}{\sqrt{\al'}}) X^0) \, \exp ((i {\bf k} + \frac{{\bf
V}}{\sqrt{\al'}}) \cdot {\bf X}) \rangle_{\msc{disk}}$ with the
D$p$-brane boundary condition,\footnote{We only consider the case
when the D-brane has Neumann boundary condition in the space-like
linear dilaton direction.} where the vertex operator is separated
into temporal and spatial parts as indicated. The two parts are
factorized in the gauge that no oscillator in temporal direction is
allowed. Consequently, the transition probability ${\cal P}(\omega)$
of the radiative process is governed entirely by the temporal part
(see (3.29) in \cite{Karczmarek:2003xm}):
\begin{align} {\cal P}(\omega) = \Big| \left\langle e^{ (- i \omega
+\frac{Q}{\sqrt{\al'}}) X^0} e^{(i {\bf k} + \frac{{\bf
V}}{\sqrt{\al'}})\cdot{\bf X}} \right\rangle_{\msc{disk}} \Big|^2
&= \Big| {1 \over \beta} \Gamma(1 + i \omega\sqrt{\alpha'} \beta)
\Gamma(- i \omega \sqrt{\alpha'} /\beta)\Big|^2 \nn
&={\pi^2/\beta^2 \over \sinh (\pi \omega\sqrt{\alpha'}  \beta)
\sinh (\pi \omega\sqrt{\alpha'}  / \beta)}~. \end{align}
Then, at leading order in string perturbation theory, the total
number of emitted closed strings from the decay of a D$p$-brane
($p\ge 1$) extended along ${\bf V}$-direction is computed as
\begin{align} \overline{\cal N} = N_p^2 V_p\sum_M \sqrt{\rho^{(c)}(M)}
\int_{-\infty}^\infty \!{\rmd^{D-1-p} {\bf k} \over (2 \pi)^{D-1-p}}
\, {1 \over 2 \omega} {\cal P}(\omega)~, \end{align} \label{closed}
\noindent where the overall coefficient abbreviates $N_p =
\pi^{\frac{D-4}{4}} (2 \pi)^{\frac{D-2}{4}-p}$ and $V_p$ is the
D$p$-brane volume. In (\ref{closed}), the sum is over all final
closed string states of mass $M$ and of oscillator excitations
symmetric between left- and right-moving sectors. Such oscillator
excitations are equivalent in combinatorics to open string
excitation, so the density of the final states is given by
square-root of \eqref{dos}.

Attributed to the Hagedorn growth of the density of states
$\rho^{(c)}(M)$, the total emission number $\overline{\cal N}$ in
(\ref{closed}) (or higher spectral moment) is ultraviolet convergent
so long as linear dilaton has a nonzero spatial component, ${\bf V}
\ne 0$, first observed in \cite{Karczmarek:2003xm}. Notice also that temporal
component of the linear dilaton does not alter the ultraviolet
behavior. This is most readily seen for small ${\bf V}$ by expanding
the density of states. To study anatomy of the ultraviolet behavior,
we shall now perform Fourier transformation and re-express
$\overline{\cal N}$ in the open string channel.

\subsubsection{open string channel viewpoint}\label{sec:5-2-2}

Physical observables such as $\overline{\cal N}$ ought to be
well-defined under the Fourier transform from the closed string
channel to the open string one because
\begin{enumerate}
 \item We start with defining expression of $\overline{\cal N}$,
 consistent with the optical theorem in the closed string channel.
 \item The expression is closed in the Euclidean signature.
Hence we are free from any subtlety that may arise from analytic
continuations between Euclidean and Lorentzian signature of the
space-time.
\end{enumerate}
As in \cite{Karczmarek:2003xm}, we expand the transition probability ${\cal
P}(\omega)$ in convergent power series, whose terms can be
interpreted as D-instantons arrayed along imaginary time
coordinate:
\begin{align} {\cal P}(\omega) = {4 \pi^2\over \beta^2} \sum_{n,m=0}^\infty
e^{-2 \pi \alpha' \omega W(m,n)} \end{align}
where the location of the D-instantons is denoted as
\begin{align} \alpha'  W(m,n) = \sqrt{\alpha'}\Big[\Big(n+{1 \over 2} \Big)
\beta + \Big(m+{1 \over 2} \Big){1 \over \beta}\Big] \ge
\sqrt{\alpha'}~. \end{align}
Thus, we take
\begin{align} \overline{\cal N} = \Big({2 \pi N_p\over \beta}\Big)^2 V_p
\sum_M \int_{-\infty}^\infty {\rmd^{D-1-p} {\bf k} \over
(2\pi)^{D-1-p}} \sum_{m,n=0}^\infty {1 \over 2 \omega} e^{ - 2 \pi
\alpha' \omega W(m,n)} \end{align}
and rewrite each D-instanton contribution parametrically via the
closed string channel modulus $t_c$ as
%
%
\begin{align} {1 \over 2 \omega} e^{ - 2 \pi \alpha' \omega W(m,n)} &=
%
\frac{\pi \alpha'}{2} \int_{-\infty}^\infty {\rmd k_0 \over 2 \pi}
\int_0^\infty \rmd {t_c} \, e^{-2\pi t_c \cdot
\frac{1}{4}\alpha'(k_0^2 + {\bf k}^2 + M^2)} e^{2 \pi i \alpha' k_0
W(m,n)} \ , \end{align}
which gives
\begin{align} \overline{\cal N} \!\!\! = \cr & \!\!\!\! \Big({2 \pi N_p\over
\beta}\Big)^2 V_p \frac{\pi \alpha'}{2} \sum_{m,n=0}^\infty
\int_0^\infty \rmd {t_c} \int_{-\infty}^\infty {\rmd k_0 \over 2
\pi} \int_{-\infty}^\infty {\rmd^{D-1-p} {\bf k} \over (2
\pi)^{D-1-p}} \, e^{-2\pi t_c \cdot \frac{1}{4}\alpha'(k_0^2 + {\bf
k}^2)} e^{2 \pi i \alpha' k_0 W(m,n)}\nn
& \hskip3.7cm \times \sum_M \sqrt{\rho^{(c)}(M)} e^{-2\pi t_c \cdot
\frac{1}{4}\alpha' M^2}. \end{align}
Here, we exchanged order of summations and integrations, and first
performed integrals over off-shell momenta $(k_0, {\bf k})$ and sum
over mass level $M$. The sum over $M$ yields modular covariant
partition function $Z^{(c)}(q_c)$ in terms of the Dedekind eta
function:
\begin{align} Z^{(c)} (q_c) &\equiv \sum_M \sqrt{\rho^{(c)}(M)} \, \,
q_c^{\frac{1}{4}\alpha'M^2} \qquad \mbox{where} \qquad q_c \equiv
e^{-2\pi t_c} \nn &= \eta^{-(D-2)}(q_c)~. \end{align}
Integrations over the $(D-p)$-dimensional momenta $(k_0, {\bf k})$ yield
$(2 \pi^4 \alpha' t_c)^{-(D-p)/2}$ times Gaussian damping factor
$e^{-2\pi \alpha' W^2(m,n)/ t_c}$. We now perform modular
transformation to the open string channel $t_c = 1/t_o$, where $t_o$
is modulus of the open string channel and $q_o \equiv e^{- 2\pi
t_o}$. Putting all these together, we finally have
\begin{align} \overline{\cal N} =  C_p \, V_p
 \sum_{m,n=0}^\infty \int_0^\infty {\rmd t_o \over t_o} t_o^{-p/2}\, e^{- 2\pi
t_o \alpha' W^2(m,n) } \, \eta^{-(D-2)} (q_o), \label{openexp}\end{align}
with $C_p = \Big({2 \pi N_p\over \beta}\Big)^2 \frac{\pi
\alpha'}{2}(2\alpha'\pi^4)^{-\frac{D-p}{2}}$, reproducing the result
reported in \cite{Karczmarek:2003xm}. As it stands, the final expression
\eqref{openexp} is at odd to the intuition based on, for example, the
Schwinger pair production in (time-dependent) electric field, since
the integral over the open string modulus $t_o$ is still intact. If
the total emission number can be interpreted as arising from on-shell
two-particle branch cut in the open string channel, the modulus
integral ought to be absent! Therefore, To understand underlying
physics better, we shall now compute the cylinder amplitude directly
and then extract the imaginary part via the optical theorem.

\subsubsection{Lorentzian cylinder amplitude}\label{sec:5-2-3}
Unitarity and optical theorem thereof, combined with the open-closed
string channel duality, should enable us to extract the emission
number $\overline{\cal N}$ of closed strings from decaying D$p$-brane
as the imaginary part of the cylinder amplitude. In the closed
string channel diagram, the computation reduces to \eqref{closed}, as
in quantum field theory. It is, however, somewhat nontrivial to
evaluate the imaginary part of the cylinder amplitude directly from
the open string channel. Here we present the {\sl ab initio}
derivation, refining that in the text of \cite{Karczmarek:2003xm}, by starting
with manifestly well-defined Lorentzian cylinder amplitude.

We begin with the cylinder amplitude in the closed string channel in
which both the world-sheet and the target space-time signatures are
taken Lorentzian. Omitting overall numerical factors for the moment,
the amplitude is given by
\begin{align}
Z_{\rm cylinder} = i \pi \alpha' V_p \int_{s_c^{\rm IR}}^{s_c^{\rm
UV}} \rmd s_c \int_{-\infty}^{\infty} \frac{\rmd \omega_L}{2\pi} \,
\frac{\pi^2/\beta^2 \cdot
q_c^{-(1-i\hat{\epsilon})^2\frac{1}{4}\alpha'\omega_L^2}
}{\sinh(\pi\beta\omega_L \sqrt{\alpha'}) \sinh(\pi\omega_L
\sqrt{\alpha'}/\beta)} \, Z_{\cM}^{(c)}(q_c) \ , \label{cl}
\end{align}
where $q_c = e^{2\pi i \tau_c}$ with $\tau_c = s_c + i\epsilon$, and
$Z_{\cM}(q_c)$ represents the contribution from the closed string
zero-modes and oscillator parts.\footnote{We are using different
normalization for modulus parameters from \cite{Karczmarek:2003xm}: $t$(KLMS) =
$(\pi/4)t$(here). In addition, they adopted $\alpha'=1$
convention.} The Lorentzian world-sheet is regularized by $i
\epsilon$ prescription, while the Lorentzian space-time is
regularized by $-i \hat{\epsilon}$-prescription. $s_c^{\rm {UV}}$
($s_c^{{\rm IR}}$) is an ultraviolet (infrared) regulator of the closed
string channel modulus. With these prescriptions, the integral over
$\om_L$ is convergent so long as $2 \hat{\ep} s_c^{\rm UV} > \ep >0$
is retained.

Defining the open string modular parameter
as $q_o = e^{-2\pi i \tau_o}$ where
$\tau_o = s_o - i \epsilon$ with $s_o = 1/s_c$, one can rewrite
\eqref{cl} in terms of open string channel energy $\omega_L'$ as
\begin{align}
Z_{\rm cylinder} &=  V_p \int_{s_o^{\rm UV}}^{s_o^{\rm IR}} {\rmd
s_o} \int_{-\infty}^{\infty} \rmd \omega'_L \left(i \pi \alpha'
\int_{-\infty}^{\infty} \rmd \omega_L \frac{\cos(\pi \alpha'
\omega_L \omega'_L)}{\sinh(\pi\beta \omega_L \sqrt{\alpha'} )
\sinh(\pi\omega_L\sqrt{\alpha'}/\beta)} \right) \nonumber \\
& \hskip4cm \times \,
q_o^{-(1+i\hat{\epsilon}')^2\frac{1}{4}\alpha'{\omega_L'}^2}
Z_{\cM}^{(o)}(q_o) \, ,
\end{align}
where $s^{\rm IR}_o\equiv 1/s^{\rm UV}_c$, $s^{\rm UV}_o\equiv
1/s^{\rm IR}_c$ are the cut-off's in the open string modulus.
As opposed to the closed string channel, we have to adopt the
$+i\hat{\ep}'$-prescription for the Lorentzian space-time, and
the above integral is well-defined as long as
$2 \hat{\ep}' s^{\rm UV}_o > \ep$.
The expression in the large parenthesis yields the open string
density of states, $\rho^{(o)}(\omega'_L)$. It is infrared divergent
at $\omega_L= 0$. To regularize it, we subtract minimally the double
pole\footnote{This subtraction does not affect the imaginary part
of the partition function we are primarily interested in.} so that
\begin{align}
\rho^{(o)}(\omega'_L)_{\mathrm{reg}} &= i \pi \alpha'
\int_{-\infty}^{\infty} \rmd \omega_L \left(\frac{\cos(\pi\alpha'
\omega_L\omega'_L)}{\sinh(\pi\beta\omega_L \sqrt{\alpha'})
\sinh(\pi\omega_L \sqrt{\alpha'}/\beta)} - \frac{1}{\pi^2
\alpha'\omega_L^2} \right) \cr &= -{2}\partial_{\omega'_L} \log
S_\beta\left({Q}_\beta + i\sqrt{\alpha'}\omega'_L\right) \ , \label{dens}
\end{align}
where the `$q$-Gamma function' $S_\beta(x)$ is defined by\footnote
  {Here the normalization of variable $x$ differs with factor 2
   from the one given in \cite{FZZ}.}
\begin{align}
-\partial_x \log S_\beta (x) = \int_{-\infty}^{\infty} \rmd t
\left(\frac{\cosh((x-Q_\beta)t)}{2\sinh(\beta t)\sinh(t/\beta)}
-\frac{1}{2t^2} \right)\
\end{align}
for $\mathrm{Re}(x) < 2 Q_\beta$ and analytically continued to the
whole complex plane.\footnote{Notice that the Lorentzian density
\eqref{dens} is well-defined without the analytic continuation. We
stress that this should be contrasted against the approach of
\cite{Karczmarek:2003xm}.} See, for example, \cite{FZZ,Nakayama:2004vk}.

Now we perform the Wick-rotation both in the target space and on the
world-sheet. First, Wick rotate the open string channel energy as
$\om'_L\, \rightarrow\, e^{i(\frac{\pi}{2}-0)} \om'_L$ and set
$\om'_L = i \om'$ $(\om' \in \br)$. Then, we can safely Wick rotate
the world-sheet Schwinger parameter as $s_o \, \rightarrow \, - i
t_o$ ($t>0$). Notice that we will need to perform the Euclidean
rotation in opposite direction for the closed and the open string
channels due to the difference of the $i\ep$-prescription.
 There is no obstruction in such contour deformation because
 $\partial_x \log S_\beta (x) $ has poles only on the real axis.
 We will see that this is specific to the decaying D-brane situation
 and do not hold generally. In fact, in section \ref{sec:8-2}
 dealing with the rolling D-branes,
 we shall show that there exist extra contributions from crossing poles
 in the course of the contour rotation and that their contributions are
 essential for maintaining the unitarity.
 After Wick rotating the world-sheet,
the cylinder amplitude in the open string sector
 is given by
\begin{align}
Z_{\rm cylinder} = -2 V_p \int_{0}^\infty \rmd t_o
\int_{(1-i0)\mathbb{R}} \rmd \omega' \partial_{\omega'} \log S_\beta
\left({Q_\beta} -\sqrt{\alpha'}
\omega'\right)q_o^{\frac{1}{4}\alpha'{\omega'}^2} Z_M^{(o)}(q_o) \ .
\end{align}

Imaginary part of the partition function comes from the simple poles
of the $q$-Gamma function $S_\beta \left(Q_\beta -\sqrt{\alpha'}
\omega'\right)$ at $\frac{1}{2}\omega' = W(m,n)$
for $n,m \in \mathbb{Z}_{\ge 0} $ and simple zeros for $n,m \in
\mathbb{Z}_{< 0}$. Therefore, collecting imaginary parts from the
contour integration over $\omega'$ and applying the optical theorem,
we finally obtain
\begin{align}
\overline{\cal N} =  \mathrm{Im}\, Z_{\rm cylinder} = C_p \, V_p
\sum_{n,m=0}^\infty \int_0^\infty {\rmd t_o \over t_o}
t_o^{-\frac{p}{2}}\, e^{-2\pi t_o \alpha' W^2 (m,n)} \,
 \eta^{-(D-2)}(q_o) \ ,
\end{align}
where we have evaluated the free oscillator part explicitly and
reinstated overall numerical factors.
This is in perfect agreement with \eqref{dilaton}, and it may be
interpreted as a nontrivial check of unitarity and open-closed
duality in the Lorentzian signature.

\subsubsection{D-brane decay in two-dimensional string theory}\label{sec:5-2-4}
In a similar method, one can compute the spectral observables from
the D-brane decay in two-dimensional string theory \cite{Klebanov:2003km}. The boundary
state for the unstable D-brane in two-dimension is given by the
ZZ-brane boundary state \cite{Zamolodchikov:2001ah}:
\begin{align}
\langle e^{(i k + 2/\sqrt{\alpha'})\phi} \rangle_{\msc{disk}} =
\mu^{-\frac{i}{2} \sqrt{\alpha'} k} \frac{2
\sqrt{\pi}}{\Gamma(1-ik\sqrt{\alpha'})\Gamma(ik\sqrt{\alpha'})} \ .
\end{align}
Combining it with the rolling tachyon boundary states, the total
emission number of closed string is given by
\begin{align}
\overline{\cal N} = N_o^2 \int^{\infty}_{0} \rmd k \int^{\infty}_{0}
\frac{\rmd \omega}{2\omega}  {\cal P} (\omega, k) \delta(\omega-k) \
,
\end{align}
where the on-shell condition $\omega = k$ is imposed, and the
transition probability is
\begin{align}
{\cal P} (\omega, k)  = \left| \langle e^{- i\omega X^{0}} e^{
(ik+2/\sqrt{\alpha'})\phi} \rangle_{\msc{disk}} \right|^2 =
\frac{\sinh^2(\pi k\sqrt{\alpha'})}{\sinh^2(\pi \omega
\sqrt{\alpha'})} \ .
\end{align}
We see that, after performing the $k$-integration, the resultant
total emission number is ultraviolet divergent.

To express $\overline{\cal N}$ in open string channel, we repeat the
analysis of section \ref{sec:5-2-3} and expand the transition probability in
arrays of imaginary D-instantons. The result is
\begin{align}
\overline{\cal N} &= N_o^2 \sum_{m,n = 0}^\infty \int_0^\infty \rmd
k \int_0^\infty \rmd t_c \int_{-\infty}^{\infty} \frac{\rmd
k_0}{2\pi} e^{-2\pi t_c \cdot \frac{1}{4}\alpha' (k_0^2 + k^2)}
e^{2\pi i \alpha' k_0 W(m,n)} \sinh^2(\pi k\sqrt{\alpha'}) \Big\vert_{\beta \rightarrow 1} \cr
&= N_o^2 \sum_{m,n = 0}^\infty \int_{0}^\infty \frac{\rmd t_o}{t_o}
\Big({1 \over q_o} - 1 \Big) q_o^{ \alpha' W^2(m,n)}
\Big\vert_{\beta \rightarrow 1}
 \ , \label{twodim}
\end{align}
where we have reinstated $W(m,n)$ for the purpose of
regularization.\footnote{Because of the subtraction of singular
vector in $(1/q_o - 1)$, the resultant amplitude is {\it
non-unitary}.} The expression exhibits ultraviolet divergence as
$t_o \to \infty$.

On the other hand, it is possible to obtain the same radiation rate
from the direct evaluation of the imaginary part of the Lorentzian
cylinder amplitude in the open-string channel as was done in section
 \ref{sec:5-2-3}:
\begin{align}
Z_{\rm cylinder} = i N_o^2 \int_0^\infty \rmd s_c
\int_{-\infty}^{\infty} \frac{\rmd \omega_L}{2\pi} \int_0^{\infty}
\frac{\rmd k}{2\pi} \, \frac{\sinh(\pi \sqrt{\alpha'} k)^2
}{\sinh(\pi \sqrt{\alpha'} \omega_L)^2}\,
q_c^{\frac{1}{4}\alpha'(-\omega^2_L + k^2)}  \ .
\end{align}
After rewriting the open string density by the $q$-Gamma function as
in section \ref{sec:5-2-3}, we obtain open string channel expression of the
partition function. We then find the imaginary part from the poles
located at $\frac{1}{2}\omega' = W(m,n)$, and reproduce
\eqref{twodim}. This confirms that the partition function is
manifestly unitary, obeying the optical theorem. Here again, the
regularization $\beta \to 1$ is implicit.

\subsubsection{ZZ brane decay in various dimensions}\label{sec:5-2-5}
It is possible to generalize the discussion of section \ref{sec:5-2-4} for ZZ branes in various dimensions by introducing the time-like linear dilaton theory. If we write the dilaton slope of the Liouville and time-like linear dilaton direction as $V_{\phi} = b+{b}^{-1}$ and $V_{t} = \beta - \beta^{-1}$ respectively, the criticality condition is given by
\begin{align}
26 = D+6V_{\phi}^2 - V_{t}^2 \ . \label{crirr}
\end{align}
We can combine the one-point function for the ZZ brane (with general $b$) and the decaying D-brane boundary states for the boundary time-like Liouville theory to compute the radiation rate as was studied in \cite{He:2006bm}.
A similar cancellation as we discussed in section \ref{sec:5-2-1} gives the UV power-like structure of the closed string radiation rate.\footnote{Technically speaking, for $D>26$, we encounter a closed string IR divergence \cite{He:2006bm}.} The result suggests again the universality of the decaying D-brane spectrum.

In this construction, owing to the criticality condition \eqref{crirr}, the existence of the time-like dilaton is unavoidable. In the following, we study the decay of the ZZ brane in $\mathcal{N}=2$ Liouville theory (or D0-brane Euclidean $SL(2;\br)/U(1)$ coset model. See section \ref{sec:6} for more details) to study more realistic models, where $\beta = 1$ (i.e. flat limit) is feasible (see \cite{Israel:2006ip} for a particular case. This subsection is based on a generalization of their results).

The absolute square of the boundary wavefunction for the $\mathcal{N}=2$ ZZ-brane is given by
\begin{align}
|\Psi(p,m)|^2 = \delta_{m,\bar{m}} \frac{\sinh(2\pi p)\sinh(2\pi p/k)}{\cosh(2\pi p) + \cos(\pi m)} \ , \label{zzwv}
\end{align}
while that for the ($\mathcal{N}=1$ supersymmetric) rolling D-brane with time-like linear dilaton $V_{t} = \beta - \beta^{-1}$ is 
\begin{align}
|\Psi(E)^2| = \frac{1}{\sinh(\beta E) \sinh(\beta^{-1}E)} \ . \label{rrd}
\end{align}
We can evaluate the on-shell ($E^2 = M^2 + \frac{p^2}{2k} + \frac{m^2}{2k}$) emission with fixed transverse mass $M$ as 
\begin{align}
N(M) &= \int \dd p \sum_m |\Psi(p,m)^2||\Psi(E(p,m))|^2 \cr
     &\sim \int \dd p e^{\frac{\pi}{k}p - \pi(\beta + \beta^{-1})\sqrt{M^2+\frac{p^2}{2k}}} \cr
	& \sim e^{-2\pi M \sqrt{(\frac{\beta+\beta^{-1}}{2})^2 - \frac{1}{2k}}} \ ,
\end{align}
where in the last line we have used the saddle point approximation. 

On the other hand, the density of states for the emitted closed string for large $M$ is given by
\begin{align}
\sqrt{\rho^{(c)}} \sim e^{4\pi M \sqrt{\frac{c_{\mathrm{eff}}}{24}}} = e^{2\pi M \sqrt{(\frac{\beta+\beta^{-1}}{2})^2 - \frac{1}{2k}}} \ .
\end{align}
Thus, we see an exact cancellation of the exponential part of the closed emission rate, leaving us with a familiar power-like universal closed string emission rate.

We have several comments here
\begin{itemize}
	\item We can analyse the bosonic case in the same way. The first difference is $k$ in \eqref{zzwv} should be replaced with $\kappa-2$. The second difference is $E$ in \eqref{rrd} should be replaced with $\sqrt{2}E$. The final closed string emission rate changes, as a consequence, to $N(M) \sim e^{-2\pi \sqrt{(\beta+\beta^{-1})^2/2 - \frac{1}{2(\kappa-2)}}}$, which will cancel against the bosonic Hagedorn density of states $\sqrt{\rho^{(c)}} \sim e^{4\pi M \sqrt{\frac{c_{\mathrm{eff}}}{24}}} \sim e^{2\pi \sqrt{(\beta+\beta^{-1})^2/2 - \frac{1}{2(\kappa-2)}}}$.
	\item For simplicity, we studied the emission rate from the closed string perspective. The open string computation like we did in section \ref{sec:5-2-2}, \ref{sec:5-2-3} is straightforward, and we will not repeat it here.
	\item The conclusion here is independent of the level $k$ of the $SL(2;\br)/(1)$, which, on one hand, suggests a universality of the D-brane decay. On the other hand, it seems curious to observe that nothing special happens at $k=1$, where we expect a ``black hole - string transition". As we will see in section \ref{sec:7}, \ref{sec:8} in detail, the rolling (or Euclidean hairpin) D-brane captures or probes the ``black hole - string transition". We will return to this question in section \ref{sec:9}.
\end{itemize} 
\subsubsection{electric field and long string formation}\label{sec:5-2-6}
One simple generalization of the rolling D-brane was, as we studied in section 5.1.3, the inclusion of the linear dilaton. Another simple generalization is to introduce constant electric field on the D-brane, i.e. we introduce the fundamental string charge \cite{Mukhopadhyay:2002en,Rey:2003xs}.

In order to introduce the constant electric field (say $F^{01} = \epsilon$) on the D-brane, we can use the stringy version of the Lorentz boost. The successive applications of T-duality, Lorentz boost and inverse T-duality, we end up with the boundary states with electric flux. Operationally, the transformation is
\begin{align}
|0 \rangle \to \gamma |0\rangle \ , \ \  t\to \gamma^{-1} t \ , \ \ \omega \to \gamma \omega  \cr
\begin{pmatrix} \alpha^0 \\ \alpha^1 \end{pmatrix} \to \Lambda^{-1} \begin{pmatrix} \alpha^0 \\ \alpha^1 \end{pmatrix} \ , \ \ \ \begin{pmatrix} \bar{\alpha}^0 \\ \bar{\alpha}^1 \end{pmatrix} \to \Lambda \begin{pmatrix} \bar{\alpha}^0 \\ \bar{\alpha}^1 \end{pmatrix} \ , 
\end{align}
where 
\begin{align}
\Lambda = \gamma \begin{pmatrix}  1& \epsilon \\ \epsilon & 1 \end{pmatrix} \ , \ \ \gamma = \frac{1}{\sqrt{1-\epsilon^2}} \ .
\end{align}

From this transformation law, the energy momentum tensor can be easily read as
\begin{align}
T_{00} & \sim E \gamma \cr
T_{01} & \sim -Ee^2\gamma -E \gamma^{-1} \exp(-\gamma^{-1}t) \cr
T_{11} & \sim -E \gamma^{-1} \exp(-\gamma^{-1}t) \ ,
\end{align}
in the $t\to \infty$ limit. The study of the closed string radiation from the boundary states is straightforward. When $x^1$ direction is noncompact, the result is 
\begin{align}
\langle N \rangle = \sum_M \int \dd k \frac{|\Psi(\omega_{k,M})|^2}{2\omega_{k,M}} \simeq \int^{\infty} \dd M \sqrt{\rho^{(c)}(M)}e^{-2\pi \gamma M} = \int^{\infty}\dd M e^{-2\pi (\gamma-1) M}
\end{align}
and the total emission rate is exponentially suppressed essentially due to the Lorentz time delay \cite{Nagami:2003yz,Nagami:2003mr}.

Now let us suppose that $x^1$ direction is compactified with the radius $R$. In this case, we have to sum over the winding mode:
\begin{align}
N(M) = \sum_w \int \dd k \frac{|\Psi(\omega_{k,M})|^2}{2\omega_{k,M}} \simeq \sum_w\int \dd k e^{- 2\pi\gamma (\sqrt{(wR)^2 + k^2 + M^2}-\epsilon Rw)} \ .
\end{align}
For large $M$, the summation over $w$ can be evaluated by the saddle point methods, which leads to 
\begin{align}
\langle N \rangle 
\sim \int^\infty \dd M \sqrt{\rho^{(c)}(M)} N(M) \sim \int^\infty \dd M M^{\beta}
\end{align}
We recover the power-like behavior of the emission rate \cite{Gutperle:2004be}.\footnote{The power dependence $\beta$ is determined from the details of the model e.g. dimensionality of the D-brane and the details of the internal CFT etc.} The computation reveals that the winding mode dominates the emission rate in the electrified D-brane decay. Mathematically, this is due to the (T-dualized) Lorentz invariance in the $R\to 0$ limit. Physically, the decay of the D-brane produces many long macroscopic strings as a final decay product, which has a cosmological significance as we will review in section \ref{sec:5-5}.

\subsection{Classical correspondence}\label{sec:5-3}
The Dirac-Born-Infeld form of the rolling tachyon effective action \eqref{tDBI} suggests a possible geometrical interpretation of the open string tachyon condensation. Such a geometrical interpretation of the rolling tachyon process would shed a new light upon our understanding of the nature of the open string tachyon and its condensation. It would also provide a guiding principle for a geometrical interpretation of the closed string tachyon condensation, for qualitative properties of the closed string tachyon condensation are poorly understood compared with the open string tachyon condensation.

In \cite{Kutasov:2004dj}, an interesting connection between the D-brane motion in the (near horizon) NS5-brane background and the rolling tachyon dynamics was pointed out. Since the NS5-brane has a tension proportional to $1/g_s^2$, in perturbative string theories, we can regard it as a fixed background, in which the D-brane, whose tension is proportional to $1/g_s$ moves. In other words, in the perturbative string theories, the probe D-brane approximation is good and trustful.

The effective action for the D-brane motion in NSNS-background (i.e. without any R-R fields), is given by the Dirac-Born-Infeld action
\begin{align}
S = -T_p\int \dd^{p+1} \sigma e^{-\Phi}\sqrt{-\det(X^*[G+B]_{\mu\nu})} \ ,
\end{align}
where $X^*[G+B]$ denotes the pullback to the D$p$-brane world-volume.

As proposed in \cite{Kutasov:2004dj}, let us consider the D0-brane motion in near horizon NS5-brane geometry \eqref{NH ext NS5}.\footnote{If one considers a homogeneous motion of the D-brane, the net result does not depend on the spacial dimension of the D-brane. We also assume that D-brane sits at a point in the internal space $\mathbb{S}^3$.} Let us fix the world-sheet reparametrization invariance by taking the static gauge $\sigma^0 = t$. In this gauge, the DBI action reduces to
\begin{align}
S = -T_0 \int \dd t e^{\frac{\rho}{\sqrt{2k}}}\sqrt{1-\dot{\rho}^2} \ , \label{DBIr}
\end{align}
where dot denotes the derivative with respect to $\sigma_0 = t$, and we have rescaled the radial direction $\rho$ so that we have a canonical kinetic term.

Let us compare the effective action for the radion field $\rho$ \eqref{DBIr} with the open string tachyon effective action \eqref{tDBI}. It is almost clear in the large (negative) region of $\rho$, these two expressions essentially coincide with each other.\footnote{If one take $k=2$, the coincidence becomes exact including the numerical factor in the tachyon (radion) potential.} This is the classical ``tachyon - radion correspondence": one can identify the effective action for the rolling tachyon problem with the effective action for the rolling D-brane in the NS5-brane, or linear dilaton, background. The ``radion field" $\rho$ plays the role of the tachyon field $T$ here.
Note, however, that the radion field is actually not tachyonic, although it has run-away potential, nor has an unstable extremum in the potential because it is a massless field at the tree level. 

One can readily solve the classical equation of motion based on the action \eqref{DBIr} as
\begin{align}
e^{-\frac{\rho}{\sqrt{2k}}} = c \cosh\left(\frac{t}{\sqrt{2k}}\right) \ , \label{seoms}
\end{align}
which agrees with the late time behavior of the rolling tachyon problem \eqref{seom}. The energy momentum tensor can be read as 
\begin{align}
T_{00} &= E\delta(\rho-\rho_0(t)) \cr
T_{0\rho} &= E \tanh\left(\frac{t}{\sqrt{2k}}\right) \delta(\rho-\rho_0(t)) \cr
T_{ij} &= -E \mathrm{sech^2}\left(\frac{t}{\sqrt{2k}}\right) \delta(\rho-\rho_0(t)) \delta_{ij} \ \ \ (i,j = 1,\cdots,p) \ ,
\end{align}
where $\rho_0(t)$ is the classical solution of the radion motion \eqref{seoms}. As expected, the energy momentum tends to that for a pressure-less dust as $t\to \infty$. The $(0\rho)$ component has a natural interpretation as the momentum transfer in the $\rho$ direction because the decaying D-brane moves in the $\rho$ direction almost at the speed of light as $t\to \infty$.

What is the end point of the ``radion condensation"? In the case of the open string tachyon condensation, Sen's conjecture states that we end up with the closed string vacuum, where the open string excitation becomes infinitely massive and disappear from the physical spectrum. From the effective field theory approach taken here, it is difficult to establish this statement in a satisfactory manner because in the large $\rho$ regime, the effective string coupling becomes larger due to the linear dilaton gradient. One way to study this might be to uplift the system to M-theory (e.g. by using the interpolating metric proposed in \cite{Aharony:1998ub}). The subsequent physics, however, is intuitively clear: the D-brane will be absorbed into the NS5-brane and form a non-threshold bound states. The open string spectrum on the D-brane should be modified so that it matches with the excitation on the bound states.\footnote{It would be an interesting open problem to study the tachyon - radion correspondence from the open string field theory and prove the analogue of Sen's conjecture.}

There are several generalizations of the problem. One interesting question is whether we can obtain the effective DBI action having the exactly identical potential with the rolling tachyon not only in the large $\rho$ region. This is possible by considering an array of the NS5-brane on $\br^3\times \mathbb{S}^1$ rather than the stack of NS5-branes in $\br^4$ \cite{Kutasov:2004ct}. Because of the oppositely-directed attractive force between two NS5-branes, the potential of the D-brane can have a local extremum:
\begin{align}
S = -T_0 \int \dd t \frac{1}{\cosh\frac{\rho}{\sqrt{2k}}}\sqrt{1-\dot{\rho}^2} \ ,\label{aDBIr}
\end{align}
which completely agrees with \eqref{tDBI}.
Unfortunately, unlike the NS5-branes on $\br^4$, the exact quantization of the rolling D-brane in this geometry is unavailable.\footnote{The exact boundary states for {\it static} (unstable) brane in a similar background has been constructed in \cite{Eguchi:2004ik}, which reproduces the mass of the geometrical tachyon (i.e. radion).}

Another interesting generalization is to consider the D-brane motion in the non-extremal black NS5-brane background. Interestingly, after a simple coordinate transformation, the classical motion of D-brane in the non-extremal NS5-brane (outside of the horizon) is identical to that in the extremal NS5-brane.
To see this we note that, by introducing `tachyon' variable $Y \equiv \log \sinh \rho$,
DBI Lagrangian of the D0-brane can be cast to that of rolling
tachyon:
\begin{align} L_{\rm D0} &= - e^{-\Phi} \sqrt{\left({\dd s \over \dd
t}\right)^2} = - V(Y) \sqrt{1 - \dot{Y}^2} \qquad \mbox{where}
\qquad
 V(Y) = M_0 \, e^Y ~, \label{nonextremal D0} \end{align}
if we restricted ourselves to the region outside of the horizon.
An important point is that,
in sharp contrast to the extremal background \eqref{NH ext NS5}, the
dilaton is finite everywhere. Thus, the strong coupling singularity
is now capped off by the horizon. The construction of the exact boundary states for the rolling D-brane in the two-dimensional black hole (or non-extremal NS5-brane) is one of the main themes of this thesis.

For another example of exactly solvable deformation, one can introduce constant electric fields as we did in the rolling tachyon example. This has been studied in \cite{Nakayama:2004ge}, where we have constructed exact boundary states and have shown the correspondence between the electrified rolling tachyon problem and the electrified rolling radion problem even with $\alpha' \sim 1/k$ corrections. As yet another generalization, the rotating D-brane solution in NS5-brane background has been also studied in \cite{Kutasov:2004dj}, which could be regarded as a rotational Lorentz boosted solution as pointed out in \cite{Nakayama:2004ge}, but the exact boundary state is yet to be constructed. Other classical studies of D-brane motion in related background include \cite{Yavartanoo:2004wb,Panigrahi:2004qr,Ghodsi:2004wn,Sahakyan:2004cq,Toumbas:2004fe,Bak:2004tp,Chen:2004vw,Kluson:2005qx,Lapan:2005qz,Thomas:2005am,Thomas:2005fw,Kluson:2005dr,Kluson:2005eb,Kluson:2005zw,Papantonopoulos:2006eg,Okuyama:2006zr,Gumjudpai:2006hg}.

Before concluding this subsection, we would like to stress again that the correspondence at the level of the effective action is only valid in the large $\rho$ or $T$ regime, where the effective action analysis loses its validity because the effective string coupling grows there. Therefore, the quantum correspondence we will prove in later sections, based on the one-loop string perturbation theory, is actually not so obvious, and we should rather regard it as a highly nontrivial statement of the universality of the properties of decaying D-branes.

\subsection{Quantum correspondence}\label{sec:5-4}

So far, we have mainly discussed the classical correspondence between the rolling tachyon problem and the rolling radion problem at the level of the effective action. Aside from the debate over the effectiveness of the rolling tachyon DBI-like action \eqref{tDBI} , we have one tunable parameter $k$ in the rolling radion problem, so it is important to analyse a possible $k$ dependence of this correspondence. 

We know that $1/k$ measures the $\alpha'$ corrections to the background geometry from the discussion in section \ref{sec:2}. When $k$ becomes larger, the classical geometry, and hence, the DBI action is more trustful. On the other hand, when $k$ becomes smaller, the geometry shows large $\alpha'$ corrections and the effective action approach may break down. Especially, the exact correspondence at the level of the effective action requires $k=2$, which is rather in a strongly coupled regime.\footnote{In the bosonic case, we need to set $k=1$.} 

In particular, if one considers the two-dimensional black hole geometry (as the non-extremal NS5-branes background), the appearance of the stretched horizon blurs the geometry. In addition, we expect a ``black hole - string transition" at $k=1$. It is of utmost interest to probe such a phase transition from the rolling D-brane. 

In the following sections, we will construct the exact boundary states for such rolling D-branes in NS5-brane background, and reveal the nature of the $1/k \sim \alpha'$ corrections to the tachyon - radion correspondence. After the construction of the exact boundary states, we study the closed string radiation rate as we did in the rolling tachyon case in section \ref{sec:5-2} and compare the results. 

For convenience, we summarize our main physical results here \cite{Nakayama:2005pk}:

\begin{enumerate}
	\item The closed string emission rate from the rolling D-brane (which will be computed in section \ref{sec:8-2}) yields exactly the same behavior as that from the rolling tachyon (which was computed in section \ref{sec:5-2}). Especially, the power-like behavior of the spectrum density does not depend on $k$ (up to an overall normalization). This is true as long as $k>1$, and confirms the tachyon - radion corresponding from the exact boundary states.
	\item Independence of the extra parameter $k$, which even governs the world-sheet (stringy) $\alpha'$ correction suggests a universal nature of the decaying D-brane: all the energy of the D-brane will be radiated as a gas of closed strings, whose dominant contribution comes from the highly massive (long) strings. If one introduces the fundamental string charge, as an electric flux, the dominant contributions from the rolling D-brane again comes from the winding strings as we have seen in the rolling tachyon problem in section \ref{sec:5-2-6}.
	\item The situation changes drastically if one studies the case $k<1$. The closed string emission rate is exponentially suppressed, and the tachyon - radion correspondence breaks down. This is in accord with the ``black hole - string transition" at $k=1$ discussed in section \ref{sec:4}. Our result is the first physical manifestation of the ``black hole - string transition" in the two-dimensional black hole probed by the rolling D-brane. 
\end{enumerate}
\subsection{Cosmological implications}\label{sec:5-5}

From the early days of its invention, the rolling tachyon system has also been studied in the context of the cosmological applications. In particular, the realization of the inflation in string theory has attracted more and more attention recently with increasing evidence for the existence of such period in the history of our universe (see \cite{Linde:2005dd} and references therein). Indeed, one of the simplest proposals for the inflation from the string theory is the tachyon inflation, where the (open string) tachyon plays the role of the inflaton \cite{Kofman:2002rh,Frolov:2002rr,Li:2002et,Fairbairn:2002yp,Shiu:2002qe}. The tachyon - radion correspondence discussed so far enables us to consider varieties of radion (or geometrical tachyon) inflation. From the classical tachyon - radion correspondence, many features of the tachyon inflation can be translated into that of the radion inflation with more generalities \cite{Thomas:2005fu}.

The starting point of the tachyon (radion) inflation is (minimal) coupling of the DBI-like action \eqref{tDBI} \eqref{DBIr} to the gravity:
\begin{align}
L_{\mathrm{eff}} = \sqrt{-g} \left(\frac{R}{16\pi G} - V(T)\sqrt{1+g^{\mu\nu}\partial_\mu T \partial_\nu T} \right) \ . \label{frwl}
\end{align}
For our realistic application, we consider the four-dimensional (non-compact) space-time, and $8\pi G = M_p^{-2}$ with the four-dimensional Planck constant $M_p$. Under the assumption of the Friedman-Robertson-Walker isotropic universe, the four dimensional metric can be written as\footnote{Since the inflation flattens the space in an exponential manner, we have assumed a flat space universe for simplicity.}
\begin{align}
\dd s^2 = -\dd t^2 + a(t)^2 \dd x^2_i  \ \ \ (i = 1,2,3) \ .
\end{align}

We begin with the equation of motion for $T$:
\begin{align}
\frac{\ddot{T}}{1-\dot{T}^2} + 3H\dot{T} + \frac{V'}{V} = 0 \ ,
\end{align}
where the prime denotes the derivative with respect to $T$ (i.e. $V' = \partial V(T)/\partial T$) and the dot denotes the time derivative. $H$ here denotes the Hubble parameter $H \equiv \dot{a}/{a}$.
In the slow-roll approximation (i.e. $\dot{T}\gg 1$), the Friedman equation reads
\begin{align}
H^2 = \left(\frac{\dot{a}}{a}\right)^2 = \frac{V(T)}{3M_p^2} \ ,
\end{align}
and the slow-roll equation reduces to
\begin{align}
3H\dot{T} = \frac{V'(T)}{V} \ .
\end{align}
For the slow-roll parameter $\eta \sim (H')^2/H^4$ to be small enough, we must require
\begin{align}
H^2 \gg \frac{(V')^2}{V^2} \ .
\end{align}

Suppose our (geometrical) tachyon potential has a local extremum as is the case with the rolling tachyon and the geometrical tachyon in the array of NS5-brane backgrounds. Inflation near the local extremum is possible if $H^2 \gg |m^2|$, where $m^2$ is the mass for the (geometric) tachyon. The condition is equivalent to
\begin{align}
 \frac{g_s}{v l_s} \gg \frac{C}{k l_s} \ , \label{gom}
\end{align}
where $v$ is the volume of the compactification,\footnote{We are assuming a direct product type compactification. If we consider the warped compactification, we can relax the condition.} and if we kept track of every numerical factor, we could find $C\sim 260$. In the original rolling tachyon problem, $k = 2$ and it is difficult to find a consistent solution in the perturbative string theory while maintaining the COBE normalization $H/M_p \sim 10^{-5}$ \cite{Kofman:2002rh,Frolov:2002rr}. In the geometric tachyon, we have one parameter $k$, and if we choose large enough $k$, it is possible to satisfy the condition \eqref{gom} consistent with the COBE normalization. We can also satisfy the slow-roll condition in the geometric tachyon. For instance, $\eta \ll 1$ is equivalent to the condition $H \gg |m|$ for $T < O(1)$. The key point here is that we have an extra tunable parameter $k$ to obtain a sustainable inflation in the case of the geometric tachyon unlike the original rolling tachyon cosmology, where such a tunable parameter is absent.

Nevertheless, we still have a serious drawback of this rolling tachyon (radion) type cosmology as pointed out in \cite{Kofman:2002rh,Frolov:2002rr,Shiu:2002qe}. The problem is related to how the inflation will end. Since the effective potential for the rolling tachyon (radion) runs away exponentially as $T \to \infty$, there is no minimum for the tachyon to oscillate. Therefore, it is allegedly impossible to reheat the universe to produce various matters, i.e. after the tachyon inflation we end up with an empty universe, which is of course unacceptable.

Again in the contex of the geometric tachyon (radion), we can avoid this reheating problem by preparing a ring of the NS5-branes and evolution of the D-brane inside the ring \cite{Thomas:2005fu}. The effective potential has global minima and the oscillation around the minima produces the reheating needed to produce matter and hence our galaxies. 

We can continue this line of reasoning and study for instance the spectral index of the cosmic microwave background etc, but this is not the main scope of this thesis. Rather, we would like to point out how inaccurate this kind of effective action analysis for the rolling D-brane  is {\it after coupling to the massive closed string sector}. Especially, the stringy treatment of the decay of the D-brane completely changes the nature of the reheating from such rolling tachyon (or D-brane) systems.

Let us begin with the following illustrative toy example. In the above discussions, couplings of the (decaying or rolling) D-brane to higher massive stringy modes (except for graviton) have been neglected. In the usual field theory, we expect corrections to the effective action of order $\sim M_{p}^2/M^2$, where $M^2$ denotes the mass of the fields integrated out. The point is that we have to sum over infinitely many massive fields: e.g. in the Kaluza-Klein theory, $M^2 \propto n^2$, where $n$ denotes the internal momenta, so the summation over $n$ schematically gives 
\begin{align}
\sum_n \frac{1}{M_n^2} \sim \sum_n \frac{1}{n^2} = \frac{\pi^2}{6} \ ,
\end{align}
which is finite. However, in string theory, the number of massive string modes grows exponentially $\rho(M) \sim \exp(\beta_{\rm Hg} M)$ as we discussed the Hagedorn temperature in section \ref{sec:4-1}. Thus the summation over all the massive modes with the coupling $\sim 1/M_s^2$ clearly diverges.

In reality, the coupling to the massive closed string sectors is much softer and the exponentially suppressed as $\exp(-\beta M)$. The case-by-case computation is needed to see  which exponential factor governs, but from our results (summarized in section \ref{sec:5-5}) it seems universal that the exponential part cancels out and the closed string backreaction is characterized by a power-like behavior irrespective of the superficial strength of the $\alpha' \sim 1/k$ corrections.\footnote{It is also interesting to note that the exponential suppression of the higher massive modes occurs when $k<1$ in the regime where the supergravity approximation is invalid.}

In this way, we can conclude that the reheating of the universe through the rolling tachyon - radion is rather effective than one might expect from the naively truncated effective action. As the direct calculation shows, almost all of the energy is radiated as the closed strings without any need for the oscillation around the extremum.\footnote{As we have discussed in section, \ref{sec:5-2}, the emission rate is power-like finite for higher dimensional branes. However, this does not mean an effectiveness of the truncated effective action (DBI+FRW such as \eqref{frwl}) to discuss the closed string backreaction. It  just means that it is more effective to decay by disconnecting patches of D-brane as D0 particles (assuming it is uncharged). Mathematically, it is just an artefact of the one-point decay and the one-point decay is no more effective than the higher-point decay.} The actual problem, therefore, is how we can transmit energy from the radiated (massive) closed string to the standard model sector. This problem is rather model dependent and we will not study it any further in detail here (see e.g. \cite{Kofman:2005yz,Frey:2005jk,Chialva:2005zy,Chen:2006ni} for recent studies).

As we have studied in section \ref{sec:5-2}, the final decay product of the rolling tachyon and rolling D-brane is highly probable to be long closed strings. This can be directly seen by assigning fundamental string charges to the unstable D-branes, but without assigning such charges, intuitively it is expected to be so by considering a pair production. This is in good agreement with the  usual Kibble mechanism of producing long macroscopic strings in the universe: causally disconnected region creates long strings and they evolve independently. It would be of great interest to study this problem quantitatively from the string theory viewpoint and determine a remaining density of cosmic strings associated with the D-brane decay (see \cite{Polchinski:2004ia} for a review of cosmic strings from the superstring theory). Such studies will verify or even exclude the geometric tachyon inflation. It is also of great importance to revisit the reheating process of various D-brane inflation scenarios to see whether the classical oscillatory contribution is really dominant over the emission of the highly massive string modes.

\newpage
\sectiono{D-branes in Two-dimensional Black Hole}\label{sec:6}
We now begin with our studies on D-branes in the two-dimensional black hole background. In this section, we review the D-branes in the Euclidean two-dimensional black hole. The organization of the section is as follows. In section \ref{sec:6-1}, we classically analyze the D-branes in the two-dimensional black hole and derive the mini-superspace boundary wavefunction. In section \ref{sec:6-2}, we review the exact boundary states describing the D-branes in the two-dimensional black hole system. 
\subsection{Classical D-branes}\label{sec:6-1}
\subsubsection{DBI analysis}\label{sec:6-1-1}
The classification of the D-brane in general curved backgrounds is given by the solution of the Dirac-Born-Infeld action coupled with Chern-Simons action.\footnote{Of course, one could imagine unstable D-branes whose effective action is {\it not} given by the DBI action + Chern-Simons, but they are outside the scope of our discussion.} The total effective action is
\begin{align}
S = -\mu_p \int \dd^{p+1}\xi e^{-\Phi}\sqrt{-\det(G_{ab}+B_{ab}+F_{ab})} +i\mu_p\int e^{F_2+B_2} \wedge \sum_q C_q \ , \label{DBICS}
\end{align}
where the summation over $C_q$ should be taken over all R-R fields in the theory we are considering. 

In this section, we study the D-branes in the Euclidean two-dimensional black hole:
\begin{align}
\dd s^2 =  k\al' (\tanh^2\rho \, \dd \theta^2 + \dd
\rho^2) \ , 
\qquad e^{2\Phi} = \frac{k}{\mu \cosh^2 \rho} \ .
\end{align}
In the Euclidean two-dimensional black hole background,
 there exist no Kalb-Ramond $B_{\mu\nu}$ field nor the R-R fields, so the effective action is simply given by the DBI term with possible electro-magnetic flux $F_{\mu\nu}$ on it. Since we are in the Euclidean signature, the DBI action \eqref{DBICS} should be Wick-rotated in an appropriate manner:\footnote{Our ``Wick rotation" here is nothing but adding a (dummy) extra decoupling time direction and set trivial Neumann boundary condition along the time. In particular, we will assume $F + B$ is real.}
\begin{align}
S^E= \mu_p \int \dd^{p}\xi e^{-\Phi}\sqrt{\det(G_{ab}+B_{ab}+F_{ab})} \ .
\end{align}

We begin with the (Euclidean) D0-brane. The D0-brane is a point particle and the DBI action on it is simply given by
\begin{align}
S_{\mathrm{D}0} = \mu_0 e^{-\Phi} \propto \cosh\rho \ .
\end{align}
It is clear that the extremum of the action is obtained when $\rho=0$. Thus we conclude that the D0-brane is localized at the tip of the cigar.

Next we study the D1-brane. The DBI action for the D1-brane is given by
\begin{align}
S_{\mathrm{D}1} = \mu_1 \int \dd\theta \cosh\rho(\theta) \sqrt{\rho'(\theta)^2 + \tanh^2\rho(\theta)} \ ,
\end{align}
where we have fixed the reparametrization invariance by using the gauge $\xi = \theta$. The equation of motion is easily solved from the ``energy" conservation:
\begin{align}
\text{const} = \cosh\rho \frac{\tanh^2{\rho}}{\sqrt{(\rho')^2 + \tanh^2\rho}} \ 
\end{align}
as
\begin{align}
\sinh(\rho) \cos(\theta -\theta_0) = \sinh\rho_0 \ . \label{trsce}
\end{align}

For later purposes, we note that if one uses the complex coordinate
\begin{align}
 u = \sinh\rho e^{i\theta} \ , \ \ \bar{u} = \sinh\rho e^{-i\theta} \ ,
\end{align}
the classical trajectory \eqref{trsce} takes the form of a straight line in the complex  $u$ plane. This can be also seen from the fact that the DBI action takes the flat form
\begin{align}
S_{D1} = \mu_1 \int \dd\xi \sqrt{\frac{du}{d\xi}\frac{d\bar{u}}{d\xi}} \ 
\end{align}
in this coordinate.

We finally examine the D2-brane. In this case, we can introduce a magnetic flux $F_{\rho\theta} = f(\rho,\theta)$. By fixing the reparametrization invariance as $\xi_1 = \rho$, $\xi_2 =\theta$, the DBI action reads
\begin{align}
S_{\mathrm{D}2} = \mu_2 \int d\theta d\rho \cosh\rho \sqrt{\tanh^2\rho + f^2(\rho,\theta)} \ .
\end{align}
From the Gauss law constraint, we have
\begin{align}
 c = \frac{\cosh\rho f(\rho,\theta)}{\sqrt{\tanh^2\rho + f^2(\rho,\theta)}} \ ,
\end{align}
which determines the magnetic flux as 
\begin{align}
f^2(\rho,\theta) = \frac{c^2\tanh^2\rho}{c^2 - \cosh^2\rho} \ . \label{magfl}
\end{align}
If $c>1$, the D2-brane partially wraps the cigar and has a boundary at $\rho = \mathrm{arccosh}(c)$ because at that value of $\rho$, the magnetic field blows up. On the other hand, if $c<1$, the D2-brane wraps the whole cigar. In the latter case, the magnetic field on the D2-brane induces a D0-brane charge near the tip of the cigar, which should be quantized. Writing $c = \sin\sigma \le 1$, we obtain the classical quantization condition as 
\begin{align}
\frac{\sigma - \sigma'}{2\pi} k \in \bz \ .
\end{align}

In quite a similar fashion, we can also study the classical D-branes in the T-dualized trumpet background:
\begin{align}
\dd s^2 = \dd\rho^2 + \frac{1}{\tanh^2\rho} \dd\tilde{\theta}^2 \ , \ \ e^{\Phi} = \frac{k}{\mu \sinh \rho} \ .
\end{align}
 Since the discussion is completely in parallel, we only present the results. 
 
 The D0-brane (probably D1-brane?) could be localized at $\rho = 0$. Since $\rho=0$ is a singularity in the trumpet geometry, the presence of such D-branes are not obvious at all. Formally, we can regard it as a T-dual of the D0-brane of the cigar geometry.
 
 The D1-brane is given by the solution of the DBI action
 \begin{align}
S_{\mathrm{D}1} = \mu_{\mathrm{D}1} \int \dd\tilde{\theta}\sinh\rho\sqrt{\frac{1}{\tanh^2\rho}+({\rho}')^2} \ ,
 \end{align}
 in the static gauge.
The solution is given by 
\begin{align}
\cosh\rho \cos(\tilde{\theta}-\tilde{\theta}_0) = \gamma \ .
\end{align}
when $\gamma>1$, the D1-brane is connected, while when $\gamma<1$, the D1-branes go through the singularity and possibly they become disconnected. Naturally, the D1-brane in the trumpet geometry is regarded as a T-dual of the D2-brane in the cigar geometry. The parameter $\gamma$ corresponds to the parameter $c$ in the cigar geometry.\footnote{The parameter $\tilde{\theta}_0$ can be T-dualized to the holonomy of the gauge field $A_0$ in the cigar. Since the D2-brane has a nontrivial fundamental group $\pi_1 = \mathbb{Z}$, different $A_0$ gives a different D-brane (for $c>1$).}

The D2-brane is classified by the solution of the DBI action
\begin{align}
S_{D2} = \mu_{D2}\int \dd\rho \dd\tilde{\theta} \sinh\rho \sqrt{\frac{1}{\tanh^2\rho} + F^2} \ .
\end{align}
The Gauss law constraint gives 
\begin{align}
 F^2 = \frac{\beta^2}{\tanh^2\rho(\sinh^2\rho-\beta^2)} \ .
\end{align}
The D2-brane always has a boundary at $\rho = \mathrm{arcsinh}(\beta)$. The D2-brane in the trumpet geometry naturally corresponds to the T-dual of the D1-brane in the cigar. The parameter identification is obviously given by $\beta = \sinh\rho_0$ appearing in \eqref{trsce}.

\subsubsection{group theoretical viewpoint}\label{sec:6-1-2}
In section \ref{sec:6-1-1}, we have studied the classical D-brane in the Euclidean two-dimensional black hole from the effective DBI action. Since the two-dimensional black hole system can be realized as the $SL(2;\br)/U(1)$ coset model, we can study the classification of the D-brane from the gauged WZNW model \cite{Maldacena:2001ky,Gawedzki:2001ye,Elitzur:2001qd,Fredenhagen:2001kw,Ishikawa:2001zu,Yogendran:2004dm}. Indeed all the D-branes discussed in section \ref{sec:6-1-1} descend from the branes in the parent $SL(2;\br)$ WZNW model. 

The starting point is the D-branes in the parent $SL(2;\br)$ WZNW model. We focus on the maximally symmetric D-branes for technical simplicity. As we proceed, we will see that the maximally symmetric D-branes are enough to obtain all the D-branes constructed from the DBI analysis done in section \ref{sec:6-1-1}. The maximally symmetric D-branes in the WZNW model are classified by the (twined) conjugacy class of the group $G$ with a possible quantization condition \cite{Kato:1996nu,Alekseev:1998mc,Birke:1999ik,Behrend:1999bn,Felder:1999ka}. We call them A-branes (conjugacy class) and B-branes (twined conjugacy class) respectively.

In our $SL(2;\br)$ group with the Euler angle parametrization $g = e^{i\sigma_2 \frac{t-\theta}{2}} e^{\rho\sigma_1}e^{i\sigma_2\frac{t+\theta}{2}}$ , the conjugacy class is given by 
\begin{align}
\mathrm{Tr} (g) = 2 \cos{t} \cosh{\rho} \equiv 2\kappa , \label{conj}
\end{align}
and the twined conjugacy class is given by
\begin{align}
\mathrm{Tr} (\sigma_1 g) = 2\cos{\theta} \sinh{\rho} \equiv 2\kappa' \ . \label{tconj}
\end{align}
up to conjugation.

The D-branes in the axial coset model is obtained by gauging $g$ by $hgh$, where $h^a= e^{i\sigma_2 a}$ in our case. For A-brane, we have to sum over the gauge orbit of the parent D-brane \label{conj} parametrized by $\kappa$ in order to obtain a gauge invariant object. The gauge transformation of the conjugacy class is given by
\begin{align}
\mathrm{Tr}(h^agh^a) = 2\cos(t+a) \cosh{\rho} ,
\end{align}
so the gauge invariant orbit of the parent D-brane is given by
\begin{align}
\cosh{\rho} \ge \kappa  \ . \label{d2wzw}
\end{align}
Projecting it down on the coset coordinate (in the gauge $t=0$) is now trivial, and we have obtained the D2-brane wrapped (partially) around the cigar whose world volume is restricted by the condition \eqref{d2wzw}. 
The shapes of the A-branes obtained here are in complete agreement with the ones obtained from the DBI analysis in section \ref{sec:6-1-1}. The precise parameter identification is $c=\kappa$ for the D2-brane.\footnote{There are several independent ways to justify this parameter identification. For instance, one can show it directly from  the detailed study on the boundary conditions of the gauged WZNW model with boundaries. In \cite{Walton:2002db}, they have shown that the parameter $\kappa$ is indeed the field strength appearing in the effective action of the D-brane at the boundary by using the T-duality technique. Their study of the $SU(2)/U(1)$ model can be translated to our Euclidean $SL(2;\br)/U(1)$ model with no essential modifications.}  We also note that A-brane is invariant under the isometry of the coset in this construction.

Similarly from the parent B-brane, we can construct the D1-brane of the coset. In this case, since the twined conjugacy class is already gauge invariant, we can directly project \eqref{tconj} down onto the coset coordinate. The resulting D1-brane trajectory is given by
\begin{align}
 \sinh{\rho}\cos{\theta} = \kappa' \ ,
\end{align}
which is nothing but the one obtained in \eqref{trsce} from the DBI analysis (with $\theta_0 =0$). The B-brane constructed in this way breaks the isometry of the coset, so it has a Nambu-Goldstone mode along the $\theta$ direction. This corresponds to the rotation of $\sigma_1$ and $\sigma_3$ in the definition of the 
 twined conjugacy class \eqref{tconj}.

We could repeat the same analysis for the vector coset ($\sim$ trumpet geometry). Since the argument is completely in parallel, we skip the detailed discussion, and simply note that the results agree with the DBI analysis.
 
\subsubsection{mini-superspace boundary wavefunction}\label{sec:6-1-3}
In the context of the string theory, D-branes can be described either from the open string viewpoint or from the closed string viewpoint (i.e. channel duality). Technically, this is achieved by the modular transformation of the cylinder amplitudes. The boundary state $|B\rangle$ is defined by
\begin{align}
Z_{\mathrm{cylinder}} \equiv \int \dd t_o \mathrm{Tr}_o e^{-\pi H_o t_o} = \int \frac{\dd t_c}{t_c} \langle B| e^{-\pi H_ct_c}| B\rangle \ ,
\end{align}
where $H_o = L_0$ is the open string Hamiltonian while $H_c = L_0 + \bar{L}_0$ is the closed string Hamiltonian. The boundary state $|B\rangle$ satisfies the gluing condition 
\begin{align}
(L_{n} -  \bar{L}_{-n}) |B \rangle = 0 , 
\end{align}
for the energy-momentum tensor (and similar gluing conditions for any other conserved currents if any: see section \ref{sec:6-2} for further details).

At the level of the minisuperspace approximation, the boundary states can be seen as the coupling of the D-brane to the closed string zero mode:
\begin{align}
\langle B |_{\mathrm{mini}} = \int_0^\infty \frac{\dd p}{2\pi} \Psi_{0}(p,n) \langle\langle p,n| \ , \label{miniwv}
\end{align}
where $|p,n\rangle\rangle$ is the so-called Ishibashi state \cite{Ishibashi:1988kg} associated with the primary states $|p,n\rangle$ (see section \ref{sec:6-2} for details), but in the mini-superspace approximation, there is no difference between the two $|p,n\rangle\rangle \sim |p,n\rangle$ because they are different only in the non-zero mode sector. The semiclassical boundary wavefunction $\Psi_{0} (p,n)$ is obtained from the overwrap between the D-brane and the primary state $|p,n\rangle$ as
\begin{align}
\Psi_0(p,n) = \langle B|p,n\rangle \ . 
\end{align}
The explicit form of the primary state $|p,n\rangle$ and the classical trajectory have been given in the form of the minisuperspace approximation as we have studied in section \ref{sec:3-2-3} and section \ref{sec:6-1-1}.

In the following, we compute the minisuperspace boundary wavefunction $\Psi_0$ for each D-branes studied in section \ref{sec:6-1-1}. The results will be compared with the proposed exact boundary states in section \ref{sec:6-2}. We expect that they will agree with each other in the semi-classical limit ($k \to \infty$), and indeed they do as we will see.

Let us begin with the D0-brane. Classically, the D0-brane is localized at the tip of the cigar $\rho = 0$, and the boundary wavefunction is simply given by the minisuperspace wavefunction for $|p,n\rangle$ evaluated at $\rho = 0$. From the explicit minisuperspace wavefunction \eqref{ef}, we can easily derive
\begin{align}
\Psi_0^{\mathrm{D}0}(p,n) = - \delta_{n,0} \frac{\Gamma^2(-j)}{\Gamma(-2j-1)} = -\delta_{n,0} \frac{\Gamma^2(\frac{1}{2}-\frac{ip}{2})}{\Gamma(-ip)} \ . \label{minico}
\end{align}
We note that D0-brane does not couple to the momentum mode along $\theta$, which is consistent with the interpretation that the D0-brane is an A-brane (see section \ref{sec:6-1-2}). The exact analysis shows that it couples to the winding mode and the discrete states localized near the tip of the cigar, but the minisuperspace analysis cannot capture them.

Next let us consider the D1-brane. Classically, the D1-brane has the shape of the hairpin. 
 A semiclassical D-brane boundary wavefunction is the weighted sum of the
wavefunction of closed string states restricted to the location of
the D-brane. In the mini-superspace approximation, as is implicit in
\cite{Ribault:2003ss}, the weighted sum equals to the overlap between the
mini-superspace wavefunction and the delta function constraint
enforcing $(\rho, \theta)$ coordinates over the hairpin trajectory
\eqref{trsce} (with respect to the volume element \eqref{vol cigar}).
The result is
\begin{align}
& \int_0^\infty \sinh \! \rho \, \dd \sinh \! \rho
\int_{-\frac{\pi}{2}+\theta_0}^{\frac{\pi}{2}+\theta_0} \dd\theta\,
\delta \Big(\cos (\theta-\theta_0) \sinh \rho-\sinh \rho_0 \Big)
\phi^p_{n}(\rho,\theta) \cr &= \int_{-\frac{\pi}{2}}^{\frac{\pi}{2}}
\!\dd \theta' \, \frac{\sinh \rho_0}{\cos^2 \theta'}
\phi^p_{n}(\widehat{\rho}(\rho_0,\theta'),\theta') e^{i n \theta_0}
, 
\end{align}
where $\theta'= (\theta-\theta_0)$ and
$\widehat{\rho}(\rho_0,\theta')$ refers to the solution of $\cos
\theta' \sinh \rho= \sinh \rho_0$. Using the decomposition
\eqref{decomp ef}, we are then to evaluate integrals:
\begin{align}
 & \int_{-\frac{\pi}{2}}^{\frac{\pi}{2}} \dd \theta \,
\frac{\sinh \rho_0}{\cos^2 \theta} \,
\phi^p_{L,n}(\hat{\rho}(\rho_0,\theta),\theta) =
\frac{2\pi\Gamma(ip)} {\Gamma\left(\frac{1}{2}+\frac{ip+n}{2}\right)
\Gamma\left(\frac{1}{2}+\frac{ip-n}{2}\right)} \, e^{-ip\rho_0} ~,
\nn
 & \int_{-\frac{\pi}{2}}^{\frac{\pi}{2}} \dd \theta \,
\frac{\sinh \rho_0}{\cos^2 \theta} \,
\phi^p_{R,n}(\hat{\rho}(\rho_0,\theta),\theta) =
\frac{2\pi\Gamma(-ip)}
{\Gamma\left(\frac{1}{2}-\frac{ip+n}{2}\right)
\Gamma\left(\frac{1}{2}-\frac{ip-n}{2}\right)} \, e^{+ ip\rho_0} ~.
\label{evaluation overlap phi}
\end{align}
Details of the computation are relegated in Appendix \ref{mini}. Using the
mini-superspace reflection amplitude \eqref{cref amp}, we then obtain
\begin{align}
&\Psi^{(0)}_{\rm D1}(\rho_0,\theta_0;p,n) =  \frac{2\pi\Gamma(ip)}
{\Gamma\left(\frac{1}{2}+\frac{ip+n}{2}\right)
\Gamma\left(\frac{1}{2}+\frac{ip-n}{2}\right)} \, e^{in\theta_0}
\left(e^{-ip\rho_0} + (-1)^n e^{+ip\rho_0}\right)~. \label{hairpin
D1 classical}
\end{align}
The D1-brane couples to the momentum mode as is clear from the geometry, which is consistent with the interpretation that the D1-brane is a B-brane (see section \ref{sec:6-1-2}).


Finally, we study the D2-brane. The D2-brane is parametrized by the parameter $c$ appearing in the amount of the magnetic flux \eqref{magfl}. Since the qualitative features of the D2-brane seem to be different for $c>1$, and $c<1$, it is natural to study them separately. Since the mini-superspace analysis for the D2-brane has not been available in the literature, we would like to present it slightly in detail here.\footnote{The author would like to thank S.~Ribault for stimulating discussions on this problem.}

Let us begin with the case when $c>1$. The D2-brane only partially wraps the cigar because at $\rho = \mathrm{arccosh}(c)$, the field strength diverges.
We parametrize $c = \cosh r_0$. 

Since the D2-brane couples to the winding states, the minisuperspace analysis is only possible for the zero winding sector.\footnote{We could avoid this problem in the T-dual picture, which will be discussed later.} The (zero momentum/winding) minisuperspace wavefunction is given by 
\begin{align}
\phi_{p,m=0}(\rho) &= -\frac{\Gamma^2(-j)}{\Gamma(-2j-1)}F(j+1,-j;1;-\sinh^2\rho) \cr &= (\sinh\rho)^{-1-ip} F\left(\frac{1}{2} + \frac{ip}{2}, \frac{1}{2} + \frac{ip}{2}; 1+ ip;-\frac{1}{\sinh^2\rho}\right) \cr &+ \frac{\Gamma(ip)\Gamma^2(\frac{1}{2}-\frac{ip}{2})}{\Gamma(-ip)\Gamma^2(\frac{1}{2}+\frac{ip}{2})}(\sinh\rho)^{-1+ip} F\left(\frac{1}{2} - \frac{ip}{2}, \frac{1}{2} - \frac{ip}{2}; 1- ip;-\frac{1}{\sinh^2\rho}\right) \ ,
\end{align}
where $j = -\frac{1}{2} + \frac{ip}{2}$.

The boundary wavefunction in the minisuperspace approximation is given by
\begin{align}
\Psi_{2}(r_0)^{\mathrm{mini}} = \int_{r_0}^{\infty} \dd\rho \cosh\rho \frac{\sinh\rho}{\sqrt{\cosh^2\rho-\cosh^2r_0}} \phi_p(\rho) \ .
\end{align}
Now we can perform the integration as follows
\begin{align}
& \int_{r_0}^\infty \dd\rho \cosh\rho  \frac{\sinh\rho}{\sqrt{\cosh^2\rho-\cosh^2r_0}} (\sinh\rho)^{-1-ip} F\left(\frac{1}{2}+\frac{ip}{2},\frac{1}{2}+\frac{ip}{2};1+ip;-\frac{1}{\sinh^2\rho}\right) \cr
&= \frac{\Gamma(ip+1)}{\Gamma(\frac{1}{2}+\frac{ip}{2})^2} \sum_{n=0}^{\infty} (-1)^n (\sinh^2 r_0)^{-n-\frac{ip}{2}}\frac{\sqrt{\pi}}{2}\frac{\Gamma(n+\frac{ip}{2})\Gamma(\frac{1}{2}+\frac{ip}{2}+n)}{\Gamma(ip+1+n)n!} \cr
&= \frac{\Gamma(\frac{ip}{2})}{\Gamma(\frac{1}{2}+\frac{ip}{2})} (\sinh^2 r_0)^{-\frac{ip}{2}} \frac{\sqrt{\pi}}{2} F\left(\frac{ip}{2},\frac{1}{2}+\frac{ip}{2};ip+1,-\frac{1}{\sinh^2 r_0}\right) \cr
&= \frac{\pi\Gamma(ip)}{\Gamma(\frac{1}{2}+\frac{ip}{2})^2} e^{-ipr_0} \ . 
\end{align}
We refer to the appendix \ref{mini} for the last equality (see also \cite{Nakayama:2005pk}).
Combining it with the second integration that can be treated in the same manner, we obtain 
\begin{align}
\Psi_2(r_0)^{\mathrm{mini}} = \frac{\pi\Gamma(ip)}{\Gamma(\frac{1}{2}+\frac{ip}{2})^2} \cos(pr_0) \ . \label{direce}
\end{align}
We can see that the boundary wavefunction for the partially wrapped D2-brane is consistent with the class 2 boundary wavefunction proposed in \cite{Fotopoulos:2004ut} at least for the zero winding sector (see section \ref{sec:6-2} for details).

We can repeat our analysis when $\beta \le 1$ and reproduces the minisuperspace limit of the class 3 boundary states in the zero-winding sector.
In the T-dual picture, (partially wrapped) D2-brane in the cigar geometry is supposed to be given by the D1-brane in the trumpet geometry.\footnote{One subtle point of the trumpet geometry is that the semiclassical limit is unclear. We regard $k\to \infty$ as the semiclassical limit for $\rho$ direction, but $\tilde{\theta}$ direction is apparently not. We will neglect this subtlety for a moment.}

Let us now move on to the minisuperspace wavefunction for the D1-brane in the trumpet geometry.
 From the semiclassical DBI action
\begin{align}
L = \sinh\rho \sqrt{\dot{\rho}^2 + k^{-2} \coth^2\rho} \ ,
\end{align}
the equation of motion is easily integrated with the help of the energy conservation:
\begin{align}
L - \dot{\rho} \frac{\partial L}{\partial \dot{\rho}} = \text{const} \ ,
\end{align}
and one can see that the classical D1-brane is described by the trajectory
\begin{align}
\cosh \rho = \frac{\cosh r_0}{\cos \left[(\tilde{\theta} - \tilde{\theta}_0)/k\right]} \ .
\end{align}
The appearance of the $1/k$ in the argument of the cosine is important. If $k$ is an even integer, the asymptotic form of the D-brane trajectory is given by the coincident two-branes, while for an odd integer $k$, it is given by the parallel two-branes placed at the anti-podal points in $\mathbb{S}^1$.\footnote{For general $k$, asymptotic position of the two branes breaks the (discrete) periodic symmetry.}

The semiclassical boundary wavefunction is obtained by integrating the classical closed string wavefunction over the classical D-brane trajectory as
\begin{align}
\Psi_{2}(\tilde{\theta}_0,r_0)^{\mathrm{mini}} &= e^{iw\tilde{\theta}_0}\int_1^{\infty} \cosh\rho \dd(\cosh\rho) \int_{-\frac{k\pi}{2}}^{\frac{k\pi}{2}} \dd\tilde{\theta} \delta(\cos(\tilde{\theta}/k)\cosh\rho-\cosh r_0) \phi_{p,w}(\rho,\tilde{\theta}) \cr
&= \int_{-\frac{k\pi}{2}}^{\frac{k\pi}{2}} \dd\tilde{\theta} \frac{\cosh r_0}{\cos^2(\tilde{\theta}/k)} \phi_{p,w}(\hat{\rho}(r_0,\tilde{\theta}),\tilde{\theta}) \ ,
\end{align}
where $\hat{\rho}(r_0,\tilde{\theta})$ is the solution of $\cos (\tilde{\theta}/k) \cosh \rho = \cosh r_0 $. The integration is feasible due to the formula
\begin{align}
\int_{-\frac{k\pi}{2}}^{\frac{k\pi}{2}} \dd\tilde{\theta} \frac{\cosh r_0}{\cos^2(\tilde{\theta}/k)} e^{iw\tilde{\theta}} (\cosh\rho)^{-1-ip} F\left(\frac{1}{2}-\frac{kw}{2}+\frac{ip}{2},\frac{1}{2}+\frac{kw}{2}+ \frac{ip}{2};1+ip;\frac{\cos^2(\tilde{\theta}/k)}{\cosh^2r_0}\right)\ \cr
= \frac{2\pi \Gamma(ip)}{\Gamma(\frac{1}{2}+\frac{ip}{2}+\frac{kw}{2})\Gamma(\frac{1}{2}+\frac{ip}{2}-\frac{kw}{2})} e^{-ip r_0} \ , \label{evaluation overlap phi2}
\end{align}
whose derivation is relegated to the appendix B (see also \cite{Nakayama:2005pk}).

In this way, we derive the minisuperspace limit of the boundary wavefunction that describes the D1-brane in the trumpet geometry:
\begin{align}
\Psi_{2}(\tilde{\theta}_0,r_0)^{\mathrm{mini}} = N(b)\frac{\Gamma(2j+1)}{\Gamma(1+j+\frac{kw}{2})\Gamma(1+j-\frac{kw}{2})} e^{iw \tilde{\theta}_0} \cos(r_0(2j+1)) \ , \label{cltrum}
\end{align}
where $j = -\frac{1}{2} + \frac{ip}{2}$. This should be identified with the boundary wavefunction for the partially wrapped D2-brane via the T-duality. Note that the for zero winding sector $w=0$, the wavefunction agrees with the direct evaluation \eqref{direce}.

In a similar manner, the classical boundary wavefunction for the totally wrapped D2-brane is given by
\begin{align}
\Psi_{3}(\sigma,\theta_0) = \Gamma(2j+1)e^{i\theta_0\omega} \left[\frac{\Gamma(-j+\frac{kw}{2})}{\Gamma(j+1+\frac{kw}{2})}e^{i\sigma(2j+1)} + \frac{\Gamma(-j-\frac{kw}{2})}{\Gamma(j+1-\frac{kw}{2})}e^{-i\sigma(2j+1)} \right] \ . \label{clth}
\end{align}
The parameters $r_0$ and $\sigma$ are supposedly related to the magnetic flux on the D2-brane:
\begin{align}
F = \frac{\beta^2 \tanh^2\rho}{\cosh^2\rho - \beta^2} d\theta d\rho \ ,
\end{align}
where $\beta = \sin \sigma $ for $\beta \le 1$ and $\beta = \cosh r_0$ for $\beta \ge 1$. We expect that these two classes of branes coincide in the limit $\beta = 1$ (i.e. $\sigma = \pm \frac{\pi}{2}$ and $r_0 = 0$).

From the geometry of the semiclassical D-brane, we expect that if one takes a suitable limit of the boundary states, the class 2 D-brane (partially wrapped D-brane) will coincide with the class 3 D-brane (totally wrapped D-brane).
To compare these two branes, we can directly show that 
\begin{align}
\Psi_3(\pi/2,-k\pi/2) + \Psi_3(\pi/2,k\pi/2) = \Psi_{2}(0,0) \ .
\end{align}
This result also shows that the boundary wavefunction \eqref{clth} describes the half cut D2-brane.

\subsubsection{embedding into NS5-branes}\label{sec:6-1-4}
We have obtained the classical D-brane solutions in the (Euclidean) two-dimensional black hole background. As we have reviewed in section \ref{sec:2}, we can embed the two-dimensional black hole into the superstring theory as NS5-branes (or more generally little string theories on singular Calabi-Yau spaces). Here we would like to summarize some of the D-brane solutions in the NS5-brane background to see how one can construct them from those in the two-dimensional black hole system \cite{Elitzur:2000pq,Lerche:2000uy,Gava:2001gv,Eguchi:2003ik,Eguchi:2004yi,Eguchi:2004ik,Israel:2005fn}.

Let us first concentrate on the D1-brane solution in the ring-likely separated NS5-brane solution \eqref{rmet} corresponding to 
\begin{align}
\frac{\Big[ {SL(2;\br)_{k} \over U(1)} \times \frac{SU(2)_{k}}{U(1)} \Big]_\perp}{\bz_k} \ .
\end{align}
Naturally, we can combine various D-branes in the $SL(2;\br)/U(1)$ coset and $SU(2)/U(1)$ coset to construct D-branes in this background. We further focus on the two-plane $x^8=x^9 = 0$, setting $\theta =0$.

\begin{figure}[htbp]
   \begin{center}
    \includegraphics[width=0.5\linewidth,keepaspectratio,clip]{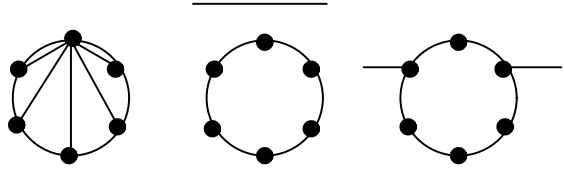}
    \end{center}
    \caption{The left figure shows W-bosons in LST. The central figure shows an uncut D1-brane. The right figure shows cut D1-branes attached to the NS5-branes.}
    \label{fig:nsbranes}
\end{figure}

The first combination is the D0-brane in the cigar and the D1-brane in the bell. The result is the D1-brane stretching between NS5-branes as in figure \ref{fig:nsbranes}. In the context of the LST, we interpret them as W-bosons. 
The second combination is the (uncut) D1-brane in the trumpet and D0-brane in the bell. The resulting geometry is the straight line on the $x^8=x^9=0$ plane as in figure \ref{fig:nsbranes}.
The third combination is the cut D1-brane in the trumpet and D0-brane in the bell. It corresponds to the semi-infinite D-brane attached to the NS5-branes as in figure \ref{fig:nsbranes}.

The geometries of the D3-brane are much more complicated. We would like to refer to \cite{Ribault:2003sg,Israel:2005fn} for detailed study of the D3-brane geometries in the NS5-brane background.

Recently, a static D-brane configuration in the black hole background has attracted much attention for a possible application to the phase transition of the fundamental matters in QCD \cite{Mateos:2006nu}. In our two-dimensional black hole setup, it amounts to the study of the D-brane in the black NS5-brane background. In the Rindler limit studied in \cite{Mateos:2006nu}, the difference between the black NS5-brane and the black D-brane does not exist. It would be interesting to study the exact boundary states for the D-branes in the black NS5-brane background to probe the $\alpha'$ corrections to the phase transition discussed there.

\subsection{Exact boundary states}\label{sec:6-2}
\subsubsection{Ishibashi states}\label{sec:6-2-1}
To construct the exact Cardy boundary states for the D-branes in the two-dimensional black hole background, we begin with the Ishibashi states. For definiteness, we first concentrate on the bosonic axial coset, which is given by the Euclidean cigar.

The coset Ishibashi states naturally descend from those for the parent current algebra. The Ishibashi state satisfies the boundary condition
\begin{align}
(L_{n} - \bar{L}_{-n}) |A \rangle \rangle &= 0  \cr 
(J_{n} -\bar{J}_{-n}) |A \rangle \rangle &= 0  \ ,
\end{align}
 for A-brane and
\begin{align}
(L_{n} - \bar{L}_{-n}) |B \rangle \rangle &= 0  \cr 
(J_{n} + \bar{J}_{-n}) |B \rangle \rangle &= 0  \ ,
\end{align}
 for B-brane. In terms of the primary states of the coset, A-boundary condition means $m = \bar{m} = \frac{k\omega}{2}$, and B-boundary condition means $m = -\bar{m} = \frac{n}{2}$. Physically, the A-branes couple to the winding states while B-brane couple to the momentum states in the coset.

The Ishibashi state is naturally endowed with the classification via the character of the coset model. For continuous series, we have the following normalization 
\begin{align}
_B\langle \langle p',n'|e^{-\pi t(L_0 + \bar{L}_0)}| p,n \rangle \rangle_B = \left[\delta(p-p') + R(p,n)\delta(p+p')\right]\delta_{n,n'}\frac{q^{-\frac{p^2}{4(\kappa-2)}+\frac{n^2}{4k}}}{\eta(\tau)^2} \cr
_A\langle \langle p',\omega'|e^{-\pi t(L_0 + \bar{L}_0)}| p,\omega \rangle \rangle_A = \left[\delta(p-p') + R(p,\omega)\delta(p+p')\right]\delta_{\omega,\omega'}\frac{q^{-\frac{p^2}{4(\kappa-2)}+\frac{\omega^2}{4}}}{\eta(\tau)^2} \ ,
\end{align}
where the subscript denote the boundary condition (either A-type or B-type), and $R(p,n)$ (or $R(p,\omega))$ denote the reflection amplitude. The Ishibashi state is parametrized by the radial momentum $p$ and the angular momentum $n$ (or the winding number $\omega$).


For the supersymmetric coset, we impose the following boundary condition for the Ishibashi states:
\begin{align}
(L_n - \bar{L}_{-n}) | A \rangle \rangle = 0 \cr
(G^{\pm}_r - i\bar{G}^{\mp}_{-r}) | A \rangle \rangle = 0 \cr
(J_n - \bar{J}_{-n}) |A \rangle \rangle = 0 \ ,
\end{align}
for A-type boundary conditions, and
\begin{align}
(L_n - \bar{L}_{-n}) | B \rangle \rangle = 0 \cr
(G^{\pm}_r - i\bar{G}^{\pm}_{-r}) | B \rangle \rangle = 0 \cr
(J_n + \bar{J}_{-n}) |B \rangle \rangle = 0 \ ,
\end{align}
for B-type boundary conditions. Both types of the boundary conditions are compatible with the diagonal $\mathcal{N}=1$ superconformal symmetry
\begin{align}
(G_r - i \bar{G}_{-r})|A \ \mathrm{or} \ B\rangle\rangle \ ,
\end{align}
where $G_r = G^+_r + G^-_r$ that should be gauged in the fermionic string theory. Physically, A-type boundary condition corresponds to Dirichlet boundary condition along the cigar angular direction, and B-type boundary condition corresponds to Neumann boundary condition. 

The Ishibashi states for the supersymmetric coset for continuous series is parametrized by three quantum number $(p,m,s)$. Our normalization is
\begin{align}
_A\langle\langle p',\omega',s'| e^{-\pi\tau_c(L_0+\bar{L}_0)} e^{i\pi y(J_0+\bar{J_0})}|p,\omega,s\rangle \rangle_A \cr
= \delta_{\omega',\omega}(\delta(p-p')+\delta(p+p')R(j,\frac{k\omega}{2},\frac{k\omega}{2})) \mathrm{ch}_{j,\frac{k\omega}{2},s}(i\tau_c,y) \ ,
\end{align}
for the A-brane, and 
\begin{align}
_B\langle\langle p',n',s'| e^{-\pi\tau_c(L_0+\bar{L}_0)} e^{i\pi y(J_0+\bar{J_0})}|p,n,s\rangle \rangle_B \cr
= \delta_{n',n}(\delta(p-p')+\delta(p+p')R(j,\frac{n}{2},-\frac{n}{2})) \mathrm{ch}_{j,\frac{n}{2},s}(i\tau_c,y) \ ,
\end{align}
for the B-brane. Here $s$ denotes the spectral flow parameter. Note that the boundary condition demands $m=\bar{m}=\frac{k\omega}{2}$ for the A-brane and $m=-\bar{m}= \frac{n}{2}$ for the B-brane. The $\mathcal{N}=2$ character $\mathrm{ch}_{j,m,s}$ is defined as
\begin{align}
\mathrm{ch}_{j,m,s}(\tau,y) = q^{\frac{p^2}{4k}+\frac{(m+s)^2}{k}+\frac{s^2}{2}}z^{\frac{2m}{k}+s} \frac{\theta_3(\tau,y)}{\eta(\tau)^3} \ ,
\end{align}
for NS sector ($z=e^{2\pi i y}$).

\subsubsection{exact boundary wavefunction}\label{sec:6-2-2}
Let us first summarize the exact boundary wavefunction for the D-branes whose classical properties we discussed in section \ref{sec:6-1}. We relegate a (partial) derivation of the exact boundary wavefunctions based on the modular bootstrap to section \ref{sec:6-2-3}.

For the bosonic two-dimensional black hole, we expand the Cardy boundary states
as
\begin{align}
\langle B | = \int \dd p \sum_m \Psi(p,m) \langle\langle p,m | + (\text{discrete}) \ .
\end{align}
 Compared with the minisuperspace approximation \eqref{miniwv}, we have allowed the winding states (for B-brane) and a possible discrete state contribution. In the following, we focus on the continuous part. The discrete part can be read from the analytic continuation of the boundary wavefunction $\Psi(p,m)$ with respect to the parameters of the continuous series restricted to the value corresponding to the discrete series (i.e. $\Psi(j=m,m)$).

The exact boundary wavefunction for the D0-brane (class 1 A-type brane) is given by
\begin{align}
\Psi_{\mathrm{D}0}(j,\omega) = \nu_b^{2j+1} \frac{\Gamma(-j+\frac{k\omega}{2})\Gamma(-j-\frac{k\omega}{2})}{\Gamma(-2j-1)\Gamma(1-b^2(2j+1))} \ , \label{exco}
\end{align}
where $b = (k-2)^{-1/2}$, and $\nu_b = \frac{\Gamma(1-b^2)}{\Gamma(1+b^2)}$. It is easy to see that the exact boundary wavefunction \eqref{exco} reduces to the mini-superspace result \eqref{minico} in the large $k$ limit by setting $\omega = 0$ up to a $p$ independent overall normalization factor. The exact boundary state for the D0-brane couples to winding states. It also couples to the discrete series localized near the tip of the cigar.

The exact boundary wavefunction for the D1-brane (class 2' B-type brane) is given by
\begin{align}
\Psi_{\mathrm{D}1}(j,n)^{r,\theta_0} =   \nu_b^{2j+1} e^{in\theta_0} \frac{\Gamma(2j+1)\Gamma(1+b^2(2j+1))}{\Gamma(1+j+\frac{n}{2})\Gamma(1+j-\frac{n}{2})} (e^{-r(2j+1)}+(-1)^n e^{r(2j+1)}) \ . \label{clss2'}
\end{align}
The D1-brane only couples to the momentum states. In particular, it does not couple to any discrete states. In the classical limit $k\to \infty$, the boundary wavefunction reproduces that of the minisuperspace approximation \eqref{hairpin
D1 classical}.

The exact boundary wavefunction for the partially wrapped D2-brane (class 2 A-type brane) is given by
\begin{align}
\Psi_{\mathrm{D}2}(j,\omega)^{r_0,\tilde{\theta}_0} = \nu_b^{2j+1} \frac{\Gamma(2j+1) \Gamma(1+b^2(2j+1))}{\Gamma(1+j+\frac{kw}{2})\Gamma(1+j-\frac{kw}{2})} e^{iw \tilde{\theta}_0} \cos(r_0(2j+1)) \ . \label{exbt}
\end{align}
It does not couple to the discrete states localized at the tip of the cigar as is expected from the geometry. We can readily see that the classical limit ($k\to \infty$) of \eqref{exbt} reduces to the minisuperspace wavefunction \eqref{cltrum}. 

Finally, the exact boundary wavefunction of the totally wrapped D2-brane (class 3 A-type brane) is given by
\begin{align}
\Psi_{\mathrm{D}2'}(j,\omega)^{\sigma,{\theta}_0} &= \nu_b^{2j+1} \Gamma(1+b^2(2j+1))\Gamma(2j+1)e^{i\theta_0\omega}  \times \cr 
&\times \left[\frac{\Gamma(-j+\frac{kw}{2})}{\Gamma(j+1+\frac{kw}{2})}e^{i\sigma(2j+1)} + \frac{\Gamma(-j-\frac{kw}{2})}{\Gamma(j+1-\frac{kw}{2})}e^{-i\sigma(2j+1)} \right] \ ,\label{exbthr}
\end{align}
with the relative quantization condition $\sigma-\sigma' = 2\pi\frac{m}{k-2}$ , $m \in \bz$.

The other possible exact boundary states for the two-dimensional black hole have been proposed in the literatures \cite{Ahn:2004qb,Hosomichi:2004ph,Ribault:2005pq}. However, they do not possess sensible open string spectra\footnote{Most of them contain tachyon in their spectra. Furthermore, they often have imaginary conformal weights when we study overlaps with class 2 (class 2') branes.} nor corresponding semiclassical limits, so we will not discuss them in detail here. Their properties are in many sense similar to the generalized ZZ branes proposed in \cite{Zamolodchikov:2001ah}. As such they could be important so as to understand the nonperturbative contributions to the partition function of the two-dimensional Euclidean black hole.

The boundary wavefunctions for the supersymmetric two-dimensional black hole are essentially the same as those for the bosonic one. In the NS sector, the only difference is to replace $b^2 = \frac{1}{k-2}$ with $\frac{1}{k}$. The boundary wavefunction for the other sector is obtained by the spectral flow.
\subsubsection{Cardy condition and modular bootstrap}\label{sec:6-2-3}
There are several different ways to derive the boundary wavefunctions for the D-branes in the $SL(2;\br)/U(1)$ coset model. One of the simplest ways to obtain them is to descend them from the branes in the parent $SL(2;\br)$  WZNW model (or $\mathbb{H}^3_+$ model). This method has a small drawback for our purposes because we should derive the boundary states for $SL(2;\br)$ WZNW model (or $\mathbb{H}_3^+$ model) first \cite{Ponsot:2001gt}. In this section, we take another root, which uses the so-called ``modular bootstrap" method to derive all the A-branes in the Euclidean two-dimensional black hole.\footnote{As we will see, we cannot derive the boundary wavefunctions for B-branes in this approach. We need a more technically involved strategy such as the conformal bootstrap to derive them.}

The modular bootstrap method is intimately related to the Cardy condition for boundary states \cite{Cardy:1989ir}. The Cardy condition is the physical constraint on the open string spectra between two different D-branes. Let us denote the boundary states for any pair of these two branes as $|a\rangle$ and $|b\rangle$. The Cardy condition says that the open string spectra between these two D-branes should have open string characters with positive multiplicities:
\begin{align}
Z_{a,b} = \mathrm{Tr}_{a,b} q^{L_o} = \sum_i n^i_{a,b}\chi_i(q) \ , \label{cardyc}
\end{align}
where $\chi_i(q)$ is the open string character, and $n^i_{a,b}$ should be positive integers from the unitarity of the theory.\footnote{Implicitly here we are assuming that the open string between $|a\rangle$ and $|b\rangle$ are bosonic. Otherwise the negative multiplicity is allowed as fermions. For NS-NS overlap, we expect that the overlap should contain bosonic excitations.} Now we modular transform the open string character $\tau \to -1/\tau$ in order to obtain the closed string description:
\begin{align}
\langle a|e^{i\tau \pi H_c}| b\rangle = Z_{ab}(q) = \sum_{i,j} n^{i}_{a,b}S_{ij} \chi_j(\tilde{q}) \ ,
\end{align}
with $\tilde{q} \equiv e^{-2\pi i/\tau}$, where $S_{ij}$ is the modular $S$-matrix for characters $\chi_i$. 

At this point, it is not immediately obvious whether the multiplicities $n^{i}_{a,b}$ are all positive integers if one introduces an arbitrary set of boundary states $|a\rangle$. This integrality condition is the Cardy condition for boundary states. The Cardy condition guarantees a physical interpretation of the open string spectra from the open-closed duality.

Actually, there is a canonical solution of the Cardy condition based on the Verlinde formula \cite{Verlinde:1988sn}. We assume that the boundary state is a superposition of the Ishibashi states with normalization
\begin{align}
\langle\langle i|e^{i\tau \pi H_c}|j\rangle\rangle = \delta_{ij}\chi_i(q) \ .
\end{align}
We then assume the existence of the simplest boundary states (identity brane) $|\hat{0}\rangle$ whose self overlap gives an identity representation in the open string sector: $n^i_{\hat{0},\hat{0}} = \delta^i_0$. Since the fusion of the identity operator is itself, we have a relation
\begin{align}
|\langle \hat{0}|j\rangle|^2 = S_{0,j} \ . 
\end{align}
As a consequence, the state $|\hat{0}\rangle$ can be written as
\begin{align}
|\hat{0}\rangle = \sum_j\sqrt{S_{0,j}}|j\rangle\rangle \ , \label{sob}
\end{align}
up to an overall phase factor (possibly dependent on $j$). 

Now we {\it define} Cardy boundary states as
\begin{align}
|a\rangle = \sum_j \frac{S_{aj}}{\sqrt{S_{0j}}}|j \rangle\rangle \label{sobb}
\end{align}
which contains the open string spectrum $n^{i}_{\hat{0}a} = \delta_a^i $ in the overlap with the identity brane. These Cardy states satisfy the Cardy condition
 \eqref{cardyc} 
\begin{align}
\langle a|j \rangle \langle j| b \rangle = \frac{S_{aj}S_{jb}}{S_{0j}} = \sum_i S_{ij} n^{i}_{ab} \ ,
\end{align}
where $n^{i}_{ab}$ is given by the Fusion coefficient $\mathcal{N}^{i}_{ab}$ that is a positive integral matrix. The last equality is due to a remarkable identity under the name of the Verlinde formula. The Verlinde formula can be shown for unitary compact CFTs by studying the monodromy constraint for the torus amplitudes.

Let us move on to the boundary wavefunction for A-branes in the two-dimensional black hole. Our first assumption is the existence of the identity brane, which will be identified as the D0-brane at the tip of the cigar. We assume that the self-overlap of this identity brane yields the identity representation summed over the spectral flow in the open string spectrum. Summation over the spectral flow is needed in order to guarantee the quantization of the $U(1)_R$ charge in the closed string spectrum.\footnote{Otherwise, we obtain the D-brane in the decompactified theory, which has been studied in \cite{Ahn:2003tt,Ahn:2004qb}, in the context of the $\mathcal{N}=2$ Liouville theory. We also note that our summation over the spectral flow is different from \cite{Eguchi:2003ik}, where the summation is taken over $n\in k \mathbb{Z}$ for integral $k$. For A-brane, the latter summation leads to the integral $U(1)_R$ charge (i.e. fractional $\omega$ quantum number).}

For definiteness, we study the NS-sector of the supersymmetric $SL(2;\br)/U(1)$ coset. Our assumption mentioned above is 
\begin{align}
Z_{00} = \langle 0|e^{-\pi\tau_c(L_0+\bar{L}_0 -\frac{c}{12})+iy(J_0+\bar{J}_0)}|0\rangle = \sum_n \frac{\mathrm{ch}_{0,n,n}(\tau_o)(1-q_o)}{(1+yq_o^{\frac{1}{2}+n})(1+y^{-1}q_o^{\frac{1}{2}-n})} \ . \label{ajka}
\end{align}
The modular S-transformation of the extended character of the identity representation (the identity character summed over the spectral flow) is given by \cite{Eguchi:2003ik}: 
\begin{align}
Z_{00} = \int_{-\frac{1}{2}+i\br_+} \dd j \sum_{m\in \frac{k}{2}Z} \frac{i\sin(\pi(2j+1))\sin\frac{\pi}{k}(2j+1)}{2\sin(j+m)\pi\sin(j-m)\pi} \mathrm{ch}_{j,m,0}(\tau_c) + \text{(discrete)} \ . \label{idmod}
\end{align}
The discrete terms are a little bit trickier to obtain, and we refer the complete form to original papers \cite{Eguchi:2003ik}.

The boundary wavefunction is essentially obtained by taking the square root of the modular S-matrix in analogy with \eqref{sob}. Expanding the identity boundary state as
\begin{align}
\langle 0|= \int_{-\frac{1}{2}+i\br_+} \dd j \sum_m \Psi(j,m)_0 \langle\langle j,m|\ ,
\end{align}
we obtain
\begin{align}
\Psi(j,m)_0 = \nu_b^{2j+1} \frac{\Gamma(-j+\frac{k\omega}{2})\Gamma(-j-\frac{k\omega}{2})}{\Gamma(-2j-1)\Gamma(1-b^2(2j+1))} \ .
\end{align}
We should note that the condition does not determine the $(j,m)$ dependent phase factor of the boundary wavefunction. The ambiguity of the phase, however, is completely fixed by the reflection relation
\begin{align}
\Psi(-p,m) = R(p,m) \Psi(p,m) \ ,
\end{align}
together with the Hermiticity condition $\Psi(p,m)^\dagger = \Psi(-p,-m)$. 

The next assumption is the overlap between the identity brane $|0\rangle$ and the general brane $|a \rangle$ labeled by the character of the $SL(2;\br)/U(1)$ coset model in analogy with \eqref{sobb}:
\begin{align}
\langle 0|e^{-\pi\tau_c(L_0+\bar{L}_0 -\frac{c}{12})+iy(J_0+\bar{J}_0)}|a\rangle = \chi_a(i\tau_o,y) \ . \label{anz}
\end{align}
We assume that the open string character appearing in the right hand side is given by the continuous series summed over the spectral flow (class 2 brane) and the discrete series summed over the spectral flow (class 3 brane). 

The S-modular transformation of the extended character for the continuous series (parametrized by $J$ and $M$) is particularly easy:
\begin{align}
\sum_{n\in \bz} \mathrm{ch}_{J,M+n,n}(\tau_o,y) = -i \sum_{m\in \frac{k}{2}Z} \int_{-\frac{1}{2}+i\br_+} \dd j \mathrm{ch}_{j,m,0}(\tau_c,y)e^{-\frac{4\pi iMm}{k}}\cos[\frac{\pi}{k}(2j+1)(2J+1)] \ .
\end{align}
From the modular bootstrap ansatz \eqref{anz}, we obtain the boundary wavefunction corresponding to the continuous series as
\begin{align}
\Psi(p,m)_{J,M} &= \Psi(-p,-m)_{0}^{-1} e^{-\frac{4\pi iMm}{k}} \cos[\frac{\pi}{k}(2j+1)(2J+1)] \cr
&= \nu_b^{2j+1}\frac{\Gamma(2j+1)\Gamma(1+b^2(2j+1))}{\Gamma(1+j+\frac{kw}{2})\Gamma(1+j-\frac{kw}{2})} e^{-\frac{4\pi iMm}{k}}\cos[\frac{\pi}{k}(2j+1)(2J+1)] \ .  
\end{align}
With the parameter identification $ r_0 = \frac{\pi}{k}(2J+1)$, $\theta_0 = -2\pi M$, we have obtained the boundary wavefunction for the partially wrapped D2-brane.

The S-modular transformation of the extended character for the discrete series are more involved:
\begin{align}
&\sum_{n\in \bz} \frac{y^{-\frac{2}{k}(M+n)}\mathrm{ch}_{J,M+n,n}(\tau_o,y)}{1+y^{-1}q_o^{\frac{1}{2}+J-M-n}}  \cr
&= -i \sum_{m\in \frac{k}{2}Z} \int_{-\frac{1}{2}+\br_+} \dd j \mathrm{ch}_{j,m}(\tau_c,y) e^{2\pi i M \omega} \left[\frac{e^{i(2J+1) (2j+1)}}{\sin\pi(j-\frac{kw}{2})} - \frac{e^{-(2J+1)(2j+1)}}{\sin(\pi(j+\frac{kw}{2}))}\right] \cr &+ \text{(discrete)} \ .
\end{align}
From the modular bootstrap ansatz \eqref{anz}, we obtain the boundary wavefunction corresponding to the discrete series as
\begin{align}
&\Psi(p,m)_{J,M} \cr
&= \nu_b^{2j+1}\Gamma(1+b^2(2j+1))\Gamma(2j+1)e^{2\pi i M \omega}  \times \cr 
&\times \left[\frac{\Gamma(-j+\frac{kw}{2})}{\Gamma(j+1+\frac{kw}{2})}e^{i\pi(m-j)+i(2J+1)(2j+1)} - \frac{\Gamma(-j-\frac{kw}{2})}{\Gamma(j+1-\frac{kw}{2})}e^{i\pi(m+j)-i(2J+1)(2j+1)} \right]  \ ,
\end{align}
where the parameter identification with the class 3 brane is $\theta_0 = 2\pi M + \frac{k\pi}{2} $ , and $\sigma = -\frac{\pi}{2} + \frac{\pi}{k}(2J+1)$.

So far, we have obtained as many branes as the extended character of the $\mathcal{N}=2$ supersymmetry (or $SL(2;\br)/U(1)$ coset). We, however, have to check whether the obtained D-branes satisfy the Cardy condition among themselves. Namely, we have to compute the cylinder amplitudes and decompose them into the open string characters and verify the positive definiteness of the density of states for continuous series and the positive integral multiplications for the discrete series. This was automatically guaranteed for the compact unitary CFTs thanks to the Verlinde formula. In our noncompact case, it is not a trivial problem. Indeed, although, almost all overlaps are consistent with the Cardy condition, the self-overlaps between class 3 branes for irrational value of $k$, we encounter negative multiplicities of discrete series in their spectra \cite{Ribault:2003ss}.\footnote{For integral value of $k$, this subtlety is avoided \cite{Israel:2005fn}. For general fractional value of $k$, the situation is more involved and the results depend on the combination of the other sectors embedded in the full string theory and the appropriate GSO condition we impose. The case by case study of these cases can be found in \cite{Eguchi:2003ik}.}

One might wonder what goes wrong with the modular bootstrap for the B-branes. The gist is that there is no identity brane for the B-boundary conditions. One can formally write down the modular bootstrap equations like \eqref{idmod}, but there does not exist any analytic solution compatible with the reflection amplitudes for B boundary conditions. Due to this lack of the identity B-brane, the whole construction of the modular bootstrap breaks down.
The coset construction from the descent of branes in $\mathbb{H}_3^+$ model was given in  \cite{Ribault:2003ss}. The conformal bootstrap for the dual $\mathcal{N}=2$ Liouville theory can be found in \cite{Ahn:2003tt,Hosomichi:2004ph}

\newpage
\sectiono{Rolling D-brane in Two-dimensional Black Hole}\label{sec:7}
In this section, we study the D-branes in Lorentzian two-dimensional black hole. The organization of the section is as follows. In section \ref{sec:7-1}, we study the classical D-branes in the Lorentzian two-dimensional black hole. In section \ref{sec:7-2}, we construct the boundary states for the rolling D-brane from the Wick rotation of the class 2 brane in the Euclidean two-dimensional black hole system.\footnote{This part of the thesis is based on \cite{Nakayama:2005pk}.} In section \ref{sec:7-3}, we study some properties of our boundary wavefunction focusing on $1/k$ corrections.
\subsection{Classical D-branes}\label{sec:7-1}
\subsubsection{DBI analysis}\label{sec:7-1-1}
The classical D-branes in Lorentzian two-dimensional black hole
 is classified by the solution of the equation of motion coming from the DBI action
\begin{align}
S^L = \mu_{p+1} \int \dd^{p+1}\xi e^{-\Phi} \sqrt{-\det(G_{ab}+B_{ab}+F_{ab})} \ . 
\end{align}
The classical background is given by
\begin{align}
\dd s^2 =k\al'(  - \tanh^2\rho \, \dd t^2 +  \dd
\rho^2) ,
\qquad e^{2\Phi} = \frac{k}{\mu \cosh^2 \rho} \ ,
\end{align}
or when we are interested in the global structure of the solution, we use the Kruscal coordinate
\begin{align}
\dd s^2 = -2k\frac{\dd u\dd v}{1-uv} \ , \ \ e^{2\Phi} = \frac{k}{\mu(1-uv)} \ . \label{krscc}
\end{align}
by the coordinate transformation: $u = \sinh \rho e^{t}$, $v=-\sinh\rho e^{-t}$.

We begin with the D(-1) instanton. A physical meaning of such D-brane is a little bit unclear in the Lorentzian signature, but the ``effective action"
\begin{align}
S_{-1} \propto e^{-\phi} = \sqrt{1-uv}  
\end{align}
could be extremized at $u=v=0$ or $\rho = 0$. 

Next we study the D0-brane. In the local coordinate outside the horizon, we can write down the DBI action as
\begin{align}
S_0 = \mu_0 \int \dd t \cosh\rho(t)\sqrt{-\dot{\rho}(t)^2 + \tanh^2\rho(t)} \ , \label{dbidzero}
\end{align}
where we have fixed the reparametrization invariance by taking the temporal gauge $\xi_0 = t$. From the energy conservation, we obtain
\begin{align}
\text{const} = \cosh\rho\frac{\tanh^2\rho}{\sqrt{-\dot{\rho}^2 + \tanh^2\rho}} \ ,
\end{align}
which can be integrated to
\begin{align}
\sinh(\rho) \cosh(t-t_0) = \text{const} \ . \label{mot}
\end{align}
The D0-brane motion \eqref{mot} also follows from the Wick rotation $\theta \to it$ to the hairpin brane \eqref{trsce} in the Euclidean two-dimensional black hole. As we mentioned in section \ref{sec:5-3}, The action \eqref{dbidzero} can be rewritten in the same form as the rolling D-brane in the linear dilaton background by
introducing `tachyon' variable $Y \equiv \log \sinh \rho$:
\begin{align} L_{\rm D0} = - V(Y) \sqrt{1 - \dot{Y}^2} \qquad \mbox{where}
\qquad
 V(Y) = M_0 \, e^Y ~, \label{nonextremal D0} \end{align}
which leads us to the ``tachyon - radion correspondence" discussed in section \ref{sec:5}.

To study the global structure, we use the Kruscal coordinate \eqref{krscc}. In this coordinate system, the DBI action takes the flat form
\begin{align}
S = \mu_0  \int \dd\xi \sqrt{\frac{\dd u}{\dd \xi}\frac{\dd v}{\dd\xi}} \ .
\end{align}
The equation of motion is solved by a straight line in the $(u,v)$ plane. It is interesting to note that the D0-brane does not feel the existence of the singularity at $uv=1$. The classical trajectory is analytically continued inside the singularity in a trivial way. This is because the curvature singularity is cancelled against the dilaton singularity, which appears in the DBI action in the opposite way. The coupling to the dilaton is a crucial difference between the D-brane and a usual particle (such as a point like F-string or folded string solution discussed in section \ref{sec:3-3-2}) in the two-dimensional black hole background. 

Let us finally consider the D1-brane. The DBI action in the Kruscal coordinate is
\begin{align}
S_1 = \mu_1 \int \dd u\dd v \sqrt{1-uv}\sqrt{\frac{1}{(1-uv)^2} - F_{uv}^2} \  
\end{align}
in the gauge $\xi_0 = u$, $\xi_1 = v$. The Gauss law constraint
\begin{align}
f = \frac{\sqrt{1-uv}F_{uv}}{\sqrt{\frac{1}{(1-uv)^2}-F^2_{uv}}} 
\end{align}
is solved by
\begin{align}
F_{uv}^2 = \frac{f^2}{(1-uv+f^2)(1-uv)^2} \ .
\end{align}
When $f^2>0$, the world-sheet of the D1-string covers the whole physical region of the two-dimensional black hole, and possibly it has a boundary inside the singularity at $uv=1+f^2$. When $f^2 = -\kappa^2 <0$, the D1-string has a boundary at $uv = 1-\kappa^2$, and describes a long folded string. However, the DBI action for the long folded string becomes imaginary, so the solution is overcritical and unphysical.\footnote{This is a general feature of the Lorentzian solution for the D-brane with the boundary coming from the blow-up of the field strength. In the Lorentzian signature, the terms inside the square root of the DBI action is bounded, or in other words the field strength has a critical value. Thus any D-brane that has a boundary due to the divergence of the field strength is overcritical and hence unphysical unlike the case in the Euclidean signature.}
\subsubsection{group theoretical viewpoint}\label{sec:7-1-2}
As we have done in section \ref{sec:6-1-2}, we can also study the classical D-brane in the Lorentzian two-dimensional black hole from the coset construction. We parametrize the parent $SL(2;\br)$ element $g$ as
\begin{align}
g = \begin{pmatrix} a \ u \cr -v \ b \end{pmatrix} \ , \ \ uv + ab = 1 \ .
\end{align}
Under the axial gauge transformation $\delta g = \epsilon(\sigma_3 g + g \sigma_3) $, the $(u,v)$ is invariant, and serves as a gauge invariant coordinate describing the two-dimensional black hole (i.e. we can identify them as $(u,v)$ in the Kruscal coordinate \eqref{krscc}).
 
The maximally symmetric D-brane in the parent $SL(2;\br)$ WZNW model is classified by the (twined) conjugacy class of the group. The conjugacy class is given by
\begin{align}
\mathrm{Tr}(g) = a + b 
\end{align}
and the twined conjugacy class is given by
\begin{align}
\mathrm{Tr}(\sigma_1 g) = u-v \ , \label{twcl}
\end{align}
up to a conjugation (i.e. Lorentz boost between $\sigma_1$ and $\sigma_2$). 
To derive the D-branes in the coset model, we have to project the twined conjugacy class to the coset variables for A-branes. For B-branes, we first take a superposition of the gauge orbit of the conjugacy class so that we obtain a gauge invariant object as a D-brane. 

Let us begin with the A-branes. The twined conjugacy class \eqref{twcl} is invariant under the axial gauge transformation, so the D-brane (D0-brane) is classified by the equation
\begin{align}
 2\kappa = \mathrm{Tr}(\sigma_1 g ) = u- v \ ,
\end{align}
which gives a straight line in the Kruscal coordinate of the two-dimensional black hole as we observed in section \ref{sec:7-1-1} from the DBI analysis. More general branes are obtained by the Lorentz boost: 
\begin{align}
ue ^{t_0} - ve^{-t_0} = 2\kappa \ .
\end{align}
The existence of such boosted D-branes are consistent with the fact that the existence of the Nambu-Goldstone modes associated with the symmetry breaking due to the A-brane as we have reviewed in section \ref{sec:6-1-2}.

Let us move on to the B-branes. In this case, the conjugacy class is not invariant under the axial gauge transformation. In order to obtain an invariant object that can be projected down to the coset, we need to sum over the gauge orbit. With fixing the trace as $a+b = 2\kappa$, the determinant constraint reads
\begin{align}
uv = 1-\kappa^2 + (a-\kappa)^2 \ge 1-\kappa^2 \ .
\end{align}
It is not difficult to see that the gauge orbit of the conjugacy class precisely agrees with the domain bounded by the last inequality. The string configuration reproduces the folded D1-string obtained from the DBI analysis in section \ref{sec:7-1-1}. We note, however, that the solution is overcritical and unphysical as we have seen in section \ref{sec:7-1-1}.\footnote{It is not uncommon that the group theoretical classification of the D-branes in the Lorentzian coset gives unphysical D-branes (see e.g. \cite{Hikida:2005vd}). Our identification of the parameter is different from the one given in \cite{Yogendran:2004dm}, which solves a small puzzle raised there. The extra $i$ comes from a (hypothetical) time-like T-duality \cite{Hull:1998vg} which we need to perform to obtain the parameter identification according to the discussion given in \cite{Walton:2002db}. } We also see that the D-string that covers the whole physical region of the two-dimensional black hole cannot be obtained from a simple descent from the D-branes in the $SL(2;\br)$ WZNW model without an analytic continuation.

\subsection{Boundary states from Wick rotation}\label{sec:7-2}

\subsubsection{analytic continuation of boundary states}\label{sec:7-2-1}

In this section, we shall construct the exact boundary state
describing the D0-brane moving in the Lorentzian two-dimensional
black hole background. Recall that the Lorentzian two-dimensional
black hole (`Lorentzian cigar') background is obtainable by the Wick
rotation $\theta = it$ of the Euclidean one \eqref{Euclidean cigar}
\begin{equation}
\dd s^2 = 2k(\dd \rho^2 - \tanh^2\! \rho  \, \dd t^2) \qquad
\mbox{and} \qquad e^{\Phi} = \frac{e^{\Phi_0}}{\cosh\rho} ~.
\label{Lorentzian cigar}
\end{equation}
Wick-rotating the geodesic of the Euclidean D1-brane, we found the
geodesic of the Lorentzian D0-brane in \eqref{mot} as
\begin{align}
 \cosh(t-t_0) \sinh \rho = \sinh \rho_0~,
\label{trajectory D0}
\end{align}
where $t_0$, $\rho_0$ are free parameters. Notice that the D0-brane
reaches the horizon $\rho = 0$ at $t \rightarrow \pm \infty$
irrespective of the values of $\rho_0$ and $t_0$. Thus, formally, the
Lorentzian D0-brane boundary state is obtainable by Wick rotation of
the Euclidean D1-brane boundary state \eqref{clss2'} if we are interested in the physics outside the event horizon.\footnote{Some
classical analysis of D-brane dynamics was attempted in
\cite{Yogendran:2004dm} within the Dirac-Born-Infeld approach.}

Reconstructing boundary states of the Lorentzian D-brane from those
of the Euclidean D-brane is generically not unique. Rather, the
following potential subtleties need to be faced:
\begin{itemize}
 \item The Euclidean momentum $n$ along the asymptotic circle
of cigar is quantized, while the corresponding quantum number in the
Lorentzian theory ({\em i.e.} the energy) takes a continuous value.
 \item The Wick rotations of primary states are not necessarily
unique. Often, appropriate boundary conditions should be specified.
\end{itemize}
As for the first point, which has to do with Matsubara formulation,
we can formally avoid the difficulty of quantized momentum by the
following heuristic consideration. Suppose the boundary wave
function $\hat{f}(n,\al)$ ($n\in \bz$ is the quantized Euclidean
energy, and $\al$ denotes the remaining quantum numbers not touched
here) is given by the Fourier transform of a periodic function
$f(x+2\pi, \al)=f(x, \al)$. We then obtain
\begin{align}
\hspace{-1cm} \bra{B}=
\sum_{\al}\sum_{n\in\bsz}\,\widetilde{f}(n,\al)\dbra{n,\al} &=
\sum_{\al}\sum_{n\in\bsz}\, \frac{1}{2\pi}\int_{-\pi}^{\pi} \dd x\,
f(x,\al) e^{in x} \, \dbra{n,\al} \nn &=  \sum_{\al}
\int_{-\infty}^{\infty} \frac{\dd q}{2\pi}\, \int_{-\infty}^{\infty}
\dd x\, f(x,\al) e^{iqx} \, \dbra{q,\al}~, \label{formal extension}
\end{align}
where we used the identity $ \sum_{n\in \bsz}\, \delta(q-n) =
\sum_{m\in \bsz}\, e^{2\pi i m q} $ in obtaining the last
expression. Assuming that $f(x, \al)$ is analytic along the entire
real $x$ axis, the Wick rotation can be performed. Often, $f(x,
\al)$ is non-analytic over the real $x$ axis, and the integral in
the last expression is ill-defined. This turns out to be the case
for the boundary wave function of the Euclidean D1-brane
\eqref{clss2'}: in the coordinate space, the wave function has
branch cuts and singularities along the real $x$-axis. In such
cases, the best we can do is to adopt the slightly deformed
integration contour $\cC$ in $x$-space\footnote
   {To be more precise, we should allow to use
  some decomposition
$$
 f(x,\al) = f_1(x,\al)+ f_2(x,\al)+\cdots~,
$$
and to take the different contours for each piece $f_i(x,\al)$. } to
render the Fourier integral well-defined:
\begin{align}
&& \bra{B'} \Big\vert_{\rm Euclidean} :=
\sum_{\al}\int_{-\infty}^{\infty} \frac{\dd q}{2\pi}\, \int_{\cC}
\dd x\, f(x,\al) \, e^{iqx} \, \dbra{q,\al} ~. \label{formal
extension 2}
\end{align}
Likewise, disk one-point function of vertex operator $\Phi^{\rm
Euclidean}_{q,\al}$ (associated with the Ishibashi state
$\dbra{q,\al}$) is evaluated as the deformed contour integral:
\begin{align}
& \Big< \Phi^{\rm Euclidean}_{q,\al} \Big>_{\msc{disk}}
   = {}_{\rm E}\!\langle B' \vert q, \al \rangle\rangle = \int_{\cC} \dd x\, f(x,\al) e^{iqx}~.
\label{formal disk amp 1}
\end{align}
Assuming sufficient analyticity, one then defines Wick rotation of
the states \eqref{formal extension 2} by the contour deformation of
$\cC$ accompanied by the continuation $q\,\rightarrow\, i\om, x
\rightarrow \, i t$;
\begin{align}
\bra{B'} \Big\vert_{\rm Lorentzian} :=
\sum_{\al}\int_{-\infty}^{\infty} \frac{i \dd \om}{2\pi}\,
\int_{-\infty}^{\infty} i\dd t\, f(it, \al) e^{-i\om t} \,
\dbra{i\om,\al} ~. \label{Wick rotation 0}
\end{align}
This is essentially the procedure taken in \cite{Nakayama:2004yx}. Of course, we
potentially have an ambiguity in the choice of the contour $\cC$,
and the correct choice should be determined by the physics under
study.

In the present case $\bra{B}$ corresponds to \eqref{clss2'} and
$\bra{B'}$ is given by
\begin{align}
{}_{\rm D1}\bra{B';\rho_0,\theta_0} = \int_0^{\infty} \frac{\dd
p}{2\pi}\, \int_{-\infty}^{\infty} \frac{\dd q}{2\pi}\, \Psi'_{\rm
D1} (\rho_0,\theta_0;p,q) \, \dbra{p,q} ~, \end{align}
where
\begin{align} &\Psi'_{\rm D1}(\rho_0,\theta_0;p,q)
\cr
 &= \frac{\sinh(\pi p)}
{\left|\cosh\Big(\pi\frac{p+iq}{2}\Big)\right|^2} \,
\frac{\pi\Gamma(ip)\Gamma\Big(1+\frac{ip}{k}\Big)}
{\Gamma\Big(\frac{1}{2}+\frac{ip+q}{2}\Big)
\Gamma\Big(\frac{1}{2}+\frac{ip-q}{2}\Big)} \, e^{iq\theta_0}
\left[e^{-ip\rho_0} +\frac {\cosh\left(\pi\frac{p-i|q|}{2}\right)}
{\cosh\left(\pi\frac{p+i|q|}{2}\right)}
 e^{ip\rho_0}\right] \cr
 &\equiv  B\left(\frac{1}{2}-\frac{ip-q}{2},
\frac{1}{2}-\frac{ip+q}{2}\right) \Gamma\left(1+\frac{ip}{k}\right)
\, e^{iq\theta_0}\left[e^{-ip\rho_0}
+\frac{\cosh\left(\pi\frac{p-i|q|}{2}\right)}
{\cosh\left(\pi\frac{p+i|q|}{2}\right)} e^{ip\rho_0}\right].
\label{D1'}
\end{align}
\begin{figure}[htbp]
    \begin{center}
   \includegraphics[width=13cm,height=10cm]
      {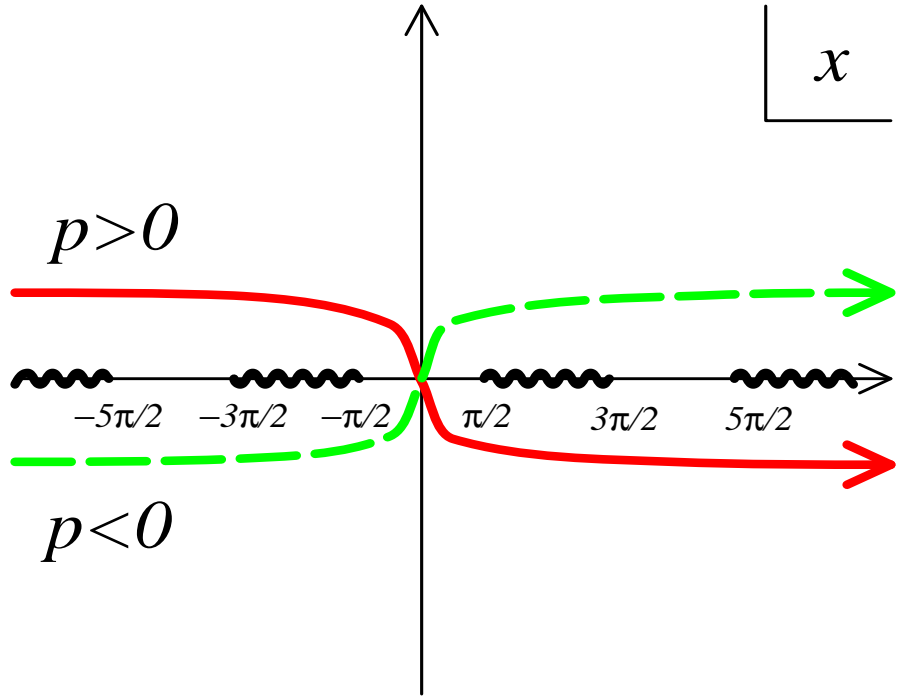}
    \end{center}
    \caption{The red (green broken) line
is the contour $\cC^+$ for $p > 0$
     to the Lorentzian time. Notice that an infinite number
     of branch cuts repeats in the Euclidean time: $\frac{\pi}{2}+2n\pi < x <
      \frac{3\pi}{2} +2n\pi$, $(n\in \bz)$ along the real $x$-axis.}
    \label{c-array}
\end{figure}
Here $B(p,q) \equiv \Gamma(p)\Gamma(q)/\Gamma(p+q)$ denotes Euler's
beta function. The integration contour $\cC$ we choose is shown in
Figure \ref{c-array} \cite{Nakayama:2004yx}. As in \eqref{evaluation overlap phi},
we separately evaluated the integrals of $\phi^{p}_{L,q}$ and
$\phi^p_{R,q}$ based on the decomposition \eqref{decomp ef}. For the
convergence of integrals, we choose the contour $\cC^+$ for
$\phi^{p}_{L,q}$ ($p>0$ sector) and $\cC^-$ for  $\phi^{p}_{R,q}$
($p<0$ sector). Such choice of integration contours rendered an
extra damping factor $\sinh(\pi p) /
{|\cosh\left(\pi\frac{p+iq}{2}\right)|^2}$, which improves the
ultraviolet behavior of the wavefunction and makes it possible to
take the Wick rotation sensibly. The non-trivial phase factor $
{\cosh\left(\pi\frac{p-i|q|}{2}\right)}/
{\cosh\left(\pi\frac{p+i|q|}{2}\right)} $ in the second term
originates from the reflection amplitude,
and it reduces to $(-1)^n$ when $q=n\in \bz$.

The second subtlety implies that $\dbra{i\om, \al}$ is not uniquely
defined in \eqref{Wick rotation 0}. This is the issue that arises in a
background with horizon, equivalently, non-existence of globally
definable timelike Killing vector. As such, this subtlety did not
arise for the extremal NS5-brane geometry (described asymptotically
by free linear dilaton theory) considered in
\cite{Nakayama:2004yx}. In section \ref{sec:7-2-4}, within the mini-superspace analysis
for the Lorentzian two-dimensional black hole, we shall clarify this
subtlety.

An alternative, sensible prescription of the analytic continuation
is to define the disk one-point correlator {\em directly\/} via the
Lorentzian Fourier transform:
\begin{align}
& \Big< \Phi^{\msc{Lorentzian}}_{\om,\al} \Big>_{\msc{disk}}
   = \int_{-\infty}^{\infty} \dd t\, f(it,\al) e^{-i\om t}~.
\label{formal disk amp 2}
\end{align}
This is {\em not\/} always equivalent to the the former method
elaborated above. In fact, the latter method does not necessarily
assert that the boundary state constructed so is expandable in terms
of the Lorentzian Ishibashi states that are analytically continued
from the Euclidean ones.\footnote{Recently, we have succeeded the direct evaluation of the overlap integral. See appendix \ref{direct} for details.}

In section \ref{sec:3-3-4}, we have reviewed the primary states for the Lorentzian two-dimensional black hole.
Having obtained the Lorentzian primary states, we shall now
construct several interesting class of boundary states for a
D0-brane propagating in the black hole background. We have seen that
the D0-brane propagates along the trajectory \eqref{trajectory D0}.
The two-dimensional black hole is eternal, so, in addition to the
past and the future asymptotic infinities, the causal propagation
region has the past horizon ${\cal H}^-$ surrounding the white hole
singularity and the future horizon ${\cal H}^+$ surrounding the
black hole singularity. As such, by taking variety of possible
boundary conditions, we can construct interesting class of boundary
states.

\subsubsection{boundary state of D0-brane absorbed to future horizon}\label{sec:7-2-2}

Consider first the boundary state obeying the boundary condition
$\psi(\rho,t)\,\rightarrow\, 0$ at the past horizon ${\cal H}^-$,
viz. the primary states $\ket{U^p_{\om}}$.
This boundary condition is relevant for scattering of a D0-brane off
the black hole, since the condition represents absorption only and
no emission of the D0-brane by the black hole. D0-brane boundary
state obeying such absorbing boundary condition is then expanded
solely by the Ishibashi states ${}^{\widehat{U}}\dbra{p,\om}$,
$\dket{p,\om}^U$ that are associated with the primary states
$\widehat{\bra{U^p_{\om}}}$, $\ket{U^p_{\om}}$:
\begin{align}
 &
{}_{\msc{absorb}}\!\bra{B;\rho_0,t_0} = \int_0^{\infty}\frac{\dd
p}{2\pi} \int_{-\infty}^{\infty}\frac{\dd \om}{2\pi}\,
  \Psi_{\msc{absorb}}(\rho_0,t_0;p,\om) \,
{}^{\widehat{U}}\!\dbra{p,\om}~, \nn
  &
\ket{B;\rho_0,t_0}_{\msc{absorb}} = \int_0^{\infty}\frac{\dd
p}{2\pi} \int_{-\infty}^{\infty}\frac{\dd \om}{2\pi}\,
  \Psi^*_{\msc{absorb}}(\rho_0,t_0;p,\om)
\, \dket{p,\om}^U~. \label{falling D0 0}
\end{align}
The boundary wavefunction $\Psi_{\msc{absorb}}(\rho_0,t_0;p,\om)$
is then interpreted as the disk one-point correlators:
\begin{align}
\Psi_{\msc{absorb}}(\rho_0,t_0;p,\om) &= \langle U^p_{\om}
\rangle_{\msc{disk}} \equiv {}_{\msc{absorb}}\!\bra{B;\rho_0,t_0}
U^p_{\om} \rangle~,
\label{falling disk}
\end{align}

The boundary wavefunction \eqref{falling disk} is then obtained by
taking the Wick rotation $q\,\rightarrow\, i\om$ ($q\,\rightarrow\,
-i\om$) for $q<0$ ($q>0$)
in \eqref{D1'} (recall \eqref{ac UVLR}):\footnote
        {In reality, there is a further overall factor $i$,
          but, for notational simplicity, we will absorb it to the definition
          of the Ishibashi states.}
\begin{align}
& \hspace{-1cm} \Psi_{\msc{absorb}}(\rho_0,t_0;p,\om) = B(\nu_+,
\nu_-) \Gamma\Big(1+\frac{ip}{k}\Big) \, e^{-i\om t_0}\left[ e^{-ip
\rho_0} -  \frac{\cosh\left(\pi \frac{p-\om}{2}\right)}
{\cosh\left(\pi \frac{p+\om}{2}\right)} e^{ip\rho_0 } \right]~,
\label{falling D0}
\end{align}
The relative minus sign in the second term of
$\Psi_{\msc{absorb}}(\rho_0,t_0;p,\om)$ originates from the fact
that the contour rotation defining the Wick rotation has opposite
directions for $\cC^+$ (suitable for $p>0$) and $\cC^-$ (suitable
for $p<0$). See figure \ref{c-array}. This boundary wavefunction
\eqref{falling D0} satisfies the exact reflection relation
\begin{align}
\Psi_{\msc{absorb}}(\rho_0,t_0;-p,\om)= \cR(-p,\om) \,
\Psi_{\msc{absorb}}(\rho_0,t_0;p,\om)~. \label{ref falling D0}
\end{align}

With such boundary condition, the boundary wavefunction
\eqref{falling D0} would have no overlap with D0-brane's trajectory
\eqref{trajectory D0} in the far past region $t\ll t_0$. In fact, the
trajectory \eqref{trajectory D0} starts from the past horizon ${\cal
H}^-$ at $t=-\infty$, reaches the time-symmetric point $\rho =
\rho_0$ at $t = t_0$, and then falls back the future horizon ${\cal
H}^+$ at $t=+\infty$, while the wavefunction $U^p_{\om}$ does not
have any component outgoing from ${\cal H}^-$. We thus interpret
that the boundary state \eqref{falling D0} describes the future half
of the classical trajectory \eqref{trajectory D0}. We shall hence call
it the `absorbed D-brane'.

By utilizing the radion-tachyon correspondence, the rolling radion
(as described by the boundary state \eqref{falling D0}) can be also
  interpreted as the rolling tachyon. In the latter interpretation,
the D0-brane absorbed to the future horizon is the counterpart of
the future-half S-brane \cite{Gutperle:2002ai,Strominger:2002pc,Larsen:2002wc}, in which the
tachyon rolls down the potential hill at asymptotic future $t
\rightarrow + \infty$ and emits radiation.

\subsubsection{boundary state of D0-brane emitted from past horizon}\label{sec:7-2-3}
Consider next the boundary condition: $\psi(\rho,t)\,\rightarrow\,
0$ at ${\cal H}^+$, viz. use the basis $\dket{p,\om}^V$,
${}^{\widehat{V}}\!\dbra{p,\om}$ instead of $\dket{p,\om}^U$,
${}^{\widehat{U}}\!\dbra{p,\om}$. Utilizing the reflection relation,
we can first rewrite \eqref{D1'} as the form which only includes the
$p<0$ Ishibashi states by means of the reflection relation. Then, we
can analytically continue the states $\ket{\phi^{-p}_q}$ ($p>0$)
into $\ket{V^p_\om}$.
The resultant boundary state is obtained by
simply replacing $p\,\rightarrow\,-p$,
$\om\, \rightarrow\, -\om$
in \eqref{falling D0};
\begin{align}
& {}_{\msc{emitted}}\!\bra{B;\rho_0,t_0} = \int_0^{\infty}\frac{\dd
p}{2\pi} \int_{-\infty}^{\infty}\frac{\dd \om}{2\pi}\,
  \Psi_{\msc{emitted}}(\rho_0,t_0;p,\om) \,
{}^{\widehat{V}}\!\dbra{p,\om}~. \nn
& \ket{B;\rho_0,t_0}_{\msc{emitted}} = \int_0^{\infty}\frac{\dd
p}{2\pi} \int_{-\infty}^{\infty}\frac{\dd \om}{2\pi}\,
  \Psi^*_{\msc{emitted}}(\rho_0,t_0;p,\om) \,
\dket{p,\om}^V ~. \label{emitted D0} \end{align}
where
\begin{align} \Psi_{\msc{emitted}}(\rho_0,t_0;p,\om) = B(\nu^*_+, \nu^*_-)
\Gamma\left(1-\frac{ip}{k}\right) \, e^{-i\om t_0}\left[
e^{ip\rho_0} - \frac{\cosh\left(\pi \frac{p-\om}{2}\right)}
{\cosh\left(\pi \frac{p+\om}{2}\right)} e^{-ip\rho_0} \right]~.
\nonumber
\end{align}
Obviously, the emitted D0-brane wavefunction is the time-reversal
of the absorbed D0-brane wavefunction \eqref{falling D0}:
\begin{align} \Psi_{\msc{emitted}}(\rho_0,t_0;p,\om) =
\Psi^*_{\msc{absorb}}(\rho_0,-t_0;p,\om) ~. \nonumber \end{align}
Namely, it describes the D0-brane emitted from the past horizon at
asymptotic past $t=-\infty$. By the choice of the boundary
condition, this boundary state \eqref{emitted D0} describes only the
past half of the classical D0-brane trajectory \eqref{trajectory D0}.

The exact reflection relation has the form
\begin{align}
& \Psi_{\msc{emitted}}(\rho_0,t_0;-p,\om)= \cR^*(-p,\om) \,
\Psi_{\msc{emitted}}(\rho_0,t_0;p,\om)~. \label{ref emiited D0}
\end{align}

Again, in light of the radion-tachyon correspondence, the D0-brane
emitted from the past horizon is the counterpart of the past-half
S-brane in tachyon rolling. The radiation creeps up the tachyon
potential hill from past infinity and forms an unstable D-brane.

\subsubsection{boundary state of time-symmetric D0-brane}\label{sec:7-2-4}

The third possible boundary state is obtainable by {\em directly\/}
taking the analytic continuation in the disk one-point amplitudes,
as we already mentioned. Recalling \eqref{ac UVLR}, we shall
analytically continue the disk amplitudes as (assume $p>0$)
\begin{align}
  \langle \phi^{+p}_q \rangle_{\msc{disk}}~\longrightarrow~
  \langle U^p_{\om} \rangle_{\msc{disk}} \qquad \mbox{and} \qquad
  \langle \phi^{-p}_q \rangle_{\msc{disk}}~\longrightarrow~
  \langle V^p_{\om} \rangle_{\msc{disk}}~.
\label{ac disk amp}
\end{align}
The Euclidean one-point amplitudes $\langle \phi^{\pm p}_q
\rangle_{\msc{disk}}$ are given in \eqref{D1'}, and can be expressed
in contour integrals as in \eqref{formal disk amp 1}. Recall that
$\langle \phi^p_{L,q}   \rangle_{\msc{disk}}$, $\langle
\phi^{p}_{R,q} \rangle_{\msc{disk}}$ are prescribed by the contour
integrals over $\cC^+$, $\cC^-$ in figure \ref{c-array}. We shall
thus analytically continue them to the real time axis (imaginary
$x$-axis). In this way, we extract the Lorentzian disk one-point
amplitudes as
\begin{align}
& \langle U^p_{\om} \rangle_{\msc{disk}} = \langle U^p_{\om}
\rangle_{\msc{disk}}^{(\msc{absorb})} \qquad \mbox{and} \qquad \langle V^p_{\om}
\rangle_{\msc{disk}} = \langle V^p_{\om} \rangle_{\msc{disc}}^{(\msc{emitted})}~,
~~~ \label{rel disc amp}
\end{align}
where the right-hand sides are simply the amplitudes associated with
the `absorbed' and `emitted' D0-branes considered in the previous
subsections and explicitly given in \eqref{falling D0} and
\eqref{emitted D0}. Since $U^p_{\om}$ and $V^p_{\om}$ constitute the
complete set of basis for Lorentzian primary fields, the amplitudes
\eqref{rel disc amp} would yield yet another Lorentzian D0-brane
boundary states. As is obvious from the above construction, this
state keeps the time-reversal symmetry manifest and reproduces the
entire classical trajectory \eqref{trajectory D0}, that is, it
describes a D0-brane emitted from the past horizon and reabsorbed to
the future horizon. From the viewpoint of the boundary conformal
theory, this would be considered the most natural one since it
captures the entire classical trajectory of the D0-brane. In the
radion-tachyon correspondence, this state is the counterpart of the
full S-brane \cite{Gutperle:2002ai,Sen:2002nu,Sen:2002in,Sen:2002an}.

Explicitly, the time-symmetric boundary states are given by
\begin{align}
&{}_{\msc{symm}}\!\bra{B;\rho_0,t_0} \cr
&={}_{\msc{absorb}}\!\bra{B;\rho_0,t_0} +
{}_{\msc{emitted}}\!\bra{B;\rho_0,t_0} \cr &=
\int_0^{\infty}\frac{\dd p}{2\pi} \int_{-\infty}^{\infty}\frac{\dd
\om}{2\pi}\, \left[
  2\Psi_{\msc{symm}}(\rho_0,t_0;p,\om)
\, {}^L\!\dbra{p,\om}
  +
  2\Psi^*_{\msc{symm}}(\rho_0,-t_0;p,\om) \, {}^R\!\dbra{p,\om}
\right]  \cr
& \ket{B;\rho_0,t_0}_{\msc{symm}} \cr &=
\ket{B;\rho_0,t_0}_{\msc{absorb}} +
\ket{B;\rho_0,t_0}_{\msc{emitted}}  \cr  &=
\int_0^{\infty}\frac{\dd p}{2\pi} \int_{-\infty}^{\infty}\frac{\dd
\om}{2\pi}\, \left[
  2\Psi^*_{\msc{symm}}(\rho_0,t_0;p,\om)
\, \dket{p,\om}^L
  +
  2\Psi_{\msc{symm}}(\rho_0,-t_0;p,\om) \, \dket{p,\om}^R
\right] ~, \label{symmetric D0} 
\end{align}
where
\begin{align} \Psi_{\msc{symm}}(\rho_0,t_0;p,\om) = B(\nu_+, \nu_-)
\Gamma\left(1+\frac{ip}{k}\right) \, e^{-ip\rho_0-i\om t_0}
\end{align}
and ${}^L\!\dbra{p,\om}$, $\dket{p,\om}^L$, ${}^R\!\dbra{p,\om}$,
$\dket{p,\om}^R$ are the Ishibashi states constructed over the
primary states $\bra{L^p_{\om}}$, $\ket{L^p_{\om}}$,
$\bra{R^p_{\om}}$, $\ket{R^p_{\om}}$,\footnote{The extra factor of
`2' was introduced for convenience. Recall \eqref{inner product
UVLR}.} respectively. One can readily check that the second lines
in \eqref{symmetric D0} are indeed correct by evaluating the disk
one-point amplitudes from them. For instance, using \eqref{inner
product UVLR}, we obtain
\begin{align}
 \langle U^p_{\om} \rangle_{\msc{disk}}^{(\msc{symm})}
&= {}_{\msc{symm}}\!\bra{B;\rho_0,t_0} U^p_{\om} \rangle \nn &=
\Psi_{\msc{symm}}(\rho_0,t_0;p,\om) +\cR(p,\om)
\Psi^*_{\msc{symm}}(\rho_0,-t_0;p,\om) \nn &=  B(\nu_+,\nu_-)
\Gamma\left(1+\frac{ip}{k}\right) e^{-i\om t_0} \, \left[
e^{-ip\rho_0} - \frac{\cosh\left(\pi \frac{p-\om}{2}\right)}
{\cosh\left(\pi \frac{p+\om}{2}\right)} e^{ip\rho_0} \right] \nn &=
\langle U^p_{\om} \rangle^{(\msc{absorb})}_{\msc{disk}} \equiv
{}_{\msc{absorb}}\!\bra{B;\rho_0,t_0}U^p_{\om}\rangle ~.
\end{align}
Other one-point amplitudes can be checked analogously.

Two remarks are in order. First, notice that, though the disk
one-point amplitudes are, the symmetric boundary states
\eqref{symmetric D0} by themselves are {\em not\/} analytically
continuable to the Euclidean boundary state \eqref{D1'}. This should
not be surprising as the Lorentzian Hilbert space is generated by
{\sl twice} as many generators as the Euclidean theory.
In other words,
the Lorentzian bases $\ket{U^p_{\om}}$, $\ket{V^p_{\om}}$ correspond
to $\ket{\phi^p_n}$, $\ket{\phi^{-p}_n}$ in the Euclidean theory,
which were however linearly dependent due to the reflection
relation.
Nevertheless, the
boundary state \eqref{symmetric D0} is a consistent one and yields
disk one-point amplitudes that can be correctly continued to the
Euclidean ones. Second, the full Lorentzian Hilbert space is
decomposed as
\begin{align}
 \cH = \cH^U \oplus \cH^V  \qquad \mbox{and} \qquad
 \widehat{\cH} = \widehat{\cH^U} \oplus \widehat{\cH^V}~,
\label{decomp Hilb}
\end{align}
where $\cH^U$ ($\cH^V$) is spanned by $\ket{U^p_{\om}}$ , ($\,
\ket{V^p_{\om}}\, $) and their descendants. The dual space
$\widehat{\cH^U}$ ($\widehat{\cH^V}$) is similarly spanned by
$\widehat{\bra{U^p_{\om}}}$, ($\widehat{\bra{V^p_{\om}}}$). Here,
the Hilbert subspaces $\cH^{U}$, $\widehat{\cH^{U}}$ ($\cH^{V}$,
$\widehat{\cH^{V}}$) correspond to the boundary condition
$\psi(\rho,t)\,\rightarrow\, 0$ at ${\cal H}^{-}$ (${\cal H}^+$).
The `absorbed' and `emitted' D0-brane boundary states \eqref{falling
D0}, \eqref{emitted D0} are consistent {\sl only} in the subspaces
$\cH^U$, $\cH^V$ ($\widehat{\cH^U}$, $\widehat{\cH^V}$), while the
`symmetric' D0-brane boundary state \eqref{symmetric D0} is
well-defined in the entire Hilbert space $\cH$ ($\widehat{\cH}$). We
thus have simple relations
\begin{align}
\ket{B;\rho_0,t_0}_{\msc{absorb}} &= P_U \,
\ket{B;\rho_0,t_0}_{\msc{symm}} & \mbox{and} \qquad 
{}_{\msc{absorb}}\! \bra{B;\rho_0,t_0} &=
{}_{\msc{symm}}\!\bra{B;\rho_0,t_0}\,\widehat{P_U}~,\cr
\ket{B;\rho_0,t_0}_{\msc{emitted}} &= P_V \,
\ket{B;\rho_0,t_0}_{\msc{symm}} &  \mbox{and} \qquad
 {}_{\msc{emitted}}\!\bra{B;\rho_0,t_0} &=
{}_{\msc{symm}}\!\bra{B;\rho_0,t_0}\,\widehat{P_V}~,
\label{proj symmetric D0}
\end{align}
where $P_{U, V}$ ($\widehat{P_{U,V}}$) denotes projection of the
Hilbert space ${\cal H}$ to $\cH^{U,V}$ ($\widehat{\cH^{U, V}}$).

\subsection{Rolling D-brane gathers moss}\label{sec:7-3}
As we have discussed in section \ref{sec:4}, it is of critical importance to study the $1/k$ corrections to the boundary states in order to understand the ``black hole - string transition" probed by our rolling D-brane. We first note that the boundary wavefunction itself is an analytic function with respect to $k$,\footnote{A possible exception is the factor $\nu_b^{2j+1}$. However, this factor can be absorbed (renormalized) into the cosmological constant operator of the $\mathcal{N}=2$ Liouville theory, or the mass of the two-dimensional black hole, so we will neglect this small subtlety.} so the boundary wavefunction itself is a well-defined quantity even for $k<1$. An alternative way to confirm this is to note that, at least in the Euclidean signature, the boundary wavefunction satisfies the conformal bootstrap equation for the dual $\mathcal{N}=2$ Liouville theory whose description is more reliable for $k<1$.

To see the effect of the $1/k$ corrections clearly, it is convenient to go to the coordinate space representation rather than the momentum space representation. For simplicity, we take the linear dilaton (extremal) limit of the boundary wavefunction (see section \ref{sec:8-5} for details of this limit). In the momentum space we have,
\begin{align}
 \Psi(\rho_0,t_0;p,\om) = \frac{1}{2} B(\nu_+,\nu_-)
\Gamma\left(1+i\frac{p}{k}\right)\, e^{-ip\rho_0-i\om t_0}~,
 \quad \mbox{where} \quad \nu_{\pm}
\equiv \frac{1}{2}- i\frac{p\pm \om}{2}~.
\end{align}

We can Fourier transform this boundary wavefunction to obtain the boundary wavefunction in the coordinate space:
\begin{align}
\Psi(\rho,t) = \frac{\sqrt{k}}{\pi(2\cosh t)^{k+1}} \exp\left[-k\rho- \frac{e^{-k\rho}}{(2\cosh t)^{k}} \right] \ .
\end{align}
It will be localized along the classical trajectory:
\begin{align}
\rho_0(t) = - \log(2\cosh t) \ 
\end{align}
in the semiclassical limit (i.e. $k\to \infty$). For finite $k$, the classical trajectory is smeared. 

To go further, we study the energy momentum distribution for finite $k$. Expanding the boundary states and reading the coupling to the gravity, we obtain (see \cite{Nakayama:2004ge} for details of the computation)
\begin{align}
T_{00} = \left(\frac{e^{-\rho}}{2\cosh t}\right)^{k-1} \exp\left[-\left(\frac{e^{-\rho}}{2\cosh t}\right)^{k}\right] \ .
\end{align}
The distribution of the energy is Poisson type and the maximum of the energy density is now located at
\begin{align}
\frac{e^{-\rho}}{2\cosh t} = 1-\frac{1}{k} \ .
\end{align}
The variance of the distribution is computed as
\begin{align}
\Delta\rho \simeq \sqrt{\frac{1}{2(k-1)}} \ , \label{ditvar}
\end{align}
which can be regarded as the smearing factor for the classical trajectory due to the $\alpha'$ corrections. One might say that the rolling D-brane gathers moss in the $\alpha'$ corrected black hole background. The moss could be identified with the analytic continuation of the winding tachyon \cite{Kutasov:2005rr}. Indeed the similar smearing factor in the Euclidean hairpin brane can be understood from the open string winding tachyon condensation near the tip of the Euclidean hairpin.

We again emphasize that the coordinate space wavefunction itself is an analytic function with respective $k$. However, below $k=1$, the variance of the smeared D-brane trajectory \eqref{ditvar} diverges, which means that the boundary wavefunction does not have a sensible interpretation as a rolling D-brane in the classical two-dimensional black hole any more. The transition point exactly coincides with the ``black hole - string transition" point we discussed in section \ref{sec:4}. The classical black hole appears no more black hole at this point, and the D-brane cannot role down into the hole as a probe. In section \ref{sec:8}, we compute the closed string radiation rate from the rolling D-brane, and see explicitly that the radiation rate also reveals such a phase transition as expected. As a consequence, we will see that the ``tachyon - radion correspondence" and the universality of the decaying D-brane breaks down.

As a generalization of the construction, we can introduce the fundamental string charge (electric flux) along the rolling D-brane boundary states. The construction is based on the Lorenz boost technique reviewed in section \ref{sec:5-2-6}. The corresponding boundary states have been studied in \cite{Nakayama:2004ge,Chen:2004vw}.
\newpage
\sectiono{Black Hole - String Transition from Probe Rolling D-brane}\label{sec:8}In this section we compute the closed string radiation rate from the rolling D-brane. The organization of this section is as follows. In section \ref{sec:8-1}, we compute the closed string radiation rate from the closed string perspective. In section \ref{sec:8-2}, we study the same closed string radiation rate from the open string perspective, establishing the consistency between unitarity and channel duality. In section \ref{sec:8-3}, we discuss the black hole - string transition from the probe rolling D-brane. In section \ref{sec:8-4}, the boundary states and radiation in R-R sector are discussed. In section \ref{sec:8-5}, we study the extremal NS5-brane limit. Finally in section \ref{sec:8-6}, we present the physical interpretations of Hartle-Hawking states for rolling D-branes.\footnote{This section is based on \cite{Nakayama:2005pk,Nakayama:2006qm}.}

\subsection{Radiation out of rolling D-brane from closed string viewpoint}\label{sec:8-1}
In the background of the black hole, the D0-brane moves along the
geodesic and we have constructed a variety of boundary states
describing the geodesic motion, specified by appropriate boundary
conditions.

Both by gravity and by strong string coupling gradient, the
D$p$-brane is pulled in and finds its minimum energy and mass at the
location of the NS5-brane. The D$p$-brane is supersymmetric in flat
space-time, but preserves no supersymmetry in black NS5-brane
background. Even in extremal NS5-brane background, until the
D$p$-brane dissociates into the NS5-brane and form a non-threshold
bound-state, the space-time supersymmetry is completely broken. In
these respects,the D$p$-brane propagating in the NS5-brane
background is much like excited D$p$-brane (many excited open
strings attached on it) in flat space-time. Decay of the latter via
closed string emission was studied extensively for $p=1$
\cite{Callan:1996dv,Das:1996wn}: the decay spectrum was found to match exactly
with the Hawking radiation of the non-extremal black hole made out
of these excited D-branes, and the effective temperature of excited
open string modes agrees exactly with the Hawking temperature. In
this section, we shall find certain analogous results for the closed
string radiation off the rolling D0-brane, though special features
also arise.

As the D0-brane is pulled in, acceleration would grow and radiates
off the binding energy into closed string modes. Details of the
radiation spectra would differ for different choice of the boundary
conditions, viz. for different boundary states of the D0-brane. In
this section, as a probe of the black hole geometry and D-brane
dynamics therein, we shall analyze spectral distribution of the
closed string radiation off the rolling D0-particle.

By applying the optical theorem, the radiation rate during the
radion-rolling process is obtainable as the imaginary part of the
annulus amplitude in the closed string channel.\footnote{For the
tachyon rolling process in flat space-time background, the amplitude
was evaluated first in \cite{Lambert:2003zr,Karczmarek:2003xm}.} Denote the differential
number density $\dd {\cal N}(p, M)$ of the radiation at a fixed
value of the radial momentum $p$ and the mass-level $M$. By the
definition of the D-brane boundary state, the radiation number
density $\dd \cN$ is then given in terms of the boundary wave
functions:
\begin{align}
\dd \cN (p,M)   &:= {\dd p \over 2 \pi} {\dd M \over (2 \pi)^d}
\int{\dd \om} \, \Big< \Psi(\om, p, M) \Big\vert \delta(L_0 +
\overline{L}_0) \Big\vert \Psi (\om, p, M) \Big>
\nonumber \\
&= \frac{\dd p}{2 \pi} \frac{\dd M}{(2 \pi)^d} \frac{1}{2
\om(p,M)}\, \Big| \Psi(p,\om(p,M))\Big|^2 ~. \label{radiation rate
0}
\end{align}
Here, $\om, p$ are the energy and the radial momentum in
two-dimensional Lorentzian background, $M$ is the total mass
(conformal weight) of the remaining subspaces of dimension $d$
(including mass gap), $\Psi(\om, p, M)$ is the boundary wave
function (including that of the remaining subspace), and
$\om(p,M)(>0)$ is the on-shell energy of the radiated closed string
state determined by the on-shell condition $L_0 + \overline{L}_0 =
0$ including the ghost contribution. From the kinematical
consideration, it is obvious that the differential number density
\eqref{radiation rate 0} is nonzero only when the D-brane is rolling.
Of particular physical interest is the spectral distribution in the
phase-space, as measured by the independent moments, {\em e.g.}
\begin{align} \Big< \om^m M^n \Big> &= \int \frac{\dd p}{2 \pi} \frac{\dd
M}{(2 \pi)^d} \om^m(p, M) M^n \frac{1}{2 \om(p, M)}  \Big|\Psi(p,
\om(p,M))\Big|^2 \nonumber \end{align}
for $m, n =0, 1, 2, \cdots$. We shall evaluate these spectral
observables by first evaluating the integral over the radial
momentum $p$ by saddle-point approximation. In doing so, we pay
particular attention to the asymptotic behavior as the mass-level
$M$ becomes asymptotically large. We shall then evaluate the
integral over the mass-level (conformal weight) $M$, and extract the
spectral observables.

Consider the
boundary state \eqref{falling D0} describing a D0-brane absorbed by
the future horizon. The radiation emitted by the D0-brane is
decomposable into `incoming' (toward the horizon) and `outgoing'
(toward the null infinity) components in the far future. The
positive energy sector is expanded by the wavefunction $U^p_{\om}$,
and has the following asymptotic behavior at $t\rightarrow +\infty$:
\begin{align}
 & U^p_{\om}(\rho,t) \sim e^{- i\om \ln \rho -i \om t}
  + d(p, \om) e^{-\rho} e^{+ ip \rho -i\om t} \qquad \mbox{where}
  \qquad  |d(p, \om)| \sim e^{-\pi p}~.
\label{as U}
\end{align}
Here, we assumed $\om \sim M \gg 0$. The first and the second terms
correspond to the incoming wave supported around $\rho=0$ and the
outgoing wave supported in the region $\rho\sim +\infty$,
respectively. The damping factor $d(p)$ originates from the exact
reflection amplitude $\cR(p,\om)$. (See \eqref{decomp ef 2},
\eqref{decomp ef 2-2}.) To obtain the radiation number density, we
need to evaluate $\left|\Psi(p,\om)\right|^2 \times
|U^{p}_{\om}(\rho,t)|^2$. At far future infinity, the interference
term in $|U^p_\om|^2$ drops off upon taking the $p$-integral.
Therefore, after integrating over the radial momentum $p$, the
partial radiation distribution is seen to consist of the `incoming'
and `outgoing' parts:
\begin{align}
 \cN(M)_{\msc{in}} &\equiv \int_0^M \dd M {\dd \cN_{\msc{in}}
 \over \dd M} = \int_0^{\infty} \frac{\dd p}{2 \pi} \frac{1}{2\om(p,M)}
 \Big|\Psi(p,\om(p,M))\Big|^2 \nn
\cN(M)_{\msc{out}} &\equiv \int_0^M \dd M {\dd \cN_{\msc{out}}
\over \dd M} = \int_0^{\infty} \frac{\dd p}{2
\pi}\frac{1}{2\om(p,M)} \Big|d(p) \Big|^2
\Big|\Psi(p,\om(p,M))\Big|^2~. \label{radiation rate 1}
\end{align}
We shall now evaluate the branching ratio between the two radiation
rates \eqref{radiation rate 1} with emphasis on possible string
world-sheet effects. To this end, consider the conformal field theory
defined by $SL(2;\br)/U(1) \times \cM$, where $SL(2;\br)/U(1)$
denotes the (super)coset model and $\cM$ denotes a unitary
(super)conformal field theory of central charge $c_\cM$. Such
(super)conformal field theory covers a variety of interesting string
theory backgrounds. For the fermionic string, superconformal
invariance asserts that the central charge ought to be critical:
\begin{align}
 3\Big(1 + \frac{2}{k}\Big) + c_\cM =15~, \nonumber
\end{align}
where $k$ denotes the level of the super $SL(2;\br)$ current
algebra. If the background describes a stack of black NS5-branes,
$\cM= SU(2)_{k} \times \br^5$ where $k$ equals to the NS5-brane
charge. Likewise, for the bosonic string case, conformal invariance
asserts that the central charge should take the critical value:
\begin{align}
 2 + \frac{6}{\kappa-2} + c_{\cM} =26~,
\end{align}
where now $\kappa$ refers to the level of the bosonic $SL(2;\br)$
current algebra. For the background describing the black hole in
two-dimensional string theory, $\cM$ is empty and $\kappa$ should be
set to $9/4$.

It would be illuminating to analyze the branching ratio for the `rolling closed
string', viz. a closed string state of fixed transverse mass $M$ and
radial momentum $p$ propagating in black hole geometry. The
branching ratio is simply given by the reflection amplitude (see
\eqref{exactra}):
\begin{align} \left. {{\cal N}_{\rm out}(p, \om) \over {\cal N}_{\rm in } (p,
\omega)} \right|_{\rm closed \, string} = |{\cal R}(p, \om)|^2 =
{\cosh^2 \pi \left(\om-p \over 2 \right) \over \cosh^2 \pi \left(
{\om+p \over 2}\right)}~. \label{tachyon} \end{align}
As emphasized below \eqref{exactra}, string world-sheet effects are
present for the reflection amplitude ${\cal R}$ itself but, being an
overall phase, it drops out of \eqref{tachyon}. The $k$-dependence
enters in the branching ratio \eqref{tachyon} only through the
on-shell dispersion relation $\omega = \sqrt{p^2 + 2 k M^2}$. For
two-dimensional case, first studied in \cite{Dijkgraaf:1992ba} and \cite{Giveon:2003wn},
$k=1/2$, $M=0$ and $\omega = p$, so the scattering probability is
exponentially suppressed as the energy increases.

For a fixed transverse mass $M$ {\sl and} the forward radial
momentum $p$, the reflection probability of the infalling D0-brane
is given precisely by the same result as \eqref{tachyon}:
\begin{align} \left. {{\cal N}_{\rm out}(p, \om) \over {\cal N}_{\rm in } (p,
\omega)} \right|_{\rm D0-brane} = |{\cal R}(p, \om)|^2 = {\cosh^2
\pi \left(\om-p \over 2 \right) \over \cosh^2 \pi \left( {\om+p
\over 2}\right)}~. \end{align}
This is simply because back-scattering of the boundary wave function
originates from that of the closed string wave function: roughly
speaking, the boundary wave function is defined by overlap of the
closed string wave function with the classical trajectory of the
D0-brane.

Radiation out of the falling D0-brane is coherent, so we integrate
over the radial momentum $p$ as in \eqref{radiation rate 1} in
extracting the branching ratio. We shall first analyze the partial
radiation distribution at large mass-level, $M \rightarrow \infty$.
More precisely, we shall examine asymptotic behavior of ${\cal N}
(M)$ multiplied by the phase-space `degeneracy factor' $\rho(M)\sim
e^{\frac{1}{2}M \beta_{\rm Hg}}$, where $\beta_{\rm Hg}$ denotes
inverse of the Hagedorn temperature. The closed string states that
couple to the boundary states are left-right symmetric, so we need
to take the square root of the usual degeneracy factor in the closed
string sector. Here, inverse of the Hagedorn temperature is given by
\begin{align}
\beta_{\rm Hg} = 4\pi \sqrt{1-\frac{1}{2k}}~, ~~~ \label{Hagedorn
super}
\end{align}
for the superstring theory, and
\begin{align}
\beta_{\rm Hg} = 4\pi \sqrt{2-\frac{1}{2(\kappa-2)}}~, ~~~
\label{Hagedorn bosonic}
\end{align}
for the bosonic string theory, where the $1/k $
$(1/\kappa)$-correction is interpreted as the string world-sheet
effects of the two-dimensional background. These results are
derivable from the Cardy formula with the `effective central charge'
$c_{\msc{eff}}=c-24 h_{\msc{min}}
$ \cite{Kutasov:1990ua}, where $h_{\msc{min}}$ refers to the lowest conformal
weight of normalizable primary states.

\subsubsection{radiation distribution in superstring theory}\label{sec:8-1-1}
\label{radiation super}
Let us begin with the spectral distribution in
superstring theories. We shall focus exclusively on the NS-NS sector
of the radiation and defer the analysis of the R-R sector to section
 \ref{sec:8-4}. The on-shell condition of closed string state in NS-NS sector is
given by
\begin{align}
& -\frac{\om^2}{4k} + \frac{p^2}{4k}+ \frac{1}{4k}+ \Delta_\cM =
\frac{1}{2}~, \label{on-shell super}
\end{align}
where $\Delta_\cM$ denotes the conformal weight of the $\cM$-part.
The on-shell energy is given by
\begin{align}
\om \equiv \om(p,M) = \sqrt{p^2+2k M^2} \qquad \mbox{where} \qquad
M^2 \equiv 2 \left(\Delta_\cM + \frac{1}{4k} - \frac{1}{2} \right)~.
\nonumber \end{align}
Consider now a D0-brane propagating outside the black hole and
absorbed into the future horizon. The relevant boundary wave
function was constructed in \eqref{falling D0} and, from them, the
differential radiation number distributions \eqref{radiation rate 1}
can be computed. At large $\omega$ and $p$, using Stirling's
approximation, we find that
\begin{align}
\cN (M)_{\msc{in}} &= \int_0^{\infty}\frac{\dd p}{2 \pi} \frac{1}
{2\om(p,M)}\, \Big| \Psi_{\rm absorb} (\rho_0,t_0;p,\om(p,M))\Big|^2
\nonumber
\\
&\sim {1 \over M} \int_0^{\infty}{\dd p}\,
e^{+\pi\left(1-\frac{1}{k}\right)p - \pi\sqrt{p^2+2k M^2}}~ \label{N
M super in}\\
\nonumber \\
\cN (M)_{\msc{out}} &= \int_0^{\infty}\frac{\dd p}{2 \pi} \,
\Big|d(p,\omega(p,M))\Big|^2 \, \frac{1} {2 \om(p,M)} \Big|
\Psi_{\rm absorb} (\rho_0,t_0;p,\om(p,M))\Big|^2 \nonumber \\
&\sim {1 \over M} \int_0^{\infty} {\dd p} \, e^{-\pi
\left(1+\frac{1}{k}\right)p - \pi\sqrt{p^2+ 2k M^2}}~. \label{N M
super out}
\end{align}
In the second lines, we have taken $M$ large, viz. $\omega \gg p \gg
1$, and keep the leading terms only. Thus, for each fixed but large
$M$, the partial number distributions take the forms:
\begin{align}
\cN (M)_{\msc{in}} \sim \int_0^{\infty} \dd p \,
\sigma_{\msc{in}}(p) e^{-\frac{1}{2}\beta_{\rm Hw} M} \qquad
\mbox{and} \qquad N(M)_{\msc{out}} \sim \int_0^{\infty} \dd p \,
\sigma_{\msc{out}}(p) e^{-\frac{1}{2}\beta_{\rm Hw} M}~, 
\label{grey body}
\end{align}
where
\begin{align}
 \beta_{\rm Hw} = 2\pi \sqrt{2k}~,
\label{Hawking temp super}
\end{align}
is the inverse Hawking temperature of the fermionic two-dimensional
black hole. As discussed above, the radiation off the D-brane in
NS5-brane background is analogous to the decay of excited D-brane in
flat ambient space-time. 
Indeed, asymptotic expression \eqref{grey
body} suggests that open string 
excitations of energy $M$ on 
the rolling D0-brane are 
populated as the distribution function $\exp
(-{1 \over 2} \beta_{\rm Hw} M)$
and decay into closed string
radiation. 
In this interpretation, the distribution function encodes
change of available states for open string excitations on the
D0-brane after emitting radiations of energy $M$. 
Curiously,
`effective temperature' 
of the excited closed strings is set by the
Hawking temperature of the nonextremal NS5-brane, not that of a
black hole that would have been made 
of the D0-brane. It is tempting
to interpret this as indicating 
that the D0-brane represents a class
of possible excitation modes of the black NS5-brane. The closed
string states of energy $M$ emitted 
by the D0-brane are certainly coherent, 
but according to this interpretation, they still can be
recasted in effective thermal distribution set by the Hawking
temperature of the two-dimensional black hole. 
We will later discuss again the origin of such effective
thermal behavior of the rolling D0-brane 
from the viewpoints of Euclidean cylinder 
amplitudes, extending the argument of \cite{Dijkgraaf:1992ba} 
about the Hawking radiation in the purely closed string 
background. 

The functions
$\sigma_{\rm in}$ and $\sigma_{\rm out}$ are 
interpretable as the black hole `greybody' factors 
for incoming and outgoing parts of the
radiation. The factor 1/2 in the exponent of the Boltzmann
distribution function reflects the fact that only left-right
symmetric closed string states can appear 
in the boundary states and the radiated closed string modes.

The `greybody factors' $\sigma_{*}(p)$ depend on the radial momentum
$p$ exponentially, so the radiation distribution would be modified
{\sl once} the radial momentum $p$ is integrated out. Below, we
shall show this explicitly. We are primarily interested in keeping
track of string world-sheet effects set by the value of the level
$k$. We shall consider different ranges of the level $k$ separately,
and focus on the asymptotic behaviors at large $M$ via the saddle
point methods.

\begin{description}
 \item[(i) \underline{$k > 1$}: ] \hfill\break

This is the case for the black NS5-brane background. Consider first
the incoming part. Since $1- \frac{1}{k}> 0$, the dominant
contribution in the $p$-integral arises from the saddle point:
\begin{align}
p \sim p_* = \frac{k-1}{\sqrt{1-\frac{1}{2k}}} M~.\nonumber
\end{align}
Substituting this to \eqref{N M super in}, we obtain
\begin{align}
\cN (M)_{\msc{in}} \sim e^{-2\pi M \sqrt{1-\frac{1}{2k}}} =
e^{-\frac{1}{2}M \beta_{\rm Hg}}~, \label{eq:in}
\end{align}
up to pre-exponential powers of $M$. Taking account of the density
of states $\rho(M) \sim e^{\frac{1}{2}M \beta_{\rm Hg}}$, we find
that $\rho(M) \cN(M)_{\msc{in}}$ scales with powers of $M$, and is
independent of $k$. More explicitly, for the black NS5-brane $\cM=
SU(2)_{k} \times \mathbb{R}^5$, the incoming radiation
distribution of the D$p$-brane parallel to the NS5-brane yields
\begin{align}
\cN(M)_{\msc{in}} &\sim {1 \over M} \int {\dd^{5-p} {\bf k}_{\perp}
\over (2 \pi)^{5-p}} \int_0^\infty {\dd p} \, e^{\pi
(1-\frac{1}{k})p-\pi\sqrt{p^2+2k(M^2+{\bf k}_{\perp}^2)}}
\nonumber \\
&\sim M^{2-\frac{{p}}{2}}\, e^{-2\pi M\sqrt{1-\frac{1}{2k}}} \ . \nonumber
\end{align}
Taking  account of the density of states $\rho(M) \sim
M^{-3}e^{2\pi M\sqrt{1-\frac{1}{2k}}}$, the average radiation number
distribution is given by
\begin{align}
\frac{\overline{\cN}_{\msc{in}}}{V_p} \sim \int^{M_{\rm D}} {\dd M
\over M} \, M^{-\frac{p}{2}} \qquad \mbox{where} \qquad M_{\rm D} \sim {\cal
O}({1 \over g_{\rm st}}) \ . \label{conL}
\end{align}
This result coincides with the computations of \cite{Lambert:2003zr,Karczmarek:2003xm}, and
corroborates with the radion-tachyon correspondence. Interestingly,
the incoming part of the radiation number distribution in the the
nonextremal NS5-brane background is exactly the same as the
distribution in the extremal NS5-brane background. Later, we shall
examine carefully taking the extremal limit and its consequence in
section \ref{sec:8-6}. As in the extremal case, \eqref{conL} implies that
nearly all the D0-brane potential energy is released into closed
string radiations before it falls into the black hole.

On the other hand, for the outgoing radiation, the far infrared $p
\sim 0$ dominates the momentum integral. We thus obtain
\begin{align}
\cN (M)_{\msc{out}} \sim e^{-2\pi M \sqrt{\frac{k}{2}}} =
e^{-\frac{1}{2}M \beta_{\rm Hw}}~, \nonumber
\end{align}
displaying effective thermal distribution set by the Hawking
temperature. Taking account of the density of states,
\begin{align}
\rho(M) \cN (M)_{\msc{out}} \sim e^{\frac{1}{2}M \left(\beta_{\rm
Hg}-\beta_{\rm Hw}\right)} = e^{2\pi M
\left(\sqrt{1-\frac{1}{2k}}-\sqrt{\frac{k}{2}} \right)}~. \nonumber
\end{align}
This is ultraviolet finite for any $k$ since
\begin{align}
\left(1-\frac{1}{2k}\right) -\frac{k}{2} = -\frac{1}{2k}
\left(k-1\right)^2 < 0~. \label{eq:ini}
\end{align}

We thus conclude that the radiation number distribution is mostly in
the incoming part:
\begin{align} \left. {{\cal N}_{\rm out}(M) \rho (M) \over {\cal N}_{\rm
in}(M) \rho (M)} \right\vert_{\rm falling \,\, D0} \sim {e^{-{1
\over 2} \beta_{\rm Hg} M } \over e^{-{1 \over 2} \beta_{\rm Hw} M}}
= e^{2 \pi M \left( \sqrt{1 -{1 \over 2 k}} - \sqrt{k \over 2}
\right)} \ll 1. \nonumber \end{align}
Intuitively, this may be understood as follows: for the absorbed
boundary state, the boundary condition is such that the D0-brane
flux is directed from past null infinity to the future horizon. This
also corroborates the observation that $T_{t\rho}$-component of
D0-brane's energy-momentum tensor is nonzero and increases
monotonically as the D0-brane approaches the future horizon. The
outgoing part of the distribution is exponentially small compared to
the incoming part and exhibits effective thermal distribution at the
Hawking temperature. Notice that, despite being so, this outgoing
part has nothing to do with the Hawking radiation of the black hole.
The latter is the feature of the background by itself. A priori, the
outgoing radiation could be in a distribution characterized by a
temperature different from the Hawking temperature. As mentioned
above, it is tempting to interpret coincidence of the two
temperatures as a consequence of maintaining equilibrium between the
black NS5-brane and the D0-brane.

\item[(ii) \underline{$\frac{1}{2} < k  \leq 1$}: ]
\hfill\break

This is the regime which includes the conifold geometry at $k=1$.
Since $1-\frac{1}{k} \leq 0$, the dominant contribution to the
momentum integral is from $p \sim 0$, not only for the outgoing
radiation but also for the incoming one. We thus obtain
\begin{align}
\cN(M)_{\msc{in}} \sim \cN (M)_{\msc{out}} \sim e^{-2\pi M
\sqrt{\frac{k}{2}}} \equiv e^{-\frac{1}{2}M \beta_{\rm Hw}}~,
\end{align}
viz. both are in effective thermal distribution set by the Hawking
temperature. All spectral moments are manifestly ultraviolet finite
since, at large $M$, exponential growth of the density of the final
closed string states is insufficient to overcome the suppression by
the distribution. Thus,
\begin{align} {{\cal N}_{\rm out}(M) \rho(M) \over {\cal N}_{\rm in} (M) \rho
(M)} \Big|_{\rm falling \, D0} \sim 1. \nonumber \end{align}
We interpret this as indicating that the D0-brane does not radiate
off most of its energy before falling into the horizon.

\item[(iii) \underline{$k=\frac{1}{2}$} : ]
\hfill\break

This special case corresponds to empty $\cM$. The two-dimensional
background permits no transverse degrees of freedom of the string.
The physical spectrum includes massless tachyon only, with $M=0$ and
$\rho(M)=1$. We now have a crucial difference from the previous
cases for the on-shell configurations. The radial momentum $p$ is
fixed by the on-shell condition as $\om = \pm p$, so it should not
be integrated over for the final states. Consequently, we cannot
decompose the radiation distribution into incoming and outgoing
radiations,
and only the total distribution is physically relevant.

We thus obtain the following large $\om$ behavior of the radiation
distribution:
\begin{align}
\cN(\om)\sim  e^{-2\pi \om} \equiv e^{-\om \beta_{\rm Hw}}~.
\label{radiation 2D BH super}
\end{align}
Again, we have found effective thermal distribution at the Hawking
temperature! Notice the absence of extra 1/2-factor in contrast to
the previous regimes. This is not a contradiction. In the present
case, the transverse oscillators are absent and the string behaves
as a point particle. Again, the D0-brane does not radiate off most
of its energy before falling across the black hole horizon.
In the linear dilaton regime, the boundary states and possible connection with the matrix model for the two-dimensional type 0A/0B string theory have been discussed in \cite{Lapan:2005qz}. Given the phase transition we observed, however, the classical intuition of such ``rolling D-brane" in the two-dimensional noncritical string theory is rather questionable. It would be interesting to give an interpretation from the dual matrix models.
\end{description}


\subsubsection{radiation distribution in bosonic string theory}\label{sec:8-1-2}

The analysis for the bosonic string case proceeds quite the same
route. The boundary state for the infalling D0-brane includes the
string world-sheet correction factor
$\Gamma\left(1+i\frac{p}{\kappa-2}\right)$, where again $\kappa$
refers to the level of bosonic $SL(2;\br)/U(1)$ coset model. The
on-shell condition now reads
\begin{align}
& -\frac{\om^2}{4\kappa} + \frac{p^2}{4(\kappa-2)}+
\frac{1}{4(\kappa-2)} + \Delta_\cM = 1~, \label{on-shell bosonic}
\end{align}
where $\Delta_\cM$ denotes the conformal weight in the $\cM$-sector.
This is solved by
\begin{align}
& \om \equiv \om(p,M) = \sqrt{\frac{\kappa}{\kappa-2}p^2+2\kappa
M^2} \qquad \mbox{where} \qquad M^2 \equiv 2 \left(\Delta_\cM +
\frac{1}{4(\kappa-2)} - 1\right)~.
\end{align}
The partial radiation number distribution at large $M$ limit is
given by:
\begin{align}
& \cN(M)_{\msc{in}}
 \sim  {1 \over M} \int_0^{\infty} {\dd p} \,
e^{+\pi \left(1- \frac{1}{\kappa-2}\right)p - \pi
\sqrt{\frac{\kappa}{\kappa-2} p^2 + 2\kappa M^2}}~.
\label{N M bosonic in} \\
& \cN(M)_{\msc{out}}
 \sim  {1 \over M} \int_0^{\infty} {\dd p} \,
e^{-\pi \left(1+ \frac{1}{\kappa-2}\right)p
- \pi \sqrt{\frac{\kappa}{\kappa-2}p^2+ 2\kappa M^2}}~.
\label{N M bosonic out}
\end{align}
Thus, as in the superstring case, there can arise several distinct
behaviors depending on how stringy the background is.


\begin{description}
 \item[(i) \underline{$\kappa > 3$}: ]
\hfill\break
Consider first the incoming radiation part. Since
$1-\frac{1}{\kappa-2}
> 0$, the dominant contribution to the momentum integral in \eqref{N M
bosonic in} is from the saddle point
\begin{align}
p \sim p_* = \frac{\kappa-3} {\sqrt{2-\frac{1}{2(\kappa-2)}}} M~.
\nonumber
\end{align}
We thus obtain, up to pre-exponential powers of $M$,
\begin{align}
\cN(M)_{\msc{in}} \sim e^{-2\pi M \sqrt{2-\frac{1}{2(\kappa-2)} } }
\, = \, e^{-\frac{1}{2} M \beta_{\rm Hg}}~, \nonumber
\end{align}
where $\beta_{\rm Hg}$ denotes the Hagedorn temperature of the
bosonic string theory \eqref{Hagedorn bosonic}.
In this way, we again find the power-law behavior of $\rho(M) \cN
(M)_{\msc{in}}$ at large $M$, independent of the level $\kappa$.

For the outgoing radiation part, again the $p\sim 0$ dominates the
momentum integral in \eqref{N M bosonic out}. The result is
\begin{align}
\cN (M)_{\msc{out}} \sim e^{-2 \pi M \sqrt{\frac{\kappa}{2}}} =
e^{-\frac{1}{2} M \beta_{\rm Hw}}~. \nonumber
\end{align}
Here,
\begin{align}
 \beta_{\rm Hw} \equiv 2\pi \sqrt{2\kappa} \nonumber
\end{align}
is the Hawking temperature of the bosonic two-dimensional black
hole. We then obtain
\begin{align}
\rho(M) \cN (M)_{\msc{out}} \sim e^{\frac{1}{2}\left(\beta_{\rm
Hg}-\beta_{\rm Hw}\right)M} = e^{2\pi M
\left[\sqrt{2-\frac{1}{2(\kappa-2)}} -\sqrt{\frac{\kappa}{2}}
\right] }~. \nonumber
\end{align}
As in the superstring case, the exponent is always negative
definite:
\begin{align}
\left(2-\frac{1}{2(\kappa-2)}\right)
-\frac{\kappa}{2} =
-\frac{\left(\kappa-3\right)^2}{2(\kappa-2)}
 \leq  0~. \nonumber
\end{align}
so the outgoing radiation distribution (as well as spectral moments)
is manifestly ultraviolet finite.

Physical interpretation of the above results is the same as the
superstring case: The D0-brane falling into the black hole has
nonzero component $T_{t\rho}$ of the energy-momentum tensor, and
entails that dominant part of the closed string radiation is
incoming toward the future horizon. The outgoing part of the
radiation is exponentially suppressed, and is in effective thermal
distribution set by the Hawking temperature. Again, this
distribution is distinct from the Hawking radiation of the
two-dimensional black hole. As for the fermionic string, the
branching ratio is exponentially suppressed.

\item[(ii) \underline{$\frac{9}{4} < \kappa \leq 3$}: ]
\hfill\break
In this regime, $1- \frac{1}{\kappa-2} < 0$ and the momentum
integrals for both incoming and outgoing radiation distributions are
dominated by $p \sim 0$:
\begin{align}
\cN(M)_{\msc{in}} \sim \cN(M)_{\msc{out}} \sim  e^{-2 \pi M
\sqrt{\frac{\kappa}{2}}} \equiv e^{-\frac{1}{2}M \beta_{\rm Hw}}~.
\nonumber
\end{align}
Both are in effective thermal distribution at the Hawking
temperature, and all spectral moments are manifestly ultraviolet
finite since, at large $M$, the growth of the density of state does
not overcome the suppression by the distribution. The branching
ratio remains order unity.

\item[(iii) \underline{$\kappa=\frac{9}{4}$} : ]
\hfill\break
This is the most familiar situation: black hole in two-dimensional
bosonic string theory, originally studied in \cite{Witten:1991yr,Elitzur:1991cb,Mandal:1991tz,Bars:1990rb}. The
physical spectrum of closed string consists only of the massless
tachyon, so we again need to set $M=0$ and $\rho(M)=1$. The
calculation is slightly more complicated than the supersymmetric
case: The canonically normalized energy is
\begin{align}
 E = \frac{\sqrt{2}}{3}\om = \sqrt{2}p~, \nonumber
\end{align}
so we obtain
\begin{align}
& \cN (E) \sim  e^{\pi \left(1-\frac{1}{1/4}\right) p -\pi
\sqrt{\frac{9/4}{1/4}}p} = e^{-3 \sqrt{2}\pi E} \equiv e^{- E
\beta_{\rm Hw}}~. \nonumber
\end{align}
It again shows effective thermal distribution of the radiated closed
string modes at the Hawking temperature: $\beta_{\rm Hw} = 2\pi
\sqrt{2\kappa} = 3\pi \sqrt{2}$.

\end{description}


\subsubsection{radiation distribution for emitted or time-symmetric boundary states}\label{sec:8-1-3}

The closed string radiations for the other types boundary states,
viz. the `emitted' \eqref{emitted D0} or the `symmetric'
\eqref{symmetric D0} D0-branes, can be studied analogously.

For the emitted D0-brane boundary state \eqref{emitted D0}, by the
time-reversal, we should observe the radiation distribution at the
far past: $t\sim -\infty$. The relevant decomposition corresponding
to \eqref{as U} is given by (assuming $\om>0$, $p>0$)
\begin{align}
 && V^p_{\om}(\rho,t) \sim e^{i\om \ln \rho-i\om t}
  + d^*(p,\om) e^{-\rho} e^{-ip\rho -i\om t} ~,
\label{as V}
\end{align}
where the first term is supported near the past horizon
and the second term corresponds to the incoming wave from the null
infinity.
Obviously we find precisely the same behavior of the radiation
distribution as the absorbed D0-brane once the role of `in' and
`out' states are reversed. So, for $k>1$, ${\cal N}(M)_{\rm in} \sim
\exp ( - {1 \over 2} \beta_{\rm Hw} M)$ while ${\cal N}(M)_{\rm out}
\sim \exp (-{1 \over 2} \beta_{\rm Hg} M)$ and, for $ 1 \ge k >
1/2$, ${\cal N}(M)_{\rm in}$, ${\cal N}(M)_{\rm out} \sim \exp (-{1
\over 2} \beta_{\rm Hw} M)$.

Consider next the boundary state describing D0-brane in symmetric
boundary condition \eqref{symmetric D0}. Recalling the relations
\eqref{rel disc amp}, one finds that the radiation rates are simply
obtained by adding contributions from `absorbed' and `emitted'
D0-brane boundary states. Thus, the radiation distributions behave as
${\cal N}(M)_{\rm in}$, ${\cal N}(M)_{\rm out} \sim \exp (-{1 \over
2} \beta_{\rm Hg} M)$ for $k>1$ and the dependence on Hawking
temperature disappeared.\footnote{Dependence on the Hawking
temperature exponentially suppressed, so completely negligible
compared to other power-suppressed subleading terms.} We then find
that the `detailed balance' $\cN(M)_{\msc{in}} = \cN(M)_{\msc{out}}$
is obeyed. This is as expected since the boundary state
\eqref{symmetric D0} is defined so that it keeps the time-reversal
symmetry and the one-particle state unitarity manifest.

\subsubsection{revisit to the radiation distribution from 
thermal sting propagator}\label{sec:8-1-4}

To close this section we discuss the radiation 
distribution from a different angle. 
Although the argument given here would be somewhat 
heuristic, it is quite helpful to grasp physical intuition 
and to understand where the thermal-like behavior 
of closed string radiation comes from. This argument 
is much like the one given in \cite{Dijkgraaf:1992ba}, where 
the Hawking radiation of 2-dimensional black-hole is discussed
by the closed string thermal propagator. In a sense, 
our discussion is an extension of it to the open string sector.

We start with the (thermal) cylinder amplitude 
for the D1-brane on the Euclidean cigar \eqref{clss2'}
   \footnote{To be more precise, we consider
   the fermionic black-hole of level $k$ 
    and focus on the space-time bosons. 
    If considering the space-time fermions, the thermal KK 
    momentum should be half integer $n\in 1/2 + \bz$, rather
    than $n\in \bz$, which leads to the fermionic distribution 
   $1/(e^{\beta_{\msc{Hw}}\om_{p,M}}+1) $ instead of 
    $1/(e^{\beta_{\msc{Hw}}\om_{p,M}}-1)$ in the following 
    argument.}. 
This is approximately evaluated as (we omit the parameters
$\rho_0$, $\theta_0$ for simplicity)
\begin{align}
& \cA_{\msc{cylinder}}^{(E)} \cr &= \int_0^{\infty} dT\, 
 {}_{D1} \bra{B} e^{-\pi T H^{(c)}} \ket{B}_{D1} 
 \approx  \sum_M \sum_{n\in \bsz} \int dp \, \frac{1}
{p^2+ \left(\frac{2\pi n}{\beta_{\msc{Hw}}}
\right)^2+M^2} \, \sqrt{\rho(M)}
  \left|\Psi_{D1}(p,n)\right|^2~ \cr
& = \frac{\beta_{\msc{Hw}}}{2\pi} \sum_M  
\int dpdq \, \sqrt{\rho(M)}
\frac{\left|\Psi_{D1}\left(p,\frac{\beta_{\msc{Hw}}q}{2\pi}
\right)\right|^2}{p^2+q^2+M^2}\,
\left(1+ \sum_{m\in \bsz_{>0}}e^{i \beta_{\msc{Hw}} m q}
 + \sum_{m\in \bsz_{>0}}e^{-i \beta_{\msc{Hw}} m q}
\right)~.
\label{thermal cylinder}
\end{align}
Here $p$ is the radial momentum and $n$ is the KK
momentum along the asymptotic circle of cigar 
(thermal circle). 
$M$ is again the transverse mass in the $\cM$-sector and 
$\rho(M)$ is the density of closed string states.
$\beta_{\msc{Hw}}\equiv 2\pi \sqrt{2k}$  again denotes the 
inverse Hawking temperature. 
Let us try to Wick rotate it by 
the contour deformation of $q$-integration 
in the similar manner to \cite{Dijkgraaf:1992ba}. 
Setting $q=i \om$\footnote
    {Here,  $\omega$,  $p$ are  normalized 
    as $L_0 = - \frac{1}{2} \om^2 + \frac{1}{2}p^2
   + \cdots$, rather than
    $L_0 = -\frac{1}{4k} \om^2 + \frac{1}{4k} p^2 + \cdots$.},
$\cA^{(L)}_{\msc{cylinder}} = -i \cA^{(E)}_{\msc{cylinder}}$,
we formally obtain 
\begin{align}
 & \cA^{(L)}_{\msc{cylinder}} \cr & \approx
\frac{\beta_{\msc{Hw}}}{2\pi} \sum_M \int dp\, 
\sqrt{\rho(M)}\, \left\lb 
\int d\om \, \frac{
\left|\Psi_{D1}\left(p,\frac{i \beta_{\msc{Hw}}\om}{2\pi}
\right)
\right|^2
}{p^2+M^2-\om^2+i\ep}
-\frac{2\pi i }{\om_{p,M}} 
\frac{\left|\Psi_{D1}\left(p,
\frac{i \beta_{\msc{Hw}}\om_{p,M}}
{2\pi}\right)\right|^2}
{e^{\beta_{\msc{Hw}} \om_{p,M}}-1}
\right\rb ~, 
 \label{evaluation AL}
\end{align}
where $\om_{p,M} \equiv \sqrt{p^2+M^2}$ 
is the on-shell energy  and we here used 
$$
\left|\Psi_{D1}\left(p,-\frac{i \beta_{\msc{Hw}}\om_{p,M} }
{2\pi}\right)\right|^2 = 
\left|\Psi_{D1}\left(p,\frac{i \beta_{\msc{Hw}}\om_{p,M} }
{2\pi}\right)\right|^2~. 
$$
Because we have 
$ \left|\Psi_{D1}\left(p,\frac{i \beta_{\msc{Hw}}\om}{2\pi}
\right)
\right|^2  \propto 
e^{\frac{1}{2}\beta_{\msc{Hw}} |\om|}$, 
the first term (including the Feynman propagator) 
gives a UV divergent contribution. 
This is not surprising and shows the reason why the naive 
Wick-rotation of \eqref{clss2'} does not work. 
The second term shows a `thermal-like' form and 
actually contributes to the imaginary part 
of cylinder amplitude we are interested in. 
It gives the expected behavior;
\begin{align}
\Im \, \cA_{\msc{thermal}}^{(L)} &\propto 
  \frac{1}{\om_{p,M}} 
\frac{1}{e^{\beta_{Hw}\om_{p,M}}-1} \, \sqrt{\rho(M)} 
 \left|\Psi_{D1}\left(p,\frac{i \beta_{\msc{Hw}}\om}{2\pi}
 \om\right)\right|^2 \cr
&\sim \frac{\sqrt{\rho(M)} \sigma(p)}{\om_{p,M}}\,
 e^{-\frac{1}{2}\beta_{\msc{Hw}} \om_{p,M}}~,
\label{thermal behavior}
\end{align}
which reproduces the previous results \eqref{grey body},
including the correct grey body factor $\sigma(p)$.
Recall that, in our construction of Lorentzian boundary 
states, the presence of the damping factor was crucial, 
which reads as 
$\frac{\sinh \pi \sqrt{2k} p}
{\cosh \left\lb \pi \sqrt{\frac{k}{2}}(p+\om) \right\rb
\cosh \left\lb \pi \sqrt{\frac{k}{2}}(p-\om)\right\rb} $ 
in the convention here. 
This factor shows the same asymptotic behavior 
in the large $\om$ or large $p$ region as 
the Boltzmann distribution functions
$1/(e^{\beta_{\msc{Hw}}\om}\pm 1)$.
In this sense, our Wick-rotation of boundary states 
would be roughly identified with the procedure 
which keeps only the second term in \eqref{evaluation AL}, 
suggesting the origin of the thermal-like distribution
derived from our Lorentzian boundary states.

Another helpful argument is achieved by starting with 
the open string channel of thermal cylinder amplitude 
\eqref{thermal cylinder}. 
Let us focus on the asymptotic region $\rho \gg 0$
for simplicity, 
in which the hairpin D1-brane \eqref{clss2'} appears just
as two halves of $D1$-$\bar{D}1$ system, which are Dirichlet
along the thermal circle, 
(so, identified as the `$sD$-$s\bar{D}$ system' 
\cite{Strominger:2002pc}) as pointed out in \cite{Nakayama:2004yx}. 
In this set up, by a simple kinematical reason, 
we find {\em on-shell} closed string states in the cylinder
amplitude, while only {\em off-shell} states in the open 
string channel.  As discussed {\em e.g.} 
in \cite{Sugawara:2002rs,Sugawara:2003xt}, using the 
modular transformation, we can show the thermal distribution of 
{\em physical} closed string states emitted/absorbed by 
the $sD$-$s\bar{D}$ system is captured by   
the {\em unphysical} open string winding modes along 
the thermal circle\footnote
   {This is a simple extension of the standard argument 
    of the thermal toroidal partition functions 
    \cite{Polchinski:1985zf,Sathiapalan:1986db,Kogan:1987jd,O'Brien:1987pn,Atick:1988si}. For instance, the Hagedorn behavior
    is interpretable as the tachyonic instability due to the 
     {\em unphysical} winding modes along the thermal circle.
     }. 
Especially, the unit of winding energy should 
determine the temperature of thermal distribution 
of closed string states coupled with 
the $sD$-$s\bar{D}$ system. 
In the present case it is identified with the interval 
of the hairpin 
$(= \frac{1}{2}\beta_{\msc{Hw}})$, which is just associated 
to the $D1$-$\bar{D}1$ open string. 
(Note that, taking suitably the GSO projection 
into account, we can find the zero winding
modes, {\em i.e.} the $D1$-$D1$ or $\bar{D}1$-$\bar{D}1$ 
strings, are canceled out. See \cite{Nakayama:2004yx}.)
This is the simplest explanation of why we get 
the thermal-like distribution 
$\propto e^{-\frac{1}{2}\beta_{\msc{Hw}}\om_{p,M}}$
from the cylinder amplitude \eqref{thermal cylinder}.

Curiously, all the {\em regular} solutions 
of $D0$-brane motion are just straight lines 
in the Kruscal coordinates, and thus Wick-rotated 
to the hairpin profiles with the {\em same} 
interval; $\frac{1}{2}\beta_{\msc{Hw}}$. This fact
leads us to the same thermal-like behaviors \eqref{grey body} 
characterized by the Hawking temperature (before integrating
$p$ out)\footnote
   {One might ask why the D0-brane motion with different 
    `temperature' is not considered. However,  
    such D0-branes correspond to singular 
    hairpin profiles and thus 
    do to singular Lorentzian trajectories. 
    These cannot be the solutions of DBI action of D0-branes
    by the divergence of the velocity at the singular points. 
    Quite interestingly, this feature is similar to 
    the original Hawking's idea : requiring the smooth 
    Euclidean geometry, we can fix the particular asymptotic 
    periodicity of Euclidean time, which yields 
    the temperature characterizing the radiation 
    from black-hole.}, 
as is already pointed out.


\subsection{Radiation out of rolling D-brane from open string viewpoint}\label{sec:8-2}
\subsubsection{open string channel viewpoint}\label{sec:8-2-1}
What is the nature of the ultraviolet behavior of the emission
number $\overline{\cal N}$ and how is it compared to the decay of
rolling D-brane? To answer these, we shall now recast \eqref{ImZ 0} in
the open string channel, following technical procedures considered
in section \ref{sec:5-2} and appendix of \cite{Karczmarek:2003xm}.

In the closed string channel, the closed string radiation has been computed as  (see section \ref{sec:8-1})
\begin{align} \overline{\cal N} = N^2_{\rm NS} \sum_M \sqrt{\rho^{(c)}(M)}
\int_0^{\infty} {\rmd p \over 2 \pi} \, \frac{1}{2\om} {\cal P}(p,
\omega) ~, \label{ImZ 0}\end{align}
where $N_{\rm NS}$ is an appropriate numerical factor, $\om =
\sqrt{p^2 + 2kM^2}$ is the on-shell energy of the emitted closed
string, and
\begin{align} {\cal P}(p, \omega) \equiv \left|\Psi(p,\om)\right|^2
 = \frac
{\sinh\pi\sqrt{\frac{\alpha'}{2}} p} {\left(\cosh \pi
\sqrt{\frac{\alpha'}{2}} p + \cosh \pi \sqrt{\frac{\alpha'}{2}} \om \right)
\sinh \frac{\pi}{k}\sqrt{\frac{\alpha'}{2}}  p }~ \label{powerspec}
\end{align}
is the transition probability.

We begin with expanding the transition probability ${\cal P}(p,\om)$ 
of the D0-brane \eqref{powerspec} in power series of contribution of
imaginary branes:
\begin{align}
 & {\cal P}(p,\om) = \sum_{n=1}^{\infty}
a_n(p) e^{-\pi n\omega\sqrt{\frac{\alpha'}{2}}} ~,
\label{expansion}
\\
 & a_n(p)= 2(-1)^{n+1}
\frac{\sinh\left(\pi n \sqrt{\frac{\alpha'}{2}} p \right)} {\sinh
(\frac{\pi}{k}\sqrt{\frac{\alpha'}{2}} p)}~. \label{a n}
\end{align}
%
%
As before,
we parametrically rewrite \eqref{ImZ 0} as
\begin{align}
\overline{\cal N} &=
 N_{\rm NS}^2 \sum_M \sqrt{\rho^{(c)}(M)} \nn
 &\times \int_0^{\infty} {\rmd p
\over 2
 \pi\sqrt{2k}} \sum_{n=1}^{\infty} \int_{-\infty}^{\infty}
{\rmd k_0 \over 2 \pi \sqrt{2k}}\,
 \int_0^{\infty} \frac{\alpha'}{2} \rmd t_c \,
  a_n(p) e^{\pi i n \sqrt{\frac{\alpha'}{2}} k_0}
 e^{-2\pi t_c \frac{1}{4}\alpha'\left(\frac{k_0^2+p^2}{2k}+M^2\right)}~, \qquad
\label{ImZ 1}
\end{align}
by introducing the Schwinger parameter $t_c$
in the closed string channel.\footnote
  {Strictly speaking, we could have the closed string tachyon
   $M^2<0$, and the rewriting \eqref{ImZ 1} would not be
   completely correct due to the infrared divergence. We can
   avoid this difficulty by considering the GSO projected
   amplitude. We are concerned with the large $M$ asymptotics,
   so shall go on ignoring it to avoid unessential complexity.
   }

We now evaluate each contribution separately. Begin with the sum
over the transverse mass $M$. By definition, the sum gives modular
invariant cylinder amplitude of the $\cM$-sector:
\begin{align}
\sum_M \sqrt{\rho^{(c)}(M)} e^{-2\pi t_c \frac{\alpha'}{4} M^2} &=
Z^{(c)}_{\cM}(q_c) \qquad \mbox{where} \qquad q_c = e^{- 2 \pi t_c}
\nn &= Z^{(o)}_{\cM}(q_o) \qquad \mbox{where} \qquad q_o = e^{- 2
\pi t_o} \quad (t_o \equiv 1/t_c)
\end{align}
by applying the standard open-closed duality and expressing the
result in terms of the open string Schwinger parameter $t_o$.

The amplitude $Z^{(o)}_{\cM}(t_o)$ asymptotes at large $t$ to
(corresponding to the ultraviolet behavior in the closed string
channel):
\begin{align}
 Z_{\cM}^{(o)}(t_o) \sim t_o^{\gamma}\,
e^{2\pi t_o \cdot \frac{c_{\cM}}{24}}
 = t_o^{\gamma} \, e^{\pi t_o \left(1-\frac{1}{2k}\right)} \qquad
 \mbox{for} \qquad t_o \, \rightarrow\, +\infty.
\end{align}
Here, the exponent $\gamma$ is determined by the number of
non-compact Neumann directions
in the $\cM$-sector. Such details,
however, are not relevant for our discussions.

The Gaussian integral over $k_0$ is readily evaluated,
resulting in
\begin{align}
 & \overline{\cal N}
= N^2_{\rm NS} \sqrt{\frac{\alpha'}{2}} \int_0^{\infty} {\rmd p \over 2 \pi\sqrt{2k}}
\sum_{n=1}^{\infty}
  \int_0^{\infty} \frac{\rmd t_o}{t_o^2} \sqrt{t_o} \,
   a_n(p) \, e^{-\pi t_o \frac{n^2k}{2} -\frac{2\pi}{t_o} \frac{\alpha'}{4}\frac{p^2}{2k}}
  \cdot Z^{(o)}_{\cM}(t_o)~.
\end{align}
The $k_0$-integral yields the Boltzmann factor with the temperature
determined by the Euclidean periodicity
($1/Q$ in our case) for the `hairpin brane'
\cite{Ribault:2003ss,Eguchi:2003ik,Ahn:2003tt,Lukyanov:2003nj}, which is the Euclidean rotation of the rolling
D-brane, as clarified in \cite{Nakayama:2005pk,Kutasov:2005rr}.
This is essentially the same as the standard argument
for thermal tachyon in the thermal string theory
\cite{Polchinski:1985zf,Sathiapalan:1986db,Kogan:1987jd,O'Brien:1987pn,Atick:1988si}.

Our goal is to re-express the rate \eqref{ImZ 0} in the open string
channel, so we shall Fourier transform the closed string momentum
$p$ to the open string momentum $p'$. This requires a careful
treatment, because the momentum-dependent coefficients $a_n(p)$ in
\eqref{a n} could be exponentially growing functions. In such cases,
the Fourier transform may not exists in a naive sense. We start with
the identity:
\begin{align}
 & e^{-2\pi t_c \cdot \frac{1}{4} \alpha' \frac{p^2}{2k}} = \sqrt{t_o} \int_{\mathbb{R}+i\xi}
 \sqrt{\frac{\alpha'}{2}}\frac{\rmd p'}{\sqrt{2k}}\,
e^{-2\pi t_o \cdot \frac{1}{4} \alpha' \frac{p^{'2}}{2k}+2\pi i \cdot
\frac{1}{2} \alpha' \frac{p p'}{2k}} \qquad \mbox{for} \qquad \xi \in \br~.
\label{gauss xi}
\end{align}
In the $p$-integral, the function $e^{2\pi i \frac{1}{2} \alpha' \frac{p
p'}{2k}}$ works as a damping factor and renders the integral finite if
the parameter $\xi$ is chosen suitably. For later convenience, we
shall decompose $a_n(p)$ as
\begin{align}
 & a_n(p) = a_n^+(p)-a_n^-(p)~, \nn
 & a^{\pm}_n(p) \equiv (-1)^{n+1}
\frac{e^{\pm\pi n \sqrt{\frac{\alpha'}{2}} p}}
   {\sinh (\pi \frac{\pi}{k} \sqrt{\frac{\alpha'}{2}})}~.
\label{a pm}
\end{align}
%
Observing the asymptotic behavior of the coefficients $a^+_n(p)$, we
readily find that the closed string channel momentum integral $\dsp
\int \rmd p\, a^+_n(p) \, e^{2\pi i \cdot \frac{1}{2} \alpha' \frac{p p'}{2k}}$
is well-defined as long as $\xi^+_n$ is chosen within the range
$(nk-1)<\sqrt{\frac{\alpha'}{2}}\xi^+_n <
(nk+1)$. We can then safely exchange the order
of the integrals.
Carrying out the $p$-integral first, we find~\footnote
  {Here, we are temporarily shifting
  the contour as $\br \, \rightarrow\, \br-i0$
  to avoid the pole $p=0$.
   We eventually restore it back to $\br$ {\sl after}
  taking the difference
  $a_n(p) \equiv a_n^+(p)-a_n^-(p)$.
  The final result \eqref{ImZ final} remains intact,
  even if another contour shift $\br+i0$ is taken, as is
  easily checked.}
\begin{align}
  & \int_{\mathbb{R}-i0}\frac{\rmd p}{\sqrt{2k}}\, a^+_n(p)
  e^{-2\pi t_c \frac{\alpha'}{4} \frac{p^2}{2k}}
= (-1)^{n+1}\frac{i\sqrt{tk}}{\sqrt{2}} \int_{\mathbb{R} + i\xi^+_n} \frac{\rmd
p'}{\sqrt{2k}}\, \frac {e^{\pi \left(\sqrt{\frac{\alpha'}{2}}\frac{p'}{2}-i
\frac{nk}{2}\right) -2\pi t_o \frac{\alpha'}{4}\frac{p^{'2}}{2k}}} {\cosh \pi
\left(\sqrt{\frac{\alpha'}{2}}\frac{p'}{2}-i\frac{nk}{2}\right)}~.
\qquad  \label{evaluation 2}
\end{align}
Finally, we shift the contour back:
$\br+i\xi_n^+ \, \rightarrow\,
\br$ so that  the open string momentum $p'$ is real-valued. In this
step, we cross the poles so need to take care of pole contributions.
(See Figure 1.)

\begin{figure}[htbp]
    \begin{center}
  \includegraphics[width=13cm,height=10cm]
     {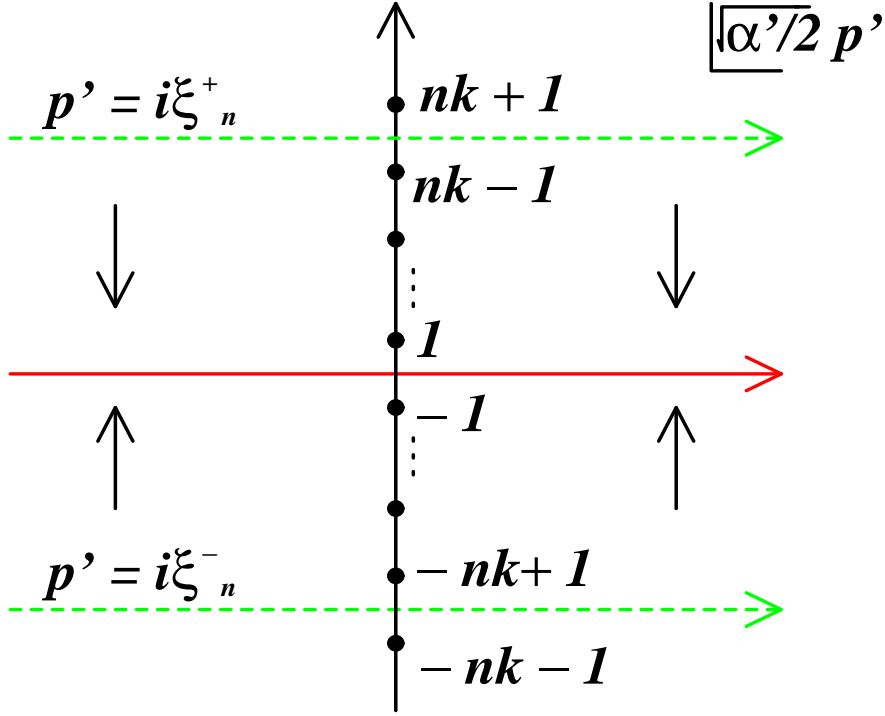}
    \end{center}
    \caption{ Deformation of the contour from the broken line
    to the solid line picks up pole contributions.}
    \label{contour1}
\end{figure}

The relevant poles are located at
\begin{align}
 & \sqrt{\frac{\alpha'}{2}}p' = i {\al_m }~, ~~~ \al_m \equiv nk
-2\left(m+\frac{1}{2}\right) \quad \mbox{where} \quad m=0,1, \ldots,
\left\lb \frac{nk}{2}-\frac{1}{2}\right\rb~, \qquad
\quad\label{poles}
\end{align}
where $\lb ~~ \rb$ denotes the Gauss symbol, and their residues are
evaluated as $(-1)^{n+1}\frac{i}{\pi} e^{\pi t_o \frac{\al_m^2}{2k}}$.
We thus obtain
\begin{align}
   \int_{\mathbb{R}-i0}\sqrt{\frac{\alpha'}{2}}\frac{\rmd p}{\sqrt{2k}}\, a^+_n(p)
   e^{-2\pi t_c \frac{\alpha'}{4} \frac{p^2}{2k}}
&=  (-1)^{n+1}\frac{i\sqrt{kt_o}}{\sqrt{2}} \int_{-\infty}^{\infty}
\sqrt{\frac{\alpha'}{2}}\frac{\rmd p'}{2k}\, \frac {e^{\pi
\left(\sqrt{\frac{\alpha'}{2}}\frac{p'}{2}-i \frac{nk}{2}\right)
-2\pi t_o \frac{\alpha'}{4} \frac{p^{'2}}{2}}} {\cosh \pi
\left(\sqrt{\frac{\alpha'}{2}}\frac{p'}{2}-i\frac{nk}{2}\right)}
\nn &+ 2(-1)^{n+1} \sqrt{t_o} \sum_{m=0}^{\left\lb
\frac{nk}{2}-\frac{1}{2}\right\rb}\, e^{\frac{\pi t_ok}{2} \left[n-\frac{2}{k}\left(m+\frac{1}{2}\right) \right]^2}~.
\label{evaluation 3}
\end{align}
The integral of $a^-_n(p)$ is calculated in a similar
way. This time, we should start with the contour $\br+i \xi^-_n$
with $(-nk- 1) < \sqrt{\frac{\alpha'}{2}}\xi^-_n <
(-nk+1)$ and, after performing the $p$-integral
first, again shift it back to $\br+i\xi^-_n\,\rightarrow\,\br$. The
relevant pole contributions come from $\sqrt{\frac{\alpha'}{2}}p'=-i\al_m$
($m=0,1,\ldots, \left\lb \frac{nk}{2}-\frac{1}{2}\right\rb$), and
we obtain
\begin{align}
\int_{\mathbb{R}-i0} \sqrt{\frac{\alpha'}{2}} \frac{\rmd p}{\sqrt{2k}}\, a^-_n(p)\,
    e^{-\pi t_c \alpha' \frac{p^2}{2k}}
&=  (-1)^{n+1}\frac{i\sqrt{t_ok}}{\sqrt{k}} \int_{-\infty}^{\infty}
\sqrt{\frac{\alpha'}{2}}\frac{\rmd p'}{\sqrt{2k}}\, \frac {e^{\pi
\left(\sqrt{\frac{\alpha'}{2}}\frac{p'}{2}+i \frac{nk}{2}\right)
-2\pi t_o \frac{\alpha'}{4} \frac{p^{'2}}{2k}}} {\cosh \pi
\left(\sqrt{\frac{\alpha'}{2}}\frac{ p'}{2}+i\frac{nk}{2}\right)}
\nn &- 2(-1)^{n+1} \sqrt{t_o} \sum_{m=0}^{\left\lb
\frac{nk}{2}-\frac{1}{2}\right\rb}\, e^{\frac{\pi t_o k}{2} \left[
n-\frac{2}{k}\left(m+\frac{1}{2}\right) \right]^2}~.
\label{evaluation 4}
\end{align}
Notice that the relative sign change in the pole term compared to
$a^+_n(p)$ integral originates from the orientation
of integration contour surrounding each pole. Therefore,
we find
\begin{align}
\int_0^{\infty}\sqrt{\frac{\alpha'}{2}}\frac{\rmd p}{\sqrt{2k}}\, a_n(p)e^{-2\pi t_c \frac{\alpha'}{4}
\frac{p^2}{2k}} &= \frac{1}{2} \int_{-\infty}^{\infty} \sqrt{\frac{\alpha'}{2}}\frac{\rmd p}{\sqrt{2k}}\,
\left(a_n^+(p) - a_n^-(p)\right)e^{-2\pi \alpha' \frac{t_c}{4} \frac{p^2}{2k}} \nn
&= (-1)^{n+1}\frac{\sqrt{t_ok}}{\sqrt{2}} \int_{-\infty}^{\infty}\sqrt{\frac{\alpha'}{2}}\frac{\rmd p'}{\sqrt{2k}}\,
\frac{\sin \left(\pi n k \right) e^{-2\pi t_o \frac{\alpha'}{4}
\frac{p^{'2}}{2k}}} {\cosh \left(\pi \sqrt{\frac{\alpha'}{2}} p'\right) +\cos
\left(\pi n k\right)} \nn
&+ 2 (-1)^{n+1} \sqrt{t_o}
\sum_{m=0}^{\left\lb \frac{nk}{2}-\frac{1}{2} \right\rb} \,  e^{\frac{\pi t_o k}{2} \left[
n-\frac{2}{k}\left(m+\frac{1}{2}\right) \right]^2}~.
\label{evaluation 5}
\end{align}
In this way,
we derive the desired open string channel expression of
the total radiation rate;
\begin{align}
 \overline{\cN} &= N^2_{\rm NS} \int_0^{\infty} \frac{\rmd t_o}{t_o}\,
\Big( F_{\rm naive} (t_o) + F_{\rm pole} (t_o) \Big)~, \nn
 F_{\rm naive} (t_o) &= \sqrt{\frac{k}{2}}\int_{-\infty}^{\infty} \sqrt{\frac{\alpha'}{2}} \frac{\rmd p'}{\sqrt{2k}}
\sum_{n=1}^{\infty} \, (-1)^{n+1} \frac{\sin \left(\pi nk\right) e^{-\pi t_o \left( \frac{\alpha'}{2} \frac{p^{'2}}{2k} +
\frac{2n^2}{k}\right)}} {\cosh \left(\pi \sqrt{\frac{\alpha'}{2}}
p' \right) + \cos \left(\pi n k\right)} \,
Z_{\cM}^{(o)}(t_o) \nn F_{\rm pole}(t_o) &= 2 \sum_{n=1}^{\infty}
(-1)^{n+1} \sum_{m=0}^{\left\lb \frac{nk}{2}-\frac{1}{2}\right\rb}
\,e^{\pi t_o \left[\frac{2}{k}\left(m+\frac{1}{2}\right)^2- 2n
\left(m+\frac{1}{2}\right)\right]}\, Z_{\cM}^{(o)}(t_o)~. \label{ImZ
final}
\end{align}
The first term in \eqref{ImZ final} coincides with the total radiation
claimed by \cite{Okuyama:2006zr} modulo inessential numerical factor~\footnote
  {It differs slightly from the one
   given in \cite{Okuyama:2006zr} in that we study
   the fermionic string, while \cite{Okuyama:2006zr} studies the bosonic string.}.
   It remains finite as $t_o\,\rightarrow\, + \infty$. The second
term, which the analysis of \cite{Okuyama:2006zr} missed altogether, is of
crucial importance. It is evident that the $m=0$ term is the leading
contribution for each $n$.\footnote{Here we have assumed that $k>1$.} Recalling $Z^{(o)}_{\cM}(t_o)\, \sim \,
e^{\pi t_o \left(1-\frac{1}{2k}\right)}$ asymptotically (up to
pre-exponential power corrections), each $m=0$ term behaves as
\begin{align}
 \sim e^{\pi t_o \left(\frac{1}{2k}- n\right)
+ \pi t_o \left(1-\frac{1}{2k}\right)} = e^{\pi t_o (1-n)} \qquad
\mbox{as} \qquad t_o\,\rightarrow\, + \infty~. \label{evaluation m=0
term}
\end{align}
Therefore, we get the leading contribution from the $n=1$ term,
which shows a massless behavior. Hence, we have reproduced
the Hagedorn-growth behavior expected in \cite{Nakayama:2004yx,Sahakyan:2004cq,
Nakayama:2004ge,Nakayama:2005pk}. Notice that all the $n>1$
contributions are massive, and thus are not relevant in the
ultraviolet regime of closed string radiations.


\subsubsection{Lorentzian cylinder amplitude}\label{sec:8-2-2}
In the previous section, we recasted the total emission number
$\overline{\cN}$ of the rolling D0-brane, defined as the sum over
the on-shell states of emitted closed string \eqref{ImZ 0}, in the
open string channel. Now, by the optical theorem and the channel
duality, we ought to be able to obtain $\overline{\cN}$ equally well
from the cylinder amplitude evaluated in the open channel. In this
section, we shall compute explicitly the cylinder amplitude in the
open string channel and show that its imaginary part reproduces
precisely the result \eqref{ImZ final}. This would serve as a
non-trivial check-point of our previous analysis for the consistency
with unitarity and the open-closed channel duality. Notice in
particular that the channel duality is far from being obvious in the
world-sheet in Lorentzian signature. For definiteness, we continue to
focus on the NS sector.

We start with the cylinder amplitude with Lorentzian
world-sheet~\footnote
  {Here, we stress the importance of
   taking the world-sheet Lorentzian.
   The Fourier transformation from the closed to
   open channel is well-defined only for
   the Lorentzian $\om_L$ in space-time. Accordingly,
    we need to take
   the Lorentzian world-sheet so that the cylinder amplitude
   becomes well-defined.} $Z_{\rm cylinder}$:
%
%
\begin{align}
& Z_{\rm cylinder} \nn
&= i \frac{\alpha'}{2} \int_{s_c^{\rm
UV}}^{s_c^{\rm IR}} \rmd s_c \int_{0}^{\infty}\frac{\rmd p}{\sqrt{2k}}
\int_{-\infty}^{\infty} \frac{\rmd \om_L}{\sqrt{2k}} \, \frac {\sinh
\left(\pi \sqrt{\frac{\alpha'}{2}}p\right)}
{\left[\cosh\left(\pi \sqrt{\frac{\alpha'}{2}}
\om_L\right) +\cosh\left(\pi \sqrt{\frac{\alpha'}{2}}p
\right)\right] \sinh(\frac{\pi}{k} \sqrt{\frac{\alpha'}{2}} p)} \nn
& \hskip3cm \times
\frac{q_c^{\frac{1}{4}\alpha'(\frac{p^{2}}{2k}-(1-i\hat{\ep})\frac{\om_L^{2}}{2k})}}
{\eta(q_c)^2} \frac{\th_3(q_c)}{\eta(q_c)}
 \cdot Z_{\cM}^{(c)}(q_c) \cdot \eta(q_c)^2
 \frac{\eta(q_c)}{\th_3(q_c)}~.
 \label{zl}
\end{align}
Here, we again adopt the $i\ep$-prescription for the Lorentzian
world-sheet, while the $-i \hat{\ep}$-prescription for the
Lorentzian space-time. The integration is well-defined so long as
$2 \hat{\ep} s_c^{\rm UV} > \ep >0$ is retained.


The second line in \eqref{zl} combines contributions of the
SL(2)$_k$/U(1), ${\cal M}$, and the world-sheet ghosts. The ghost
contribution $\eta(q_c)^2 \frac{\eta(q_c)}{\th_3(q_c)}$ is seen to
cancel out the contribution of longitudinal oscillators. Thus, the
amplitude simplifies to
\begin{align}
& \hskip0.5cm Z_{\rm cylinder} \nn &= i\frac{\alpha'}{2} \int_{s_c^{\rm
UV}}^{s_c^{\rm IR}} \rmd s_c \int_{0}^{\infty} \frac{\rmd p
}{\sqrt{2k}}\int_{-\infty}^{\infty} \frac{\rmd \om_L}{\sqrt{2k}} \, \frac {\sinh
\left(\pi \sqrt{\frac{\alpha'}{2}}p \right)
q_c^{\frac{1}{4}\alpha'(\frac{p^{2}}{2k}-(1-i\hat{\ep})^2\frac{\om_L^{2}}{2k})} \cdot
Z_{\cM}^{(c)}(q_c) } {\left[\cosh\left(\pi \sqrt{\frac{\alpha'}{2}}
\om_L\right) +\cosh\left(\pi \sqrt{\frac{\alpha'}{2}} p
\right)\right] \sinh(\frac{\pi}{k} \sqrt{\frac{\alpha'}{2}} p)}~. \nn
%
&
\label{L cylinder closed}
\end{align}

We now modular transform \eqref{L cylinder closed} to the open string
channel.
Define again the open string modulus as $q_o=e^{-2\pi i \tau_o}$,
where $\tau_o = s_o -  i \ep$ and $s_o = 1/s_c$. Using the Fourier
transform identity:
\begin{align}
 & \int_{-\infty}^{\infty}\rmd x\,
\frac{\sin(\pi a x)}{\sinh(\pi x)} e^{-2\pi i kx} = \frac{\sinh (\pi
a)}{\cosh(2\pi k)+ \cosh (\pi a)} ~, \qquad
(\left|\mbox{Im}\,a\right| < 1) \label{FT formula}
\end{align}
we then obtain
\begin{align}
& \hskip+0.5cm Z_{\rm cylinder} = \nn
& \hskip-0.8cm \frac{i\alpha'}{4} \int_{s_o^{\rm UV}}^{s_o^{\rm IR}}
\frac{\rmd s_o}{s_o} \int_{-\infty}^{\infty}\frac{\rmd p'}{\sqrt{2k}}
\int_{-\infty}^{\infty}\rmd \frac{\om_L'}{\sqrt{2k}}\, \frac {\sinh
\left(\pi \sqrt{\frac{\alpha'}{2}} \om'_L \right)
q_o^{\frac{1}{4}\al'(\frac{p^{'2}}{2k}-(1+i\hat{\ep}')^2\frac{\om_L^{'2}}{2k})} \cdot
Z_{\cM}^{(o)}(q_o) } {\left[\cosh\left(\pi \sqrt{\frac{\alpha'}{2}}
\om'_L \right) +\cosh\left(\pi \sqrt{\frac{\alpha'}{2}}p'
\right)\right]\sinh(\frac{\pi}{k} \sqrt{\frac{\alpha'}{2}} \om_L')}.\,\,\,\,\,\,
\nn
%
& \label{L cylinder open}
\end{align}
Again
$s_o^{\rm UV} \equiv 1/s_c^{\rm IR}$, $s_o^{\rm IR}\equiv
1/s_c^{\rm UV}$ are the cut-off's
and the expression \eqref{L cylinder
open} is well-defined so long as
$2 \hat{\ep}' s_o^{\rm UV}
> \ep$.

\subsubsection{analytic continuation}\label{sec:8-2-3}
We shall now analytically continue both the space-time and the
world-sheet to the Euclidean signature. We have to carefully make the
continuation so that keeping the original amplitude \eqref{L cylinder
open} unchanged (up to cut-off's). As in the previous section, we
should first Wick rotate in space-time $\om'_L\, \rightarrow\,
e^{i(\frac{\pi}{2}-0)} \om'_L$ with $\om'_L = i \om'$ $(\om' \in
\br)$, and then rotate the world-sheet $s_o \, \rightarrow \, - i
t_o$ ($t>0$).
%
%
%
We shall omit the cutoff's from now on.
We reach the expression
\begin{align} Z_{\rm cylinder} = Z_{\rm naive} + Z_{\rm pole}~, \end{align}
where the first part is the contribution from naive continuation,
while the second parts originates from the poles passed over by the
rotated contour: $\om'_L\, \rightarrow\, e^{i(\frac{\pi}{2}-0)}
\om'_L$. See Figure 2.
\begin{figure}[htbp]
    \begin{center}
   \includegraphics[width=13cm,height=10cm]
     {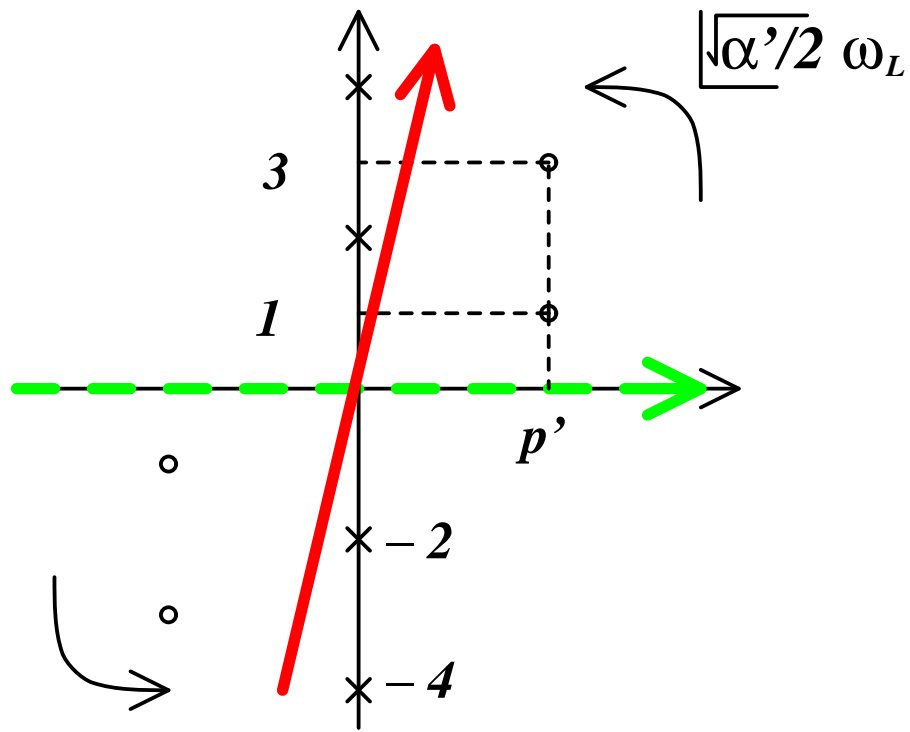}
    \end{center}
    \caption{ The $\om_L$-integral with Lorentzian contour (broken line)
    and the Euclidean contour (solid line).}
    \label{contour2}
\end{figure}
The first part $Z_{\rm naive}$ is given by
\begin{align} & \hspace{+0.2cm} Z_{\rm naive} = \nn
& \int_{0}^{\infty} \frac{\rmd t_o}{t_o}
\int_{-\infty}^{\infty}\!\!\frac{\rmd p'}{\sqrt{2k}}
\int_{(1-i0)\mathbb{R}}\frac{\rmd \om'}{\sqrt{2k}}\, \frac {-
\frac{1}{4}\alpha' \sin \left(\pi
\sqrt{\frac{\alpha'}{2}}\om' \right)
q_o^{\frac{1}{4}\alpha'(\frac{p^{'2}}{2k}+\frac{\om^{'2}}{2k})} \cdot Z_{\cM}^{(o)}(q_o) }
{\left[\cos\left(\pi \sqrt{\frac{\alpha'}{2}} \om' \right)
+\cosh\left(\pi \sqrt{\frac{\alpha'}{2}} p'\right)\right]
\sin(\frac{\pi}{k}\sqrt{\frac{\alpha'}{2}} \om')}~.  \quad \label{Z naive}
%
\end{align}
The second part $Z_{\rm pole}$ arises from the poles located at
\begin{align}
 & \sqrt{\frac{\alpha'}{2}}\om'_L =
\left\{
\begin{array}{ll}
 \sqrt{\frac{\alpha'}{2}}|p'| + {i}
 \left(2m+1\right) & ~~ m\in \mathbb{Z}_{\geq 0}  \\
 -\sqrt{\frac{\alpha'}{2}}|p'| + {i}
 \left(2m+1\right) & ~~ m\in \mathbb{Z}_{<0 }
\end{array}
\right.
\end{align}
whose residues are (after taking the open string channel modulus
Euclidean, $q_o=e^{-2\pi t}$)
\begin{align}
 & \frac{i}{2} \sqrt{\frac{2}{\alpha'}}\cdot
\frac{\sqrt{2}}{2 \pi\sqrt{k}} \frac{e^{\pm \frac{i\pi}{k}t(2m+1)\sqrt{\frac{\alpha'}{2}} |p'| -
\pi t \frac{2}{k}\left(m+\frac{1}{2}\right)^2}} {\sinh \frac{\pi}{k} \left(\pm
\sqrt{\frac{\alpha'}{2}} |p'|+i \left(2m+1\right)\right)} ~.
\end{align}
We thus obtain
\begin{align}
 &
Z_{\msc{pole}} = \int_0^{\infty} \frac{\rmd t_o}{t_o}\, \left\lb
\sum_{m\geq 0} \int_{0}^{\infty}\sqrt{\frac{\alpha'}{2}}\frac{\rmd p'}{\sqrt{2k}}
 -\sum_{m<0} \int_{-\infty}^{0}\sqrt{\frac{\alpha'}{2}}\frac{\rmd p'}{\sqrt{2k}}
\right\rb \,  
{e^{\frac{i\pi}{k} t  (2m+1)\sqrt{\frac{\alpha'}{2}} p' - \pi t \frac{2}{k}
\left(m+\frac{1}{2}\right)^2}}
\nn & \hskip3cm \times 2\pi i \cdot \frac{i\sqrt{2}}{4 \pi\sqrt{k}} \left\lb
\frac{1}{\sinh \frac{\pi}{k}  \left(\sqrt{\frac{\alpha'}{2}} p'+i
\left(2m+1\right)\right)} + (p' \leftrightarrow -p')
\right\rb Z_{\cM}^{(o)}(q_o)\nn
& \hspace{1cm} = - 2\sqrt{\frac{2}{k}}  \int_0^{\infty} \frac{\rmd t_o}{t_o}\,
\sum_{m=0}^{\infty} \int_{0}^{\infty}\sqrt{\frac{\alpha'}{2}}\frac{\rmd
p'}{\sqrt{2k}}\, \frac{e^{\frac{i\pi}{k} t (2m+1)\sqrt{\frac{\alpha'}{2}} p' - \pi t
\frac{2}{k}\left(m+\frac{1}{2}\right)^2}} {\sinh \frac{\pi}{k}  \left(
\sqrt{\frac{\alpha'}{2}} p'+i \left(2m+1 \right)\right)}
\cdot Z_{\cM}^{(o)}(q_o) ~. \nn & \label{Z pole}
\end{align}

We thus obtained manifestly convergent open string channel
expressions \eqref{Z naive}, \eqref{Z pole} for the cylinder amplitude
in Lorentzian signature of the space-time.

\subsubsection{optical theorem at work}\label{sec:8-2-4}
With the Lorentzian (in space-time) cylinder amplitude \eqref{Z naive},
\eqref{Z pole} available, we now apply the unitarity and obtain total
emission number $\overline{\cal N}$ via imaginary part of $Z_{\rm
cylinder}$. In the analysis of \cite{Okuyama:2006zr} only the naive contribution
$Z_{\msc{naive}}$ was considered. Taking the imaginary part picks up
infinite poles located at the real $\om'$-axis (the imaginary
$\om'_L$-axis), depicted in Figure 2. Their contributions yield
\begin{align}
  & \mbox{Im}\, Z_{\msc{naive}} \nn
  & = - \frac{1}{2}
\int_{0}^{\infty} \frac{\rmd t_o}{t_o} \int_{-\infty}^{\infty} \sqrt{\frac{\alpha'}{2}}\frac{\rmd p'}{\sqrt{2k}} \sum_{\stackrel{n\neq 0}{n\in \bsz}} \, \pi \mbox{sgn}\,(n) \,
\frac{(-1)^n\sqrt{k}}{\pi \sqrt{2}} \frac {\sin\left(\pi n k\right)
e^{-\pi t_o \left( \frac{\alpha'}{2} \frac{p^{'2}}{2k}+\frac{n^2 k}{2}\right)}}
{\cos\left(\pi n k \right) +\cosh \left(\pi \sqrt{\frac{\alpha'}{2}} p' \right)} \cdot Z_{\cM}^{(o)}(q_o) \nn &=
-\sum_{n=1}^{\infty}\int_{0}^{\infty} \frac{\rmd t_o}{t_o}
\int_{-\infty}^{\infty} \sqrt{\frac{\alpha'}{2}} \frac{\rmd p'}{\sqrt{2k}} \, \frac{(-1)^n\sqrt{k}}{\sqrt{2}} \frac
{\sin\left(\pi n k \right) e^{-\pi t_o \left(\frac{\alpha'}{2}
\frac{p^{'2}}{2k}+\frac{n^2k}{2}\right)}} {\cos\left(\pi n k\right)
+\cosh \left(\pi \sqrt{\frac{\alpha'}{2}}p'\right)} \cdot
Z_{\cM}^{(o)}(q_o)~, \label{ImZ 1st}
\end{align}
reproducing the first term in \eqref{ImZ final}.

We next evaluate the contribution from the pole contribution
$Z_{\msc{pole}}$ \eqref{Z pole}. As is easily seen, taking the
imaginary part just amounts to extending the integration region of
$p'$ in \eqref{Z pole}
to the whole real axis $(-\infty, \infty)$. By closing the
$p'$-contour in the upper half plane, we thus obtain
\begin{align}
& \mbox{Im}\, Z_{\msc{pole}} \nn
&=
 i\sqrt{\frac{2}{k}} \int_0^{\infty} \frac{\rmd t_o}{t_o}\,
\sum_{m=0}^{\infty} \int_{-\infty}^{\infty}\sqrt{\frac{\alpha'}{2}}\frac{\rmd p'}{\sqrt{2k}}\, \frac{e^{\frac{i\pi}{k} t_o
 (2m+1)\sqrt{\frac{\alpha'}{2}} p' - \pi t_o \frac{2}{k}\left(m+\frac{1}{2}\right)^2}}
{\sinh \frac{\pi}{k}  \left(\sqrt{\frac{\alpha'}{2}} p'+i
\left(2m+1\right)\right)} \cdot Z_{\cM}^{(o)}(q_o) \nn
&=
 2\pi i \cdot i\sqrt{\frac{2}{k}}
\int_0^{\infty} \frac{\rmd t_o}{t_o}\,   \sum_{m=0}^{\infty}
\sum_{\stackrel{n> \frac{1}{k}\left(2m+1\right)} {n\in \bsz_{>0}}}
\, \frac{(-1)^n\sqrt{k}}{\pi \sqrt{2}} e^{-\pi t_o n (2m+1)+\pi t_o
\frac{2}{k}\left(m+\frac{1}{2}\right)^2} \cdot Z_{\cM}^{(o)}(q_o)
 \nn
&= -2  \int_0^{\infty} \frac{\rmd t_o}{t_o} \,
\sum_{n=1}^{\infty} \sum_{m=0}^{\left\lb
\frac{nk}{2}-\frac{1}{2}\right\rb} \, (-1)^n e^{-\pi t_o n (2m+1)+\pi
t_o \frac{2}{k}\left(m+\frac{1}{2}\right)^2} \cdot Z_{\cM}^{(o)}(q_o)~.
\label{ImZ pole}
\end{align}
In the last line, we exchanged order of the double summations. The
final result agrees perfectly with the total emission number
$\overline{\cal N}$ in \eqref{ImZ final} evaluated via direct
computation of the transition amplitudes in Euclidean world-sheet.

\subsubsection{imaginary D-instantons: decaying versus rolling}
In the previous sections, we studied spectral observables in causal
processes involving decay of unstable D-brane and rolling of
accelerated D-brane. The main result of this work is that
transformation of the total emission number $\overline{\cal N}$ and
the cylinder amplitude $Z_{\rm cylinder}$ from the closed string
channel to the open string channel require careful analytic
continuation on the world-sheet and that, unlike other results
claimed in the literatures, the analytic continuation we adopt gives
results consistent with the unitarity via the optical theorem
$\overline{\cal N} = \mbox{Im}\, Z_{\rm cylinder}$. 
In particular, we
found that the cylinder amplitude consists in general of two parts
$Z_{\rm cylinder} = Z_{\rm naive} + Z_{\rm pole}$, and the second
part is crucial for ensuring the unitarity through its imaginary
part. While we dealt with decaying or rolling process of the
D-brane, the rules we developed ought to extend to other real-time
processes such as open string and D-brane dynamics in electric field
or plane-wave field background.

In this section, we highlights several important steps we noted in
establishing consistency between the channel duality and the
unitarity.

Throughout this work, the strategy for recasting the closed string
emission spectra in open string channel was to expand the transition
probability ${\cal P}(\omega, {\bf p})$ in power series 
of `imaginary D-instantons' \cite{Maloney:2003ck,Lambert:2003zr,Gaiotto:2003rm}, viz. 
contributions of localized states at time $2 \pi
i \alpha' W(m,n)$ for decaying D-branes and at time $(2 \pi i \sqrt{\frac{k}{2}}) n
\sqrt{\frac{\alpha'}{2}}$ for rolling D-branes, respectively.

A crucial difference we noted for the rolling D-brane in NS5-brane
background, $k>1$, that weight of the $n$-th imaginary
D-instanton, $a_n(p)$, is a non-trivial function of $p$. We
emphasized above that the momentum dependence came about because
accelerated D-brane rolls in the two-dimensional subspace
$\mathbb{R}_t \times \mathbb{R}_\phi$. Being process dependent, it
could be that, in general, {\em the weights are exponentially
growing functions of momentum, and their Fourier transformations are
not necessarily well-defined.}
This was indeed the case for the rolling D-brane case. We thus
prescribed the Fourier transform of the D-instanton weight by
analytic continuation via a deformed integration contour. The
prescription then yielded in the open string channel the
contribution $Z_{\rm pole}$ beyond the naive one $Z_{\rm naive}$.
Moreover, whereas the naive contribution is always ultraviolet
finite, the pole contribution exhibited ultraviolet divergence.
Since $\overline{\cal N}$ (or higher spectral moment) is ultraviolet
divergent, we concluded that the presence of ultraviolet divergent
$Z_{\rm pole}$ is crucial for consistency with the unitarity and the
channel duality.

From mathematical viewpoint, we found that the pole contribution
$Z_{\rm pole}$ in \eqref{ImZ final} is present in so far as we adopt
mathematically well-posed prescription of the Fourier transform.
From physics viewpoint, we can also argue that the first term
$Z_{\rm naive}$ by itself cannot be the correct answer and the
second term $Z_{\rm pole}$ ought to dominate over the first one.

In the range $k>1$, it is easy to see that 
$\mbox{Im}\,Z_{\rm
naive}$ can take a negative value if we tune the value $k$ suitably
within this range. If $Z_{\rm naive}$ is all there is for the
cylinder amplitude, the negative value of its imaginary part
contradicts with the fact that the total emission number
$\overline{\cal N}$ is positive by definition. Moreover, for
 integral value of $k$, which corresponds to rolling
D-branes in $k$ coincident NS5 backgrounds, we observe that the
first term $Z_{\rm naive}$ vanishes identically since the integrand
vanishes. The above observations indicate that extra contribution
ought to be present to the cylinder amplitude beyond the naive
contribution, $Z_{\rm naive}$.

On other other hand, we do not have any contradiction of the
cylinder amplitude with the unitarity once the contribution $Z_{\rm
pole}$ is taken into account. This is because $Z_{\rm pole}$ is
dominant (generically divergent) over $Z_{\rm naive}$ and always
positive. We conclude that our prescription for the cylinder
amplitude renders the total emission number, as extracted from the
optical theorem as $\overline{\cal N} = \mbox{Im}\,Z$ always positive
and well-defined.

The situation is in sharp contrast to that for decaying D-brane
case. There, as recapitulated in section 2, the D-instanton weights
were constant ($a_{n,m} = 1$), so the issue of Fourier transform was
void from the outset. Again, as explained in section 2, the momentum
independence came about because unstable D-brane decays at rest (or
trivially Lorentz boosted). The situation in NS5-brane phase $k > 1$ is also in contrast to that in extreme string phase 
\cite{Giveon:2005mi},
$1/2 \le k \leq 1$, or in `out-going' radiation in nonextremal
NS5-brane background (which involves two-dimensional black hole
geometry) \cite{Nakayama:2005pk}. For these, 
the leading weight $a_1(p)$ is a bounded
function and have well-defined Fourier transformation. Thus, there
does not arise any extra contribution beyond $Z_{\rm naive}$. We
thus obtain via optical theorem an ultraviolet finite total emission
number.\footnote
  {Even in the deep stringy phase $1/2 \le k \leq 1$, $a_n(p)$ is
  exponentially divergent for sufficiently large $n$.
  Therefore, the formula given in \cite{Okuyama:2006zr} have to be still
  corrected. However, only the $n=1$ term
  could cause the Hagedorn divergence as noted above.
   Hence, this correction does
   not modify ultraviolet behavior of the emission number density.}

In the previous work \cite{Nakayama:2005pk}, we also noted that the first
D-instanton weight $a_1(p)$ is identifiable with the `grey body
factor' $\sigma(p)$ in the total emission number $\overline{\cal
N}$. There, the identification was based on saddle-point analysis
valid at large mass $M \rightarrow \infty$ in the closed string
channel. The present result in the open string channel, where the
leading ultraviolet divergence arises from the weight $a_1(p)$, then
supports the identification.\footnote{Footnote 3 of \cite{Okuyama:2006zr} claims
the saddle-point approximation used in our earlier works is invalid.
We disagree with their claim: the relevant 
integral is of the type
\begin{align} \int^\infty \rmd p \, \exp \Big[-M f\Big({p \over
M}\Big)\Big]. \nonumber \end{align}
As $M \rightarrow \infty$, the saddle point approximation is well
justified in so far as  
$$
f(p_*/M) \sim \cO(1)~, ~~~ f''(p_*/M) >0~, ~~~ 
f^{(2n)}(p_*/M) \sim \cO(1)~, ~~ (n \geq 2)~, ~~~ (p_*~:~
\mbox{saddle})~,
$$
and this is indeed our case.}

\subsubsection{comparisons}

From our analysis, it became clear the reason why \cite{Karczmarek:2003xm}
obtained the correct result for the decay of unstable D-brane is
because the contour rotation in Fourier transform did not encounter
any pole (since the D-instanton weights $a_n(p)$ were
$p$-independent constants), and the naive manipulation yielded the
correct result. In \cite{Okuyama:2006zr}, the prescription of \cite{Karczmarek:2003xm} was
taken literally also for the rolling of accelerated D-brane. It was
then concluded that $Z_{\rm naive}$ refers to the total cylinder
amplitude. We showed throughout this work that this is incorrect
since it overlooked the pole contribution $Z_{\rm pole}$. After all,
only after taking this extra contribution into account, we showed
that the cylinder amplitude is consistent with the channel duality
and the unitarity.

Finally, we find it illuminating to understand why $\overline{\cal
N}$ exhibited Hagedorn divergence in the two-dimensional string
theory studied in \cite{Klebanov:2003km}, whereas it is ultraviolet finite in
the linear dilaton background studied in \cite{Karczmarek:2003xm} in
two-dimensional space-time (that is, $c_{\msc{eff}}=0$). The reason
is because the boundary wave function (D-brane transition amplitude)
of the former has non-trivial $p$-dependence that exponentially
diverges, whereas the latter does not.

We finish this subsection with a comment on the origin of the ``black hole - string transition" from the open string viewpoint. Operationally, the transition occurs because in the double summation of \eqref{ImZ pole}, the lightest contribution $(m=0, n=1)$ is outside the rage of summation for $ k<1$. In other words, the lightest open string exchange mode, contributing to the power-like divergence  of the imaginary part is projected out. This suggests that the long range interaction between the brane is drastically different between $k<1$, and $k>1$. It would be of great interest to uncover this phenomenon and explain the intuitive reason for the breaking of the tachyon - radion correspondence from the open string viewpoint.


\subsection{Black hole - string transition}\label{sec:8-3}
It has been a recurrent theme \cite{'tHooft:1987tz,Holzhey:1991bx,Horowitz:1996nw,Susskind:1993ws,Sen:1995in} that an elementary particle
or a string is a black hole: a configuration consisting of
(multiple) strings with high enough total mass is equivalent to a
black hole of the same mass and other conserved charges as we have reviewed in section \ref{sec:4}. This brings
a question whether a given configuration is most effectively
described in terms of strings or black holes. By the black hole - string transition, we will refer to such change of the effective
description for a configuration involving massive string
excitations. Roughly speaking, the string is dual to the black hole
and vice versa.

An immediate, interesting question is whether the two-dimensional
black hole geometries is also subject to the black hole - string
transition and if so what precisely the dual of the geometries would
be. In this section, we shall investigate this transition by
studying rolling dynamics of a D0-brane placed on the background. If
the background undergoes the transition between the black hole and
the string configurations, propagation of a probe D0-brane would be
affected accordingly. The transition is triggered by $k$ or
$\kappa$, which measures characteristic curvature scale of the
background measured in sting unit and hence string world-sheet
effects. We shall explore a signal of the transition by examining
spectral distribution of the closed string radiation out of the
rolling D0-brane. Other physical observables associated with
D0-brane would certainly be equally viable probes. Though
straightforward to analyze, in this work, we shall not consider
them.

\subsubsection{probing black hole - string transition via D-brane}\label{sec:8-3-1}

In section \ref{sec:8-2}, we observed that $\cN(M)_{\msc{in}} \gg
\cN(M)_{\msc{out}}$ for both the supersymmetric and bosonic string
theories in case the string world-sheet effects are weak enough, viz.
$k > 1$ and $\kappa > 3$, respectively. Obviously, such behavior can be interpreted as indicating that the background on which the
radiative process takes place is indeed a black-hole: D0-brane falls
into the horizon and subsequent radiation is mostly absorbed by the
black hole. On the other hand, the behavior that $\cN
(M)_{\msc{in}}\sim \cN (M)_{\msc{out}} \gg \rho(M)^{-1}$ for $k<1$
or $\kappa < 3$ does not seem to bear features present in the black
hole background: while D0-brane falls inward, subsequent radiation
is not mostly absorbed by the black hole but disperse away. Since
this is the regime where the string world-sheet effects are
significant, the background may be described most effectively in
terms of strings. We are thus led to conclude that the background,
whose stringy effects are controlled by the parameter $k$ or
$\kappa$, would make a phase-transition between the black hole and
the string across $k=1$ or $\kappa = 3$. In a different physical
context, this so-called ``black hole - string transition" was studied
recently \cite{Karczmarek:2004bw,Giveon:2005mi}. What distinguishes our consideration and
result from \cite{Karczmarek:2004bw,Giveon:2005mi} is that we are probing possible
phase-transition of the (closed string) background by introducing a
D0-brane in it and studying open string dynamics.


Possible existence of such a phase transition was first hinted in
\cite{Kutasov:1990ua} in the closed string sector, where they
observed that the $\cN=2$ Liouville superpotential becomes
normalizable once $k>1$ and it violates the Seiberg bound. Recall
that the marginal interaction term is
\begin{align}
S^{\pm} = \psi^{\mp} e^{-\frac{1}{\cQ}(\phi \pm iY)}~,
~~~(\cQ=\sqrt{2/k})
\end{align}
for the $\cN=2$ Liouville theory, and
\begin{align}
S^{\pm} =  e^{-\frac{1}{\cQ}(\phi \pm \sqrt{1+\cQ^2}iY)}
\equiv e^{-\sqrt{\frac{\kappa-2}{2}} \phi \mp
\sqrt{\frac{\kappa}{2}} iY}~,
~~~ (\cQ=\sqrt{2/(\kappa-2)})~,
\end{align}
for the bosonic sine-Liouville theory, respectively. Both
interactions are normalizable (exponentially falling off in the
asymptotic far region) if the curvature is sufficiently small that
$k>1$ or $\kappa
>3$ is satisfied. As is well-known, $\cN=2$ Liouville or
sine-Liouville theory is T-dual to the $SL(2;\br)/U(1)$ coset theory
\cite{FZZ,Giveon:1999px,Giveon:1999tq,Hori:2001ax}, so the condition on the level $k$ or $\kappa$
ought naturally to descend to the two-dimensional black hole
description. Indeed, such aspect was discussed in \cite{Karczmarek:2004bw}
purely in the language of the $SL(2;\br)/U(1)$ coset theory (see
also \cite{Hori:2001ax}). Their reasoning is closely related to the
non-formation of the black hole in two-dimensional string theory
(see also \cite{Friess:2004tq} for the discussion concerning this
issue from the matrix model viewpoint).
In the strong curvature regime, $k<1$, the background is described
more effectively in terms of the $\cN=2$ Liouville theory as it is
weakly coupled. Evidently, the black hole interpretation of the
$SL(2;\br)/U(1)$ theory is less clear in this region, because the
classical $\cN=2$ Liouville theory does not admit an interpretation
in terms of black hole geometry in any obvious way.

We emphasize that such black hole - string transition is not likely
to arise perturbatively and could arise only from nonperturbative
string world-sheet effects as we have reviewed in section \ref{sec:3} and \ref{sec:4}. For instance, tree-level closed string
amplitudes are manifestly analytic with respect to the level $k$.
These amplitudes exhibit a finite absorption rate (thus displaying
the non-unitarity of the reflection amplitudes) regardless of the
value of $k$. In fact, finite-$k$ correction to the amplitudes yield
an irrelevant phase-factor \cite{Dijkgraaf:1992ba,Giveon:2003wn}.


However, as was first observed in \cite{Nakayama:2004ge}, situation changes
drastically if we consider the closed string radiation from the
rolling D-brane in such a background. In \cite{Nakayama:2004ge}, it was shown
that the distribution of radiation off D0-brane in extremal
NS5-brane background becomes ultraviolet finite for $k < 1$. In the
previous section, extending the analysis of \cite{Nakayama:2004ge}, we have
shown that the $k=1$ transition shows up manifestly in the open
string sector in the sense that branching ratio between the incoming
and the outgoing radiation distribution (as well as spectral
moments) behaves very differently across $k=1$.
Remarkably, retaining finite $1/k$-correction, which originated from
consistency with the exact reflection relations, was crucial in
obtaining physically sensible results {\sl even for} $k\gg 1$.
Cancellation between the radiation distribution and the exponential
growth of the density of states at large $M$ is quite nontrivial,
and relied crucially on precise functional dependence on $k$.

An `order-parameter' of the transition is thus provided by the
radiation distribution of rolling D-brane. The phase transition
across $k=1$ is that while the radiation distribution from the
falling D-brane exhibits powerlike ultraviolet divergence for $k>1$,
it becomes finite for $k<1$. Thus, the rolling D-brane in the $k<1$
regime does {\it not} yield a large back-reaction unlike the $k>1$
case. This is also consistent with the assertion that black hole
cannot be formed in the two-dimensional string theory: It seems
difficult to construct two-dimensional black hole by injecting
D-branes to the linear dilaton (or usual Liouville) theory.\footnote
{Such a possibility was proposed in \cite{Karczmarek:2004bw}.}

It is also worth mentioning that the radion-tachyon correspondence
is likely to fail in the two-dimensional string theory ($k=1/2$). In
fact, had we have such a correspondence, the rolling radion of the
D0-brane could be identified with the rolling tachyon of the
ZZ-brane in the Liouville theory. On the other hand, it is known
that the radiation distribution of the-ZZ brane exhibits a powerlike
ultraviolet divergence \cite{Klebanov:2003km} at leading order in string
perturbation theory, while that of the falling D0-brane does not.

\subsubsection{holographic viewpoint}\label{sec:8-3-2}

The  black hole - string transition across $k=1$ also has a natural
interpretation in terms of the holographic principle, as recently
discussed in \cite{Giveon:2005mi}. Adding $Q_1$ fundamental strings to $k$
NS5-branes, one obtains the familiar bulk geometry of the
$AdS_3/CFT_2$-duality. In this context, the density of states of the
dual conformal field theory is given by the naive Cardy formula
$S=2\pi\sqrt{\frac{cL_0}{6}}+2\pi\sqrt{\frac{\bar{c}\bar{L}_0}{6}}$
with $c = 6 k Q_1$ for $k>1$, but not for $k<1$. Rather, the central
charge that should be used in the Cardy formula is replaced by an
effective one $c_{\rm eff}= 6Q_1(2-\frac{1}{k})$ \cite{Kutasov:1990ua}. The
similar effects also showed up in the double scaling limit of the
`little string theory'(LST) \cite{Giveon:1999px,Giveon:1999tq}.\footnote
   {Even though the original `little string theory' is the theory of
   NS5-brane, so $k$ should be positive integer-valued,
one can also consider models with fractional value of the level $k$,
which is less than 1 generically. This is achieved by considering
the {\em wrapped\/} NS5-brane backgrounds, or compactifications on a
Calabi-Yau threefold having rational singularity \cite{Giveon:1999zm}.  From
the regularized torus partition function, one can prove that there
is no normalizable massless states (corresponding to the
`Lehmann-Symanzik-Zimmerman-poles' \cite{Aharony:2004xn}) in such string vacua
if $k<1$, as was discussed in {\em e.g.} \cite{Eguchi:2004ik,Eguchi:2004yi}. }
We shall now show that such change of the central charge is also
imperative for reproducing the closed string radiation distribution
correctly from the dual holographic picture.

It is an interesting attempt to reproduce the phase transition in
the radiation distribution of rolling D-brane across
$k=1$ from the holographic viewpoint. In \cite{Sahakyan:2004cq}, it was
proposed that the rolling D-brane should correspond to the decay of
a certain defect in the dual LST. We shall now extend that analysis
to the $k<1$ case and explore the phase-transition. The relevant
holographic description is based on the following two assumptions.
\begin{enumerate}
\item \underline{fixed radiation number distribution}: The radiation distribution for a fixed mass $M$ is determined
by large $k$ behavior of the pressure in the far future (past). This
is equivalent to the statement that the decay of the radion is
described by a `holographic tachyon condensation'. We assume that
there is no phase transition at $k=1$ for a fixed mass $M$.\footnote
   {Theoretically, there is no reason to exclude
    a finite $1/k$ correction here.
     We only need this assumption phenomenologically
    in order to reproduce the ten-dimensional calculation
    even for $k>1$. A priori,
     the tachyon condensation (in the critical bosonic string)
    itself may receive large string world-sheet corrections. In the
    Dirac-Born-Infeld action analysis, such potential corrections
    were completely dropped out.}
In our convention, the distribution is given by
\begin{equation}
\cN(M)_{\rm LST} \sim e^{-2\pi M \sqrt{\frac{k}{2}}} \ .
\label{asum}
\end{equation}

\item \underline{change of density of states}: The final density of closed little string states
in the `holographic tachyon condensation' is given by the square
root of the full nonperturbative density of states in LST. As is
discussed in \cite{Giveon:2005mi}, the full nonperturbative density of states
of the LST is believed to exhibit a phase transition at $k=1$: for
$k>1$, the density of states is related to the Hawking temperature
as
\begin{equation}
n(M)_{\rm LST} \sim e^{{4\pi} M \sqrt{\frac{k}{2}}} \ .
\label{asuma}
\end{equation}
In other words, the Hagedorn temperature in LST should
be equated with the Hawking temperature \cite{Aharony:1998ub}
(see also, {\em e.g.} \cite{Harmark:2000hw,Berkooz:2000mz,Kutasov:2000jp}).

On the other hand, for $k<1$, because of the non-normalizability of
the black hole excitation, the nonperturbative density of states of
the LST is equivalent to the density of states of the (dual)
perturbative string theory \cite{Giveon:2005mi}:
\begin{equation}
n(M)_{\rm LST} \sim e^{4\pi M \sqrt{1-\frac{1}{2k}}} \ .
\label{asumb}
\end{equation}
\end{enumerate}
With these assumptions, we can estimate the average radiation number
of the `holographic tachyon condensation' to be
\begin{align}
\overline{\cN}_{\rm LST} = \int_0^\infty \dd M \cN(M) \,
\sqrt{n(M)_{\rm LST}} \ . \nonumber
\end{align}
Note that, in contrast to the bulk string theory calculation, we
have no integration over the radial momentum. Substituting
\eqref{asum} and \eqref{asuma} or \eqref{asumb} according to the
value of $k$, we obtain
\begin{align}
\overline{\cN}_{\rm LST} \sim \int^\infty \dd M \, e^{-{2\pi}M
{\sqrt{\frac{k}{2}}} +{2\pi}M \sqrt{\frac{k}{2}}} \nonumber
\end{align}
for $k>1$, showing powerlike ultraviolet divergent behavior because
of the complete cancellation in the exponent, and
\begin{align}
\overline{\cN}_{\rm LST} \sim \int^\infty \dd M \, e^{-{2\pi M }
{\sqrt{\frac{k}{2}}}+2\pi M \sqrt{1-\frac{1}{2k}}} \ , \nonumber
\end{align}
for $k<1$, showing exponential suppression in the ultraviolet. It is
easy to see that this holographic dual computation reproduces the
bulk computation presented in section \ref{radiation super} up to a
subleading power dependence \eqref{eq:in}, \eqref{eq:ini}.\footnote
   {The exact determination of the pre-exponential power part
   is beyond the scope of the rough estimate presented here.
    It requires the full computational ability in the LST.}

It should be noted, however, that the cancellation between the
radiation distribution and the density of states has a different
origin in the dual holographic description as compared to the bulk
side. In the holographic description, the origin of the phase
transition is the nonperturbative density of the states in LST while
the radiation distribution at a fixed mass-level $M$ keeps its
functional form unchanged. On the other hand, in the bulk theory,
origin of the cancellation was that the radiation distribution
changes at $k=1$ due to the disappearance of the non-trivial saddle
point in the integration of the radial momentum $p$, while the
density of states is always given by the same formula. Thus the
agreement between the two descriptions is quite non-trivial and we
believe that our results provide yet another evidence of the
holographic duality for the NS5-brane and black hole physics.

Though we presented the dual description based on some assumptions,
we can turn the logic around and regard our results as a support for
such assumptions. In particular the quantum gravity phase transition
at $k=1$ in the dual theory proposed in \cite{Giveon:2005mi} is crucial for
understanding the radiation distribution out of a defect decay in
the dual LST. We thus propose our discussion in this section as a
strong support for black hole - string transition.

\subsection{Boundary states and radiation in Ramond-Ramond sector}\label{sec:8-4}

In the case of fermionic black hole background, the rolling D0-brane
would also radiate off closed string states in the Ramond-Ramond
(R-R) sector. In this section, we shall construct R-R boundary state
of the D0-brane and compute radiation rates. Since the world-sheet
theory corresponds to $\cN =2$ superconformal field theory,
correlation functions of the R-R sector and boundary states are
readily obtainable by performing the standard $\cN=2$ spectral flow.

We shall begin with discussion regarding properties of reflection
amplitudes for the R-R sector (see \cite{Giveon:2003wn} in the context of two-dimensional black hole). Recall that the reflection relation was given in the
NS-NS sector as
\begin{align}
 U^{-p}_{\om}(\rho,t)^{\rm NS}= \cR^{\rm NS}(-p,\om)
U^{p}_{\om}(\rho,t)^{\rm NS} \quad \mbox{and} \quad
V^{-p}_{\om}(\rho,t)^{\rm NS}= \cR^{\rm NS *}(-p,\om)
V^{p}_{\om}(\rho,t)^{\rm NS}~, \nonumber \end{align}
where the exact reflection amplitude $ \cR^{\rm NS}(-p,\om)$ was
defined by
\begin{align}
\cR^{\rm NS}(p,\om) = \frac{\Gamma(1+\frac{ip}{k})\Gamma(+ip)
\Gamma^2(\frac{1}{2}-i\frac{p+\omega}{2})}
{\Gamma(1-\frac{ip}{k})\Gamma(-ip)
\Gamma^2(\frac{1}{2}+i\frac{p-\omega}{2})} \ . \nonumber
\end{align}
To obtain the reflection relation of the R-R sector, we shall
perform the spectral flow by half unit of the $\cN=2$ $U(1)$
current.

In sharp contrast to the $\cN=2$ Liouville theory,
the reflection amplitude now depends on the spin structure of the
R-R sector.\footnote
  {This is because, in the $\cN=2$ Liouville theory,
   the reflection amplitudes for the momentum modes have
   a symmetry under $\omega \to -\omega$.}
Explicitly, the spectral flow is defined as $\omega \to \omega \pm
i$, where the $+$ sign corresponds to spin ($+,-$) states and $-$
sign corresponds to spin ($-,+$) states (in the
$(\frac{1}{2},\frac{1}{2})$ picture): in the $\rho \to \infty$
limit, they are described by $S^{\pm} e^{-\rho} e^{-ip \rho-i\omega
t}$ and the conformal weight is given by $h =
\frac{p^2-\omega^2+1}{4k} + \frac{1}{8}$.

Therefore, for the R-R states with spin ($+,-$), the exact
reflection amplitudes become
\begin{equation}
\cR^{\rm R+}(p,\om) = \frac{\Gamma(1+\frac{ip}{k})\Gamma(+ip)
\Gamma^2(1-i\frac{p+\omega}{2})} {\Gamma(1-\frac{ip}{k})\Gamma(-ip)
\Gamma^2(1+i\frac{p-\omega}{2})} \ . \label{refRp}
\end{equation}
Equivalently, if we take spin ($-,+$) R-R states, the exact
reflection amplitudes become
\begin{equation}
\cR^{\rm R-}(p,\om) =
\frac{\Gamma(1+\frac{ip}{k})\Gamma(+ip)\Gamma^2(-i\frac{p+\omega}{2})}
{\Gamma(1-\frac{ip}{k})\Gamma(-ip)\Gamma^2(+i\frac{p-\omega}{2})} \
. \label{refRm}
\end{equation}
It is important to notice that the latter amplitudes have a second
order zero in the light-cone direction $p = \omega >0$ (recall that
$p>0$ in our convention). Similarly, we could derive the reflection
relation for $(\pm,\pm)$ spin structure, but the resultant
amplitudes are compatible only with the analytic continuation to the
`winding time' (in the interior of the singularity), so we would not
delve into details anymore.

Consider next the boundary wave function of the R-R sector. For
definiteness, we shall take the absorbed D0-brane \eqref{falling D0}
(We focus on the $t_0=0$ case for simplicity.)
\begin{align}
{}_{\msc{absorb}}\!\bra{B,{\rm NS};\rho_0} =
\int_0^{\infty}\frac{\dd p}{2\pi} \int_{-\infty}^{\infty}\frac{\dd
\om}{2\pi}\,
  \Psi_{\msc{absorb:NS}}(\rho_0;p,\om) \,
{}^{\widehat{U}}\!\dbra{p,\om} \nonumber \end{align}
where
\begin{align} \Psi_{\msc{absorb:NS}}(\rho_0;p,\om) =
\frac{\Gamma(\frac{1}{2}-i\frac{p+\omega}{2})
\Gamma(\frac{1}{2}-i\frac{p-\omega}{2})}{\Gamma(1-ip)}
\Gamma\left(1+\frac{ip}{k}\right) \, \left[ e^{-ip\rho_0} -
\frac{\cosh\left(\pi \frac{p-\om}{2}\right)} {\cosh\left(\pi
\frac{p+\om}{2}\right)} e^{+ip\rho_0} \right]~. \nonumber
\end{align}
The boundary wave functions of the R-R sector are then derived by
applying the ${\cal N}=2$ spectral flow $\omega \to \omega \pm i$:
\begin{align}
 \Psi_{\msc{absorb:R}+}(\rho_0;p,\om)
\frac{\Gamma(-i\frac{p+\omega}{2})
\Gamma(1-i\frac{p-\omega}{2})}{\Gamma(1-ip)}
\Gamma\left(1+\frac{ip}{k}\right) \, \left[ e^{-ip\rho_0} +
\frac{\sinh\left(\pi \frac{p-\om}{2}\right)} {\sinh\left(\pi
\frac{p+\om}{2}\right)} e^{+ip\rho_0} \right]~, \nonumber
\end{align}
and
\begin{align}
\Psi_{\msc{absorb:R}-}(\rho_0;p,\om) =
\frac{\Gamma(1-i\frac{p+\omega}{2})
\Gamma(-i\frac{p-\omega}{2})}{\Gamma(1-ip)}
\Gamma\left(1+\frac{ip}{k}\right) \, \left[ e^{-ip\rho_0} +
\frac{\sinh\left(\pi \frac{p-\om}{2}\right)} {\sinh\left(\pi
\frac{p+\om}{2}\right)} e^{+ip\rho_0} \right]~, \nonumber
\end{align}
for the two opposite spin structures. These boundary wave functions
are of course consistent with the exact reflection amplitudes
\eqref{refRp},\eqref{refRm}.

From these boundary wave functions, we can deduce some physical
properties of the boundary states in the R-R sector:
\begin{itemize}
\item For $k > {1 \over 2}$, in the saddle point approximation of
the radial momentum integral,
radiation distribution of the R-R sector behaves the same as that of
the NS-NS sector. In particular, the absolute value of the
reflection amplitudes behave in the similar manner. Thus, the
radiation distribution of the R-R sector is the same as that of the
NS-NS sector.
\item For $k = {1 \over 2}$, viz. the two-dimensional black hole,
considerable differences arise. Both boundary wave function and
reflection amplitudes show singularity (or zero) when we take
particular spin structure. It is not clear what the origin of these
singularities of lightlike on-shell states $p = \omega$ would be. We
note that some related discussions were given in \cite{Giveon:2003wn}.
\item In the mini-superspace limit $k \to \infty$,
the mass gap in the R-R sector vanishes. Therefore, it is well-posed
to question radiation of the massless R-R states off the R-R charge.
From the boundary states given above, we observe that, assuming $p,
\omega > 0$, there is no lightlike pole in $R+$ state while there is
a pole at $p = + \omega$ in the $R-$ state. It is also interesting
to note that, in the subleading contribution proportional to
$e^{+ip\rho_0}$, the pole from the gamma function is cancelled by
the zero in the $\sinh(\pi\frac{p-\omega}{2})$ factor.

A possible interpretation is that, roughly speaking, R-R charge is
localized on the incoming light-cone $p = \omega$.\footnote
        {This is true only in the asymptotic region
       $\rho \to \infty$ since the distribution
   near $\rho = 0$ is further related
   to the basis of Ishibashi states
   used in the expansion. In the case of `absorbed'
   basis, there is no contribution from the past horizon.
In addition, because the reflection amplitude vanishes in the $R-$
sector, an observer at $\rho \to \infty$ do not detect any outgoing
wave.}

\end{itemize}
\subsection{Back to extremal NS5-brane background}\label{sec:8-5}

By tuning off $\mu \rightarrow 0$, we are back to the extremal
NS5-brane background. Roughly speaking, the extremal background is
described by the free linear dilaton theory, but crucial differences
from the non-extremal counterpart studied in this work are the
followings:
\begin{itemize}
 \item We have no reflection relation, and the $p>0$ and $p<0$
 states should be treated as independent states.\footnote
   {In this sense, the arguments given in \cite{Nakayama:2004yx}
    are not completely precise, although the main part of
    physical results, say, the closed string
    radiation rates, are not altered.
    }
 \item The conformal field theory description is not effective in the
 entire space-time: the string coupling diverges at the location of the
 NS5-brane. We cannot completely trace the classical trajectory of
 the D0-brane \eqref{trajectory D0} without facing strong coupling problem.
\end{itemize}
We thus have to keep it in mind that the validity of the conformal
field theory description of extremal NS5-brane is limited to the
sufficiently weak string coupling region.

For the extremal NS5-brane, since the relevant conformal field
theory involves linear dilaton and hence is a free theory, we can
introduce the basis of the Ishibashi states as $\dket{p,\om}$,
$(p,\om \in \br)$ associated with the wave function
$\psi^p_{\om}(\rho,t)\propto e^{-\rho} e^{-ip \rho-i\om t}$. Another
non-trivial difference from the non-extremal case is the volume form
of the space-time. Since we have the linear dilaton $\Phi =
\mbox{const}-\rho$ and a flat metric $G_{ij}=\eta_{ij}$, the
relevant volume form becomes
\begin{align}
 \dd \mbox{Vol}= e^{-2\Phi}\sqrt{G}\dd \rho \dd t = e^{2\rho} \dd \rho \dd t~.
\label{vol linear dilaton}
\end{align}

Now, the classical trajectory of D0-brane in the extremal NS5-brane
is given by \cite{Kutasov:2004dj}:
\begin{align}
 2\cosh(t-t_0) e^{\rho} = e^{\rho_0}~.
\label{trajectory 2}
\end{align}
The boundary state describing the D0-brane moving along
\eqref{trajectory 2} ought to have the following form:
\begin{align}
 \bra{B;\rho_0,t_0} = \int_{-\infty}^{\infty}\frac{\dd p}{2\pi}\,
\int_{-\infty}^{\infty} \frac{\dd \om}{2\pi}\, \Psi
(\rho_0,t_0;p,\om) \dbra{p,\om}~. \label{symmetric D0 extremal}
\end{align}
The boundary wave function is evaluated as
\begin{align}
 \Psi(\rho_0,t_0;p,\om) &\sim \int \dd v\, \delta\Big(
2\cosh(t-t_0)e^{\rho}-e^{\rho_0} \Big)\, e^{-\rho-ip\rho-i\om t} \nn
&= \int_{-\infty}^{\infty} \dd t \, e^{-ip\rho_0} e^{-i\om t}
\Big[2 \cosh(t-t_0)\Big]^{ip-1} \nn & =
\frac{1}{2}B\left(\frac{1}{2}-i\frac{p+\om}{2},
\frac{1}{2}-i\frac{p-\om}{2} \right) \, e^{-ip \rho_0-i \om t_0}~.
\nn \end{align}
In the last expression, we used the formula \eqref{formula 2}. This is
essentially the calculation given in \cite{Nakayama:2004yx}. Finally, by
restoring the important `world-sheet correction factor'
$\Gamma\left(1+i\frac{p}{k}\right)$,\footnote
  {Since in this case we do not have the reflection relation,
   the inclusion of the factor
   $\Gamma\left(1+i\frac{p}{k}\right)$ may sound less affirmative than
   the nonextremal NS5-brane background.
   We argue that the procedure is actually justified by
   considering the limit from the non-extremal case.}
we obtain the boundary wave function
\begin{align}
 \Psi(\rho_0,t_0;p,\om) = \frac{1}{2} B(\nu_+,\nu_-)
\Gamma\left(1+i\frac{p}{k}\right)\, e^{-ip\rho_0-i\om t_0}~.
 \qquad \mbox{where} \qquad \nu_{\pm}
\equiv \frac{1}{2}- i\frac{p\pm \om}{2}~,\nonumber
\end{align}
This is the extremal counterpart of the `symmetric D0-brane' in the
non-extremal NS5-brane background \eqref{symmetric D0}.

We can also consider the `half S-brane' counterpart by
taking the Hartle-Hawking contours depicted in the Figures
\ref{HH-future} and \ref{HH-past}. Namely, for the `absorbed brane',
we obtain
\begin{align}
 {}_{\msc{absorb}}\bra{B;\rho_0,t_0}
= \left( \int_0^{\infty}\frac{\dd p}{2\pi}
\int_0^{\infty}\frac{\dd \om}{2\pi} + \int_{-\infty}^0\frac{\dd
p}{2\pi} \int_{-\infty}^0\frac{\dd \om}{2\pi}\right)\,
\Psi(\rho_0,t_0;p,\om)\, \dbra{p,\om} ~, 
\label{falling D0 extremal}
\end{align}
and for the `emitted brane',
\begin{align}
 {}_{\msc{emitted}}\bra{B;\rho_0,t_0}
= \left( \int_0^{\infty}\frac{\dd p}{2\pi}
\int^0_{-\infty}\frac{\dd \om}{2\pi} + \int_{-\infty}^0 \frac{\dd
p}{2\pi} \int^{\infty}_0\frac{\dd \om}{2\pi}\right)\,
\Psi(\rho_0,t_0;p,\om)\, \dbra{p,\om} ~. \label{emitted D0 extremal}
\end{align}
They are regarded as the counterparts of \eqref{HH symm D0 1} and
\eqref{HH symm D0 2}.

The radiation rates were already evaluated in \cite{Nakayama:2004yx,Sahakyan:2004cq}.\footnote{
In this paper, we scaled energy and momentum differently
from \cite{Nakayama:2004yx}. In light of normalization as in \eqref{on-shell
super}, $\om, p$ in this work should be read as $2 \sqrt{k}$ times
$\om, p$ in \cite{Nakayama:2004yx}.} Crucial differences from the non-extremal
case are the followings: We have the `forward radiations' ({\em
e.g.}, the incoming radiation for the absorbed D-brane \eqref{falling
D0 extremal}) only and no `backward radiations' ({\em e.g.}, the
outgoing radiation for the absorbed D-brane). This is because there
is no reflection relation in the extremal case. The forward
radiations behave in the completely same way as the non-extremal
case (that is, in a fermionic two-dimensional black hole with $k >1
$), giving rise to the Hagedorn-like ultraviolet divergence again.
At fixed but large $M$ before integrating over $p$, the partial radiation number distribution
takes again exactly the same asymptotic form as in \eqref{grey body}
except that now the coefficient $2 \pi \sqrt{2k}$ is {\sl not}
interpretable as the inverse Hawking temperature of the black hole.\footnote
{An obvious alternative interpretation could be that, even
for extremal background, the falling D0-brane excites the NS5-brane
above the extremality.} Again, this has to do with the peculiarity
that the Hawking temperature of the two-dimensional black hole is
set by the level $k$, not by the nonextremality $\mu$. On the other
hand, the absence of the backward radiation matches with the
extremality of the background; there is no Hawking radiation.


\subsection{More on physical interpretations :
Hartle-Hawking states}\label{sec:8-6}

We shall now revisit the boundary states we constructed in this work
and elaborate further on their physical interpretations with
particular emphasis on analogy with the rolling tachyon problem via
the radion-tachyon correspondence. We also elaborate on the fate of
R-R charge carried by the D0-brane. To be concrete, we shall focus
on the cases $k \geq 2$ admitting interpretation in terms of near
horizon geometry of black NS5 branes.

The boundary state \eqref{falling D0} describes the late-time rolling
($t \gg t_0$) of the D0-brane rolling into the black NS5 branes.
The relevant D0-brane has the initial condition $\rho=\rho_0$,
$\frac{d\rho}{dt}=0$ at $t=t_0$ and starts to roll down toward the
black hole. After sufficiently long coordinate time elapsed, the
D0-brane gets close to the future horizon (${\cal H}^+$). As
examined in section 4, almost all energy of the D0-brane is absorbed
by the black hole in the form of incoming radiation.
The incoming radiation is dominated by very massive, and hence
highly non-relativistic closed string excitations. Via the
radion-tachyon correspondence, these states are identifiable with
the `tachyon matter' in the rolling tachyon problem in flat
space-time. On the other hand, we have seen that a small part of
energy escapes to the spatial infinity (${\cal I}^+$) as the
outgoing radiation. We have seen that the spectral distribution is
characterized by the Hawking temperature, and is necessarily
dominated by light modes. This interpretation is quite natural from
the viewpoint of the radion-tachyon correspondence for the extremal
NS5-brane background \cite{Kutasov:2004dj}. Since we are now working with
the non-extremal NS5-brane background, our analysis may be
considered as an evidence that the correspondence is valid even at
finite temperature.


What about evolution in the far past $t < t_0$? Here, we face a
subtlety. Recall that the boundary condition defining \eqref{falling
D0} does not allow contributions from the past horizon (${\cal
H}^-$), namely, the basis of Ishibashi states $\dket{p,\om}^U$
does not reproduce the past half of the classical trajectory
\eqref{trajectory D0}. Rather, the NS-NS sector of the D0-brane
boundary wave function appears widely distributed in the space-time
in the far past. This may be interpreted as radiations imploding to
$\rho = \rho_0$ from spatial infinity, but then it is subtle to
trace the R-R charge carried by the D0-brane, created out of the
imploding radiation. Classically, the D0-brane charge density ought
to be localized along the classical trajectory \eqref{trajectory D0}
and hence emanates from the past horizon. Once stringy effects are
taken into account, the charge appears to originate from asymptotic
infinity along the light-cone coordinate. Complete understanding of
this curious feature is highly desirable but we shall relegate it to
future study. Here, instead, we present a simple prescription of
avoiding this subtlety: a version of `Hartle-Hawking' boundary
condition.

\begin{figure}[htbp]
    \begin{center}
   \includegraphics[width=0.5\linewidth,keepaspectratio,clip]{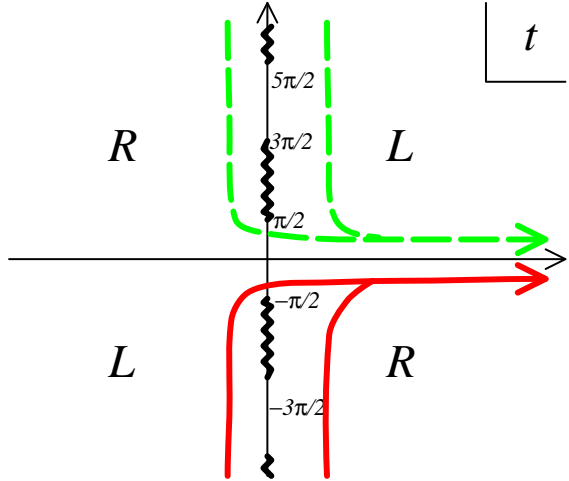}
    \end{center}
    \caption{`future Hartle-Hawking contour' : the red (green broken)
        line is the contour $\cC^+_{\msc{future}}$ for $\om > 0$
        ($\cC^-_{\msc{future}}$ for $\om <0$). The `$L$' (`$R$') contour
         should be used if calculating the overlap with
        $L^p_{\om}(\rho,t)$ ($R^p_{\om}(\rho,t)$) for
         the convergence of integral.}
    \label{HH-future}
\end{figure}
\begin{figure}[htbp]
   \begin{center}
    \includegraphics[width=0.5\linewidth,keepaspectratio,clip]{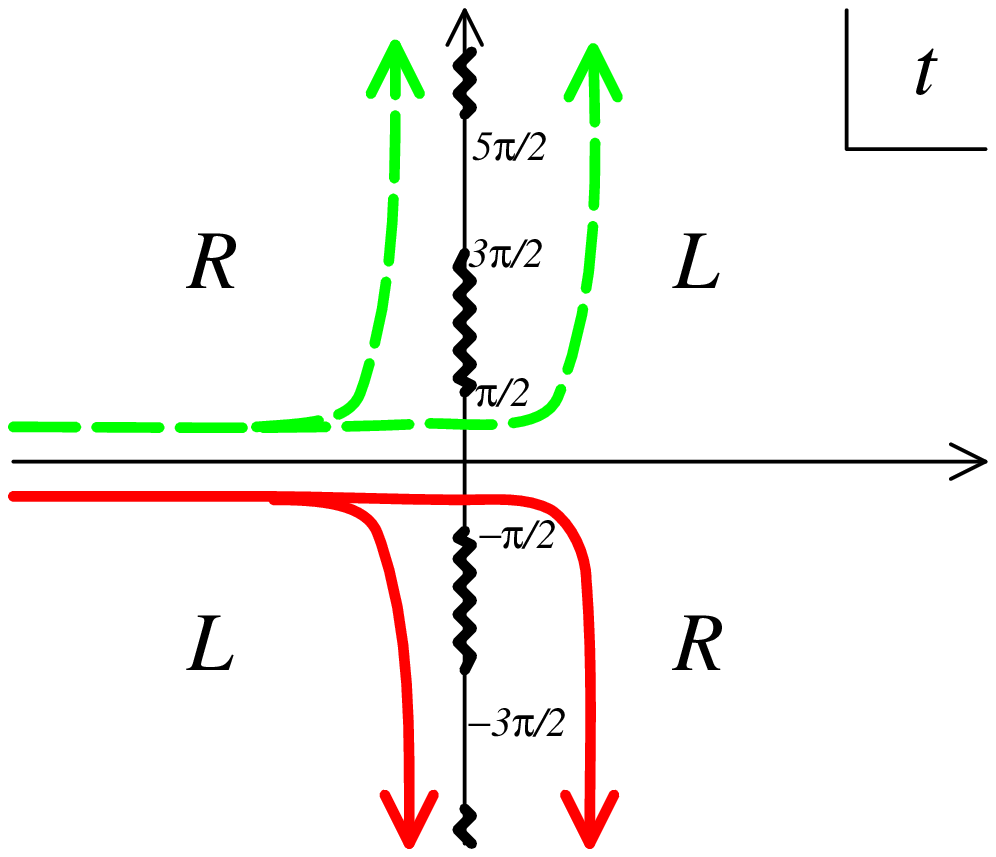}
    \end{center}
    \caption{`past Hartle-Hawking contour' : the red (green broken)
        line is the contour $\cC^+_{\msc{past}}$ for $\om > 0$
         ($\cC^-_{\msc{past}}$ for $\om <0$).}
    \label{HH-past}
\end{figure}

We shall first focus on the absorbed D0-brane boundary state
\eqref{falling D0}. Formally, by construction, we can regard
the boundary wave function specified by the time-integration over
the `real contour' $\cC= \br$ as in \eqref{Wick rotation 0}. Now, let
us discuss what happens if we choose the `Hartle-Hawking' type
contour instead of the real contour, which connect the Euclidean
time with the future or past half of real time axis at $t=t_0$:
\begin{align}
\cC_{\msc{future}}^{\pm}
= \left(t_0+i\br_{\mp}\right) \cup \left(t_0+\br_{+}\right)~,~~~
\cC_{\msc{past}}^{\pm} = \left(t_0+i\br_{\mp}\right) \cup
\left(t_0+\br_{-}\right)~.
\end{align}
More precisely, we should avoid suitably the branch cuts on
$t_0+i\br$ to render the integral convergent. See Figures
\ref{HH-future} and \ref{HH-past} for details. The superscript $+$
$(-)$ is associated with the positive (negative) energy sector. Note
that the phase-factor $e^{-i\om t}$ behaves well on the lower
(upper) half of complex $t$-plane if $\om $ is positive (negative).
Let us pick up $\cC_{\msc{future}}$. Following the traditional
interpretation of the Hartle-Hawking type wave function, we may
suppose that both the D0-brane and black NS5-brane are created from
`nothing' at $t=t_0$, and then the D0-brane starts to fall down
toward the future horizon along the classical trajectory
\eqref{trajectory D0}. In this prescription, the subtlety we mentioned
above is completely circumvented.

One may paraphrase the prescription as follows: choosing the
Hartle-Hawking contour $\cC_{\msc{future}}$, we explicitly obtain
\begin{align}
& \hspace{-5mm} {}_{HH +,\,\msc{absorb}}\!\bra{B;\rho_0,t_0} \cr 
&=
\int_0^{\infty}\frac{\dd p}{2\pi}\, \left[
\int_{0}^{\infty}\frac{\dd \om}{2\pi}\,
\Psi_{\msc{symm}}(\rho_0,t_0;p,\om) + \int_{-\infty}^{0}\frac{\dd
\om}{2\pi}\, \cR(p,\om)\Psi^*_{\msc{symm}}(\rho_0,-t_0;p,\om)
\right] \, {}^{\widehat{U}}\!\dbra{p,\om}~, \cr \label{HH falling
D0}
\end{align}
where $\Psi_{\msc{symm}}(\rho_0,t_0;p,\om)$ is defined in
\eqref{symmetric D0}. In fact, by taking $\cC_{\msc{future}}$, we are
only left with the $L^p_{\om}$ ($R^p_{\om}$)-part of the one-point
function for the $\om>0$ ($\om<0$) sector. See figure
\ref{HH-future}. This boundary wave function is formally regarded as
the limit of \eqref{falling D0} under $t_0\,\rightarrow\, -\infty$,
$\rho_0\,\rightarrow\,+\infty$ while keeping $|\rho_0|/|t_0|$
finite. Note that the second (first) term $\propto e^{ip\rho_0-i\om
t_0}$ ($\propto e^{-ip\rho_0-i\om t_0}$) in \eqref{falling D0}
oscillates very rapidly in this limit for $\om >0$ ($\om <0$) and
hence drops off.\footnote
   {More precise argument would be as follows:
    The disk amplitude for a wave packet {\em e.g.}
    $\int \frac{\dd p}{2\pi} \int \frac{\dd \om}{2\pi}\, f(p,\om) \ket{L^p_{\om}}$ is evaluated
    as $
    \lim_{\rho_0\,\rightarrow\,+\infty , \,
    t_0\,\rightarrow\,-\infty}\,
    \int \frac{\dd p}{2\pi} \int \frac{\dd\om}{2\pi} f(p,\om) \Psi(\rho_0,t_0;p,\om)$.
    Then, the rapidly oscillating term in the boundary wave
     function $\Psi(\rho_0,t_0;p,\om)$ cannot contribute for any
     $L^2$-normalizable wave packet $f(p,\om)$ due to the Riemann-Lebesgue theorem.
} The limit just means that the D0-brane moving along the
trajectory \eqref{trajectory D0} is coming from the past infinity
$({\cal I}^-)$, and falling into the future horizon (${\cal H}^+$).
Everything is supposed to be localized over the classical trajectory
in this case.


Adopting the past Hartle-Hawking contour $\cC_{\msc{past}}$ for the
boundary state of emitted D0-brane \eqref{emitted D0} is completely
parallel. We take the time-reversal of the above:
\begin{align}
& {}_{HH -,\,\msc{emit}}\!\bra{B;\rho_0,t_0} \cr &=
\int_0^{\infty}\frac{\dd p}{2\pi} \left[ \int_{0}^{\infty}\frac{\dd
\om}{2\pi}\, \Psi^*_{\msc{symm}}(\rho_0,-t_0;p,\om) +
\int_{-\infty}^{0}\frac{\dd \om}{2\pi}\,
\cR^*(p,\om)\Psi_{\msc{symm}}(\rho_0,t_0;p,\om) \right]
 {}^{\widehat{V}}\!\dbra{p,\om}~, \nn \label{HH emitted D0}
\end{align}
which is regarded as the $t_0\,\rightarrow\,+\infty$,
$\rho_0\,\rightarrow\,+\infty$ limit of \eqref{emitted D0}. It
describes the trajectory of D0-brane emitted from the past horizon
${\cal H}^-$ and escaping to the future infinity ${\cal I}^+$.


Let us turn to the `symmetric' D0-brane \eqref{symmetric D0}. Naively,
it appears that the prescription is that
\begin{align}
& {}_{HH +,\,\msc{symm}}\!\bra{B;\rho_0,t_0}' \cr &=
\int_0^{\infty}\frac{\dd p}{2\pi} \left[ \int_{0}^{\infty}\frac{\dd
\om}{2\pi}\, 2 \Psi_{\msc{symm}}(\rho_0,t_0;p,\om) \,
{}^L\!\dbra{p,\om} +\int_{-\infty}^{0}\frac{d\om}{2\pi}\, 2
\Psi^*_{\msc{symm}}(\rho_0,-t_0;p,\om) {}^R\!\dbra{p,\om} \right]
\nn \label{HH symm D0 1}
\end{align}
for the future Hartle-Hawking contour $\cC_{\msc{future}}$, and
\begin{align}
& {}_{HH -,\,\msc{symm}}\!\bra{B;\rho_0,t_0}' \cr &=
\int_0^{\infty}\frac{\dd p}{2\pi} \left[ \int_{0}^{\infty}\frac{\dd
\om}{2\pi}\, 2 \Psi^*_{\msc{symm}}(\rho_0,-t_0;p,\om)
{}^R\!\dbra{p,\om} +\int_{-\infty}^{0}\frac{d\om}{2\pi} 2
\Psi_{\msc{symm}}(\rho_0,t_0;p,\om) {}^L\!\dbra{p,\om} \right] \nn
\label{HH symm D0 2}
\end{align}
for the past Hartle-Hawking contour $\cC_{\msc{past}}$.
However, this cannot be the whole story. The existence of Euclidean
part of the Hartle-Hawking path-integral enforces the boundary
states to be expanded by the basis smoothly connected to the
Euclidean ones, while $\ket{L^p_{\om}}$, $\ket{R^p_{\om}}$ do not
possess such a property. Consequently, to achieve the correct
Hartle-Hawking states, we ought to make further the projection to
$\cH^U$, ($\widehat{\cH^U}$) for the contour $\cC_{\msc{future}}$,
and to $\cH^V$, ($\widehat{\cH^V}$) for $\cC_{\msc{past}}$. We thus
obtain as the correct Hartle-Hawking states:
\begin{align}
& {}_{HH +,\,\msc{symm}}\!\bra{B;\rho_0,t_0} =  {}_{HH
+,\,\msc{symm}}\!\bra{B;\rho_0,t_0}' \widehat{P_U} \equiv {}_{HH
+,\,\msc{absorb}}\!\bra{B;\rho_0,t_0} ~, \nn & {}_{HH
-,\,\msc{symm}}\!\bra{B;\rho_0,t_0} =  {}_{HH
-,\,\msc{symm}}\!\bra{B;\rho_0,t_0}' \widehat{P_V} \equiv {}_{HH
-,\,\msc{emitted}}\!\bra{B;\rho_0,t_0} ~, \label{relation HH}
\end{align}
where the right-hand sides are already given in \eqref{HH falling D0},
\eqref{HH emitted D0}.


Remarkably, this feature resembles much that of the S-branes
discussed in \cite{Lambert:2003zr}. Namely, it was shown there that
\begin{equation}
\mbox{half S-brane} ~ \cong ~ \mbox{full S-brane with the
Hartle-Hawking contour} ~. \label{half full S}
\end{equation}
In our case, \eqref{symmetric D0} corresponds to the full S-brane,
while the Hartle-Hawking state \eqref{HH falling D0} (\eqref{HH emitted
D0}) is identifiable as the analogue of the half S-brane describing
unstable D-brane decay (creation) process. The equalities
\eqref{relation HH} suggest that we have roughly identical relation to
\eqref{half full S}.

Notice that the parameters $\rho_0$, $t_0$ appear just as phase
factors of boundary wave functions in \eqref{HH falling D0}, \eqref{HH
emitted D0} contrary to \eqref{falling D0}, \eqref{emitted D0}. Namely,
the choice of parameters $\rho_0$, $t_0$ does not cause any physical
difference for the Hartle-Hawking type states : They all can be
regarded as describing the D0-brane moving from ${\cal I}^-$ to
${\cal H}^+$ (from ${\cal H}^-$ to ${\cal I}^+$) for \eqref{HH falling
D0} (for \eqref{HH emitted D0}) irrespective of $\rho_0$, $t_0$. These
two parameters merely parameterize displacing the trajectory in
two-dimensional black hole background. Similar feature comes about
for the full S-brane with Hartle-Hawking contour as well: It is
equivalent to the half S-brane not depending on any shift of the
origin (the point connecting the real and imaginary times).

Finally, we remark a comment from the viewpoints of boundary
conformal field theory: in contrast to the original ones
\eqref{falling D0}, \eqref{emitted D0} and \eqref{symmetric D0}, the
Hartle-Hawking boundary states \eqref{relation HH} (or equivalently
\eqref{HH falling D0}, \eqref{HH emitted D0}) are not compatible with
the reflection relations. One may regard the boundary states
\eqref{falling D0} and \eqref{emitted D0} as the `completions' of the
Hartle-Hawking states \eqref{relation HH} so that they satisfy the
reflection relations.


\newpage
\sectiono{Conclusion and Discussions}\label{sec:9}

In this thesis, we have examined the exact boundary states describing the rolling D-brane in the two-dimensional black hole system. In this final section, we would like to summarize our main results and discuss their physical relevance.

In the introduction, we asked three fundamental questions about the nature 
of the quantum gravity, or string theory as a candidate for the theory of 
everything:
\begin{itemize}
 \item Small charge black hole v.s. large charge black hole.
 \item Analyticity v.s. non-analyticity in physical amplitudes.
 \item Unitarity v.s. open closed duality.
\end{itemize}
It would be natural to conclude this thesis by asking how far we can answer 
these questions after our studies on the rolling D-brane in two-dimensional 
black hole system. 
To answer these three fundamental questions, in this paper, we have constructed
the exact boundary states describing the rolling D-brane in the two-dimensional black hole system (section \ref{sec:7}) to probe the quantum geometry. Our  main results are
\begin{itemize}
	\item The tachyon - radion correspondence is proved for $k>1$ by studying the closed string radiation rate from the $\alpha'$ exact rolling D-brane solution (section \ref{sec:8-1}). 
	\item The black hole - string transition is observed at $k=1$ in the closed string radiation rate as a physical order parameter (section \ref{sec:8-3}). 
	\item The consistency between the unitarity and open closed duality is shown to be recovered after a careful treatment of the Wick rotation (section \ref{sec:8-2}).
\end{itemize}
 Although our model is rather a specific one, we can naturally extend our results to draw many universal features of the quantum gravity. In the rest of this section, we recapitulate our arguments and present some discussions with possible future directions to pursue.

First of all, we have shown in section \ref{sec:8} that the total emission rate of the rolling D-brane into the two-dimensional black hole system behaves exactly same as that for the rolling tachyon in flat Minkowski space studied in section \ref{sec:5}. This result strongly suggests universal features of the physics associated with the D-brane decay. 

The universality is an important concept in any physical system. In our decaying D-brane system, we have shown that the closed string radiation rate is independent of the free parameter $k$ representing the level of the current algebra. Since $1/k$ correction governs the $\alpha'$ correction to the geometry, the physical quantity observed in the decaying D-brane process is independent of the stringy corrections. The classical tachyon - radion correspondence still holds even after introducing the stringy corrections independent of its strength (as long as $k>1$).

Indeed, the universal behavior of the closed string radiation should be true from the following simple argument. The D-brane energy that should be released during the decay is always proportional to $1/g_s$, so in the perturbative string computation, we expect a divergence in the radiation rate: otherwise we have to face the missing energy problem.\footnote{At first sight, this viewpoint contradicts our computation that the higher dimensional D-brane shows a power-like {\it finite} emission rate (energy), but this is an artefact of the one-particle decay.} 

Furthermore the universality holds under almost every exactly solvable deformations of the model such as an inclusion of the time-like linear dilaton, electric field etc (see section \ref{sec:5},\ref{sec:8}). In \cite{Israel:2006ip}, a similar computation has been performed in the $AdS_3$ space, supporting the universality of the decaying D-brane systems in yet another solvable background. We would like to emphasize again that our results do not depend on the  level $k$ which governs the strength of the world-sheet $\alpha'$ corrections {\it as long as} $k>1.$ This indeed provides a strong support for the tachyon-radion correspondence even at the quantum level.

At this point, it is worthwhile mentioning Sen's open-string completeness conjecture \cite{Sen:2003iv,Sen:2003mv,Sen:2004nf}: {\it There is a quantum open string theory (OSFT) that describes the full dynamics of an unstable Dp-brane without an explicit coupling to closed strings. Furthermore, Ehrenfest theorem holds in the weakly coupled OSFT; the classical results correctly describes the time evolution of the quantum expectation values}. The tachyon-radion correspondence directly results in the same conjecture for the rolling D-brane system. The smeared trajectory (or ``moss" around the rolling D-brane) we observed in our rolling D-brane system with exact $\alpha'$ correction is an interesting twist to this conjecture.

Secondly, we have shown that the interplay between the analyticity and 
non-analyticity of the physical amplitudes are crucial to discuss the black 
hole - string phase transition. The integration over the radial momenta, 
which at the same time is crucial to prove the tachyon - radion 
correspondence, introduces the non-analyticity in the physical observables, 
resulting in the phase transition.

 More precisely, we have directly observed the black hole - string phase 
transition from the exact boundary states for the probe rolling D-brane in 
the two-dimensional black hole background. The phase transition occurs 
exactly when the Hawking temperature of the two-dimensional black hole 
coincides with the Hagedorn temperature of the string background as we 
decrease the charge of the two-dimensional black hole (level of the current 
algebra). Below the phase transition point, the physical interpretations of 
the $SL(2;\br)/U(1)$ coset model as a black hole geometry break down and 
become obscure. Our results show that the tachyon - radion correspondence 
fails at the phase transition point, and the physics associated with the 
D-brane decay changes drastically.

Indeed, as we have shown in section \ref{sec:8}, a drastic change occurs when we study the dynamics of rolling D-branes in the two-dimensional black hole with $k<1$. From the arguments given in section \ref{sec:4}, we expect ``black hole - string transition". This transition is subtle even from the exact CFT analysis because in deriving every formulae in the closed string scattering amplitudes, we assume an analyticity in $k$. How can we probe the ``black hole - string transition" with respect to $k$ when the amplitude is analytic in $k$? In the open string channel, $k$ dependence is also analytic in the amplitudes as well. However, if we compute the physical quantities such as closed string emission rate, the non-analyticity with respect to $k$ emerges. In this way, we have succeeded in probing the ``black hole - string transition", as an emergent phase transition, by studying the rolling D-brane dynamics in the background.

It would be interesting to note that not every D-brane can probe the ``black hole - string transition". As we have seen in section \eqref{sec:5-2-5}, the decay of the unstable D0-brane in the Euclidean two-dimensional black hole does {\it not} show any ``black hole - string transition" at $k=1$. The decay rate of unstable D-branes shows a universal property irrespective of the value of $k$. We do not have a good physical explanation of this phenomenon at this moment, but it would be interesting to give a further study and determine which objects can probe the phase transition.

In the Euclidean signature target-space theory after the analytic 
continuation, the phase transition is induced from the non-perturbative 
$\alpha'$ corrections related to the winding-tachyon condensation. In the 
original Lorentzian signature target-space, one might understand it as a 
thermal (winding) tachyon condensation, or in the real time picture at the 
phase transition point, we would encounter associated (local) Hagedorn 
divergence of the black hole thermodynamics.

It is natural to expect that our results on the string - black hole phase 
transition is rather robust and universal. Indeed, the transition is barely 
affected under various marginal deformations of the solvable model such as 
incorporation of the linear dilaton or the electric field. It would be 
interesting to extend our analysis to more realistic higher dimensional 
black hole systems realized in superstring theory.\footnote{Recently, the 
black hole - string transition has been studied in the context of 
$AdS_5/CFT_4$ correspondence in \cite{Alvarez-Gaume:2006jg}.}

Philosophically, the concept of the phase in the quantum gravity is rather 
obscure. We know that in smaller dimensional field theories or in finite size 
theories, there is no notion of the phase transition. What happens, then, if 
the dimensionality or size of the universe fluctuate as is supposed to be 
the case with the quantum gravity?\footnote{A good example is the de-Sitter 
space, where the quantum gravity is supposed to consist of finite degrees of 
freedom.} Our study only touches a possible hidden nature of the phase 
transition as  non-analytic behaviors of the physical quantities (not 
amplitudes themselves). It would be worth studying further the potential 
origins of the non-analyticity in physical quantities in more general 
situations.

So far, we have restricted ourselves to $g_s \to 0$ limit throughout this thesis, but finite $g_s$ effects cannot be neglected in any realistic string theory. It is natural to assume that the finite $g_s$ effect sets a cut-off for the emitted closed string energy because it cannot emit energy greater than the tension of the decaying D-brane $\sim 1/g_s$. Therefore, the emission rate roughly behaves as
\begin{eqnarray}
\mathcal{N} \sim \int^{1/g_s} \dd dM N(M) \sqrt{\rho^{(c)}(M)} \ 
\end{eqnarray}
with an explicit cut-off at $1/g_s$. This also means that the radial momentum $p/\sqrt{k}$ should be less than $1/g_s$. Does this constraint smoothen out the phase transition? The saddle point approximation is not accurate as $M$ becomes smaller, so we expect that the phase transition becomes smooth as we introduces $g_s$ corrections. This is consistent with the statement that the ``black hole - string transition" is actually a ``black hole - string crossover". 

We have constructed several boundary states for the rolling D-branes in the two-dimensional black hole system. The failure of the uniqueness is physically relevant because in the time-dependent problems in string theory, the boundary conditions should be always imposed in accordance with the physics we are interested in. Mathematically, the different choices of the contour integration give rise to different physics. The tachyon - radion correspondence beautifully connects different solutions (contours) of the rolling tachyon with those for the rolling radion.

For the absorbed D-brane boundary condition studied in section \ref{sec:7-2-2}, the dominant infalling closed string radiation (at the Hagedorn temperature) accompanies the outgoing closed string radiation (at the Hawking temperature). The existence of the anomalously small outgoing radiation originates from the boundary condition imposed at the horizon. This reminds us of the closed string Hawking radiation discussed in section \ref{sec:2-4}. Combining the discussion given in section \ref{sec:8-1-4}, we can see that the origin of the thermal-like behaviour of the rolling D-brane radiation is closely related to the boundary condition imposed on the wavefunction (Ishibashi states). It would be interesting to make more precise the relation between the boundary conditions imposed on the string theory and the apparent anomaly as Hawking-like radiation.

Finally, we have discussed the consistency between the unitarity and the 
open-closed duality in the radiative process for the decay of unstable 
D-brane and rolling of accelerated D-brane dynamics in section \ref{sec:5-2} and \ref{sec:8-2} respectively. From ``ab initio" 
derivation in the open string channel, both in Euclidean and Lorentzian 
world-sheet approaches, we have found heretofore overlooked contribution to 
the spectral amplitudes and observables. The contribution is fortuitously 
absent for decay of unstable D-brane, but is present for rolling of 
accelerated D-brane. We have shown that the contribution is imperative for 
ensuring unitarity and optical theorem.

Our notion of the unitarity is rather specific, so we have not discussed 
more fundamental questions e.g. about the unitarity of the quantum black hole 
systems associated with Hawking evaporation. The information paradox of the 
black hole system should be resolved within the contex of the string theory 
if it is really a fundamental theory of everything.

These three questions raised in this thesis are basic but profound ones that 
people might first come up with when they would like to discuss the 
fundamental properties of the quantum gravity. We have attacked them from 
the concrete examples of the exactly solvable string black hole background. 
At this moment, we do not have complete answers to these questions in the 
tremendously huge string landscape, but we believe that our little step in 
the small corner will ultimately lead to their final answers. 

\section*{Acknowledgements}
The author would like to thank all my friends, my family, and my teachers for supporting him to write up this thesis. In particular the author would like to express his sincere thanks to his supervisor Tohru Eguchi for continuous encouragement and advice. Also he would like to express his special thanks to Soo-Jong Rey and Yuji Sugawara for the fruitful collaborations. The most of the main results in this thesis is based on the collaborations with them. He also acknowledges Sylvain Ribault and Yuji Tachikawa for stimulating discussions on the type 2 boundary states and a-theorem violation for generalized conifolds. This research is supported in part by JSPS Research Fellowships for Young Scientists.

\newpage

\appendix\sectiono{Appendices I: Conventions and Useful Formulae}\label{sec:A}
\subsection{conventions}

{\bf world-sheet}

In this thesis, we use $(-,+,+,\cdots,+)$ conventions for target-space metric signature. For the world-sheet coordinate with Lorentzian signature, we use $ -\infty < \tau < \infty$ and $ 0 \le \sigma \le 2\pi$ $(\sigma+2\pi \simeq \sigma)$. The light cone coordinate is defined as
\begin{align}
\begin{cases}
\sigma_+ = \sigma + \tau \cr
\sigma_- = \sigma - \tau 
\end{cases} \ .
\end{align}
We abbreviate their derivatives as $\partial_+ = \frac{\partial}{\partial\sigma_+}$ and $\partial_- = \frac{\partial}{\partial\sigma_-}$.

On the other hand, the complex coordinate on the complex plane is defined as
\begin{align}
\begin{cases}
z = x_1 + ix_2 \cr
\bar{z} = x_1 - ix_2 \ 
\end{cases}
\end{align}
We abbreviate their derivatives as $\partial = \frac{\partial}{\partial z}$ and $\bar{\partial} = \frac{\partial}{\partial \bar{z}}$. The integration measure is given by $\dd z^2 \equiv \dd x_1 \dd x_2$.

Throughout this thesis, we use the convention $\alpha' = l_s^2 = 2$ as long as otherwise stated. However, in several places, we explicitly write down $\alpha'$ for reader's convenience to compare our results with those in literatures, where different conventions are sometimes used.

{\bf theta functions with characteristic}

\begin{align}
\theta_0(\tau,v) &=\theta_4(\tau,v) = \prod_{m=1}^\infty (1-q^m)(1-zq^{m-1/2})(1-z^{-1}q^{m-1/2}) \cr
\theta_1(\tau,v) &= -2q^{1/4}\sin\pi v \prod_{m=1}^\infty (1-q^m)(1-zq^m)(1-z^{-1}q^m) \cr
\theta_2(\tau,v) &= 2q^{1/4}\cos\pi v \prod_{m=1}^\infty (1-q^m)(1+zq^m)(1+z^{-1}q^m) \cr
\theta_3(\tau,v) &= \prod_{m=1}^\infty (1-q^m)(1+zq^{m-1/2})(1+z^{-1}q^{m-1/2})  \cr
\eta(\tau) &= q^{1/24} \prod_{m=1}^{\infty}(1-q^m) \ ,
\end{align}
where $q = \exp(2\pi i \tau)$ and $z= \exp(2\pi i v)$. Their $S$-modular transformations are
\begin{align}
\theta_0(-1/\tau,v/\tau) &= (-i\tau)^{1/2} \exp(\pi i v^2/\tau) \theta_2(\tau,v) \cr
\theta_1(-1/\tau,v/\tau) &= -(-i\tau)^{1/2} \exp(\pi i v^2/\tau) \theta_1(\tau,v) \cr
\theta_2(-1/\tau,v/\tau) &= (-i\tau)^{1/2} \exp(\pi i v^2/\tau) \theta_0(\tau,v) \cr
\theta_3(-1/\tau,v/\tau) &= (-i\tau)^{1/2} \exp(\pi i v^2/\tau) \theta_3(\tau,v) \cr
\eta(-1/\tau) &=(-i\tau)^{1/2} \eta(\tau) \ .
\end{align}

The theta functions satisfy the following Riemann quartic identity:
\begin{align}
2\prod_{a=1}^4 \theta_1(\tau,v'_a) = \prod_{a=1}^4 \theta_3(\tau,v_a) - \prod_{a=1}^4 \theta_2(\tau,v_a) - \prod_{a=1}^4 \theta_0(\tau,v_a) + \prod_{a=1}^4 \theta_1(\tau,v_a) \ , \label{rqi}
\end{align}
where 
\begin{align}
2v'_1 &= v_1 + v_2 + v_3 + v_4 \ ,& 2v_2'   &= v_1 + v_2 -v_3 -v_4  \ ,\cr
2v'_3 &= v_1 - v_2 + v_3 -v_4 \ , & 2v_4'  &= v_1 - v_2 - v_ 3 + v_4 \ .
\end{align}
As a corollary, we obtain the abstruse identity of Jacobi:
\begin{eqnarray}
\theta_0^4(\tau,v) + \theta_2^2(\tau,v) = \theta_1^2(\tau,v) + \theta_3^4(\tau,v) \ .
\end{eqnarray}

{\bf Hypergeometric functions}

Gauss's hypergeometric function is defined as
\begin{eqnarray}
F(a_1,a_2;b;z) = \  _2F_1(a_1,a_2;b;z) = \sum_{l=0} \frac{(a_1)_l (a_2)_l}{(b)_l}\frac{x^l}{l!} \ ,
\end{eqnarray}
where 
\begin{eqnarray}
(a)_l \equiv \frac{\Gamma(a+l)}{\Gamma(a)} \ .
\end{eqnarray}

The analytic continuation of the hypergeometric function is defined by
\begin{align}
F(\alpha,\beta;\gamma;z) &=
\frac{\Gamma(\gamma)\Gamma(\beta-\alpha)}
{\Gamma(\beta)\Gamma(\gamma-\alpha)}
(-z)^{-\alpha}F(\alpha,\alpha+1-\gamma;
\alpha+1-\beta;1/z) \cr
&+ \frac{\Gamma(\gamma)\Gamma(\alpha-\beta)}
{\Gamma(\alpha)\Gamma(\gamma-\beta)}(-z)^{-\beta}
F(\beta,\beta+1-\gamma;\beta+1-\alpha;1/z)
\label{eq:inv} \ ,
\end{align}
\begin{align}
F(\alpha,\beta;\gamma;z) &=
\frac{\Gamma(\gamma)\Gamma(\gamma-\beta-\alpha)}
{\Gamma(\gamma-\beta)\Gamma(\gamma-\alpha)}
F(\alpha,\beta;
\alpha+\beta+1-\gamma;1-z) \cr
&+ \frac{\Gamma(\gamma)\Gamma(\alpha+\beta-\gamma)}
{\Gamma(\alpha)\Gamma(\beta)}(1-z)^{\gamma-\alpha-\beta}F(\gamma-\alpha,\gamma-\beta;1+\gamma-\alpha-\beta;1-z)
\label{eq:inv2} \ .
\end{align}
If $\gamma = \alpha + \beta + m $ with a certain integer $m$, the above formula should be modified due to the logarithmic singularity at $z=1$. In a particular case ($m=0$), which is interesting to us, the modified formula \cite{transcendental} reads 
\begin{align}
F(\alpha,\beta;\alpha+\beta;z) = \frac{\Gamma(\alpha+\beta)}{\Gamma(\alpha)\Gamma(\beta)} \sum_{n=0}^\infty \frac{a_{(n)}b_{(n)}}{(n!)^2} [h''_n - \log(1-z)] (1-z)^n \ ,
\end{align}
where $h''_n = 2\psi(n+1)-\psi(\alpha+n) - \psi(\beta +n)$ with $\psi(z) \equiv \frac{\Gamma'(z)}{\Gamma(z)}$. 

We also introduce the generalized hypergeometric function
\begin{eqnarray}
_3F_2(a_1,a_2,a_3;b_1,b_2;z) = \sum_{l=0} \frac{(a_1)_l (a_2)_l(a_3)_l}{(b_1)_l(b_2)_l}\frac{x^l}{l!} \ ,
\end{eqnarray}
whose asymptotic expansion (as $z\to \infty$) is given by
\begin{align}
&_3F_2(a_1,a_2,a_3;b_1,b_2;z) \cr &
= \frac{\Gamma(b_1)\Gamma(b_2)}{\Gamma(a_1)\Gamma(a_2)\Gamma(a_3)} \sum_{k=1}^3 \frac{\Gamma(a_k)\prod_{j=1;j\neq k}^3(a_j-a_k)}{\prod_{j=1}^2 \Gamma(b_j-a_k)}(-z)^{-a_k}\left[1 + O(z^{-1})\right] \ .
\end{align}

\subsection{$SL(2;\br)$ current algebra}\label{a-1}
We will collect useful facts about the $SL(2;\br)$ current algebra and fix our notations. We begin with the zero mode. Our conventions for the commutation relations  of the $SL(2;\br)$ algebra are
\begin{align}
[J^1,J^2] = -i J^3 \ , [J^2,J^3] = i J^1 \ , [J^3,J^1] = iJ^2 \ .
\end{align}
We will also introduce
\begin{align}
J^{\pm} = J^1 \pm iJ^2 \ ,
\end{align}
which gives 
\begin{align}
[J^3, J^{\pm}] = \pm J^{\pm} \ , [J^+,J^{-}] = -2J^3 \ . 
\end{align}
The quadratic Casimir is defined as
\begin{align}
C_2 = (J^1)^2 + (J^2)^2 - (J^3)^2 \equiv -j(j-1) \ .
\end{align}
We summarize the unitary (irreducible) representations of the $SL(2;\br)$ algebra parametrized by $j$

\begin{enumerate}
	\item Principal discrete representations (highest or lowest weight states): they contain the state that is annihilated by $J^+$ or $J^-$ respectively. 
Their modules are generated as
\begin{align}
D^+ = \left\{ |j;m\rangle : m = -j , -j+1, -j+2, \cdots\right\} \ ,
\end{align}
and
\begin{align}
D^- = \left\{ |j;m\rangle : m = j , j-1, j-2, \cdots\right\} \ ,
\end{align}
obtained by acting $J^-$ or $J^+$ on the highest or lowest weight state.
The representation is unitary when $j \le -\frac{1}{2}$.
	\item Principal continuous series: they are realized by
\begin{align}
C_j^{\alpha} = \left\{|j,\alpha;m\rangle : m = \alpha, \alpha \pm1, \alpha \pm 2, \cdots \right\}
\end{align}
where $J^3|j,\alpha;m\rangle = m |j,\alpha;m\rangle$. Without loss of generality, we take $0\le \alpha <1$. The representation is unitary when $j=-1/2 + is $ with $s\in \br^+$.

	\item Complementary representations: similar to the continuous series but with real $j$. They are unitary when $-1 < j <-1/2$ and $-j-1/2<|\alpha-1/2|$.
	\item Identity representation: this is the trivial representation with $j=-1$.
\end{enumerate}

Let us move to the affine current algebra $\widehat{SL(2;\br)}$. The relevant OPEs are
\begin{align}
\begin{cases} j^3(z)j^3(0) \sim -\frac{\kappa}{2z^2} \cr
j^3(z) j^{\pm}(0) \sim \pm \frac{1}{z}j^{\pm}(0) \cr
j^+(z)j^{-}(0) \sim \frac{\kappa}{z^2} - \frac{2}{z}j^{3}(0) 
\end{cases} \ .
\end{align}
Corresponding affine Lie algebra is given by
\begin{align}
[J^3_n,J^3_m] &= -\frac{k}{2}n \delta_{m,-n} \ , \cr
[J^3_n,J^{\pm}_m] & = \pm J^\pm_{n+m} \ , \cr
[J^+_n,J^{-}_m] & = -2J^3_{n+m} + kn\delta_{m,-n} \ .
\end{align}
The energy momentum tensor is given by the Sugawara form:
\begin{align}
T(z) = \frac{1}{\kappa-2}(j^1j^1 + j^2 j^2 -j^3j^3)(z) \ 
\end{align}
with the central charge $c=3 + \frac{6}{\kappa-2}$.

We summarize the characters of the unitary  representations of $\widehat{SL(2;\br)}$ current algebra. The character is defined as $\mathrm{Tr} q^{L_0} y^{J_0^3} $ with $q\equiv e^{2\pi i\tau}$, $y\equiv e^{2\pi i u}$.
\begin{enumerate}
	\item Principal discrete representations (highest or lowest weight states): the characters can be written as
\begin{align}
\chi_j^{\pm}(\tau,u) = \pm i \frac{q^{-\frac{1}{\kappa-2}(j-\frac{1}{2})^2}y^{\pm (j-\frac{1}{2})}}{\theta_1(\tau,u)} \ .
\end{align}

	\item Principal continuous series: the character is given by
\begin{align}
\chi_{s,\alpha}(\tau,u) = \frac{q^{\frac{s^2}{\kappa-2}}}{\eta(\tau)^3} \sum_{n\in \bz} y^{n+\alpha} \ .
\end{align}

	\item Complementary representations: the character is given by
\begin{align}
\chi_{j,\alpha}(\tau,u) = \frac{q^{\frac{(j-\frac{1}{2})^2}{\kappa-2}}}{\eta(\tau)^3} \sum_{n\in \bz} y^{n+\alpha} \ .
\end{align}

	\item Identity representation: the character is given by
\begin{align}
\chi_0(\tau,u) = i \frac{q^{-\frac{1}{4(\kappa-2)}}y^{-1/2}(1-y)}{\theta_1(\tau,u)} \ .
\end{align}
It is possible to construct more general representations of the current algebra by using ($n$-units of) the spectral flows
\begin{align}
j_m^3 \to j_m^3 - \frac{\kappa}{2}n \delta_{m,0} \ ,  \ \ j^{\pm}_m \to j^{\pm}_{m\pm n} \ , \ \ L_m \to L_m + nj^3_m - \frac{\kappa}{4}n^2 \delta_{m,0} \ . 
\end{align}
Unlike in the case of the compact group, we actually obtain new representations, but their conformal dimensions are typically unbounded below.
\end{enumerate}

\subsection{Coordinate on $SL(2;\br)$}\label{a-1-2}
The Euler angle parametrization $g(r,t,\phi) \in SL(2;\br)$ suitable for the Euclidean coset is given by
\begin{eqnarray}
e^{i\sigma_1\frac{t-\phi}{2}} e^{r\sigma_2}e^{i\sigma_1\frac{t+\phi}{2}} 
= \begin{pmatrix} \cos t \cosh r - \sin\phi \sinh r & \cosh r \sin t + \cos \phi \sinh r \cr
-\cosh r \sin t + \cos \phi \sinh r& \cos t \cosh r + \sin\phi \sinh r 
\end{pmatrix} \ ,
\end{eqnarray}
where $0<t\le 2\pi$, $ 0\le r$ and $0<\phi \le 2\pi$ for the single cover of $SL(2;\br)$. 

On the other hand, for the Lorentzian coset, a useful Euler angle parametrization $g(r,t_L,t_R)$ is given by 
\begin{eqnarray}
 e^{\sigma_3 t_L} e^{r\sigma_2}e^{\sigma_3 t_R} = \begin{pmatrix} e^{t_L+t_R} \cosh r & e^{t_L-t_R} \sinh r \cr
e^{-t_L+t_R} \sinh r & e^{-t_L-t_R} \cosh r \end{pmatrix} \ , 
\end{eqnarray}
where $t_L$ and $t_R$ are noncompact.\footnote{The parametrization does {\it not} cover the whole $SL(2;\br)$ manifold.}

To connect it with the $AdS_3$ space, it is customary to introduce the parametrization
\begin{eqnarray}
g = \begin{pmatrix} x^0 + x^2 & x^1 + x^3 \cr x^1-x^3 & x^0-x^2 \end{pmatrix} \ 
\end{eqnarray}
 so that we can see the $SL(2;\br)$ group as a hyperbola 
\begin{eqnarray}
(x^0)^2 - (x^1)^2 - (x^2)^2 + (x^3)^2 = 0 \ 
\end{eqnarray}
embedded in Minkowski space $(x^0,x^1,x^2,x^3)$ with signature $(-,+,+,-)$.

\subsection{Frequently used formulae} 

\begin{eqnarray}
\Gamma(z+1) = z\Gamma(z) \ .
\end{eqnarray}
\begin{eqnarray}
\Gamma(z)\Gamma(1-z) = \frac{\pi}{\sin\pi z} \ .
\end{eqnarray}
\begin{eqnarray}
\Gamma(\frac{1}{2}+z)\Gamma(\frac{1}{2}-z) = \frac{\pi}{\cos\pi z} \ .
\end{eqnarray}
\begin{eqnarray}
\Gamma(2z) = (2\pi)^{-1/2} 2^{2z-1/2} \Gamma(z) \Gamma(z+1/2) \ .
\end{eqnarray}
\begin{align}
\int_0^\infty dx \frac{x^c}{\sqrt{x^2 + a^2}} = \frac{a^c \Gamma(-\frac{c}{2})\Gamma(\frac{1+c}{2})}{2\sqrt{\pi}} \ .
\end{align}
\begin{align}
& \hspace{-5mm} \int_{-\frac{\pi}{2}}^{\frac{\pi}{2}} (2 \cos
\theta)^{a-1} e^{i b \theta} \dd \theta = \pi \frac{\Gamma(a)}
{\Gamma\left(\frac{1}{2}+\frac{a+b}{2}\right)
\Gamma\left(\frac{1}{2}+\frac{a-b}{2}\right)}~,~~~ (\Re\,
a>0~,~~\left|\Re\, b \right|< \Re\, a + 1)~,
\label{formula 1} \\
& \hspace{-5mm} \int_{-\infty}^{\infty} (2 \cosh t)^{a-1} e^{i b t}
\dd t = \frac{1}{2} B\left(\frac{1}{2}-\frac{a+ib}{2},
\frac{1}{2}-\frac{a-ib}{2}\right) \equiv \frac{1}{2} \frac
{\Gamma\left(\frac{1}{2}-\frac{a+ib}{2}\right)
\Gamma\left(\frac{1}{2}-\frac{a-ib}{2}\right)} {\Gamma(1-a)}~,\nn &
\hspace{10cm} (\Re\, a <1~,~~ \left|\Im \, b \right| < 1- \Re\, a)~.
\label{formula 2}
\end{align}
The integral \eqref{formula 2} follows from the more general formula:
\begin{align}
\int_0^{\infty} \frac{\cosh(2at)}{\cosh^{2\beta} (pt)} \dd t =
4^{\beta-1} p^{-1} B\left(\beta+\frac{a}{p},
\beta-\frac{a}{p}\right)~,~~~ (p>0,
~~\Re\,\left(\beta\pm\frac{a}{p}\right)>0)~,
\end{align}
given in \cite{transcendental}.


\subsection{Proof of \eqref{evaluation overlap phi} and \eqref{evaluation overlap phi2}}\label{mini}

Here we would like to evaluate explicitly the integral
\eqref{evaluation overlap phi} for any $\rho_0$ (strictly speaking, we
need to assume $\sinh \rho_0>1$). We begin with series expansion of
the hypergeometric function in $\phi^p_n(\rho,\theta)$:
\begin{align}
&
F\left(\frac{1}{2}+\frac{ip+n}{2},\frac{1}{2}+\frac{ip-n}{2};
ip+1;
-\frac{\cos^2 \theta}{\sinh^2 \rho_0}\right)\nn
&=\sum_{\ell=0}^\infty \frac{\Gamma(ip+1)}
{\Gamma(\frac{1}{2}+\frac{ip+n}{2})
\Gamma(\frac{1}{2}+\frac{ip-n}{2})}
\frac{\Gamma(\frac{1}{2}+\frac{ip+n}{2}+\ell)
\Gamma(\frac{1}{2}+\frac{ip-n}{2}+\ell)}
{\Gamma(ip+1+\ell)}\frac{(-1)^\ell}{\ell !}
\left(\frac{\cos \theta}{\sinh \rho_0}\right)^{2\ell}.
\label{expansion}
\end{align}
Using the formula (\ref{formula 1}), we can perform, in the
$\ell$-th sector, the integral (\ref{evaluation overlap phi}) as
\begin{align}
\Psi_\ell=g(\ell)\int_{-\frac{\pi}{2}}^{\frac{\pi}{2}} \dd\theta \,
e^{in\theta}(\cos \theta)^{ip-1+ 2\ell} =\frac{g(\ell)}{2^{ip-1
+2\ell}}\cdot \frac{\pi\Gamma(ip-1+2\ell+1)}
{\Gamma(\frac{1}{2}+\ell+\frac{ip+n}{2})
\Gamma(\frac{1}{2}+\ell+\frac{ip-n}{2})}, \nonumber
\end{align}
where $g(\ell)$ refers to
\begin{equation}
g(\ell)=\frac{(-1)^\ell}{\ell !}
(\sinh \rho_0)^{-ip -2\ell}\frac{\Gamma(ip+1)}
{\Gamma(\frac{1}{2}+\frac{ip+n}{2})
\Gamma(\frac{1}{2}+\frac{ip-n}{2})}
\frac{\Gamma(\frac{1}{2}+\frac{ip+n}{2}+\ell)
\Gamma(\frac{1}{2}+\frac{ip-n}{2}+\ell)}{\Gamma(ip+1+\ell)}.
\end{equation}
Then the total integral (\ref{evaluation overlap phi}) is
\begin{equation}
\sum_{\ell=0}^\infty \Psi_\ell
=\frac{\pi}{2^{ip-1}(\sinh \rho_0)^{ip}}\frac{\Gamma(ip+1)}
{\Gamma(\frac{1}{2}+\frac{ip+n}{2})
\Gamma(\frac{1}{2}+\frac{ip-n}{2})}\sum_{\ell=0}^\infty
\frac{(-1)^\ell}{\ell !}\frac{1}{2^{2\ell}(\sinh \rho_0)^{2\ell}}
\frac{\Gamma(ip+2\ell)}{\Gamma(ip+1+\ell)}.
\end{equation}
We can rewrite the summation into a hypergeometric function
by using
\begin{equation}
\Gamma(ip+2\ell)=\frac{2^{ip-1+2\ell}}{\sqrt{\pi}}
\Gamma\left(\frac{ip}{2}+\ell\right)
\Gamma\left(\frac{1}{2}+\frac{ip}{2}+\ell\right)~.
\end{equation}
and then obtain
\begin{equation}
\sum_{\ell=0}^\infty \Psi_\ell=\frac{\sqrt{\pi}}
{(\sinh \rho_0)^{ip}}\frac{\Gamma(\frac{ip}{2})
\Gamma(\frac{1}{2}+\frac{ip}{2})}
{\Gamma(\frac{1}{2}+\frac{ip+n}{2})
\Gamma(\frac{1}{2}+\frac{ip-n}{2})}
F\left(\f{ip}{2},\frac{1}{2}+\frac{ip}{2}; ip+1 ;
-\frac{1}{\sinh^2 \rho_0}\right).
\label{calc1}
\end{equation}
Making use of the formula
\begin{align}
&F\left(2\alpha,2\beta;\alpha+\beta+\frac{1}{2};z\right)=
F\left(\alpha,\beta;\alpha+\beta+\frac{1}{2};4z(1-z)\right).
\label{calc2} \\
& \hskip5cm |z|<\frac{1}{2},\quad |z(1-z)|<\frac{1}{4}~ \nonumber
\end{align}
we find that
\begin{equation}
\sum_{\ell=0}^\infty \Psi_\ell=\frac{\sqrt{\pi}}
{(\sinh \rho_0)^{ip}}\frac{\Gamma(\frac{ip}{2})
\Gamma(\frac{1}{2}+\frac{ip}{2})}
{\Gamma(\frac{1}{2}+\frac{ip+n}{2})
\Gamma(\frac{1}{2}+\frac{ip-n}{2})}
F\left(ip,ip+1,ip+1;\frac{1}{2}-
\frac{\cosh \rho_0}{2\sinh \rho_0}\right).
\end{equation}
Note that the second and third arguments
of the hypergeometric function are the same.
The function is thus simplified as
\begin{equation}
F\left(ip,ip+1,ip+1;\frac{1}{2}-
\frac{\cosh \rho_0}{2\sinh \rho_0}\right)=
\left(\frac{\sinh \rho_0 +\cosh \rho_0}
{2\sinh \rho_0}\right)^{-ip}
=\left(2e^{-\rho_0}\sinh \rho_0\right)^{ip}~,
\end{equation}
because of the relation
\begin{equation}
(1-z)^{\alpha}=F(-\alpha,\beta;\beta;z)~.
\end{equation}
In this way, we finally obtain
\begin{align}
& \sum_{\ell=0}^\infty \Psi_\ell=\sqrt{\pi}
e^{-ip \rho_0 }2^{ip}
\frac{\Gamma(\frac{ip}{2})\Gamma(\frac{1}{2}+\frac{ip}{2})}
{\Gamma(\frac{1}{2}+\frac{ip+n}{2})\Gamma(\frac{1}{2}
+\frac{ip-n}{2})}
= \frac{2\pi \,\Gamma(ip)}
{\Gamma(\frac{1}{2}+\frac{ip+n}{2})
\Gamma(\frac{1}{2}+\frac{ip-n}{2})}e^{-ip\rho_0}~,
\end{align}
and this is the desired formula.


In a quite similar manner, we can prove
\eqref{evaluation overlap phi2}. We begin with series expansion of
the hypergeometric function in $\phi_{pw}(\rho,\theta)$:
\begin{align}
&
F\left(\frac{1}{2}+\frac{ip+kw}{2},\frac{1}{2}+\frac{ip-kw}{2};
ip+1;
\frac{\cos^2 \theta}{\cosh^2 r_0}\right)\nn
&=\sum_{\ell=0}^\infty \frac{\Gamma(ip+1)}
{\Gamma(\frac{1}{2}+\frac{ip+kw}{2})
\Gamma(\frac{1}{2}+\frac{ip-kw}{2})}
\frac{\Gamma(\frac{1}{2}+\frac{ip+kw}{2}+\ell)
\Gamma(\frac{1}{2}+\frac{ip-kw}{2}+\ell)}
{\Gamma(ip+1+\ell)}\frac{1}{\ell !}
\left(\frac{\cos \theta}{\cosh r_0}\right)^{2\ell}.
\label{expansion}
\end{align}
Using the formula (\ref{formula 1}), we can perform, in the
$\ell$-th sector, the integral (\ref{evaluation overlap phi}) as
\begin{align}
\Psi_\ell=g(\ell)\int_{-\frac{\pi}{2}}^{\frac{\pi}{2}} \dd\theta \,
e^{ikw\theta}(\cos \theta)^{ip-1+ 2\ell} =\frac{g(\ell)}{2^{ip-1
+2\ell}}\cdot \frac{\pi\Gamma(ip-1+2\ell+1)}
{\Gamma(\frac{1}{2}+\ell+\frac{ip+kw}{2})
\Gamma(\frac{1}{2}+\ell+\frac{ip-kw}{2})}, \nonumber
\end{align}
where $g(\ell)$ refers to
\begin{equation}
g(\ell)=\frac{1}{\ell !}
(\cosh r_0)^{-ip -2\ell}\frac{\Gamma(ip+1)}
{\Gamma(\frac{1}{2}+\frac{ip+kw}{2})
\Gamma(\frac{1}{2}+\frac{ip-kw}{2})}
\frac{\Gamma(\frac{1}{2}+\frac{ip+kw}{2}+\ell)
\Gamma(\frac{1}{2}+\frac{ip-kw}{2}+\ell)}{\Gamma(ip+1+\ell)}.
\end{equation}
Then the total integral (\ref{evaluation overlap phi}) is
\begin{equation}
\sum_{\ell=0}^\infty \Psi_\ell
=\frac{\pi}{2^{ip-1}(\cosh r_0)^{ip}}\frac{\Gamma(ip+1)}
{\Gamma(\frac{1}{2}+\frac{ip+kw}{2})
\Gamma(\frac{1}{2}+\frac{ip-kw}{2})}\sum_{\ell=0}^\infty
\frac{1}{\ell !}\frac{1}{2^{2\ell}(\cosh r_0)^{2\ell}}
\frac{\Gamma(ip+2\ell)}{\Gamma(ip+1+\ell)}.
\end{equation}
We can rewrite the summation into a hypergeometric function
by using
\begin{equation}
\Gamma(ip+2\ell)=\frac{2^{ip-1+2\ell}}{\sqrt{\pi}}
\Gamma\left(\frac{ip}{2}+\ell\right)
\Gamma\left(\frac{1}{2}+\frac{ip}{2}+\ell\right)~.
\end{equation}
and then obtain
\begin{equation}
\sum_{\ell=0}^\infty \Psi_\ell=\frac{\sqrt{\pi}}
{(\cosh r_0)^{ip}}\frac{\Gamma(\frac{ip}{2})
\Gamma(\frac{1}{2}+\frac{ip}{2})}
{\Gamma(\frac{1}{2}+\frac{ip+kw}{2})
\Gamma(\frac{1}{2}+\frac{ip-kw}{2})}
F\left(\f{ip}{2},\frac{1}{2}+\frac{ip}{2}; ip+1 ;
\frac{1}{\cosh^2 r_0}\right).
\label{calc1}
\end{equation}
Making use of the formula
\begin{align}
&&F\left(2\alpha,2\beta;\alpha+\beta+\frac{1}{2};z\right)=
F\left(\alpha,\beta;\alpha+\beta+\frac{1}{2};4z(1-z)\right).
\label{calc2} \\
&& \hskip5cm |z|<\frac{1}{2},\quad |z(1-z)|<\frac{1}{4}~ \nonumber
\end{align}
we find that
\begin{equation}
\sum_{\ell=0}^\infty \Psi_\ell=\frac{\sqrt{\pi}}
{(\sinh r_0)^{ip}}\frac{\Gamma(\frac{ip}{2})
\Gamma(\frac{1}{2}+\frac{ip}{2})}
{\Gamma(\frac{1}{2}+\frac{ip+n}{2})
\Gamma(\frac{1}{2}+\frac{ip-n}{2})}
F\left(ip,ip+1,ip+1;\frac{1}{2}(1-\tanh r_0)\right).
\end{equation}
Note that the second and third arguments
of the hypergeometric function are the same.
The function is thus simplified as
\begin{equation}
F\left(ip,ip+1,ip+1;\frac{1}{2}(1-\tanh r_0)\right)
=\left(2e^{-r_0}\cosh r_0\right)^{ip}~,
\end{equation}
because of the relation
\begin{equation}
(1-z)^{\alpha}=F(-\alpha,\beta;\beta;z)~.
\end{equation}
In this way, we finally obtain
\begin{align}
&& \sum_{\ell=0}^\infty \Psi_\ell=\sqrt{\pi}
e^{-ip r_0 }2^{ip}
\frac{\Gamma(\frac{ip}{2})\Gamma(\frac{1}{2}+\frac{ip}{2})}
{\Gamma(\frac{1}{2}+\frac{ip+kw}{2})\Gamma(\frac{1}{2}
+\frac{ip-kw}{2})}
= \frac{2\pi \,\Gamma(ip)}
{\Gamma(\frac{1}{2}+\frac{ip+kw}{2})
\Gamma(\frac{1}{2}+\frac{ip-kw}{2})}e^{-ipr_0}~,
\end{align}
and this is the desired formula.

\section{Appendix II: Miscellaneous Topics}\label{sec:B}

\subsection{Partition functions}\label{part}
The modular invariant torus partition function is of critical importance in closed string theory to read the spectrum of a given background.  In this appendix, we collect partition functions of several CFTs that are relevant for our discussions.

Our main focus is the partition function for the Lorentzian two-dimensional black hole. The partition function for the two-dimensional black hole, however, suffers some subtleties because of the Lorentzian signature of the target space and the divergence coming from the non-compact target space. 

With these subtleties  in mind, our starting point is the (twisted) partition function for the $\mathbb{H}_3^+$ model. For a time-being, we restrict ourselves to the bosonic CFT. Since the Euclidean action is bounded, the direct path integral computation is possible \cite{Gawedzki:1988nj,Gawedzki:1991yu}. For the vector gauging
\begin{align}
Z^{\mathbb{H}^3_+}_{(V)}(\tau,u) \equiv \int \dd g e^{-kS^{\mathbb{H}_3^+}_{(V)}(g,h^u,h^{u\dagger})} = \frac{e^{\frac{u_2^2}{\tau_2}}}{\sqrt{\tau_2}|\theta_1(\tau,u)|^2 }\ .
\end{align}
For the axial gauging
\begin{align}
Z^{\mathbb{H}^3_+}_{(A)}(\tau,u) \equiv \int \dd g e^{-kS^{\mathbb{H}_3^+}_{(A)}(g,h^u,h^{u\dagger})} = \frac{e^{\frac{u_2^2}{\tau_2}-\pi k\frac{|u|^2}{\tau_2}}}{\sqrt{\tau_2}|\theta_1(\tau,u)|^2 }\ .
\end{align}
In the $u\to0$ limit, both expressions coincide and (formally) modular invariant.\footnote{Because of the divergence as $u\to 0$, these expressions contain rather poor information to read the spectrum (e.g. the partition function is $k$ independent). Actually, the resurrection of the $k$ dependence is one of the key issues to extract contributions from the discrete states and give the improved unitarity bound. In order to do this, we should actually know the central charge of the model independently since the torus partition function does not know {\it a priori} the shift of the central charge coming from the linear dilaton coupled to the curvature of the world-sheet.} 

The partition function for the Euclidean two-dimensional black hole from the axial coset of the $\mathbb{H}^3_+$ model (denoted by ${\mathbb{H}^{3(A)}_+/\mathbb{R}}$) is 
\begin{align}
Z_{\mathbb{H}^{3(A)}_{+}/\br} = \int_{\Sigma} \frac{\dd u^2}{\tau_2} \frac{e^{\frac{u_2^2}{\tau_2}}}{\sqrt{\tau_2}|\theta_1(\tau,u)|^2 } \sqrt{\tau_2}|\eta(\tau)|^2 \sum_{m,\omega\in \bz} e^{-\frac{\pi k}{\tau_2}|\omega\tau - m + u|^2} \ , \label{hpz}
\end{align}
where the integration is over the Jacobian torus $\int_0^1 ds_1 \int_0^1 ds_2$ with $u=s_1\tau-s_2$. The appearance of the (twisted) partition of the compactified boson (with radius $R^2 = k$) 
\begin{align}
Z_{\sqrt{k}}(\tau,u) = \sqrt{\tau_2}|\eta(\tau)|^2 \sum_{m,\omega\in \bz} e^{-\frac{\pi k}{\tau_2}|\omega\tau - m + u|^2}
\end{align}
suggests an asymptotic geometry of the cigar (with the same radius).
Note that the summation over $m,\omega$ can be combined into
\begin{align}
Z_{\mathbb{H}^{3(A)}_{+}/\br} =  \int_{\br^2} \frac{\dd u^2}{\tau_2}\frac{e^{\frac{u_2^2}{\tau_2}-\frac{\pi k}{\tau_2}|u|^2}  }{\sqrt{\tau_2}|\theta_1(\tau,u)|^2 } \sqrt{\tau_2}|\eta(\tau)|^2  \ .
\end{align} 
Due to the fact that ${\mathbb{H}^{3(A)}_{+}/\br}  = SL(2;\br)^{(A)}/U(1)$, one can interpret \eqref{hpz} as the partition function of the $SL(2;\br)^{(A)}/U(1)$ axial coset model.\footnote{In the Euclidean axial coset model, there is no distinction between the single cover of $SL(2;\br)$ and the universal cover of the $SL(2;\br)$.}

The partition function for the Lorentzian coset is rather subtle, and we should resort to analytic continuations. Instead of gauging compact subgroup generated by $J^3$, we gauge the noncompact subgroup generated by $J^2$. Effectively, we can Wick rotate $J^3 \to iJ^3$. The axially $J^2$ twisted partition function for (the universal cover of) $SL(2;\br)_{(\infty)}^{(A)}$ could be obtained from the analytic continuation $(u\to iv)$ as
\begin{align}
Z^{SL(2;\br)}_{(A)}(\tau,v) \equiv \int \dd g e^{-kS^{H_3^+}_{(V)}(g,h^v,h^{v\dagger})} = \frac{e^{\frac{v_1^2}{\tau_2}}}{\sqrt{\tau_2}|\theta_1(\tau,v)|^2 } \ .
\end{align}
The partition function for the Lorentzian two-dimensional black hole (after an analytic continuation) could be written as
\begin{align}
Z_{\mathbb{H}^{3(A)}_{+}/\br} =  \int_{\br^2} \frac{\dd v^2}{\tau_2}\frac{e^{\frac{v_1^2}{\tau_2}-\frac{\pi k}{\tau_2}|v|^2}  }{\sqrt{\tau_2}|\theta_1(\tau,iv)|^2 } \sqrt{\tau_2}|\eta(\tau)|^2  \ ,
\end{align}
which agrees with the proposal made in \cite{Israel:2003ry} from a different perspective.

\subsection{T-duality and Buscher's rule}\label{busc}
We summarize Buscher's rule \cite{Buscher:1987sk,Buscher:1987qj} for T-duality:
\begin{align}
\tilde{G}_{00} &= \frac{1}{G_{00}} \cr
\tilde{G}_{0i} &= \frac{B_{0i}}{G_{00}} \cr
\tilde{B}_{0i} &= \frac{G_{0i}}{G_{00}} \cr
\tilde{G}_{ij} &= G_{ij} - \frac{G_{0i}G_{0j}-B_{0i}B_{0j}}{G_{00}} \cr
\tilde{B}_{ij} &= B_{ij} -\frac{G_{0i}B_{0j}-B_{0i}G_{0i}}{G_{00}} \cr
\tilde{\Phi} &= \Phi - \frac{1}{4}\log\left(\frac{G_{00}}{\tilde{G}_{00}}\right) \ ,
\end{align}
where $0$ is the direction to be T-dualized.

\subsection{$\mathcal{N}=2$ superconformal algebra and spectral flow}\label{SCA2}
OPEs for the $\mathcal{N}=2$ SCA are summarized as
\begin{align}
T(z) T(0) &\sim \frac{c}{2z^4} + \frac{2}{z^2}T(0) + \frac{1}{z}\partial T(0) \cr
T(z) G^{\pm}(0) &\sim \frac{3}{2z^2}G^{\pm}(0) + \frac{1}{z}\partial G^{\pm}(0) \cr
T(z) J(0) &\sim \frac{1}{z^2} J(0) + \frac{1}{z} \partial J(0) \cr
G^+(z)G^{-}(0) &\sim \frac{2c}{3z^3}+\frac{2}{z^2}J(0) + \frac{2}{z} T(0) + \frac{1}{z} \partial J(0) \cr
G^{\pm}(z)G^{\pm}(0) &\sim 0 \cr
J(z)G^{\pm}(0) &\sim \pm G^{\pm}(0) \cr
J(z)J(0) &\sim \frac{c}{3z^2} \ .
\end{align}
Correspondingly the mode expansion yields 
\begin{align}
[L_m,L_n] &= (m-n)L_{m+n} + \frac{c}{12}(m^3-m)\delta_{m,-n} \cr
[L_m,G^{\pm}_r] &= \left(\frac{m}{2}-r\right)G^{\pm}_{m+r} \cr
[L_m,J_n] &= -n J_{m+n} \cr
\{G_r^{+},G_{s}^{-}\} &= 2L_{r+s} + (r-s)J_{r+s} + \frac{c}{3}\left(r^2 -\frac{1}{4}\right) \delta_{r,-s} \cr
\{G_r^{\pm},G_s^{\pm}\} &= 0 \cr
[J_n,G^{\pm}_r] &= \pm G_{r+n}^{\pm} \cr
[J_m,J_n] &= \frac{c}{3}m\delta_{m,-n} \ ,
\end{align}
where $r,s$ are integers for the R-sector and half-integers for the NS-sector.
The $\mathcal{N}=2$ superconformal algebra admits an isomorphism known as spectral flow:
\begin{align}
 U_{\eta}L_{m}U_\eta^{-1} &\to L_{m} + \eta J_m +\frac{\eta^2 c}{6}\delta{m,0} \cr
U_{\eta}G^{\pm}_mU^{-1}_{\eta} &\to G^{\pm}_{m\pm \eta} \cr
U_{\eta}J_mU^{-1}_{\eta} &\to J_m + \frac{\eta}{3}\delta_{m,0} \ .
\end{align}
The explicit form of the unitary operator $U_\eta$ can be obtained by
\begin{align}
U_{\eta} = e^{-i\sqrt{\frac{c}{3}}\eta\phi} \ ,
\end{align}
where 
\begin{align}
J = i\sqrt{\frac{c}{3}}\partial \phi \ .
\end{align}
 
\subsection{Lichnerowicz obstruction and $a$-theorem}\label{Lic}
In this appendix, we show that an apparent violation of the $a$-theorem for the  gauge theory living on the D3-brane at the tip of the Brieskorn-Pham singularities are avoided by the Lichnerowicz obstruction we reviewed in section \ref{sec:2-2-3}. Let us consider the Brieskorn-Pham type generalized conifolds
\begin{align}
x_1^{k_1} + x_2^{k_2} + k_3^{k_3} + k_4^{k_4} = 0 \ . \label{bifs}
\end{align}

Assuming that the Reeb vector (conformal $U(1)_R$ charge) is given by the natural $\bc^*$ action induced by the charge vector $(1/k_1,1/k_2,1/k_3,1/k_4)$ on $(x_1,x_2,x_3,x_4)$, the volume of the associated Sasaki-Einstein manifold can be computed as
\begin{align}
V = \frac{\pi^3 k_1k_2k_3k_4 \left(\frac{1}{k_1}+\frac{1}{k_2}+\frac{1}{k_3}+\frac{1}{k_4}-1\right)^3}{27} \ .
\end{align}
The conjectured $a$-theorem states that $a$, which is inversely proportional to $V$ via $AdS-CFT$ correspondence, should be an decreasing function as we decrease $k_i$ as a relevant deformation. The condition is equivalent to
\begin{align}
-\frac{2}{k_1} +\frac{1}{k_2}+\frac{1}{k_3}+\frac{1}{k_4} \le 1 \ , \label{a-theo}
\end{align}
for $k_1$, and similarly for $k_2$, $k_3$ and $k_4$. 

On the other hand, the Lichnerowicz obstruction applied for the generalized conifold \label{bifs} is
\begin{align}
\sum_i \frac{1}{k_i} \le 1 + 3\omega_m \ , \label{Lich}
\end{align}
where $\omega_m$ is a charge of holomorphic functions on \eqref{bifs}. As a holomorphic function, we choose a monomial $x_i$ that has a charge $1/k_i$. Then the Lichnerowicz obstruction \eqref{Lich} directly yields the necessary and sufficient condition for the conjectured $a$-theorem \eqref{a-theo}. Physically speaking, this suggests that the unitarity of the SCFT is crucial for the establishment of the $a$-theorem as is the case with the two-dimensional $c$-theorem, where the unitarity is imperative for its proof.

\subsection{Boundary wavefunction from direct integration}\label{direct}
We would like to compute the boundary wavefunction for the rolling D-brane in the Lorentzian two-dimensional black hole from the direct integration, which was not carried out in \cite{Nakayama:2005pk}.

We focus on the overlap between the minisuperspace wavefunction
\begin{align}
U^p_{\om}(\rho,t) &= -
\frac{\Gamma^2(\nu_+)}{\Gamma(1-i\om)\Gamma(-ip)} e^{-i\om t} (\sinh
\rho)^{-i\om} F(\nu_+,\nu^*_-;1-i\om;-\sinh^2\rho) ~,
\label{U} 
\end{align}
and the classical trajectory
\begin{eqnarray}
\cosh(t) \sinh (\rho) = \sinh(\rho_0) \ .
\end{eqnarray}
Explicitly, we would like to evaluate the integral
\begin{align}
\Psi(p,\omega) &= \int_0^\infty \sinh\rho d\sinh\rho \int_{-\infty}^{\infty} dt \delta\left(\cosh(t) \sinh (\rho) - \sinh(\rho_0) \right) U^p_{\omega}(\rho,\theta) \cr
& = \int_{-\infty}^{\infty} dt \frac{\sinh\rho_0}{\cosh^2t} U^p_{\omega}(\rho(\rho_0,t),t)  \ . \label{integ}
\end{align}

After expanding the hypergeometric function as 
\begin{align}
& F(\frac{1}{2}-\frac{ip}{2}-\frac{i\omega}{2},\frac{1}{2}+\frac{ip}{2}-\frac{i\omega}{2};1-i\omega,-\frac{\sinh^2\rho_0}{\cosh^2t}) \cr
=& \sum_{l=0} \frac{\Gamma(\frac{1}{2}-\frac{ip}{2}-\frac{i\omega}{2}+l)\Gamma(\frac{1}{2}+\frac{ip}{2}-\frac{i\omega}{2}+l)}{\Gamma(1-i\omega+l)} \times \cr &\times \frac{\Gamma(-i\omega)}{\Gamma(\frac{1}{2}-\frac{ip}{2}-\frac{i\omega}{2})\Gamma(\frac{1}{2}+\frac{ip}{2}-\frac{i\omega}{2})} \frac{(-)^l\sinh^{2l}\rho_0}{l!\cosh^{2l}t} \ .
\end{align}
the integration over $t$ is possible by the formula
\begin{eqnarray}
\int_{-\infty}^{\infty} (2 \cosh t)^{a-1} e^{i b t}
\dd t = \frac{1}{2} B\left(\frac{1}{2}-\frac{a+ib}{2},
\frac{1}{2}-\frac{a-ib}{2}\right) \equiv \frac{1}{2} \frac
{\Gamma\left(\frac{1}{2}-\frac{a+ib}{2}\right)
\Gamma\left(\frac{1}{2}-\frac{a-ib}{2}\right)} {\Gamma(1-a)}~, &
\hspace{10cm} (\Re\, a <1~,~~ \left|\Im \, b \right| < 1- \Re\, a)~. 
\end{eqnarray}

After collecting terms by using the duplication formula
\begin{eqnarray}
\sqrt{\pi} \Gamma(2l+2-i\omega) = 2^{2l+1-i\omega} \Gamma(2l+1-\frac{i\omega}{2})\Gamma(2l+\frac{3}{2}-\frac{i\omega}{2}) \ ,
\end{eqnarray}
we obtain\footnote{Presumably up to a numerical factor.}
\begin{align}
\Psi(p,\omega) =
& \left(\frac{\sinh \rho_0}{2}\right)^{1-i\omega}\frac{\Gamma^2(\frac{1}{2}-\frac{ip}{2}-\frac{i\omega}{2})\Gamma(\frac{1}{2}+\frac{ip}{2}-\frac{i\omega}{2})}{\Gamma(-ip)\Gamma(1-\frac{i\omega}{2})\Gamma(\frac{3}{2}-\frac{i\omega}{2})}\cr &\times
 \ _3F_2\left(1,\frac{1}{2}-\frac{ip}{2}-\frac{i\omega}{2},\frac{1}{2}-\frac{ip}{2}-\frac{i\omega}{2};1-\frac{i\omega}{2},\frac{3}{2}-\frac{i\omega}{2};-\frac{\sinh^2\rho_0}{2}\right)\ . \label{ans}
\end{align}
Here we have introduced the generalized hypergeometric function
\begin{eqnarray}
_3F_2(a_1,a_2,a_3;b_1,b_2;z) = \sum_{l=0} \frac{(a_1)_l (a_2)_l(a_3)_l}{(b_1)_l(b_2)_l}\frac{x^l}{l!} \ ,
\end{eqnarray}
where 
\begin{eqnarray}
(a)_l \equiv \frac{\Gamma(a+l)}{\Gamma(a)} \ .
\end{eqnarray}

Let us discuss a particular limit of our boundary wavefunction \eqref{ans}. We take the limit $\rho_0 \to \infty$ to see the connection to our previous results \cite{Nakayama:2005pk}. To see this, we use the asymptotic expansion formula
\begin{align}
&_3F_2(a_1,a_2,a_3;b_1,b_2;z) \cr  &= \frac{\Gamma(b_1)\Gamma(b_2)}{\Gamma(a_1)\Gamma(a_2)\Gamma(a_3)} \sum_{k=1}^3 \frac{\Gamma(a_k)\prod_{j=1;j\neq k}^3(a_j-a_k)}{\prod_{j=1}^2 \Gamma(b_j-a_k)}(-z)^{-a_k}\left[1 + O(z^{-1})\right] \ .
\end{align}
In this limit, we see that it asymptotically approaches to our previous results
\begin{align}
\lim_{\rho_0\to \infty} \Psi(p,\omega) = B(\nu_+,
\nu_-) \Gamma\Big(1+\frac{ip}{k}\Big) \, \left[ e^{-ip
\rho_0} -  \frac{\cosh\left(\pi \frac{p-\om}{2}\right)}
{\cosh\left(\pi \frac{p+\om}{2}\right)} e^{ip\rho_0 } \right] +O(e^{-\rho_0})~. 
\label{falling D00}
\end{align}
So importantly, our previous results coincide with the direct evaluation of the overlap only in the limit $\rho_0 \to \infty$.

Another interesting limit is to take $\rho_0 \to 0$. If we further specialize in the zero energy overlap (i.e. $\omega = 0$), we have
\begin{eqnarray}
\lim_{\rho_0\to 0} \Psi(p,0) = -\sinh\rho_0 \frac{\Gamma^2(\frac{1}{2}-\frac{ip}{2})}{\Gamma(-ip)} \ ,
\end{eqnarray}
which coincides with the zero-winding sector of the D0-brane in the cigar (up to $\sinh\rho_0$), which is quite expected, given the origin of the minisuperspace calculation. It would be interesting but seem very difficult to compute the emission rate for general $\rho_0$.

\newpage
\bibliographystyle{utcaps}
\bibliography{d}

\end{document}